\documentclass[a4paper,fleqn,usenatbib]{mnras}

\usepackage[T1]{fontenc}
\usepackage{ae,aecompl}

\usepackage{graphicx}	
\usepackage{amsmath}	
\usepackage{amssymb}	
\usepackage{url}
\usepackage{hyperref}

\usepackage{etoolbox}
\makeatletter
\patchcmd\@combinedblfloats{\box\@outputbox}{\unvbox\@outputbox}{}{%
   \errmessage{\noexpand\@combinedblfloats could not be patched}%
}%
 \makeatother

\title[X-ray Flares and CO Destruction]{Non-Equilibrium Chemistry and Destruction of CO by X-ray Flares}

\author[J.~Mackey et al.]{Jonathan Mackey,$^{1,2,3}$\thanks{E-mail: jmackey@cp.dias.ie (JM)} Stefanie Walch,$^{3}$ Daniel Seifried,$^{3}$ Simon C.O. Glover,$^{4}$
\newauthor{Richard W\"unsch,$^{5}$ Felix Aharonian$^{1,2,6}$}\\
$^1$Centre for AstroParticle Physics and Astrophysics, DIAS Dunsink Observatory, Dunsink Lane, Dublin 15, Ireland\\
$^2$Dublin Institute for Advanced Studies, Astronomy \& Astrophysics Section, 31 Fitzwilliam Place, Dublin 2, Ireland\\
$^{3}$I.~Physikalisches Institut, Universit\"at K\"oln, Z\"ulpicher Str.~77, 50937 K\"oln, Germany\\
$^{4}$Zentrum f\"ur Astronomie, Institut f\"ur Theoretische Astrophysik, Universit\"at Heidelberg, Albert-Ueberle-Str. 2, D-69120 Heidelberg, Germany\\
$^5$Astronomical Institute, Czech Academy of Sciences, Bo\v{c}n\'\i\ II 1401, 141 00 Prague, Czech Republic \\
$^{6}$Max-Planck-Institut f\"ur Kernphysik, P.O. Box 103980, D-69029 Heidelberg, Germany
}

\date{Submitted 27 March 2018; accepted 25 March 2019}

\pubyear{2018}

\begin{document}
\label{firstpage}
\pagerange{\pageref{firstpage}--\pageref{lastpage}}
\maketitle

\begin{abstract}
Sources of X-rays such as active galactic nuclei and X-ray binaries are often variable by orders of magnitude in luminosity over timescales of years.
During and after these flares the surrounding gas is out of chemical and thermal equilibrium.
We introduce a new implementation of X-ray radiative transfer coupled to a time-dependent chemical network for use in 3D magnetohydrodynamical simulations.
A static fractal molecular cloud is irradiated with X-rays of different intensity, and the chemical and thermal evolution of the cloud are studied.
For a simulated $10^5\,\mathrm{M}_\odot$ fractal cloud, an X-ray flux $<0.01$\,erg\,cm$^{-2}$\,s$^{-1}$ allows the cloud to remain molecular, whereas most of the CO and H$_2$ are destroyed for a flux of $\geq1$\,erg\,cm$^{-2}$\,s$^{-1}$.
The effects of an X-ray flare, which suddenly increases the X-ray flux by $10^5\times$ are then studied.
A cloud exposed to a bright flare has 99\% of its CO destroyed in 10-20 years, whereas it takes $>10^3$ years for 99\% of the H$_2$ to be destroyed.
CO is primarily destroyed by locally generated far-UV emission from collisions between non-thermal electrons and H$_2$; He$^+$ only becomes an important destruction agent when the CO abundance is already very small.
After the flare is over, CO re-forms and approaches its equilibrium abundance after $10^3$-$10^5$ years.
This implies that molecular clouds close to Sgr A$^\star$ in the Galactic Centre may still be out of chemical equilibrium, and we predict the existence of clouds near flaring X-ray sources in which CO has been mostly destroyed but H is fully molecular.
\end{abstract}




\section{Introduction}
\label{sec:introduction}
Heating and ionization by X-rays and cosmic rays is known to be a key process in setting the temperature and ionization state of interstellar gas \citep{SpiTom68,  ShuSte85, DalMcC72}.
X-rays with energy $>1$\,keV can propagate deeper into molecular clouds than ultraviolet (UV) or optical radiation because their interaction cross-section is  smaller and decreases with increasing photon energy.
The ionizations induced by X-rays that are absorbed in a molecular cloud can strongly affect the chemical balance of the cloud by heating it and increasing the electron fraction \citep{LepMcC83, MalHolTie96}.
The sources of X-rays, especially non-thermal sources related to X-ray binaries or active galactic nuclei (AGN), tend to be strongly variable on timescales from minutes to years depending on the size of the emitting region.

Even mostly inactive black-hole sources such as Sgr A$^\star$ in the Galactic Centre occasionally have giant flares where the X-ray luminosity increases by a factor of $10^3-10^6$ for a few years at a time.
\citet{PonTerGol10} studied X-ray reflection from molecular clouds around the Galactic Centre in the iron K-shell lines.
They find that the luminosity of Sgr A$^*$ has been at $L_\mathrm{x}\lesssim 10^{35}$ erg\,s$^{-1}$ for the past 60-90 years, but that a bright flare with $L_\mathrm{x}\approx 1.4\times10^{39}$ erg\,s$^{-1}$ occurred about 100 years ago, with a duration of at least 10 years \citep[see also][]{Sunyaev1993, KoyMaeSon96, Sunyaev1998, Churazov2017a}.
The scattering of X-rays in molecular clouds has been studied using Monte-Carlo radiative-transfer simulations \citep{OdaAhaWat11, MolKhaSun16, WalCheTer16, MolKhaSun16} and shown to be a powerful diagnostic of the incident X-ray flux on a cloud.
The inferred luminosity is still far below the Eddington luminosity for Sgr A$^*$, but is $\gtrsim10^4$ times brighter than its current luminosity in X-rays.

X-ray binaries are also powerful sources during their active periods \citep[e.g.~GRS 1915+105 with $L_\mathrm{x}\approx 10^{39}$ erg\,s$^{-1}$ for $\sim$ 10 years, see][]{Punsly2013}.
This shows that molecular clouds close to black holes or luminous X-ray binaries are subject to occasional bright X-ray irradiation, which may affect their thermal and chemical state \citep{Krivonos2017, Churazov2017c}.
If these flares are frequent enough \citep{Churazov2017b}, then the clouds could spend most of their time out of chemical and thermal equilibrium \citep{Moser2016}.

Bright X-ray sources are also usually sites of efficient cosmic-ray (CR) production.
For example, the link between CR production and supernova remnants is now well established \citep{Aha13}, the Galactic Centre is a bright and diffuse source of $\gamma$-rays produced by CRs \citep{HESS17}, and \textit{FERMI} has detected hundreds of AGN at 0.1-100\,GeV energy \citep{AckAjeAll11}.
Like X-rays, CRs propagate deep into molecular clouds, but their interaction with atoms produce $\gamma$-rays as a by-product of nuclear reactions.
For both X-ray and CR interaction with matter the main ionizing and heating agents are so-called \emph{secondary electrons}, produced when high-energy photons or cosmic rays ionize a heavy element.
These electrons have large kinetic energy, comparable to that of the ionizing photon, and so they ionize and heat molecules and atoms as they lose energy through collisional interactions \citep[e.g.][]{MalHolTie96}.
This means that the effects of an elevated CR energy density and of an elevated X-ray radiation field can be difficult to distinguish, and one must either look deeply into the abundances of rare chemical species or consider the different attenuation of CRs and X-rays with column density.
X-rays propagate in straight lines at the speed of light and are simply attenuated, whereas CRs follow trajectories determined by the local magnetic field and on large enough scales their propagation follows a diffusion equation \citep{Girichidis2016, PfrPakSch17}.

Under the assumption that X-rays are unimportant for the chemistry,  \citet{CasWalTer98} showed that the electron fraction and CR ionization rate within a dense cloud can be inferred from abundance ratios of HCO$^+$, CO, and DCO$^+$ (the deuterated form of HCO$^+$).
\citet{VauHilCec14} studied a molecular cloud being impacted by the W28 supernova remnant, using the observed molecular lines to constrain the CR ionization rate to be $>100$ times the background Galactic rate.
\citet{ClaGloRag13} compared observations of the Galactic Centre cloud G0.253+0.016 with simulations using different CR ionization rates, finding that it too should have a CR energy density $>100$ times the background Galactic value.
Investigating extreme environments, \citet{BisPapVit15} studied how CO is destroyed in molecular clouds as the CR energy density increases, using chemical equilibrium calculations of photodissociation regions (PDR).
They found that the number ratio of CO to H$_2$ decreases strongly with increasing CR energy density, because CO is effectively destroyed by He$^+$ ions created by CR ionization.
This was followed up with 3D simulations of fractal clouds exposed to different CR energy densities \citep{BisVanPap17}, confirming their previous results.
\citet{GonOstWol17} also studied PDR chemistry with elevated CR energy density, finding that grain-assisted recombination of He$^+$ limits the effectiveness of CO destruction by CRs.

\citet{MeiSpaIsr06} studied X-ray dominated regions (XDR) and PDRs including elevated CR ionization and heating rates.
For a cloud exposed to high X-ray flux, the XDR is most of the cloud volume, the PDR traces the cloud surface, and CRs affect both the surface and interior of a cloud.
They found that line ratios of HCN, CO and HCO$^+$ can be used, with high-J lines of CO, to distinguisgh between X-ray- and CR-irradiated clouds.
Subsequently, \citet{MeiSpaLoe11} found that OH, OH$^+$, H$_2$O, H$_2$O$^+$, and H$_3$O$^+$ can also be used to discriminate CR and X-ray irradiation.

Most previous chemical studies of X-ray irradiated MCs assume chemical and/or thermal equilibrium, similar to PDR models \citep[e.g.][]{MalHolTie96, MeiSpa05, HocSpa10}.
The codes developed for these projects therefore cannot capture the time-dependent chemistry and thermodynamics that occurs within a molecular cloud irradiated by a time-dependent X-ray radiation field.
A recent departure from this is the study of \citet{CleBerObe17}, who investigated variable H$^{13}$CO$^+$ emission (observed in a protstellar disk) as a consequence of a time-varying X-ray irradiation.
So far there are no studies of the time-dependent chemistry of, for example, the molecular clouds near Sgr A$^\star$, which arises from the recent flare.

Here we introduce a non-equilibrium code that couples X-ray irradiation to chemistry and thermodynamics (and potentially hydrodynamics) of molecular gas, using a simplified chemical network of 17 species.
The treatment of X-ray radiation, the chemical network, and coupling to the \textsc{flash} code is described in Section~\ref{sec:methods}.
Tests of the network using one-dimensional, constant-density slabs are presented in Section~\ref{sec:tests}. 
Section~\ref{sec:fractal} introduces the modelling of a fractal cloud in 3D using the \textsc{flash} code, embedded in a homogeneous and isotropic background radiation field.
The equilibrium state of the gas for different X-ray radiation intensities is obtained, and the states compared with each other.
In Section~\ref{sec:flare} the equilibrium state is disturbed by X-ray flares of duration 1 to 100 years, and we show the time-dependent effects of the flares on the chemical abundances and gas temperature during and after the flare event.
Our results are discussed in Section~\ref{sec:discussion} and our conclusions presented in Section~\ref{sec:conclusions}.

\section{Algorithms and methods}
\label{sec:methods}

\subsection{X-ray transport and absorption}
\label{sec:xrays}
In previous works using the SILCC simulation framework \citep{WalGirNaa15, Girichidis2016, Gatto2017, Peters2017} the X-ray flux was assumed to be constant and was simply scaled with the background interstellar UV radiation field (ISRF).
Here, we develop a fully self-consistent X-ray absorption module and introduce the algorithms used for X-ray radiative transfer and absorption.
We split the X-ray radiation field into $N_E$ energy bins, equally spaced in $\log E$, and calculate a mean cross-section for each bin, $\langle\sigma_i\rangle$.
One-dimensional radiative transfer is very simple and requires little explanation.
For 3D simulations in this paper we consider only an isotropic external radiation field to study the effects of X-rays on the chemistry of molecular clouds, similar to assuming an isotropic background interstellar UV radiation field \citep[e.g.][]{Dra78}.
We use the \textsc{TreeRay/Optical depth} algorithm \citep{WunWalDin18} for three-dimensional radiative transfer, implemented in the \textsc{flash} code \citep{Fryxell2000}, described in more detail below.
Modifying \textsc{TreeRay/Optical depth} to handle anisotropic radiation fields is a relatively simple extension.

The term ``flux'' can mean different things depending on context: when we say X-ray flux, denoted $F_X$, we mean (i) uni-directional energy flux of radiation for one-dimensional slab-symmetric calculations, and (ii) $4\pi J_X$ (where $J_X$ is the angle-averaged mean intensity) for three-dimensional simulations.
In both cases it is the X-ray energy flux available to be absorbed at a point.
We integrate over a given energy range, usually 0.1-10\,keV, and so the units are [erg\,cm$^{-2}$\,s$^{-1}$].
In the nomenclature of \citet{RoeAbeBel07}, the 1D simulations have uni-directional flux, and the 3D simulations isotropic flux.
We also quote the X-ray energy density, $E_\mathrm{rad}$, for clarity and for ease of comparison with other potentially relevant energy densities, such as cosmic rays, FUV radiation, thermal energy, etc.
For the one-dimensional flux, $E_\mathrm{rad} = F_X/c$ ($c$ is the speed of light), and for three-dimensional calculations $E_\mathrm{rad} = 4\pi J_X/c$.

\subsubsection{X-ray absorption cross-section}
\label{sec:xray:xsec}
X-rays are mainly absorbed by ions of heavy elements (especially iron) because their large cross section more than makes up for their trace abundance.
However, calculating the absorption by each ion individually is computationally expensive, as it requires knowledge of the abundance and ionization stage of many heavy ions and is therefore only feasible for detailed PDR/XDR codes \citep[e.g.][]{MeiSpa05,FerPorVan13}.
\citet{PanCabPin12} used a mean cross-section that takes account of all of the heavy elements in a single analytic function, which we also use:
\begin{equation}
\sigma_\mathrm{x} = 2.27\times10^{-22} E_\gamma^{-2.485} \;\mathrm{cm}^2
\label{eqn:sigma}
\end{equation} 
per H nucleus, where $E_\gamma$ is the photon energy in keV.

This cross-section was also used by \citet{ShaGlaShu02} and is based approximately on results of \citet{MorMcC83} for a gas of solar metallicity with abundances in table 1 of this paper \citep[typically within 0.1 dex of updated values from][]{AspGreSau09}.
It assumes that the temperature is low enough that heavy atoms are not significantly ionized, and so the dominant absorbers at large energy are those heavy atoms with K and L shells and corresponding large cross sections.
The approximate formula does not capture resonances or sharp jumps in cross section at K or L shell edges \citep[e.g.][]{DeABre12}.
\citet{MorMcC83} and \citet{WilAllMcC00} show that H and He contribute significantly to the cross section up to the oxygen K-shell edge at $\sim0.5$\,keV, and that a power law with slope $\sim-2.5$ is a good approximation to the total cross section in the range $0.1-10$\,keV.
Our cross section is therefore reliable as long as the electron fraction is small, and fails first at low energies ($\lesssim0.5$\,keV) as the ionization fraction increases.
For highly ionized gas the approximate cross section becomes unreliable and a more accurate treatment would be required, but our aim here is to model molecular clouds and so this regime is not relevant.
The cross section is only valid at or near solar metallicity and does not scale simply with metallicity because H and He contribute significantly for $E_\gamma\lesssim0.5$\,keV.
\citet{MorMcC83} show that the cross section shows only marginal changes even when most heavy elements are completely depleted onto grains.

For an energy bin, $i$, in the energy range $E_a<E_\gamma<E_b$, with $E_m = 0.5(E_a+E_b)$, and defining $\sigma_m\equiv \sigma_\mathrm{x}(E_m)$, we define the mean cross section $\langle\sigma_i\rangle$ using the relation
\begin{equation}
\exp \left(-\frac{\langle\sigma_i\rangle}{\sigma_m}\right) = 
\frac{1}{E_b-E_a}\int_{E_a}^{E_b} \exp  \left(-\frac{\sigma_\mathrm{x}(E)}{\sigma_m}\right) dE \;.
\label{eqn:xsec}
\end{equation}
This formula averages the attenuation factor over the energy bin, and this is used to obtain an appropriate $\langle\sigma_i\rangle$.
This provides a better estimate of the energy absorbed than using a simple average of $\sigma_\mathrm{x}$.
The constant $\sigma_m$ is chosen so that  the exponent is of order unity over most of the integral, but in principle a different value could be used.
A similar averaging was used by \citet{MacLim10} to improve photon (and hence energy) conservation in photo-ionization calculations.

We stress that computational requirements force us to minimize the number of bins, $N_E$, and so it is always the case that $\sigma_\mathrm{x}$ changes significantly within the energy bin because of its strong scaling with energy.
There is no way to avoid some level of inaccuracy when choosing $\langle\sigma_i\rangle$ without making assumptions about the shape of the X-ray spectrum.

\subsubsection{One-dimensional radiative transfer}
For uni-directional flux the equation of radiative transfer is very simple, having a source at infinity with flux entering the simulation domain, $F_{X,0}$, and only absorption everywhere else (i.e.\ scatterings are not considered).
For an energy bin $i$, the X-ray flux, $F_{X,i}$, at a point $x$ is simply
\begin{equation}
F_{X,i}(x) = F_{X,0} \exp\left\{-\tau_{i}(x)\right\} \;,
\end{equation}
where $\tau_{i}(x)\equiv \int_{-\infty}^{x} n_\mathrm{H}(x^\prime) \langle\sigma_i\rangle dx^\prime$ is the optical depth along the ray to point $x$,
and $n_\mathrm{H}$ is the local number density of H nuclei.

\subsubsection{Three-dimensional radiative transfer}

In the 3D \textsc{flash} simulations we use the \textsc{TreeRay/Optical depth} algorithm \citep{WunWalDin18}, which is similar to the {\sc Treecol} method developed by \citet{Clark2012}.
The \textsc{TreeRay/Optical depth} algorithm computes the mean column density of any given species in every time step and for each cell of the computational domain using a \textsc{Healpix} tessellation \citep{Gorski2005} with $N_{\rm pix}$ pixels for each grid cell, using an Oct-tree method.
We modified the tree solver such that it can be used to calculate the X-ray optical depth between each grid cell and the boundary of the computational domain.
As a result, we obtain the columns and fluxes for every grid cell.
Here we use $N_{\rm pix} = 48$ and a geometric opening angle criterion \citep{Barnes1986} with an opening angle of $\theta_{\rm lim} = 0.5$.

We consider that the simulation domain is embedded in a uniform and isotropic external X-ray radiation field with mean intensity $J_\nu$, where $\nu$ is frequency.
For an isotropic 3D radiation field the intensity, $I_\nu$, is equal to $J_\nu$, and so all rays entering the simulation domain satisfy this equality.
For an X-ray energy bin, denoted $i$, the external mean intensity can be denoted $J_{0,i}$, and the fluxes $F_{0,i}\equiv4\pi J_{0,i}$ are input parameters to our calculations.

The intensity along a ray, labelled $n$, from the edge of the simulation domain to a grid cell located at $\mathbf{r}$ can be obtained by solving the equation of radiative transfer with zero emissivity, as in the 1D case above:
\begin{equation}
I_{X,i}^n(\mathbf{r}) = J_{0,i} \exp \left( -\tau_{i}^n \right) \;,
\end{equation}
where $\tau_{i}^n\equiv \int n_\mathrm{H}(\mathbf{r}^\prime) \langle\sigma_i\rangle d\mathbf{r}^\prime$ is now the optical depth along the ray.
For a given number of rays, $N$, uniformly covering $4\pi$ steradians, the mean intensity at $r$ is simply the average value of $I_{X,i}^n(\mathbf{r})$:
\begin{equation}
J_{X,i}(\mathbf{r})  = \frac{1}{N}\sum_{n=1}^N I_{X,i}^n(\mathbf{r})
 = \frac{J_{0,i}}{N} \sum_{n=1}^N \exp \left( -\tau_{i}^n \right)
\end{equation}

The local attenuated flux at $\mathbf{r}$ is then
\begin{equation}
F_{X,i}(\mathbf{r})\equiv 4\pi J_{X,i}(\mathbf{r}) = \frac{F_{0,i}}{N} \sum_{n=1}^N \exp \left( -\tau_{i}^n \right) \;.
\label{eqn:xrayflux}
\end{equation}

From this we can calculate a local rate of X-ray energy absorption, $H_\mathrm{x}$ (erg\,s$^{-1}$) per H nucleus using
\begin{equation}\label{EQ_HX}
H_\mathrm{x} = \sum_{i=1}^{N_E} F_{X,i} \langle\sigma_i\rangle \;,
\end{equation}
where the sum is over all energy bins.

We use isolated boundary conditions for the \textsc{Optical depth} module, which means that the simulation domain is bathed in a uniform and isotropic  (but potentially time-varying) X-ray radiation field.
The X-ray optical depths are calculated between the target cell and the boundary of the simulation domain, so that every cell contributes to attenuating the radiation field seen at a given point.
Such a setup is not always appropriate for X-ray radiation fields, which are often dominated by point sources \citep[e.g.][]{Ponti2015}, but it is an improvement on a 1D slab (see section \ref{sec:ms05}) because it allows us to consider a more realistic density field.
We also run our calculations in the limit of infinite speed of light.

The column densities of total gas, CO and H$_2$ are necessary to compute the (self-) shielding of gas from the ISRF, whereas the X-ray attenuation factors, $\exp (-\tau_{i}^n)$, depend only on the total gas column density. 
We therefore calculate the attenuated X-ray flux for each of the $N_E$ X-ray energy bins arriving at every cell using Eq.~\ref{eqn:xrayflux} and use it as an input for the chemical network.
The radiative transfer is completed before the chemistry update in \textsc{flash}, and so we need to store the attenuation factors
\begin{equation}
\frac{1}{N}\sum_{n=1}^N \exp (-\tau_{i}^n)
\label{eqn:attenuation}
\end{equation}
for each X-ray energy bin, $i$, at every grid cell.
This is accomplished by adding $N_E$ scalar fields to the grid.
Within the chemistry network, the local X-ray absorption rate is calculated using Eq.~\ref{EQ_HX}.

\subsection{Chemical Network}\label{sec:chem}
We use a chemical network based largely on the NL99 network of \citet{GloCla12}, which combines a model for hydrogen chemistry taken from \citet{GloMac07a, GloMac07b} and a model for CO chemistry introduced by \citet{NelLan99}.
We also include a number of modifications and updated reaction rates as suggested by more recent work \citep[e.g.][]{GonOstWol17}.
The X-ray reactions and rates are taken largely from \citet{Yan97} and \citet[][hereafter MS05]{MeiSpa05}.

The number fraction of species $\mathrm{Q}$ with respect to the total number of hydrogen nuclei is denoted $y(\mathrm{Q})$, and $Y_\mathrm{R}$ is the fractional abundance by number of nuclei of \emph{element} $\mathrm{R}$ with respect to hydrogen.
For example, $y(\mathrm{H}_2)\in[0,0.5]$ because $Y_\mathrm{H}\equiv1$, and $y(\mathrm{CO})\in[0,\mathrm{min}(Y_\mathrm{C},Y_\mathrm{O})]$.
Note in particular that the electron fraction, $y(\mathrm{e^-})$, can be larger than unity with this definition.

\begin{table}
  \centering
  \caption{Species calculated in our chemical network.}
  \label{tab:species}
  \begin{tabular}{ll} 
    \hline
    Species & Treatment  \\
    \hline
    H       & conservation eqn. \\
    H$^+$   & ODE solve         \\
    H$_2$   & ODE solve         \\
    OH$_\mathrm{x}$     & ODE solve         \\
    C       & conservation eqn. \\
    C$^+$   & ODE solve         \\
    CO      & ODE solve         \\
    CH$_\mathrm{x}$     & ODE solve         \\
    HCO$^+$ & ODE solve         \\
    He      & conservation eqn. \\
    He$^+$  & ODE solve         \\
    M       & conservation eqn. \\
    M$^+$   & ODE solve         \\
    O       & equilibrium         \\
    O$^+$   & equilibrium     \\
    H$_2^+$ & instantly reacts further \\
    H$_3^+$ & equilibrium       \\
    e$^-$       & conservation eqn. \\
    \hline
  \end{tabular}
\end{table}

The chemical species that we solve for are listed in Table~\ref{tab:species}.
The non-equilibrium species solved for are H$_2$, H$^+$, CO, C$^+$, CH$_\mathrm{x}$, OH$_\mathrm{x}$, HCO$^+$, He$^+$, and M$^+$.
Following \citet{NelLan99}, CH$_\mathrm{x}$ is a proxy species for simple hydrocarbons CH, CH$_2$, CH$_3$, etc., and similarly OH$_\mathrm{x}$ for OH, H$_2$O, etc.
Intermediate molecular ions CH$^+$, CH2$^+$, OH$^+$, etc., are also included in CH$_\mathrm{x}$ and OH$_\mathrm{x}$, as appropriate, as well as the neutral species.
We assume that each CH$_\mathrm{x}$ and OH$_\mathrm{x}$ molecule only contains one H atom for accounting purposes, but this makes no difference because the abundance of the species is very low compared to hydrogen.

M is a proxy element for metals (e.g. N, Mg, Si, S, Fe) that can be the primary source of electrons in molecular gas at large column density.
We assume that M is a two-ionization-stage atom, tracking M$^+$ as a species, and neutral M with a conservation equation.
The abundances of neutral atomic species H, He, C, are also computed using conservation equations, and we assume that the abundance of doubly (and more highly) ionized species is negligible.
Oxygen is also treated as a two-ionization-stage atom, and its ionization fraction is assumed to be the equilibrium value (after accounting for the fraction of O that is in OH$_\mathrm{x}$ and CO) because of the rapid charge exchange reactions with H and H$^+$ \citep{StaSchKim99}.
The equilibrium abundance of H$_3^+$ is calculated from the local chemical abundances and temperature, and used in the network following \citet{NelLan99}.
In total there are 9 species in the network that are solved by the ODE solver \citep{Brown1989}, 5 species tracked by conservation equations (H, He, C, M, e$^-$) and 4 species (O, O$^+$, H$_2^+$, H$_3^+$) tracked by assuming equilibrium abundances or instantaneous further reaction.
All of these contribute to gas heating and cooling.

\begin{table}
  \centering
  \caption{Elemental abundances in the gas phase by number with respect to hydrogen nuclei, $Y_\mathrm{R}$.}
  \label{tab:elements}
  \begin{tabular}{ll} 
    \hline
    Species & $Y_\mathrm{R}$  \\
    \hline
    H       & 1.0 \\
    He      & 0.1 \\
    C       & $1.4\times10^{-4}$ \\
    O       & $3.4\times10^{-4}$ \\
    M       & $1.0\times10^{-5}$ \\
    \hline
  \end{tabular}
\end{table}

The elemental abundances are listed in Table~\ref{tab:elements}.
The metal abundance can be set somewhat arbitrarily because it covers a number of different elements, although we take reaction rates appropriate for silicon throughout the paper.
\citet{MalHolTie96} considered Si, Fe, S, and Ni, with the most abundant being Si ($3.5\times10^{-6}$) and S ($1.0\times10^{-5}$).
\citet{NelLan99} used a rather low value of $Y_\mathrm{M}=2\times10^{-7}$, whereas \citet{BisPapVit15} use $Y_\mathrm{M}=4\times10^{-5}$ as the sum of the abundances of all relevant gas-phase metal abundances, and \citet{GonOstWol17} used Si as a proxy for all metals with $Y_\mathrm{Si}=1.7\times10^{-6}$.
The metal abundance is important at high column densities because it determines the electron fraction once C$^+$ has recombined.

The collisional reactions are listed in Table~\ref{tab:reactions}, and photo-reactions in Table~\ref{tab:photoreactions} in Appendix~\ref{app:network}.
An analysis of the differences between results with and without the \citet{GonOstWol17} additional reactions is also presented in App.~\ref{app:network}.
A noteworthy addition is that we follow \citet{GonOstWol17} in including grain recombination reactions for C$^+$, He$^+$, M$^+$, as well as H$^+$.
In addition, in view of the potential importance of He$^{+}$ ions in the CO chemistry of X-ray-irradiated gas, it is worthwhile highlighting the difference in our treatment of He$^{+}$ recombination.
\citet{GonOstWol17} use the case B radiative recombination rate from \citet{HumSto98}, while we attempt to account for the fact that in gas which is optically thick to ionizing photons, the actual radiative recombination rate lies between the case A and case B rates owing to absorption of helium recombination photons by atomic hydrogen \citep{Ost89}.
In addition, we also account for dielectronic recombination of He$^{+}$, a process neglected by \citet{GonOstWol17}.
At low temperatures, this process is unimportant, but in hot gas ($T \sim 10^{5}$\,K), it comes to dominate the total He$^{+}$ recombination rate. 

\subsubsection{H$_2^+$ and H$_3^+$ abundance and reactions}

There are four formation channels for H$_2^+$: cosmic ray ionization of H$_2$ (\#56 in Table~\ref{tab:photoreactions}), charge exchange between He$^+$ and H$_2$ (\#27 in Table~\ref{tab:reactions}), charge exchange between H$_2$ and H$^+$ (\#18 in Table~\ref{tab:reactions})  and X-ray ionization of H$_2$ (\#66 in Table~\ref{tab:photoreactions}).
H$_2^+$ is considered to react immediately once it is formed and, following the discussion in MS05, it has three further reaction pathways:
\begin{enumerate}
\item
dissociative recombination with an electron to 2H plus 10.9\,eV of heat (\#43 in Table~\ref{tab:reactions});
\item charge exchange with H to produce H$_2$ and H$^+$ and 0.94\,eV of heating (\#17 in Table~\ref{tab:reactions}); and
\item
further reaction with H$_2$ to produce H$_3^+$ and H (with subsequent recombination or reaction with other species), with net heating of 8.6\,eV per H$_3^+$ ion production (\#32 in Table~\ref{tab:reactions}).
\end{enumerate}
The creation rate of these products is given by the H$_2^+$ formation rate multiplied by the fraction of the H$_2^+$ ions that follow each pathway.
H$_2^+$ can also be photodissociated by the interstellar radiation field, but this process is competitive with processes (ii) and (iii) above only when $n / G_{0} < 1$ \citep{Glo03}.
Since $n / G_{0} \gg 1$ in typical molecular cloud conditions, we are justified in neglecting this process in the models presented in this paper.

We assume that H$_3^+$ has its equilibrium abundance at all times.
Its only significant creation channel\footnote{H$_{3}^{+}$ can also form via the radiative association of H$_{2}$ with H$^{+}$, but this process is slow (see e.g. the discussion in \citealt{GloSav09}), and is only competitive with formation via H$_{2}^{+}$ in gas with a very low H$_{2}$ abundance.
In these conditions, the H$_{3}^{+}$ abundance itself is very small and H$_{3}^{+}$ plays a negligible role in the gas chemistry.} is through H$_2^+$ (\#32 in Table.~\ref{tab:reactions}), and it is destroyed by:
\begin{enumerate}
\item
reaction with C to form CH$_\mathrm{x}$ (\#21 in Table.~\ref{tab:reactions})
\item
reaction with O to form OH$_\mathrm{x}$ (\#22 in Table.~\ref{tab:reactions}), and further with an electron to produce O + 3H (\#23 in Table.~\ref{tab:reactions});
\item
reaction with CO to form HCO$^+$ and H$_2$ (\#24 in Table.~\ref{tab:reactions});
\item
dissociative recombination with an electron (\#20 in Table.~\ref{tab:reactions}); and
\item
charge exchange with M to form H$_2$ + H + M$^+$ (\#19 in Table.~\ref{tab:reactions}).
\end{enumerate}
The equilibrium abundance is obtained by balancing the creation rate with the destruction rates listed.

\subsection{X-ray heating, ionization and dissociation}
\label{sec:xraychem}

X-rays are absorbed by dust and gas, affecting both components through the following processes, most of which we include. They are described in more detail below:
\begin{enumerate}
\item
Dust
heating, following \citet{Yan97}.
\item
Dust destruction and charging by X-rays.
\item
Direct ionization of an atom/molecule by X-rays.
This is generally only important for elements that have a K-shell, because these elements have much larger direct ionization cross-sections than lighter elements.
For H, H$_2$, and He it is negligible \citep[e.g.][]{DalYanLiu99}.
\item
Secondary ionization of atoms/molecules through collisions with the fast (keV) electrons that are produced by a direct X-ray ionization.
This is the main ionization channel for H, H$_2$, and He.
\item
Secondary ionization/dissociation of atoms/molecules through FUV radiation that is locally generated by H$_2$ molecules, which are collisionally excited by fast electrons.
This provides important photodissociation channels for molecules (except H$_2$) and photoionization channels for atomic species with low ionization energy (e.g.\ C).
\item
Coulomb heating of the gas arising from energy exchange between the fast electrons and other charged particles in the gas.
\item
Heating through dissociation of molecules and ionization of atoms (these rates are typically already in the chemical model, and the X-rays only increase the heating rate).
\end{enumerate}

For the dust we consider only heating (i), ignoring ionization and dust destruction (ii).
This is reasonable for the molecular clouds that we consider, but would not be suitable for strongly irradiated, hot gas.
We also do not consider direct ionization/dissociation by X-rays (iii), but only secondary ionizations through collisional (iv) and FUV (v) processes.
All of the other processes are included as described below.

\subsubsection{Dust heating}
The dust temperature, $T_\mathrm{D}$, in an X-ray irradiated gas is calculated following \citet{Yan97} and MS05 as
\begin{equation}
T_\mathrm{D} = 1.5\times10^{2}\left(\frac{H_\mathrm{x}}{10^{-18}\,\mathrm{erg\,s}^{-1}}\right)^{0.2} \;\mathrm{K}.
\end{equation}
We take the maximum of this temperature and the radiative equilibrium temperature resulting from FUV irradiation \citep[which is calculated following][]{GloCla12}.
There is evidence for dust temperatures between 125-150\,K in the circumnuclear disk of the Galactic Centre via detection of the J$=4-3$, v$_2=1$ vibrationally excited transition of HCN, which \citet{Mills2013} argue is excited by local IR radiation from hot dust grains.
In our 3D simulations described later the dust temperature ranges from 10 to 70\,K.

\subsubsection{Coulomb heating}
Secondary electrons are produced when an X-ray photon is absorbed by a heavy element, resulting in ionization and the ejection of an electron with kinetic energy comparable to the photon energy.
The absorbed X-ray power per H nucleus, $H_\mathrm{x}$ (erg\,s$^{-1}$), is transferred to these hot electrons, and subsequently goes partly into heating the gas and partly into ionizations~\citep{DalYanLiu99}.
The fraction that goes into heating is determined in part by the electron abundance in the gas, because the heating arises from energy exchange through Coulomb interactions between the hot electron and the thermal electrons (and, to a lesser extent, thermal ions).
For small electron fractions, most of the X-ray energy goes into ionizations, but the heating fraction increases towards unity as the electron fraction increases~\citep{DalYanLiu99}.
The heating fraction is also dependent on the energy of the hot electron (and hence the energy of the X-ray photon), because higher-energy electrons are much more likely to cause ionizations in a collisional interaction than lower-energy electrons.
We follow MS05 in implementing the results of \citet{Yan97} and  \citet{DalYanLiu99} to model these processes.

The Coulomb heating rate by secondary electrons is obtained from the tables of \citet{DalYanLiu99} using the local abundances of electrons, H, and H$_2$.
The local heating rate, $\Gamma_\mathrm{x}$ (erg\,cm$^{-3}$\,s$^{-1}$) is given by
\begin{equation}
\Gamma_\mathrm{x} = \eta n_\mathrm{H} H_\mathrm{x} \;,
\end{equation}
where $\eta$ is a heating efficiency obtained from tables in \citet{DalYanLiu99}.
The efficiency depends on $y(\mathrm{e^-})$, $y(\mathrm{H})$, $y(\mathrm{H}_2)$ and $y(\mathrm{He})$.

Coulomb heating becomes more efficient as $y(\mathrm{e^-})$ increases, and the fit of \citet{DalYanLiu99} becomes invalid for $y(\mathrm{e^-})>0.1$.
We therefore assume that, for $y(\mathrm{e^-})>0.1$, the fraction of absorbed X-ray energy that goes to Coulomb heating, $\eta$, scales linearly with the electron fraction, starting from the \citet{DalYanLiu99} value at $y(\mathrm{e^-})=0.1$ and reaching 100 per cent for $y(\mathrm{e^-})\geq1$, i.e.,
\begin{equation}
\eta[y(\mathrm{e^-})] = \eta(0.1) + \frac{1-\eta(0.1)}{0.9}\left(\min[1,y(\mathrm{e^-})]-0.1\right) \;,
\end{equation}
where the minimum operator ensures $\eta \leq1$ even when $y(\mathrm{e^-})>1$.
This interpolation is important for ensuring that the ODE solver converges in highly ionized gas.

\subsubsection{Secondary collisional ionization}
The hot electrons ionize and dissociate, as well as heat, the gas.
H is ionized with rate $\zeta(\mathrm{H})$ per H atom per second, and He with rate $\zeta(\mathrm{He})$ per He atom per second.
Molecular hydrogen, H$_2$, is dissociated (with rate $\zeta_\mathrm{D}(\mathrm{H}_2)$ per H$_2$ molecule per second) or ionized to H$_2^+$ (with rate  $\zeta(\mathrm{H}_2)$ per H$_2$ molecule per second).

These collisional ionization and dissociation rates by secondary electrons are calculated by interpolating the tables of \citet{DalYanLiu99} for $y(\mathrm{e^-})\leq0.1$.
As for the heating rates above, for $y(\mathrm{e^-})>0.1$ we take the \citet{DalYanLiu99} rates at $y(\mathrm{e^-})=0.1$ and make them proportional to the abundance of the neutral species being ionized (or dissociated) so that the rate has the correct limit as full ionization is approached, e.g.
\begin{equation}
\zeta(\mathrm{H})y(\mathrm{H}) = \frac{H_\mathrm{x}}{W_{\mathrm{H}}(y(\mathrm{e^-})=0.1)}y(\mathrm{H})^{1 - (0.1/y(\mathrm{e^-}))^3} \;.
\end{equation}
Here $W_{\mathrm{H}}$ is the mean energy per H ionization from \citet{DalYanLiu99}.
This is an ad-hoc extrapolation of the \citet{DalYanLiu99} tables but is not important for the results presented in this work because we are not studying highly ionized plasmas.
It does, however, ensure that the ODE solver converges for all values of $y(\mathrm{e^-})$.
The rates for reactions \#62, \#63, \#66 and \#67 from Table~\ref{tab:photoreactions} are calculated using this formula and the tables from \citet{DalYanLiu99}.

C is ionized by secondary electrons, with a rate 3.92 times that of H according to appendix~D3.2 of MS05.
We generalise their equation to the following:
\begin{equation}
\zeta(\mathrm{C})y(\mathrm{C}) = \frac{\zeta(\mathrm{H})y(\mathrm{H})+\zeta(\mathrm{H}_2)y(\mathrm{H}_2)}{y(\mathrm{H})+y(\mathrm{H}_2)}3.92 y(\mathrm{C}) \;.
\end{equation}
This has the correct limiting values when H is fully atomic and fully molecular, and is the equation used for reactions \#64, \#65, \#68-71 in Table~\ref{tab:photoreactions}.
Similarly CO, CH$_\mathrm{x}$, OH$_\mathrm{x}$, HCO$^+$ can be collisionally ionized and destroyed by secondary electrons.
For  CO, CH$_\mathrm{x}$, and HCO$^+$ we use the same scaling factor as for C (3.92), whereas for OH$_\mathrm{x}$ we use a scaling factor of 2.97 appropriate for oxygen (MS05).
For M, we use the same scaling factor as for silicon, 6.67.
For simplicity we assume that ionization of all the carbon-bearing molecules produces C$^+$, OH$_\mathrm{x}$ produces O and H$^+$, and HCO$^+$ produce C$^+$ and H$^+$ and O.
Ionization of M produces M$^+$.

These factors of 3.92 for C, 2.97 for O, and 6.67 for Si were obtained by integrating the cross sections over the range 0.1$-$10\,keV to obtain an average value (see MS05), whereas in reality they should vary as a function of energy bin.
In all of our calculations, however, these reactions are negligible compared with dissociation by the locally generated FUV field and so such an approximate treatment can be accepted.
For future work that consistently includes the transition to highly ionized plasmas one would need to improve this aspect of our chemical model \citep[cf.][]{DeAMigBre12}, ideally considering the energy-dependent cross-section of each ion.

\subsubsection{Secondary ionization by locally generated FUV radiation}
A local FUV radiation field is generated by collisional excitation of H$_2$ and H by hot electrons \citep{PraTar83, GreLepDal87, MalHolTie96}.
In our network, this contributes to the ionization of C and M \citep[rates from][]{MalHolTie96,Yan97}, and to the dissociation of CH$_\mathrm{x}$, OH$_\mathrm{x}$, HCO$^+$, and CO \citep{Yan97}.

The \citet{GreLepDal87} rate for CO destruction per second is fitted with
\begin{equation}
R^\mathrm{FUV}_\mathrm{CO}y(\mathrm{CO}) = 2.7 \sqrt{y(\mathrm{CO})\frac{T}{10^3\,\mathrm{K}}} \zeta(\mathrm{H_2}) y(\mathrm{H_2}) \;,
\end{equation}
and this is often used \citep[e.g.][MS05]{MalHolTie96}.
This does not scale linearly with $y(\mathrm{CO})$ as $y(\mathrm{CO})\rightarrow0$, which causes numerical problems for the ODE solver (the destruction timescale goes to zero as $y(\mathrm{CO})\rightarrow0$).
The physical reason for this scaling is that the process is photon limited: photons are produced at a rate that depends on $\zeta(\mathrm{H_2})$ and $n(\mathrm{H_2})$, and are then primarily absorbed by CO.

We instead use the UMIST12 \citep{McEWalMar12} rate for reaction \#74 in  Table~\ref{tab:photoreactions} because it has a more numerically stable asymptotic behaviour, although it may be less accurate for $T>50$\,K than the \citet{MalHolTie96} rate (T.~Millar, private communication), and it probably underestimates the rate at which CO is destroyed as the CO abundance goes to zero:
\begin{equation}
R^\mathrm{FUV}_\mathrm{CO}y(\mathrm{CO}) =210.0 \left(\frac{T}{300\,\mathrm{K}}\right)^{1.17}  y(\mathrm{CO}) \zeta(\mathrm{H_2}) y(\mathrm{H_2}) \;.
\label{eqn:crphot}
\end{equation}

For other species we follow previous authors \citep[][MS05]{MalHolTie96, Yan97} using the following functional form for reactions \#72, \#73, \#75 and \#76 in Table~\ref{tab:photoreactions}:
\begin{equation}
R^\mathrm{FUV}_\mathrm{x}y(\mathrm{x}) = \left[p_\mathrm{m} \zeta(\mathrm{H_2}) y(\mathrm{H_2}) + p_\mathrm{a} \zeta(\mathrm{H})y(\mathrm{H})\right] \frac{y(\mathrm{x})}{1-w} \;,
\label{eqn:xrfuv}
\end{equation}
where $p_\mathrm{m}$ relates to the cross-section of species $x$ for dissociation/ionization by Lyman-Werner photons, and $p_\mathrm{a}$ by Lyman-$\alpha$ photons.
The values of $p_\mathrm{m}$ and $p_\mathrm{a}$ used are given in Table~\ref{tab:crphot}.
The grain albedo, $w$, is taken to be 0.5 for all energies \citep{MalHolTie96, PanCabPin12}.
Cosmic rays also produce secondary electrons and a local FUV field in the same way, and so reactions \#59, \#60 and \#61 have the same form.

\citet{HeaBosVan17} have recently calculated updated rate coefficients for $p_\mathrm{m}$ (their table 20).
Their new values are similar to what we use here.
In particular their updated value for C is 520 (scaled to our normalisation) compared with our value of 510.
This and the CO rate (Equation~\ref{eqn:crphot}), for which \citet{HeaBosVan17} refer to \cite{GreLepDal87}, are the key ones for our work.
For the others, the rate for M is so large that it remains ionized to the largest column densities considered, and our treatment of CHx and OHx is very approximate and so a factor of $\sim2$ difference in $p_\mathrm{m}$ does not impact on our results.

\begin{table}
  \centering
  \caption{
    Constants for destruction of species by FUV radiation generated by hot electrons exciting molecular ($p_\mathrm{m}$) and atomic ($p_\mathrm{a}$) hydrogen (see Eq.~\ref{eqn:crphot}).
    Values for OH and CH are used for OH$_\mathrm{x}$ and CH$_\mathrm{x}$, respectively, and values for Si are used for M.
    Values for $p_\mathrm{a}$ are already multiplied by $\epsilon_\mathrm{L}=0.1$, following \citet{LepDal96}.
    Most $p_\mathrm{m}$ values are taken from the UMIST12 database \citep{McEWalMar12} and
    are multiplied by 2 because they are relative to a cosmic-ray/X-ray ionization rate per H$_2$ molecule, whereas we use an ionization rate per H nucleus.
    The $p_\mathrm{m}$ value for M is attributed to Rawlings (1992, private communication) in \citet{McEWalMar12}.
    In the fourth column, the first reference is for $p_\mathrm{m}$ and the second for $p_\mathrm{a}$.
    References: 1 \citet{GreLepDal87}; 2 \citet{McEWalMar12}; 3 \citet{LepDal96}; 4 \citet{Yan97}; 5 \citet{MalHolTie96}; 6  \citet{GreLepDal89}. }
  \label{tab:crphot}
  \begin{tabular}{lrrc} 
    \hline
    Species & $p_\mathrm{m}$  & $p_\mathrm{a}$  & Reference \\
    \hline
    C   & 510   & 0    & 1,5 \\
    M   & 4230  & 10\,500 & 2,4 \\
    OH$_\mathrm{x}$ & 508   & 87.6 & 6,3 \\
    CH$_\mathrm{x}$ & 730   & 35   & 6,4 \\
    \hline
  \end{tabular}
\end{table}

\subsection{Time-dependent solution in the FLASH code}
\label{sec:timeint}
Chemistry and cooling are operator-split from the other parts of the \textsc{flash} code \citep{Fryxell2000}, which compute e.g. the magneto-hydrodynamic evolution of the gas or the gas self-gravity. As in \citet{WalGirNaa15}, the chemistry and gas temperature are integrated simultaneously using the ODE solver \textsc{dvode} \citep{Brown1989}. We employ sub-timestepping if the chemical abundances or the internal energy are about to change significantly in a given cell. This ensures that the reaction and cooling rates are accurate even if the gas temperature changes by a large factor over a single timestep. 
The heating and cooling processes considered and a table of references for their implementation are given in Appendix~\ref{sec:cooling}.

The inputs to the ODE solver and the chemical network are the total column density $N_\mathrm{H}$, the column densities of CO and H$_2$, the attenuated ISRF, the attenuated X-ray flux in each energy bin (see section \ref{sec:xrays}), the gas density, internal energy and the chemical state at the beginning of a timestep. The ODE solver integrates the equations and returns the updated internal energy and chemical state at the end of each timestep. 

Therefore, chemistry and thermodynamics are mostly time-dependent, giving us an advantage over previous XDR calculations because we can study what happens when the X-ray radiation field varies on timescales shorter than the chemical or thermal timescale in full 3D geometry.
There are some caveats to this statement: we do use a chemical network in which we assume 
(i) that the O/O$^+$ ratio has reached its equilibrium value based on the the H$^+$ fraction;
(ii) that H$_2^+$ reacts instantly to produce further products; and
(iii) that H$_3^+$ has its equilibrium abundance; and
(iv) that the locally-generated UV radiation field is produced instantly by hot electrons in the molecular cloud.
The first three approximations are made because these reactions are usually faster than others which are calculated in a fully time-dependent way.
Regarding the fourth assumption, we note that the timescale on which the local UV field builds up is of the order of the stopping time of the hot photoelectrons (i.e. the time it takes for them to lose the bulk of their kinetic energy).
At typical molecular cloud densities this is $\ll 1$~yr \citep{DalYanLiu99}, much shorter than the timescales of interest in Section~\ref{sec:flare} and, therefore, for our purposes the approximation that the UV field appears instantly is reasonable.

\section{Test problems}
\label{sec:tests}
\begin{figure}
\includegraphics[width=0.49\textwidth]{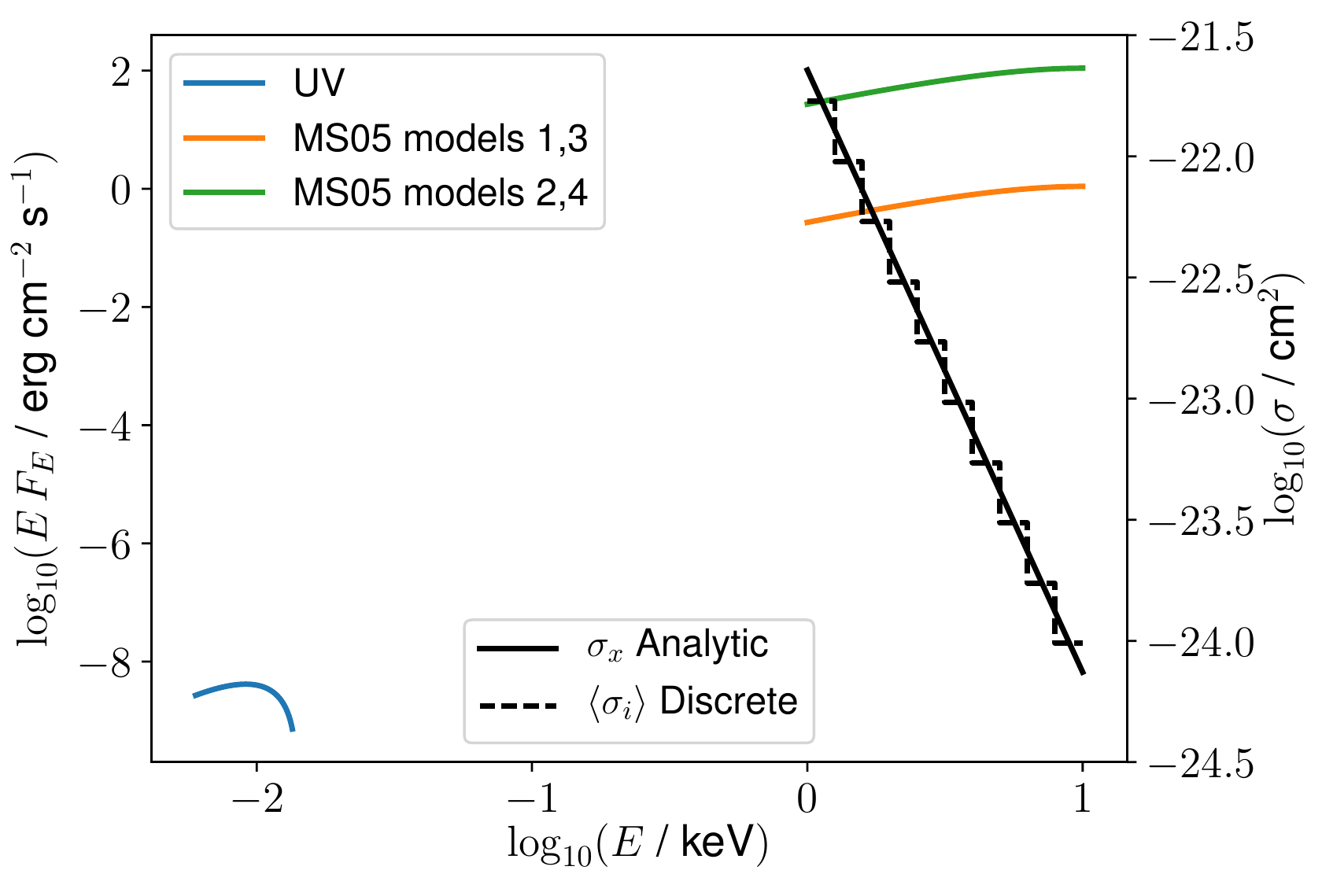}
\caption{
  UV flux (blue) and X-ray flux from Table~\ref{tab:MS05bins} for the 4 test problems considered by MS05 in section~\ref{sec:ms05} (left $y$-axis) and X-ray absorption cross-section (right $y$-axis).
  $E$ is the energy in keV and $F_E$ is the energy flux in units erg\,cm$^{-2}$\,s$^{-1}$\,keV$^{-1}$.
  For the cross-section, the continuous black line plots Eqn.~\ref{eqn:sigma} from \citet{PanCabPin12}, and the dashed black line the discrete cross-section used for each of the 10 energy bins.
  For these tests the UV flux is scaled to $G_0=10^{-6}$ to make it insignificant.
  \label{fig:ms05_spec}}
\end{figure}

\begin{figure*}
\begin{tabular}{ccc}
&$F_X=1.6$ erg\,cm$^{-2}$\,s$^{-1}$ & $F_X=160$ erg\,cm$^{-2}$\,s$^{-1}$ \\
\raisebox{7.5\normalbaselineskip}[0pt][0pt]{\rotatebox{90}{$n_\mathrm{H}=10^3$ cm$^{-3}$}} &
\includegraphics[trim = 0mm 12mm 15mm 1mm, clip, height=6.0cm]{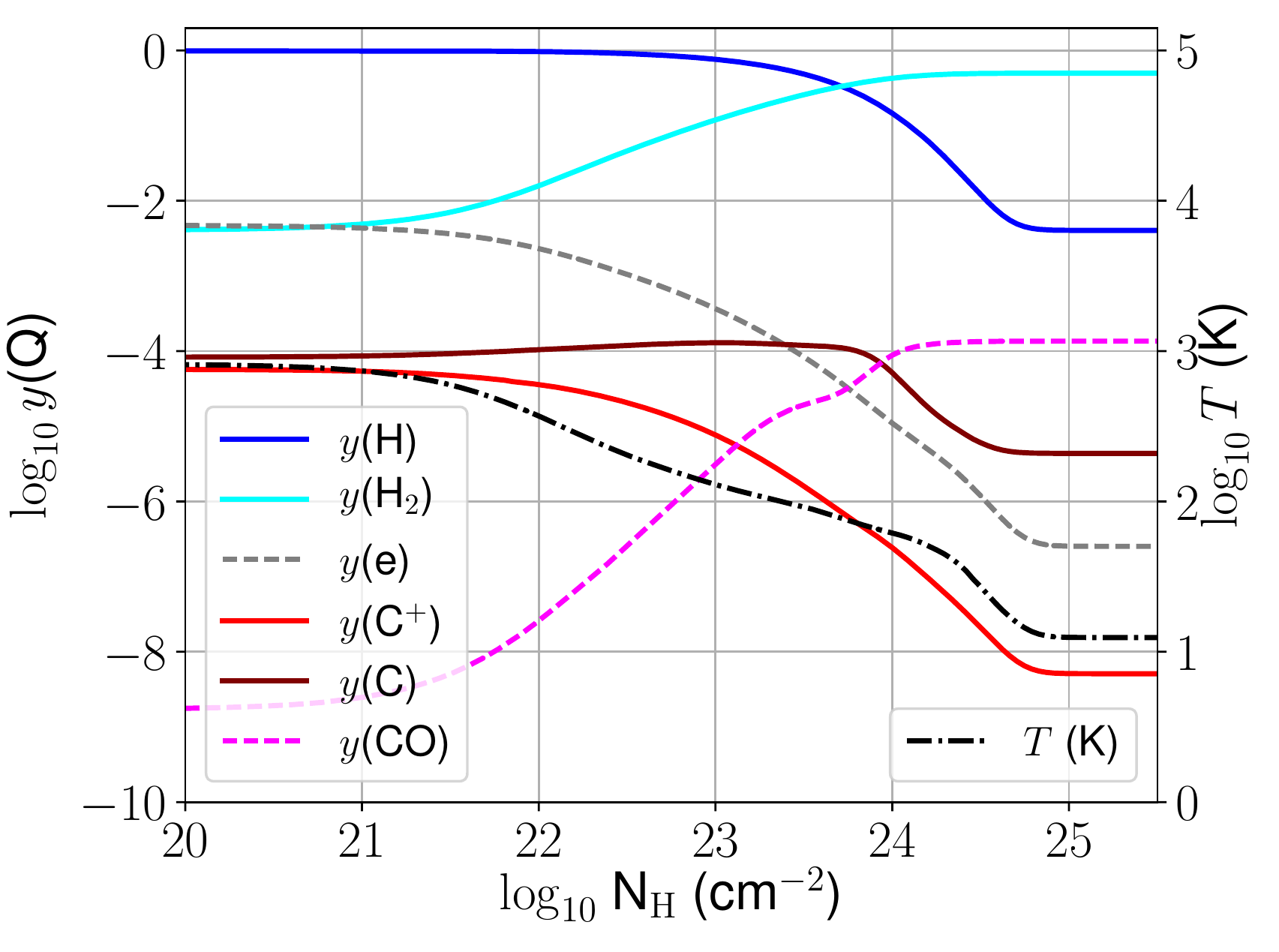} &
\includegraphics[trim = 13mm 12mm 0mm 1mm, clip, height=6.0cm]{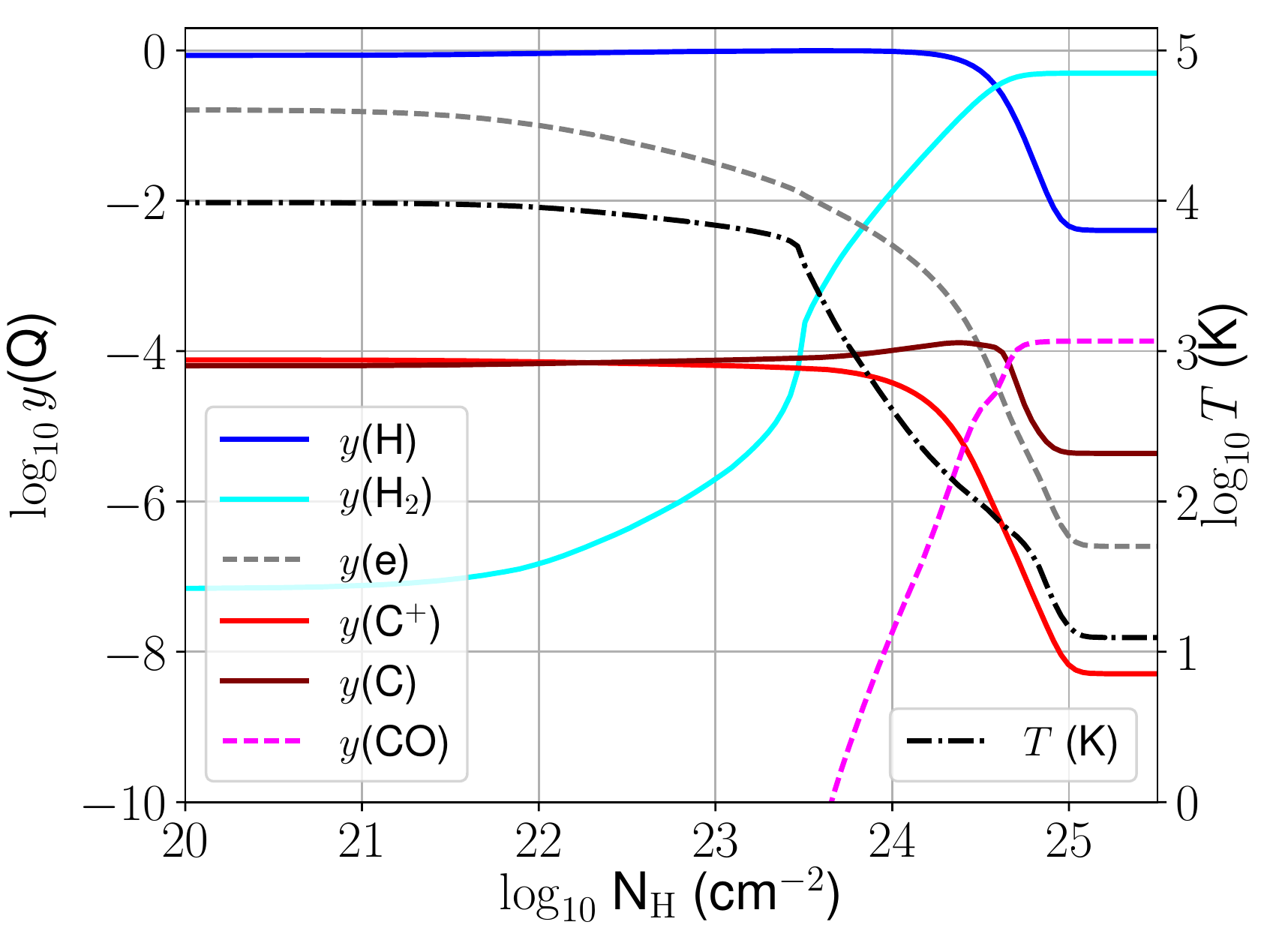} \\
\raisebox{7.5\normalbaselineskip}[0pt][0pt]{\rotatebox{90}{$n_\mathrm{H}=10^{5.5}$ cm$^{-3}$} }& 
\includegraphics[trim = 0mm 0mm 15mm 0mm, clip, height=6.7cm]{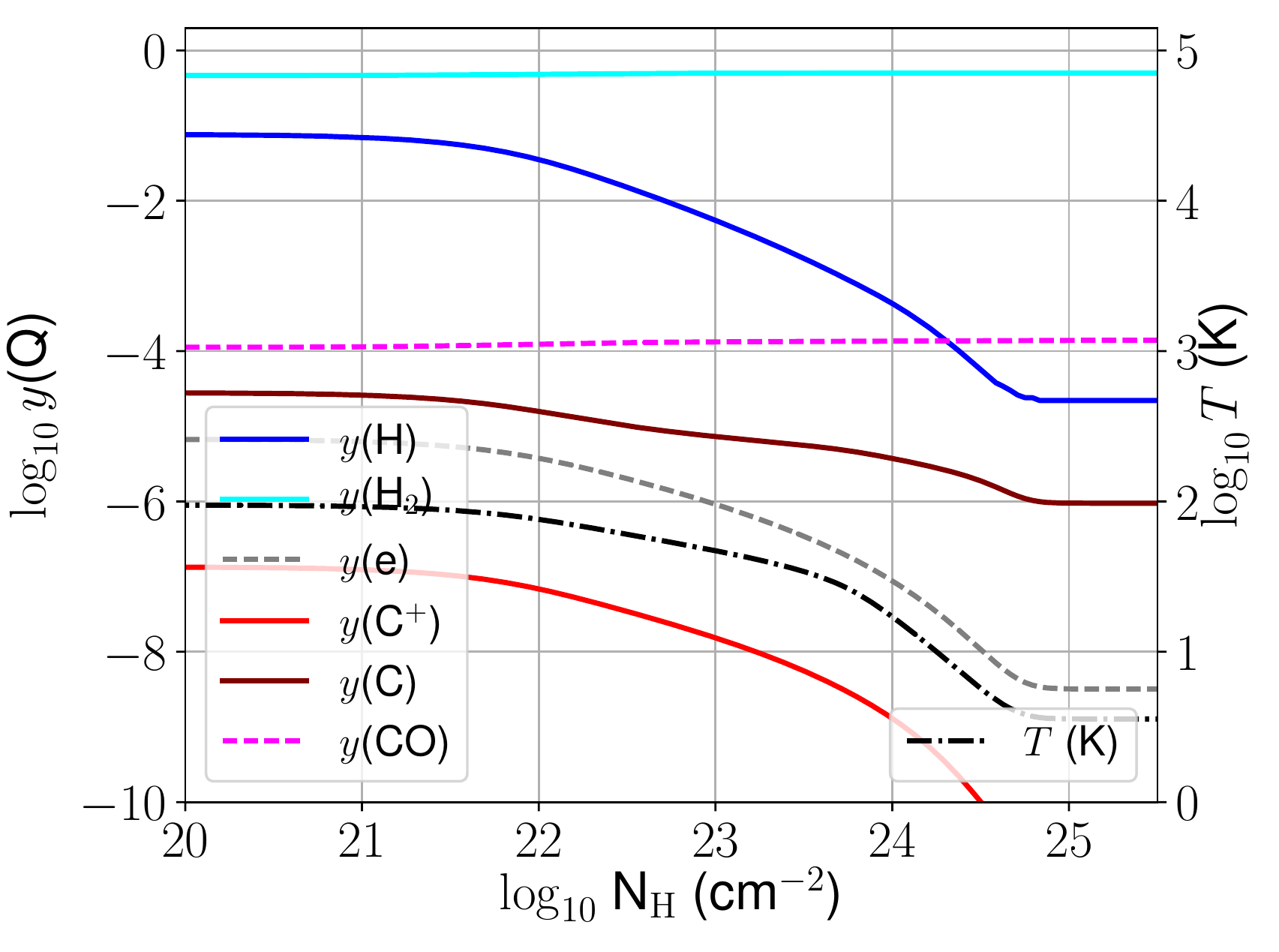} &
\includegraphics[trim = 13mm 0mm 0mm 0mm, clip, height=6.7cm]{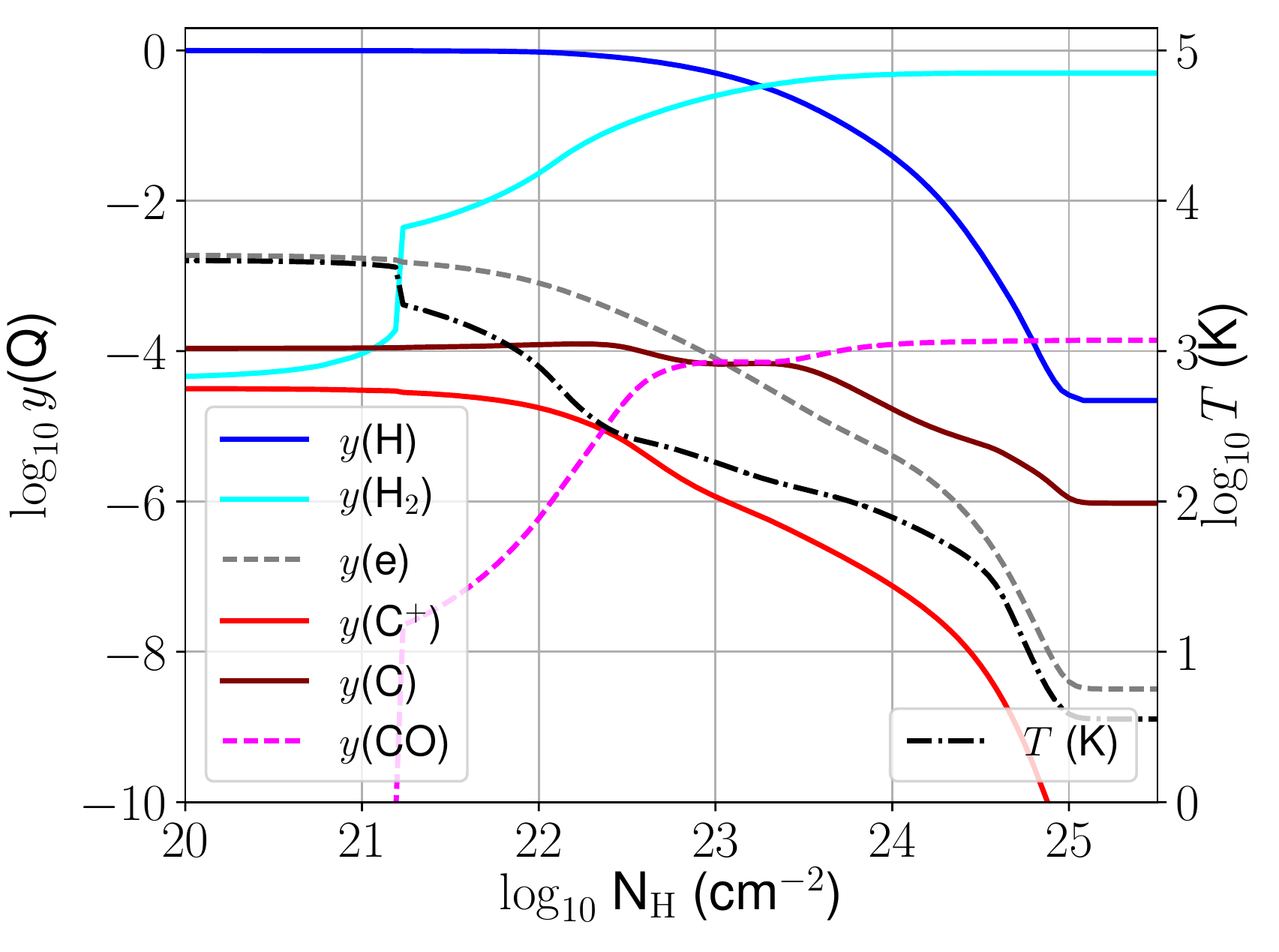}
\end{tabular}
\caption{
  Results obtained for MS05 models 1 (upper left), 2 (upper right), 3 (lower left) and 4 (lower right; see Table \ref{tab:ms05} for the run parameters). The H, H$_2$, electron, C$^+$, C and CO abundances, and gas temperature, $T$, are plotted as a function of column density, $N_\mathrm{H}$, both on log scales. The left $y$-axis refers to abundances, and the right $y$-axis to temperature.
  The results can be compared with figs.~3 and 4 of MS05.
  \label{fig:ms05_1234}}
\end{figure*}

Our chemical network is much smaller than networks used by XDR calculations in the literature that assumed chemical equilibrium \citep[e.g.][hereafter MS05]{MeiSpa05}.
This means that we have fewer potential coolants in the gas and fewer potential sources of electrons in highly shielded gas, although the inclusion of species M is intended to mimic the effects of a number of metals that are not explicitly incorporated.
Furthermore, in some cases we are using different reaction rates and cooling rates from previous authors.
These differences may be significant, so it is important to benchmark our results against other codes, and to try to understand any differences that may be present.
We begin by considering the test problems studied by MS05, and then run calculations using a large range of densities and X-ray fluxes, to make sure that our model produces sensible results for all ISM conditions.

\begin{table}
  \centering
  \caption{Simulation parameters for the 4 test problems of MS05.}
  \label{tab:ms05}
  \begin{tabular}{lccc} 
    \hline
    Model & $n_\mathrm{H}$ (cm$^{-3}$) & $F_X$ (erg\,cm$^{-2}$\,s$^{-1}$) & $E_\mathrm{rad}$ (erg\,cm$^{-3}$) \\
    \hline
    1 & $10^3$      & 1.6 & $5.34\times10^{-11}$ \\
    2 & $10^3$      & 160 & $5.34\times10^{-9}$ \\
    3 & $10^{5.5}$  & 1.6 & $5.34\times10^{-11}$ \\
    4 & $10^{5.5}$  & 160 & $5.34\times10^{-9}$ \\
    \hline
  \end{tabular}
\end{table}

\subsection{Comparison with MS05}\label{sec:ms05}
We consider the four calculations by MS05 as test problems for our XDR chemistry module, and follow these authors by referring to them as models 1-4.
They are one-dimensional XDR calculations of an infinite slab that is irradiated from one side by X-ray radiation, and follow closely the work of \citet{Yan97}.
The gas density and X-ray fluxes for models 1-4 are given in Table~\ref{tab:ms05}.
Models 1 and 2 have $n_\mathrm{H}=10^3$\,cm$^{-3}$ whereas Models 3 and 4 have a density about 300 times larger.
Models 1 and 3 have a moderate total X-ray flux of $F_X=1.6$ erg\,cm$^{-2}$\,s$^{-1}$, and models 2 and 4 have a flux 100 times larger.
MS05 considered an X-ray spectrum of the form $F\propto\exp(-E/10\,\mathrm{keV})$  (a typo in MS05 said 1\,keV in the exponential instead of 10\,keV; R.~Meijerink, private communication), and they only considered X-rays in the range 1-10\,keV.
We run the calculations with 10 energy bins, logarithmically spaced between 1 and 10 keV, shown in Table~\ref{tab:MS05bins}.
The ISRF is set to $G_0=10^{-6}$, i.e.~effectively no UV irradiation.
This radiation field is plotted in Fig.~\ref{fig:ms05_spec} together with the absorption cross-section used in each of the 10 bins.
For consistency with previous work, the radiation field is assumed to be zero from the Lyman limit up to 1 keV.
This can be justified because of the large interstellar absorption cross-section at these energies, although the abrupt switch-on of the X-rays at 1 keV is somewhat artificial.

\begin{table}
  \centering
  \caption{
    Energy bins, mean absorption cross section, X-ray flux and energy density in each bin for MS05 test models 1 and 3.
    Models 2 and 4 are identical except that the flux in each bin is multiplied by 100.
    Bin energy limits $E_\mathrm{min}$ and $E_\mathrm{max}$ are in keV, mean cross section $\langle\sigma_i\rangle$ in cm$^{-2}$, flux $F_X$ is in erg\,cm$^{-2}$\,s$^{-1}$ per bin, and energy density $E_\mathrm{rad}$ in units $10^{-12}$\,erg\,cm$^{-3}$ per bin.
  }
  \label{tab:MS05bins}
  \begin{tabular}{llllll} 
    \hline
    Bin, $i$ & $E_\mathrm{min,i}$ & $E_\mathrm{max,i}$ & $\langle\sigma_i\rangle$ & $F_{X,i}$ & $E_{\mathrm{rad}, i}$ \\
    \hline
    0 & 1.000 & 1.259 & $1.686\times10^{-22}$ & 0.069 & $2.30$ \\
    1 & 1.259 & 1.585 & $9.515\times10^{-23}$ & 0.084 & $2.81$ \\
    2 & 1.585 & 1.995 & $5.369\times10^{-23}$ & 0.102 & $3.41$ \\
    3 & 1.995 & 2.512 & $3.030\times10^{-23}$ & 0.123 & $4.10$ \\
    4 & 2.512 & 3.162 & $1.710\times10^{-23}$ & 0.146 & $4.87$ \\
    5 & 3.162 & 3.981 & $9.648\times10^{-24}$ & 0.171 & $5.69$ \\
    6 & 3.981 & 5.012 & $5.444\times10^{-24}$ & 0.196 & $6.54$ \\
    7 & 5.012 & 6.310 & $3.072\times10^{-24}$ & 0.220 & $7.33$ \\
    8 & 6.310 & 7.943 & $1.734\times10^{-24}$ & 0.239 & $7.97$ \\
    9 & 7.943 & 10.00 & $9.782\times10^{-25}$ & 0.250 & $8.35$ \\
    \hline
  \end{tabular}
\end{table}

We set up a one-dimensional grid with 200 logarithmically spaced grid-zones, without hydrodynamics and with constant gas density, and we set the grid-zones so that column densities from $N_\mathrm{H}=10^{16}$\,cm$^{-2}$ to $10^{26}$\,cm$^{-2}$ are calculated.
The initial conditions are uniform, with sound speed 10\,km\,s$^{-1}$, and partially ionized with $y$(H$^+)=0.5$, $y$(He$^+)=0.05$, $y$(C$^+)=Y_\mathrm{C}$, $y$(M$^+)=Y_\mathrm{M}$, and molecular species set to have abundance $10^{-20}$.
The column density of H, H$_2$, and CO are trivially calculated at each timestep on such a grid, and these are used as an input to the chemistry solver.
The chemical and thermodynamic properties are then integrated for each grid point over a timestep.
The initial timestep is $10^5$\,s, and this is doubled after each step.
The MS05 calculations assume chemical and thermal equilibrium, so we integrate our chemical network for $10^9$ years to ensure that equilibrium conditions are obtained in all cases.
Models 1 and 2 reach equilibrium in 5-10\,Myr, and models 3 and 4 take $<1$\,Myr because of their higher gas density.

The results obtained at the end of the integration are shown in Fig.~\ref{fig:ms05_1234}.
The effects of attenuation are negligible for $N_\mathrm{H} \lesssim 10^{21}$\,cm$^{-2}$ (which corresponds to a visual extinction $A_V \sim 0.5$), and attenuation is basically complete by $N_\mathrm{H} \gtrsim 10^{25}$\,cm$^{-2}$; the abundances and temperature tend to constant values in these limits.
Models 1 and 2 have a moderate gas density ($n_\mathrm{H}=10^3$\,cm$^{-3}$) and so weaker gas cooling (per unit volume) than the denser Models 3 and 4.
As a result they have higher equilibrium temperatures at all $N_\mathrm{H}$.
At low $N_\mathrm{H}$, Model 1 has $T\approx10^3$\,K, Model 2 has $T\approx10^4$\,K, Model 3 has $T\approx10^2$\,K, and Model 4 has $T\approx10^{3.6}$\,K.
All models are charaterised by decreasing temperature and electron fraction in the range  $N_\mathrm{H} \in [10^{22}, 10^{25}]$\,cm$^{-2}$.
Models 1, 2 and 4 have low molecular fractions at low column density, and increasing abundance with increasing column density.
Model 3 is so dense that the moderate X-ray flux cannot destroy the molecules even at low column density, and so it is mostly molecular at all column densities.
For all four calculations, the atomic-to-molecular transition happens at $T\sim100$\,K and when $y(\mathrm{e^-})\lesssim10^{-4}$, although the column density at which this occurs is strongly dependent on gas density and X-ray flux.
The C to CO transition occurs at approximately the same column density as the H to H$_2$ transition.

Our results can be directly compared with figs.~3 and 4 of MS05.
Taking each model in turn, we discuss the similarities and differences between our results and those of MS05.

\paragraph*{Model 1 (Fig.~\ref{fig:ms05_1234}, top-left panel):}
At small $N_\mathrm{H}$ we find larger $y(\mathrm{H}_2)$, larger $y(\mathrm{C}^+)$, smaller $T$ and $y(\mathrm{e^-})$ than found by MS05.
At large $N_\mathrm{H}$ we cannot see the asymptotic values that the MS05 results will tend to, but the results appear comparable.
At intermediate $N_\mathrm{H}$ some changes occur at smaller $N_\mathrm{H}$ in our calculations: the H to H$_2$ transition occurs at $N_\mathrm{H} \sim 10^{23.7}$\,cm$^{-2}$, at which point $T<100$\,K, $y(\mathrm{e^-})\sim10^{-4.5}$.
These $T$ and $y(\mathrm{e^-})$ values are consistent with MS05, except that they find the transition at $N_\mathrm{H} \sim 10^{24.2}$\,cm$^{-2}$, about 0.5 dex larger than us.
MS05 also find that $y(\mathrm{C}^+)$ remains large until $N_\mathrm{H} \sim 10^{24.2}$\,cm$^{-2}$, whereas we find a significant decrease already at $N_\mathrm{H} \sim 10^{23}$\,cm$^{-2}$.
Similarly to H$_2$, we find that the C to CO transition happens at about 0.5 dex smaller $N_\mathrm{H}$ than MS05.
Apart from the offset in $N_\mathrm{H}$ and the qualitative difference in $y(\mathrm{C}^+)$, the results are very comparable.

\paragraph*{Model 2 (Fig.~\ref{fig:ms05_1234}, top-right panel):}
At small $N_\mathrm{H}$ we find very similar results, except that $y(\mathrm{H}_2)$ is smaller than MS05.
The reason for this close agreement is probably that the gas is partially ionized and $T\sim10^4$\,K, and this convergence of electron fraction and temperature means that most quantities are comparable.
At large $N_\mathrm{H}$ we see the same trends as for Model 1, namely that the H to H$_2$ transition happens at smaller $N_\mathrm{H}$ in our calculations, by about 0.3 dex, and the same for the C to CO transition.

Model 2 has a weak discontinuity in $T$ and $y$(H$_2$) at $N_\mathrm{H} \approx 10^{23.5}$\,cm$^{-2}$, which was not found by MS05.
This is one of the more striking features of Fig.~\ref{fig:ms05_1234}, and also appears in Model 4 at $N_\mathrm{H} \approx 10^{21.2}$\,cm$^{-2}$.
Such discontinuities were also obtained by \citet{Yan97} for gas with sub-solar metallicity, and arise from a chemothermal instability that is associated with a region in $n-T$ space where H$_2$ is the dominant coolant (see also \textsc{Cloudy} results in Section~\ref{sec:cloudy}).
These discontinuities are superficially reminiscent of an ionization front, where the thermal and ionization properties of a medium change very rapidly.
In that case, however, the cross-section for absorption of ionizing photons is so large that there is very strong deposition of energy in a thin layer separating neutral from ionized gas.
In contrast, the X-ray heating rate as a function of column density is unaffected by the chemothermal instability and remains a smooth function of $N_\mathrm{H}$.

\paragraph*{Model 3 (Fig.~\ref{fig:ms05_1234}, bottom-left panel):}
This shows the largest discrepancies between our results and MS05.
We find that the gas is mostly molecular at all column densities, and at small $N_\mathrm{H}$ we find that $y(\mathrm{H})\approx0.08$ and $y(\mathrm{C})\approx3\times10^{-5}$, whereas MS05 found that H and H$_2$ should have comparable abundances and that C should be more abundant than CO.
They also found a larger electron fraction but comparable temperature.
The difference seems to arise from the treatment of C$^+$: MS05 find $y(\mathrm{C}^+)>10^{-5}$ up to $N_\mathrm{H} \approx 10^{24.5}$\,cm$^{-2}$, whereas we have $y(\mathrm{C}^+)\approx10^{-7}$ at small $N_\mathrm{H}$ and decreasing as $N_\mathrm{H}$ increases.
Consequently MS05 have a significantly larger electron fraction than we do, and this affects the chemical balance.

\paragraph*{Model 4 (Fig.~\ref{fig:ms05_1234}, bottom-right panel):}
  The asymptotic temperatures at low and high $N_\mathrm{H}$ are similar to MS05, and the run of $T$ with $N_\mathrm{H}$ is also similar, although not identical.
  As mentioned above, there is a temperature discontinuity at $N_\mathrm{H} \approx 10^{21.2}$\,cm$^{-2}$, associated with a chemo-thermal instability.
  This was not found by MS05, and it is the most striking difference between our results and theirs.
  We again find that the atomic-to-molecular transition occurs at smaller $N_\mathrm{H}$ than MS05 by about 0.5 dex, and the temperature and electron fraction also decrease more rapidly with $N_\mathrm{H}$.
  For example at $N_\mathrm{H} = 10^{23}$\,cm$^{-2}$, MS05 find $T\approx10^3$\,K and $y(\mathrm{e^-})\approx10^{-3}$, whereas we find $T=180$\,K and $y(\mathrm{e^-})=7\times10^{-5}$.

The broad agreement between our results and those of MS05 is encouraging, but there are systematic differences in the column density of the atomic-to-molecular transition, the abundance of $y(\mathrm{C}^+)$ and the occurance of temperature discontinuities.
This prompted a direct comparison with an XDR code that uses a much larger network, discussed in the next subsection.
We also present a study of the effects of the new reactions added to the NL99 network following \citet{GonOstWol17} in Appendix~\ref{app:network}.
Regarding $y(\mathrm{C}^+)$, the appendices show that the addition of new reactions following \citet{GonOstWol17} is driving the discrepancy, particularly the grain recombination reactions, without which we obtain similar C$^+$ abundances to MS05.

\begin{figure*}
\begin{tabular}{ccc}
&$F_X=1.6$ erg\,cm$^{-2}$\,s$^{-1}$ & $F_X=160$ erg\,cm$^{-2}$\,s$^{-1}$ \\
\raisebox{7.5\normalbaselineskip}[0pt][0pt]{\rotatebox{90}{$n_\mathrm{H}=10^3$ cm$^{-3}$}} &
\includegraphics[trim = 0mm 12mm 11mm 1mm, clip, height=6.0cm]{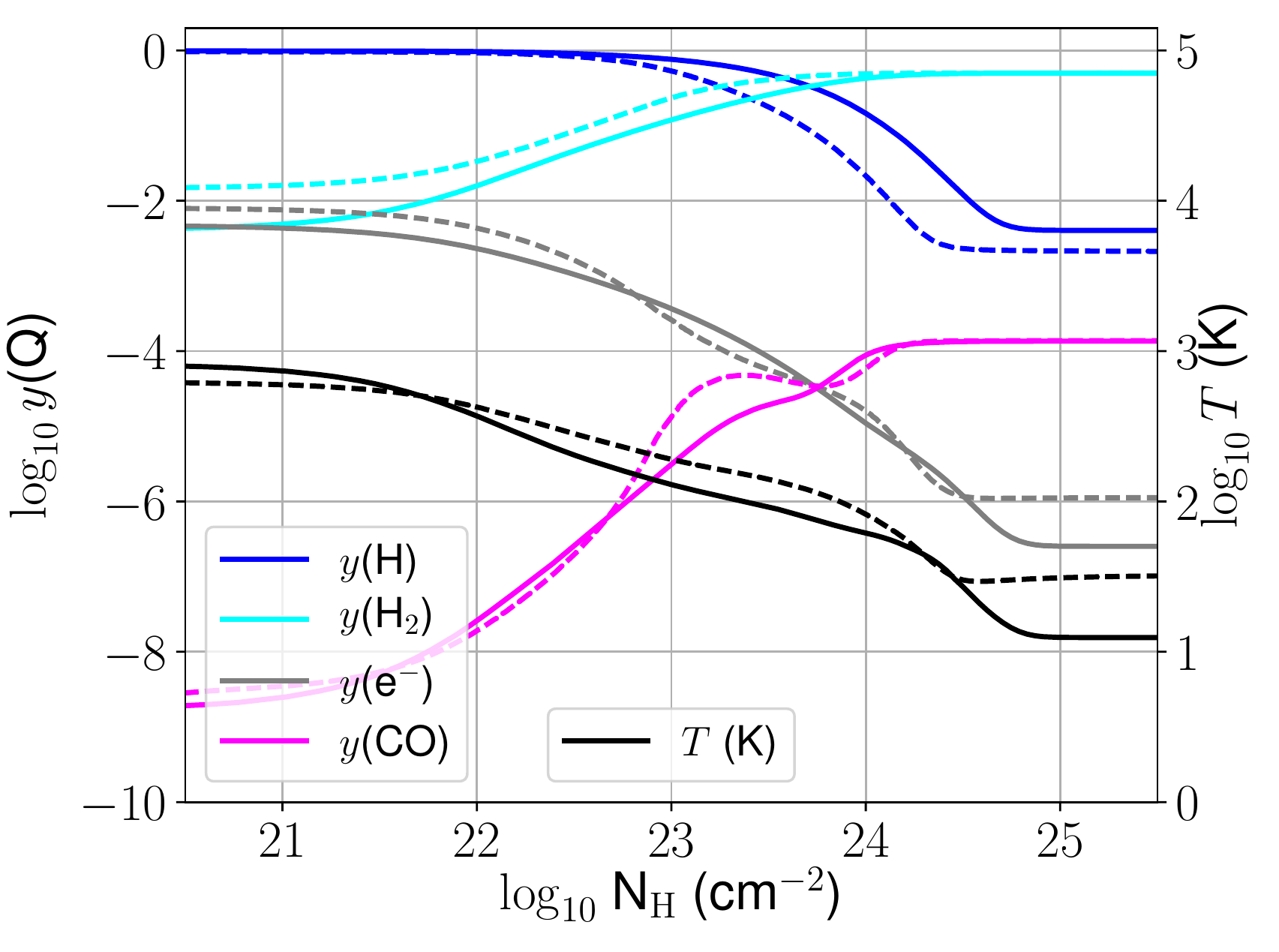} &
\includegraphics[trim = 12mm 12mm 0mm 1mm, clip, height=6.0cm]{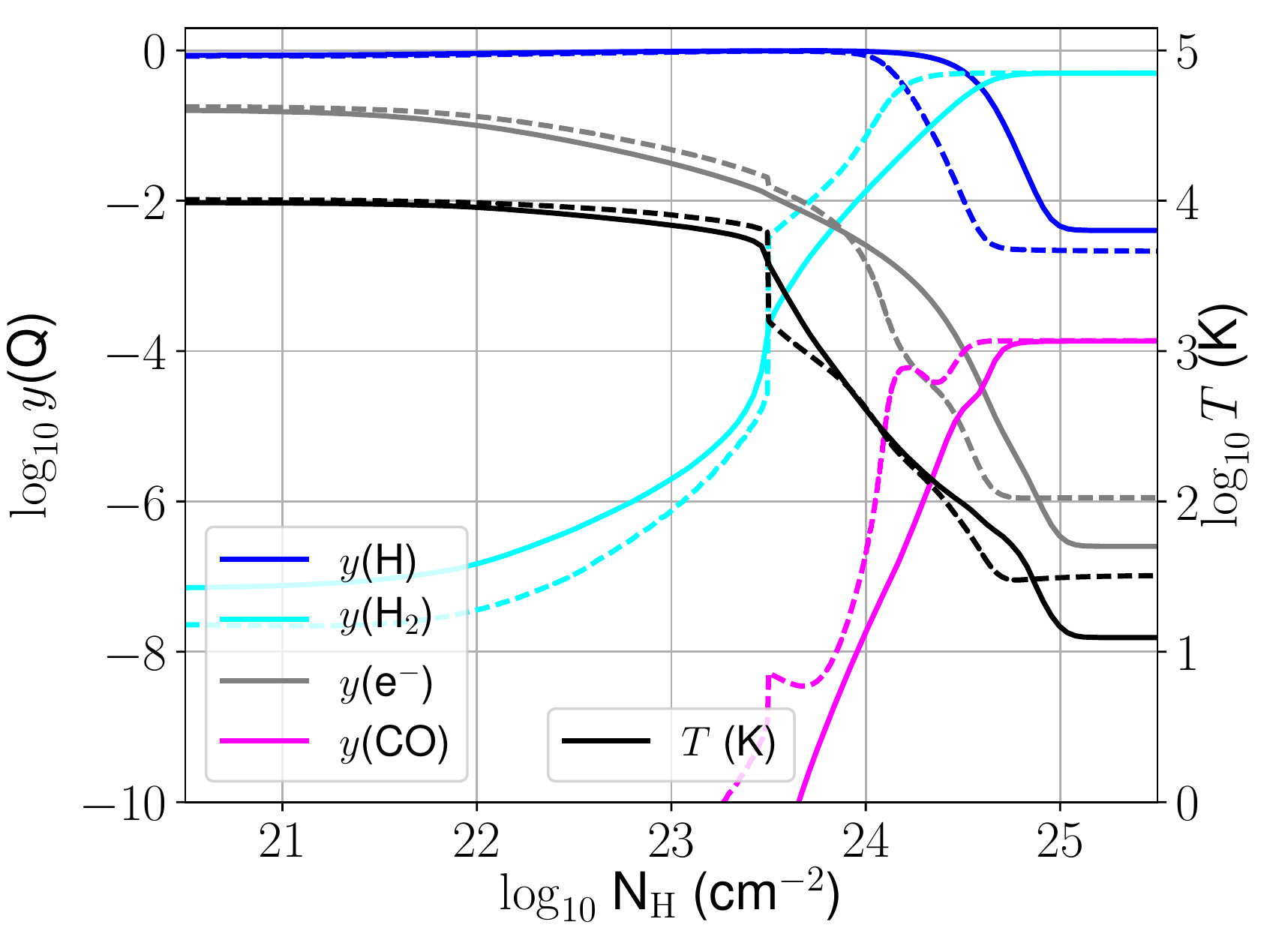} \\
\raisebox{7.5\normalbaselineskip}[0pt][0pt]{\rotatebox{90}{$n_\mathrm{H}=10^{5.5}$ cm$^{-3}$}} &
\includegraphics[trim = 0mm 0mm 11mm 0mm, clip, height=6.5cm]{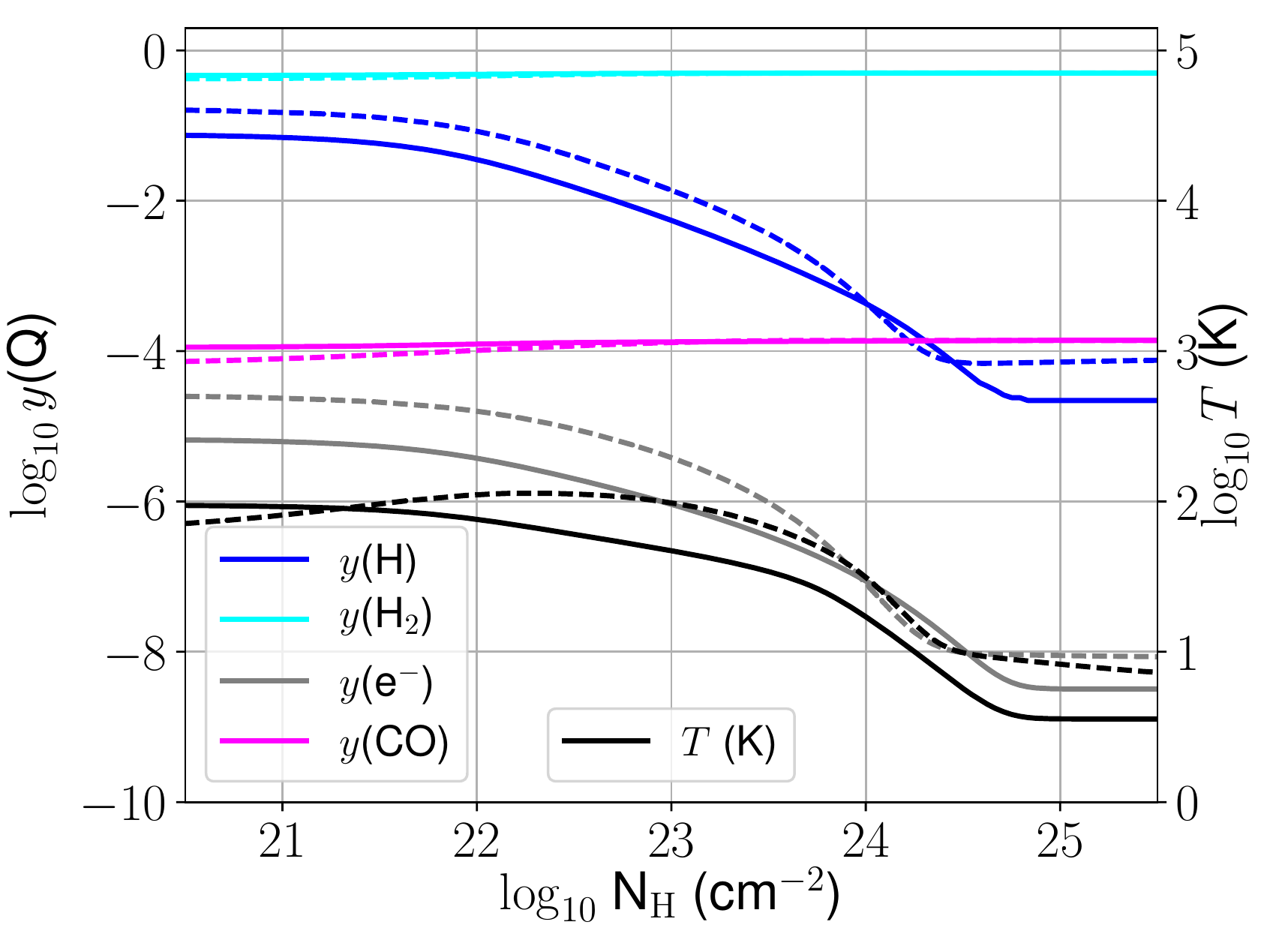} & 
\includegraphics[trim = 12mm 0mm 0mm 0mm, clip, height=6.5cm]{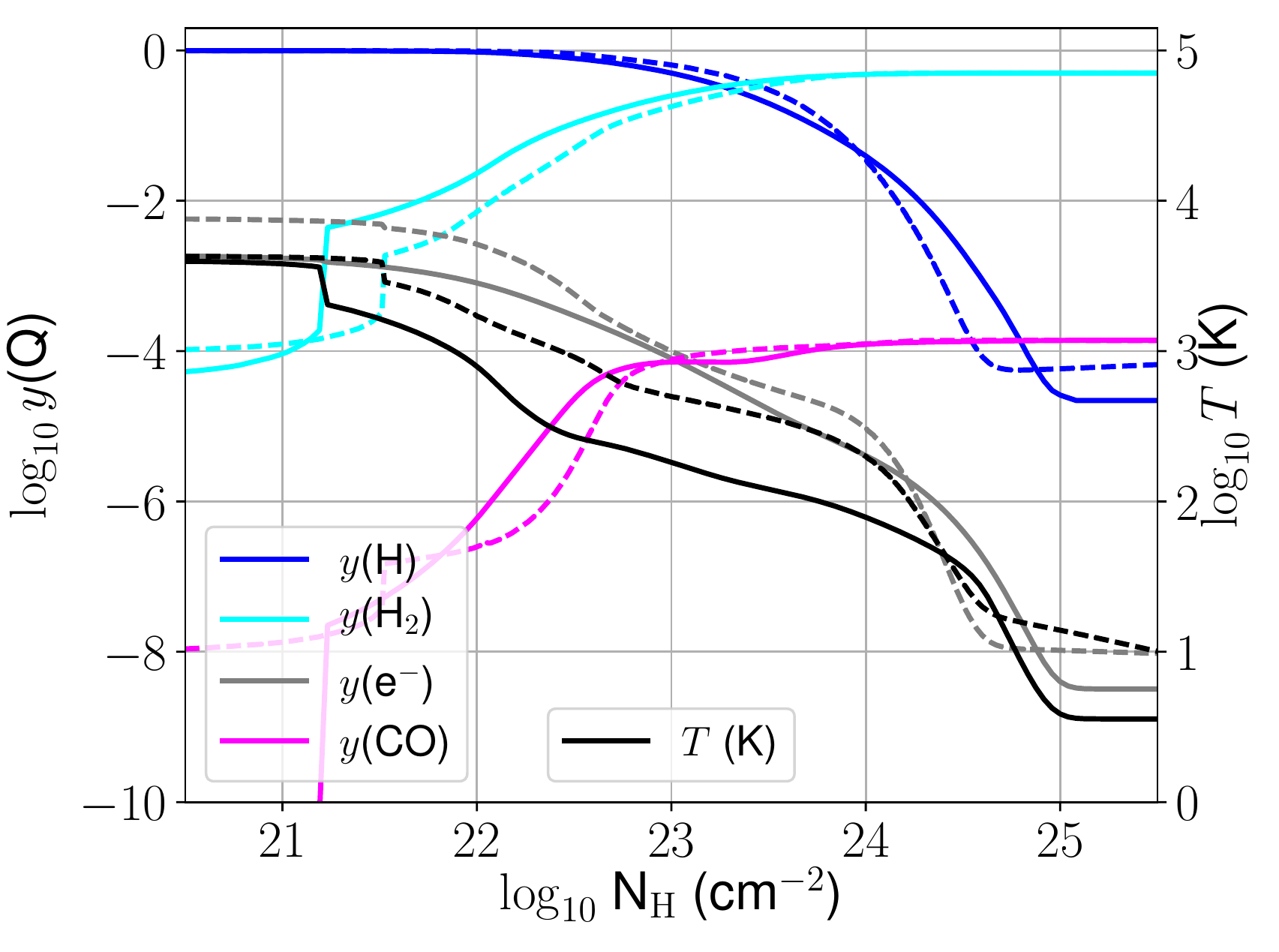} \\
\end{tabular}
  \caption{
  Abundances of H$_2$, CO, H, electrons, and gas temperature for models 1 (upper left), 2 (upper right), 3 (lower left) and 4 (lower right) calculated using \textsc{Cloudy} (dashed lines) and compared with our calculations (solid lines).
  The results are plotted as a function of column density of hydrogen.
  The left-hand vertical axis shows the fractional abundance whereas the right-hand vertical axis shows the temperature scale.
}
  \label{fig:cloudy}
\end{figure*}

\begin{figure*}
\begin{tabular}{ccc}
&$F_X=1.6$ erg\,cm$^{-2}$\,s$^{-1}$ & $F_X=160$ erg\,cm$^{-2}$\,s$^{-1}$ \\
\raisebox{7.5\normalbaselineskip}[0pt][0pt]{\rotatebox{90}{$n_\mathrm{H}=10^3$ cm$^{-3}$}} &
\includegraphics[trim = 0mm 12mm 0mm 3mm, clip, height=5.8cm]{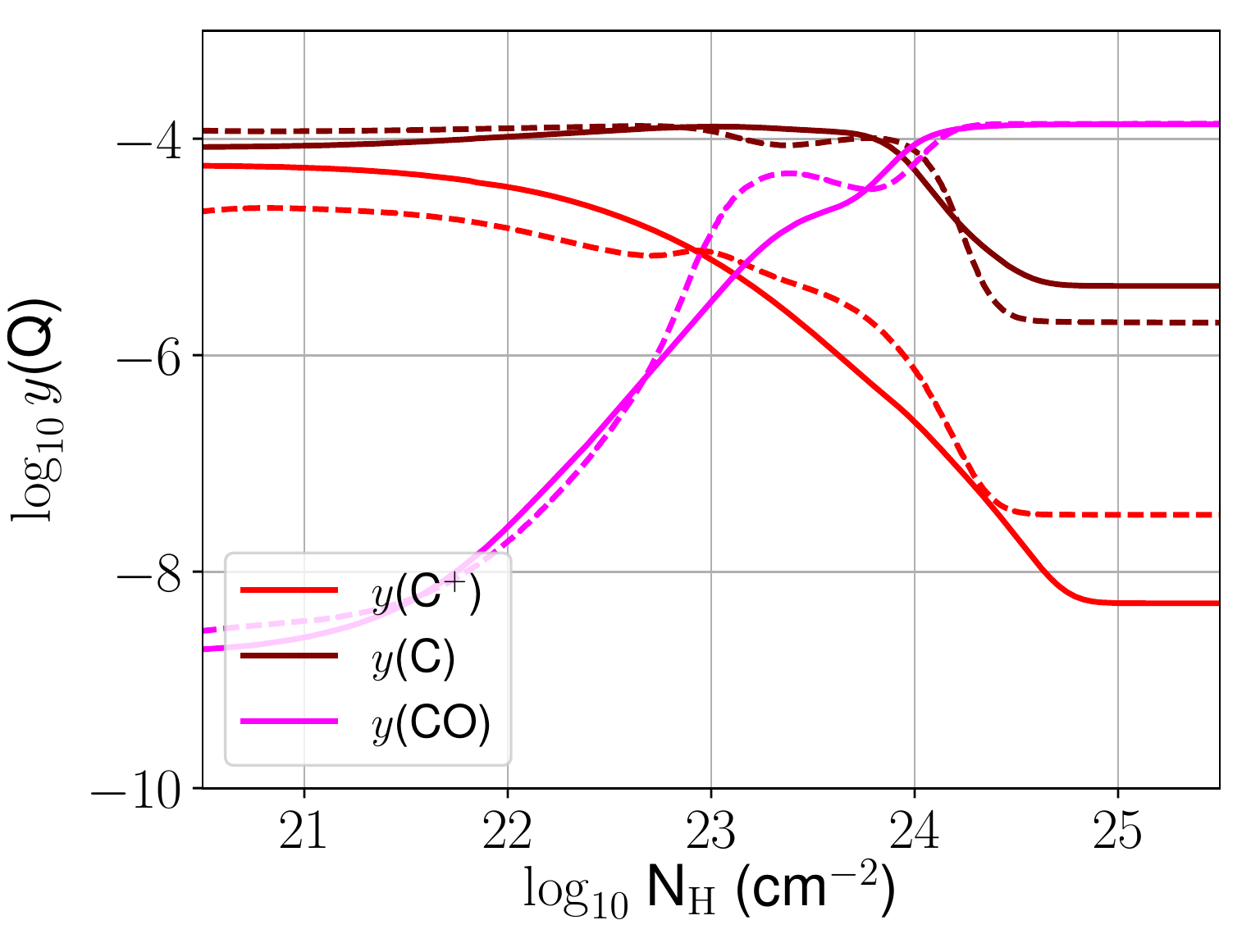} &
\includegraphics[trim = 12mm 12mm 0mm 3mm, clip, height=5.8cm]{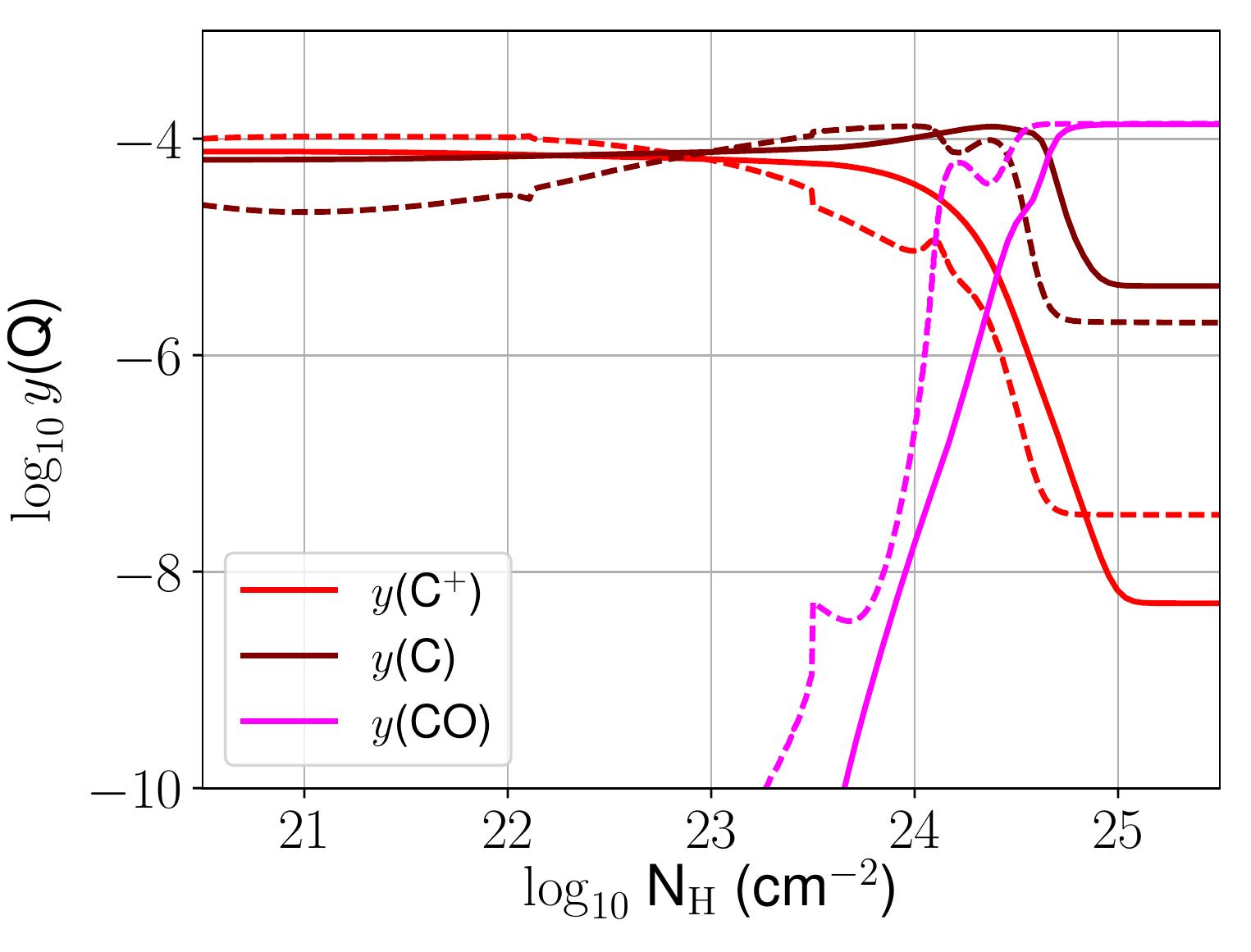} \\
\raisebox{7.5\normalbaselineskip}[0pt][0pt]{\rotatebox{90}{$n_\mathrm{H}=10^{5.5}$ cm$^{-3}$}} &
\includegraphics[trim = 0mm 0mm 0mm 3mm, clip, height=6.5cm]{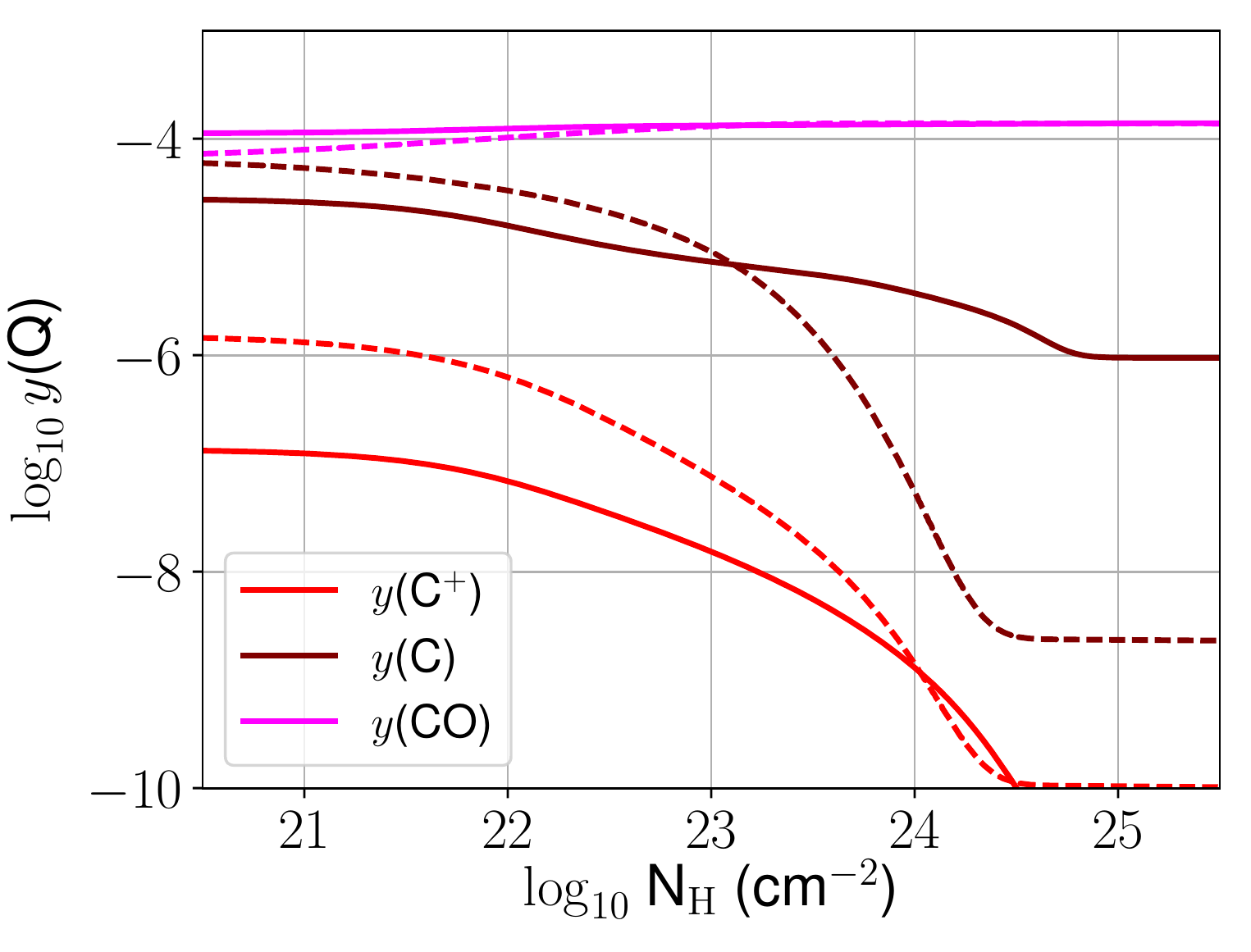} & 
\includegraphics[trim = 12mm 0mm 0mm 3mm, clip, height=6.5cm]{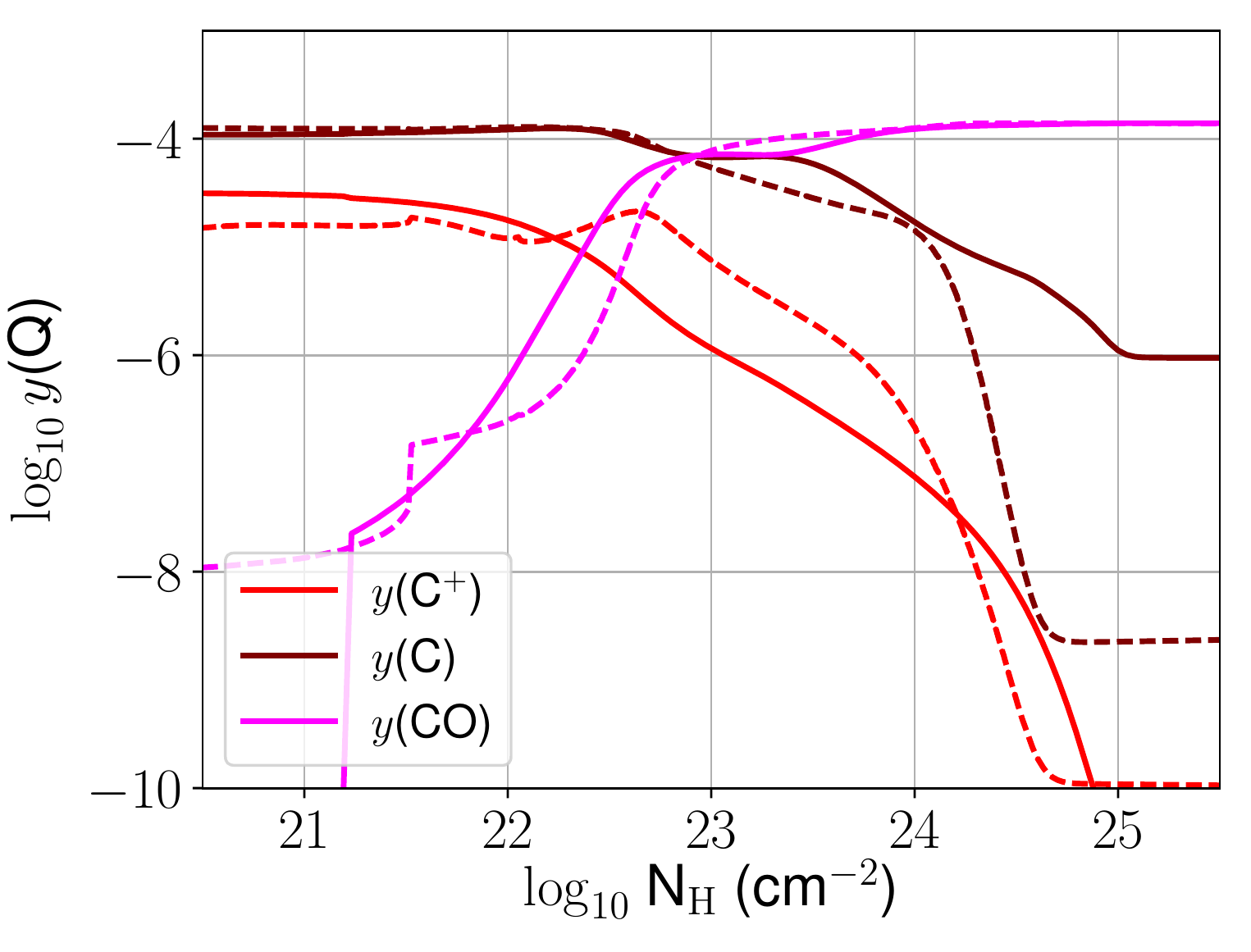} \\
\end{tabular}
  \caption{Same as Fig. \ref{fig:cloudy} but for the main carbon-bearing species included by the network: abundances of C$^+$, C, CO for the 4 models.
  The dashed lines show the results of the \textsc{Cloudy} calculation and the solid lines show our results.
  }
  \label{fig:cloudy2}
\end{figure*}

\subsection{Comparison with CLOUDY}
\label{sec:cloudy}

We also ran the same test problems with \textsc{Cloudy} \citep{FerPorVan13}, which has a more detailed treatment of X-ray absorption and ionization processes than our module and also a much larger chemical network.
The calculations were performed with version 17.00 of \textsc{Cloudy} as described by \citet{FerPorVan13} and \citet{FerChaGuz17}.
Note that even for the species that we have in common with the \textsc{Cloudy} network, the reaction and cooling rates used may not be the same.

We use the standard {\sc Cloudy} mix of of silicate and graphitic dust grains with a ratio of total to selective extinction of $R_V=3.1$, which is typical for the ISM in the Milky Way in terms of abundance and size distribution \citep{MatRumNor77}. Polycyclic aromatic hydrocarbons (PAHs) are not included.
As in previous work \citep{WalGirNaa15} we set the overall dust-to-gas mass ratio to 0.01.
We started from the full ISM model for the gas phase abundances and reduced the included heavy elements to the most important ones, i.e. the ones we find to be necessary in order to reproduce our results reasonably well (see Table \ref{table_cloudy}, left column).
All other elements were switched off but were thoroughly checked to only result in minor changes when included with their standard ISM gas phase abundances from {\sc Cloudy}.
We find that magnesium and iron are most important for setting the electron abundance.
The abundances of the elements that we do include are shown in Table \ref{table_cloudy}, middle column.

\begin{table}
\begin{tabular}{lcr}
\hline
Element & Abundance & Ionization potential \\
\hline
Sodium & $3.16\times 10^{-7}$ & 5.14 eV \\
Magnesium & $1.26 \times 10^{-5}$  & 7.65 eV \\
Iron & $6.31 \times 10^{-7}$ & 7.9 eV \\
Silicon & $3.16 \times 10^{-6}$ & 8.15 eV \\
Sulphur & $3.24 \times 10^{-5}$ & 10.36 eV \\
Carbon & $1.40\times10^{-4}$ & 11.26 eV \\
Oxygen & $3.40 \times 10^{-4}$ & 13.62 eV \\
Nitrogen & $7.94 \times 10^{-5}$ & 14.53 eV \\
Helium & $1.00 \times 10^{-1}$ & 24.59 eV \\
\hline
\end{tabular}
\caption{List of heavy elements included in the {\sc Cloudy} models (first column) sorted by their their respective ionization potential (last column). The relative abundances with respect to hydrogen are given in the middle column.}\label{table_cloudy}
\end{table}

In {\sc Cloudy} the ISRF is modeled as a blackbody with temperature 30000 K in the energy range of 0.44 to 0.99 Rydberg as suggested by the {\sc Cloudy} documentation. 
The total intensity of the ISRF is scaled to the same value as used in section \ref{sec:ms05}, i.e. corresponding to a $G_0=10^{-6}$.
All other initial conditions are also the same as in section \ref{sec:ms05}.

The results are plotted in Fig.~\ref{fig:cloudy}, in a similar manner to Fig.~\ref{fig:ms05_1234}.
\textsc{Cloudy} also obtains the chemo-thermal instability for models 2 and 4.
It occurs at the same $N_\mathrm{H}$ as what we find for model 2, but the jump in $T$ and $y(\mathrm{H}_2)$ is larger.
For model 4 \textsc{Cloudy} finds a weaker discontinuity that occurs at larger $N_\mathrm{H}$ than in our calculations.

For hydrogen, the atomic-to-molecular transition happens at similar $N_\mathrm{H}$ for \textsc{Cloudy} and our code, and the H$_2$ abundance is comparable in both calculations for all models.
The biggest difference is for model 2, where $y(\mathrm{H}_2)$ increases more rapidly with $N_\mathrm{H}$ in the  \textsc{Cloudy} calculation and the H$\rightarrow$H$_2$ transition occurs at smaller $N_\mathrm{H}$ (by $\sim0.5$ dex).
This is the opposite of what we found comparing with MS05, where they found the transition at larger $N_\mathrm{H}$ than our results by $\sim0.5$ dex.

The gas temperature from our calculations agrees well with \textsc{Cloudy} for models 1 and 2, but for models 3 and 4 \textsc{Cloudy} finds larger temperature than our module in the range $21.5 \lesssim \log N_\mathrm{H}/\mathrm{cm}^{-3} \lesssim 24.5$.
The electron fraction is also larger in this range.
The temperature discrepancy is up to 0.5\,dex for model 4.

The results for carbon-bearing species are plotted in Fig.~\ref{fig:cloudy2}.
\textsc{Cloudy} can include freeze-out of molecules onto grains, which is not in our network, so we switched this off for the comparison.
The CO abundance agrees well for all calculations in Figs.~\ref{fig:cloudy} and \ref{fig:cloudy2}.
In the \textsc{Cloudy} results, the dip in CO abundance just below $N_\mathrm{H}\approx10^{24}$\,cm$^{-2}$ in model 1 (slightly larger $N_\mathrm{H}$ in model 2) is because of CS formation at this depth, which is not in our network.
In Models 1 and 2 the \textsc{Cloudy} abundance of CO increases more rapidly with $N_\mathrm{H}$ than what we find, but the opposite is true in model 4.
The abundances of atomic C and C$^+$ generally agree well between the two networks, but the limiting $y(\mathrm{C})$ at large $N_\mathrm{H}$ is much lower in the \textsc{Cloudy} results for models 3 and 4.
We find generally smooth and monotonic curves for C$^+$, C and CO, with at most a single maximum for $y($C$)$, whereas \textsc{Cloudy} has more pronounced maxima and and other features.
This is probably due to interaction with other carbon-bearing species that are not included in our network.
Notably, the agreement with \textsc{Cloudy} is better than with MS05, suggesting that updated reaction and cooling rates over the past 13 years have a bigger impact on our results than the size of the chemical network.

In summary, our results for the H\,$\rightarrow$\,H$_2$ and C$^+$\,$\rightarrow$\,C$\,\rightarrow$\,CO transitions agree well with results obtained from \textsc{Cloudy}, with small differences in the exact value of $N_\mathrm{H}$ for each transition.
The temperature and electron fractions as a function of $N_\mathrm{H}$ also agree well with some caveats, notably the discrepancy in model 4.
Less abundant species (CH$_\mathrm{x}$, OH$_\mathrm{x}$, HCO$^+$) are poorly predicted by our simple reaction network, probably because these are primarily included in the network in order to obtain the correct relative abundances of C$^+$, C, and CO.
These trace species are not the focus of this work.

\begin{figure*}
\begin{tabular}{ccc}
&$F_X=10^{-3}$ erg\,cm$^{-2}$\,s$^{-1}$ & $F_X=10^5$ erg\,cm$^{-2}$\,s$^{-1}$ \\
\raisebox{7.5\normalbaselineskip}[0pt][0pt]{\rotatebox{90}{$E_\mathrm{rad}=0.1$\,keV}} &
\includegraphics[height=6.5cm]{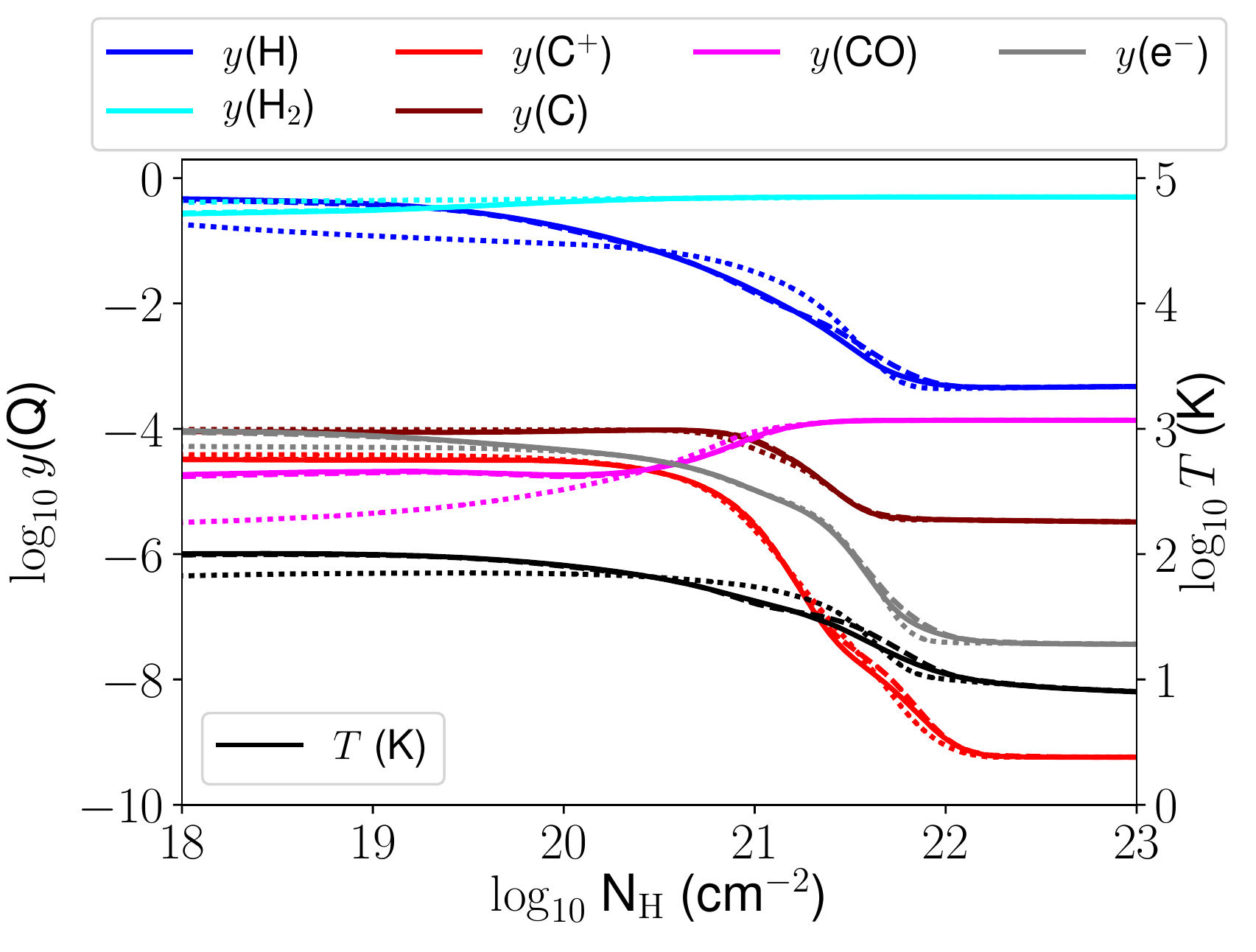} &
\includegraphics[trim = 14mm 0mm 0mm 1mm, clip, height=6.5cm]{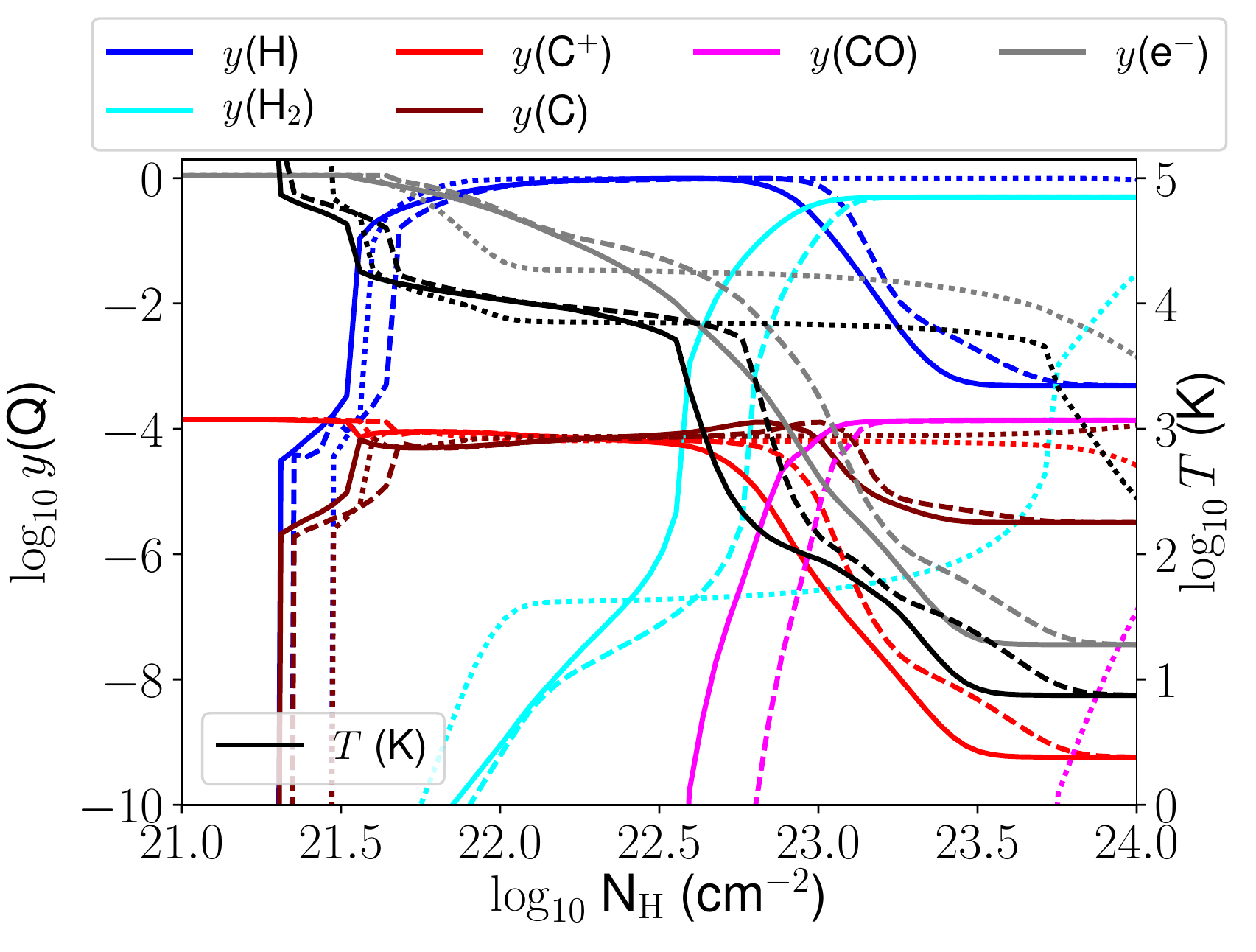} \\
\raisebox{7.5\normalbaselineskip}[0pt][0pt]{\rotatebox{90}{$E_\mathrm{rad}=1$\,keV}} &
\includegraphics[trim = 0mm 0mm 0mm 22mm, clip, height=5.5cm]{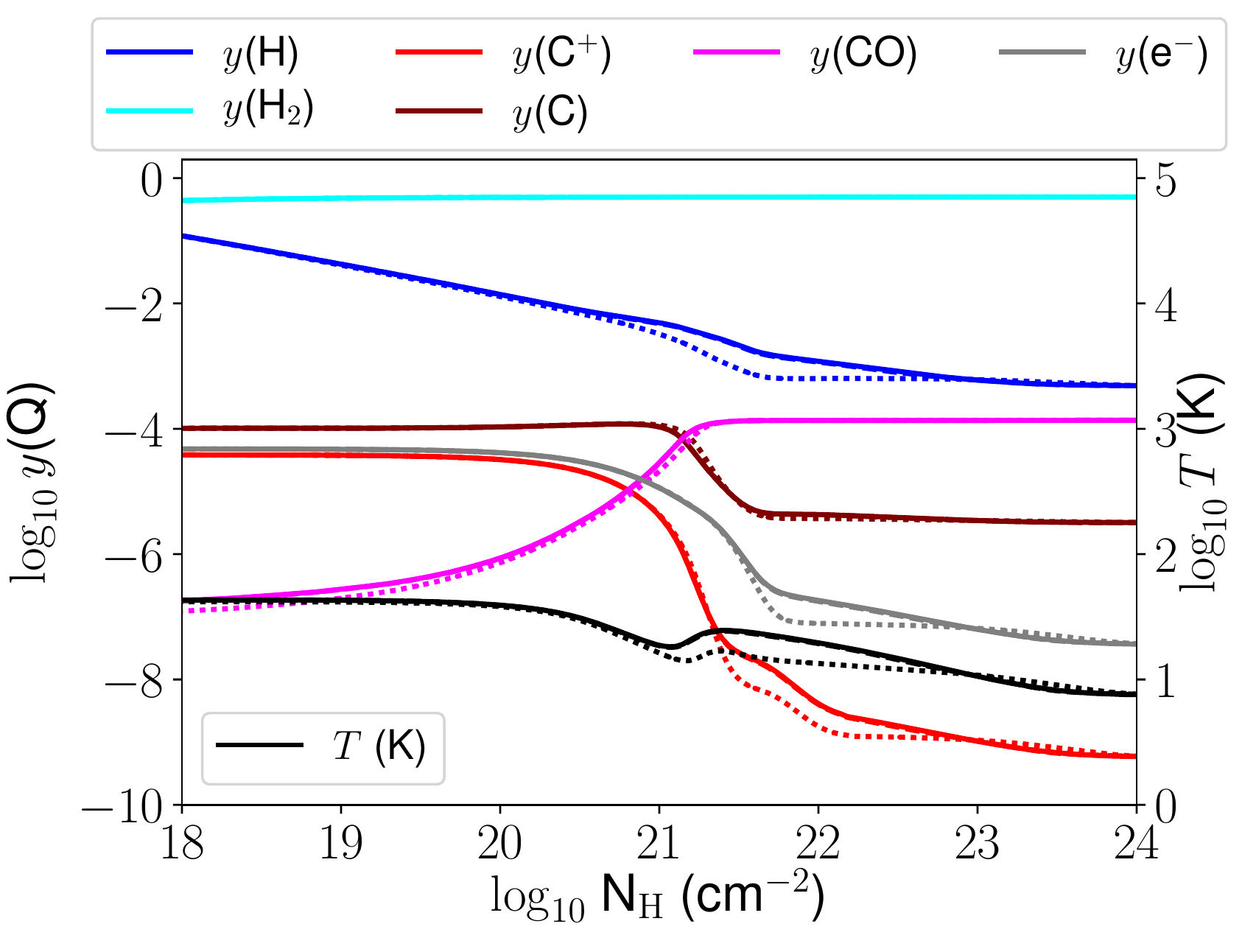} & 
\includegraphics[trim = 14mm 0mm 0mm 22mm, clip, height=5.5cm]{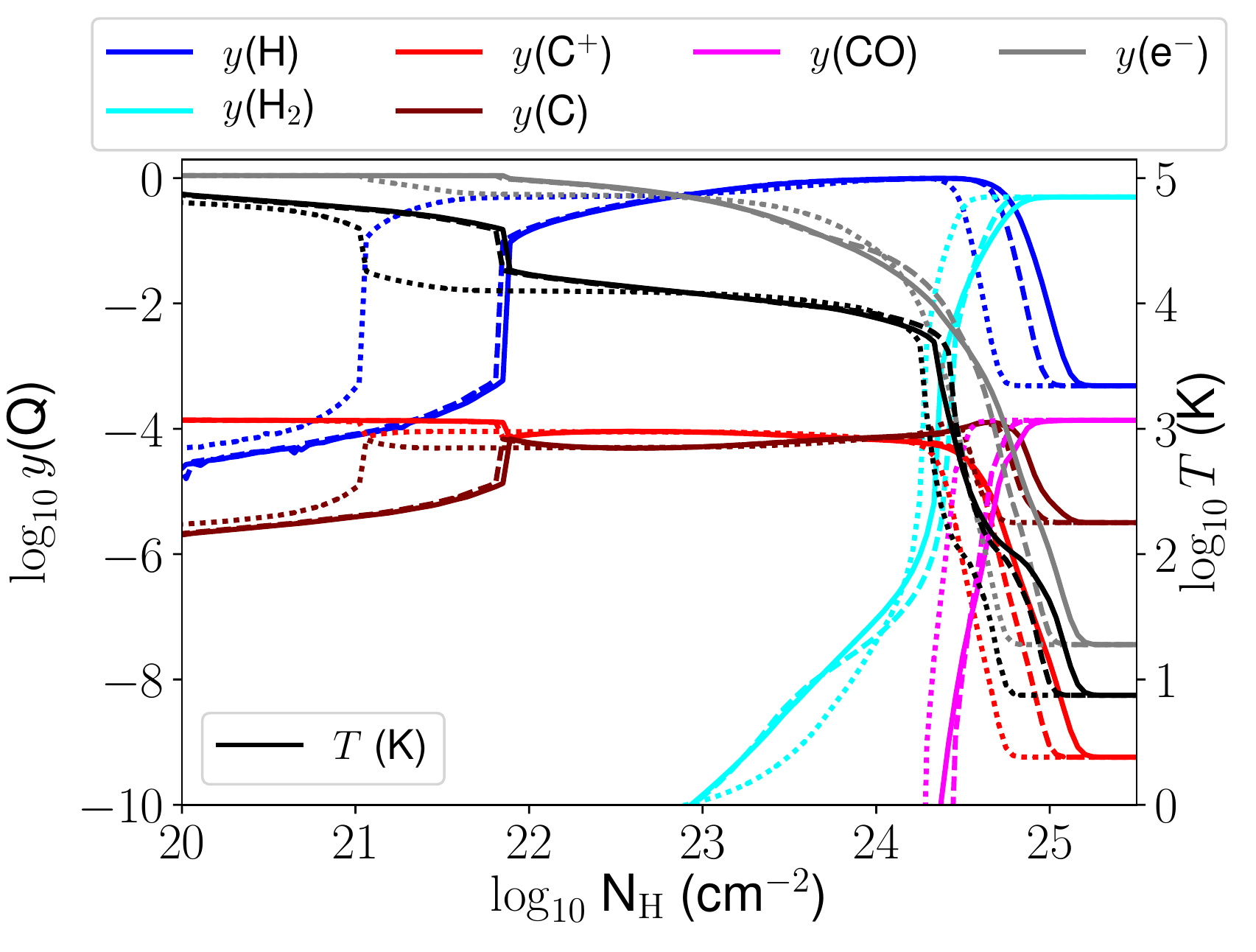} \\
\end{tabular}
  \caption{
  Abundances of H$^+$, H, H$_2$, C$^+$, C, CO, electrons, and gas temperature for a set of 1D calculations using different X-ray fluxes and radiation temperature.
  Dotted lines are results using two energy bins in 0.1-10\,keV, dashed lines using 6, and solid lines using 20.
  The results are plotted as a function of column density of hydrogen nuclei.
  In each case the gas has number density $n_\mathrm{H}=10^4$\,cm$^{-3}$.
  The left-hand vertical axis shows the fractional abundance whereas the right-hand vertical axis shows the temperature scale.
}
  \label{fig:eres}
\end{figure*}

\subsection{Tests of energy resolution}
\label{sec:eres}

We ran a large grid of 1D models with varying ISM density, X-ray flux, and X-ray spectrum.
Density varies from $n_\mathrm{H}=[0.1-10^6]$\,cm$^{-3}$, flux from $F_X=[10^{-5}-10^5]$\,erg\,cm$^{-2}$\,s$^{-1}$, and blackbody spectra with radiation temperature $E_\mathrm{rad}=[0.1-10]$\,keV.
This was used to validate the code over a large range of different conditions, find any regions of parameter space where the ODE solver fails to converge, and test how many energy bins are required for different ISM conditions.
A sample of results are shown in Fig.~\ref{fig:eres}, for a fixed gas density ($n_\mathrm{H}=10^4$\,cm$^{-3}$), two different X-ray fluxes and two different radiation temperatures, $E_\mathrm{rad}$ (for a blackbody spectrum).
All of these calculations have a UV radiation field of $G_0=1$, which is why the CO abundance is low at low column density.

The models of MS05 (Section~\ref{sec:ms05}) had $G_0=10^{-6}$, and so the gas could be fully molecular at low column density for Model 3.
Apart from this, the low-flux calculations in Fig.~\ref{fig:eres} have many similarities to Model 3.
The high-flux calculations are most similar to Model 4, but the flux is significantly higher.
In these extreme conditions the energy resolution plays a key role because the cross-section of the softest (hardest) energy bin increases (decreases) as the energy bin gets narrower.

For all plotted calculations, 2 energy bins (0.1-1 and 1-10\,keV, dotted lines) is a rather crude approximation and some atomic-to-molecular transitions happen at quite different column densities for $F_X=10^5$\,erg\,cm$^{-2}$\,s$^{-1}$.
For the low-flux calculations, 6 energy bins (dashed lines) are sufficient in all cases and seem adequate but not ideal for the high-flux calculations.
The transitions between different phases (ionized-to-atomic, atomic-to-molecular) happen at column densities differing by up to 0.1\,dex between 6 and 20 energy bins, whereas the difference can be up to 1\,dex between 2 and 20 bins.

The tradeoff between number of energy bins and computational cost (memory and cpu cycles) means that we have to accept some level of error from using discrete energy bins.
The worst case found on the grid of calculations was for $E_\mathrm{rad}=10$\,keV and $F_X=10^5$\,erg\,cm$^{-2}$\,s$^{-1}$, i.e.\ gas irradiated very strongly by a hard X-ray field.
In this case the location of the atomic-to-molecular transition differed by about 0.1\,dex between 6 and 20 energy bins.
This is because there is a lot of flux in the highest-energy bin for such a hard spectrum, and so its cross-section is key to determining the column density at which X-ray heating becomes ineffective.
For the calculations in the next sections we use a thermal spectrum with $T=1$\,keV, and so this problem is not so severe because there is very little flux in the highest energy bins.

\section{Irradiation of a Fractal Cloud}
\label{sec:fractal}

\begin{figure}
\includegraphics[width=0.49\textwidth]{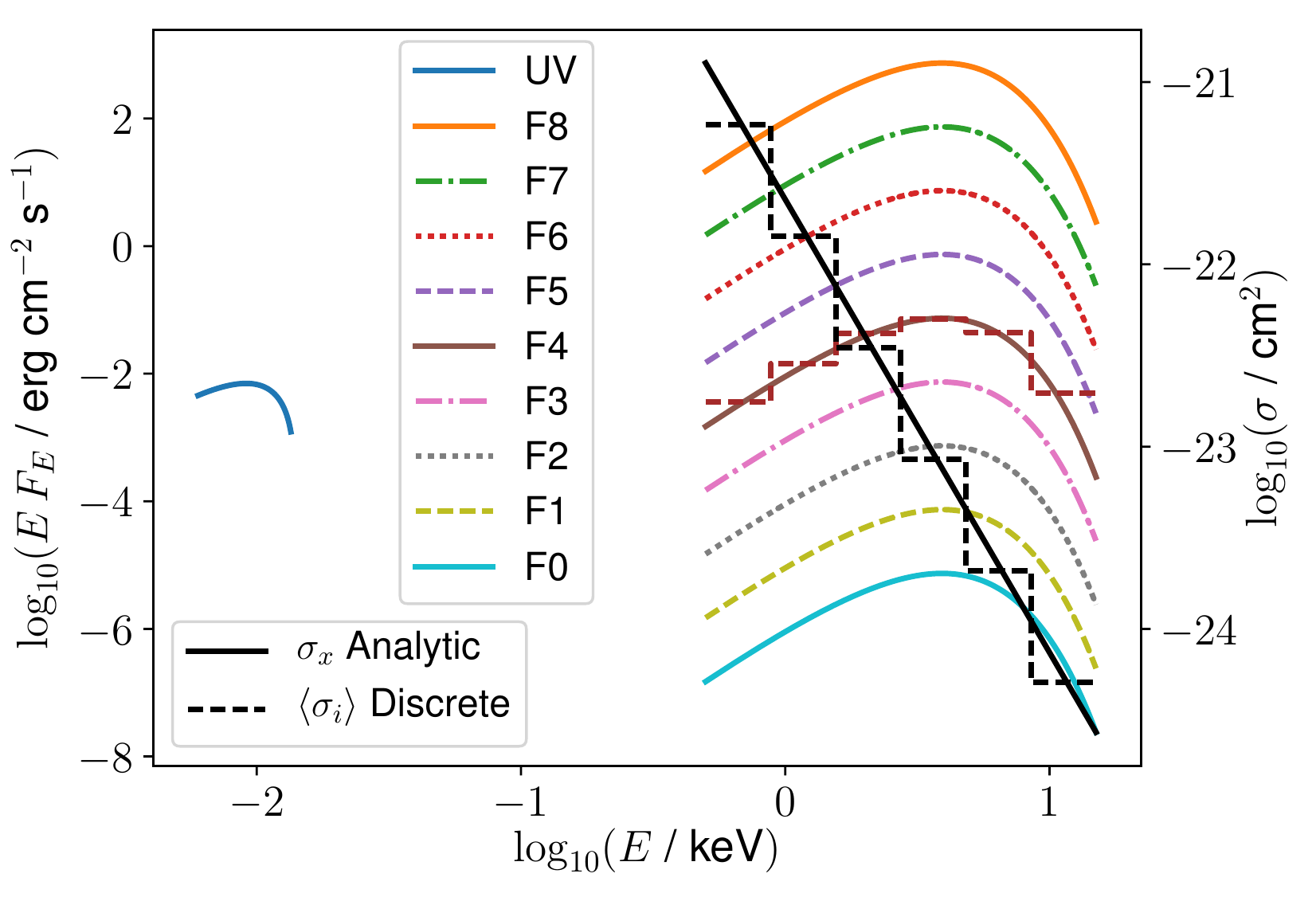}
\caption{
  UV flux (blue) and X-ray flux from Table~\ref{tab:sims} for the 9 simulations in section~\ref{sec:fractal} (left $y$-axis) and X-ray absorption cross-section (right $y$-axis).
  The continuous flux is plotted in all cases, and the discrete flux for simulation F4 using the dashed brown line.
  $E$ is energy in keV and $F_E$ is energy flux in units erg\,cm$^{-2}$\,s$^{-1}$\,keV$^{-1}$.
  For the cross-section, the continuous black line plots Eqn.~\ref{eqn:sigma} from \citet{PanCabPin12}, and the dashed black line the discrete cross-section used for each of the six energy bins.
  \label{fig:spectrum}}
\end{figure}

\begin{table}
  \centering
  \caption{X-ray fluxes and energy densities considered in each of the simulations in section~\ref{sec:fractal}.}
  \label{tab:sims}
  \begin{tabular}{lcc}
    \hline
    Simulation & Flux (erg\,cm$^{-2}$\,s$^{-1}$) & $E_\mathrm{rad}$ (erg\,cm$^{-3}$) \\
    \hline
    F0 & $10^{-5}$  & $3.3\times10^{-16}$ \\
    F1 & $10^{-4}$  & $3.3\times10^{-15}$ \\
    F2 & $10^{-3}$  & $3.3\times10^{-14}$ \\
    F3 & $10^{-2}$  & $3.3\times10^{-13}$ \\
    F4 & $10^{-1}$  & $3.3\times10^{-12}$ \\
    F5 & $10^{0}$   & $3.3\times10^{-11}$ \\
    F6 & $10^{1}$   & $3.3\times10^{-10}$ \\
    F7 & $10^{2}$   & $3.3\times10^{-9}$ \\
    F8 & $10^{3}$   & $3.3\times10^{-8}$ \\
    \hline
  \end{tabular}
\end{table}

\begin{table*}
  \centering
  \caption{Energy limits, mean absorption cross-section  $\langle\sigma\rangle$, X-ray radiation flux $4\pi J_{X,i}$, and X-ray energy density $E_\mathrm{rad}$ for the six energy bins used in the 3D \textsc{flash} simulations.
  The radiation flux and energy density are quoted for simulation F5, and are scaled up or down by powers of 10 for the other simulations.
  }
  \label{tab:bins}
  \begin{tabular}{lllccc} 
    \hline
    Bin & $E_{\mathrm{min},i}$ (keV) & $E_{\mathrm{max},i}$ (keV) & $\langle\sigma_i\rangle$ (cm$^{-2}$) & $4\pi J_{X,i}$ (erg cm$^{-2}$ s$^{-1}$) & $E_{\mathrm{rad},i}$ (erg cm$^{-3}$)\\
    \hline
0  &  0.500  &  0.881  &  $5.84\times10^{-22}$  &  $1.97\times10^{-2}$  &  $6.57\times10^{-13}$  \\
1  &  0.881  &  1.554  &  $1.43\times10^{-22}$  &  $7.85\times10^{-2}$  &  $2.62\times10^{-12}$  \\
2  &  1.554  &  2.739  &  $3.49\times10^{-23}$  &  $2.34\times10^{-1}$  &  $7.81\times10^{-12}$  \\
3  &  2.739  &  4.827  &  $8.54\times10^{-24}$  &  $3.98\times10^{-1}$  &  $1.33\times10^{-11}$  \\
4  &  4.827  &  8.510  &  $2.09\times10^{-24}$  &  $2.42\times10^{-1}$  &  $8.09\times10^{-12}$  \\
5  &  8.510  &  15.000  &  $5.10\times10^{-25}$  &  $2.76\times10^{-2}$  &  $9.20\times10^{-13}$  \\
    \hline
  \end{tabular}
\end{table*}

We added the new chemistry network to the \textsc{flash} code, as discussed in Section~\ref{sec:methods}; this was implemented in a similar way to how the NL97 network \citep{NelLan97, GloCla12} has been used for the SILCC simulations \citep{WalGirNaa15}.
Multiple chemical species are implemented using the \textsc{flash} Multispecies framework, and radiative transfer uses \textsc{TreeRay} \citep{WunWalDin18}.

We follow \citet{Shadmehri2011} and \citet{Walch2012} to set up a fractal density field with a given fractal index $D_f$ and a lognormal density probability density function (PDF).
The fractal density field is setup in Fourier space using a power-law distribution of the amplitude squared, $A_{\rho(k)}^2 \propto k^{-n}$ on all modes ranging from 1 to 128.
The power spectral index $n$ is related with $D_f$ through $D_f = 4-\frac{n}{2}$.
Here we choose $D_f = 2.5$ and hence $n=3.0$, typical for molecular clouds in the Milky Way \citep{Stutzki1998}.
The simulation box is a cube of diameter 25.6\,pc and we use a uniform grid with $256^3$ grid cells, so the grid cell-size is 0.1\,pc, sufficient to resolve the CO chemistry \citep{SeiWalGir17}.
The total mass in the box is $10^5$\,M$_\odot$ and the maximum density located at the origin of the computational domain is $\rho_\mathrm{max}=1.6\times10^{-20}$\,g\,cm$^{-3}$.

Nine different simulations were run without hydrodynamics, labelled F0-F8, each with a different X-ray flux irradiating the outer boundary given in Table~\ref{tab:sims}.  
Recall that this flux is equal to $\sum_{i=1}^{N_E}4\pi J_{X,i}$ where $J_{X,i}$ is the mean intensity of the isotropic radiation field in energy bin $i$.
The hydrodynamic boundary conditions are irrelevant for the calculation, and as noted above we use isolated boundaries for the \textsc{TreeRay} algorithm.
We consider a thermal X-ray spectrum between 0.5 and 15 keV, with a temperature of 1\,keV.
Six new scalar field variables are added to account for the attenuation of the six logarithmically spaced X-ray energy bins, with energy limits and mean cross sections in each bin given in Table~\ref{tab:bins}.
The unattenuated X-ray flux and energy density in each energy bin is also quoted for Simulation F5 in Table~\ref{tab:bins}; for other simulations these values can be scaled, e.g., F0 is scaled down by $10^5$ and simulation F8 is scaled up by $10^3$.
This Table shows that the energy range 0.5-15\,keV covers almost all of the emission for the 1\,keV blackbody that we consider; adding further energy bins above or below this range would add less than 1\% to the total X-ray energy density.
Fig.~\ref{fig:spectrum} plots the UV and X-ray flux for each of the 9 simulations, as well as the continuous and discrete cross section for X-ray absorption.
For simulation F4 the discrete flux in each bin is also shown as the brown dashed line, converted to the appropriate units by multiplying the flux by the midpoint energy of the bin.
The external UV radiation field is set to $G_0=1.7$ in units of the Habing field, corresponding to the \citet{Dra78} field, and is not scaled with the X-ray field strength but rather kept constant.

For each simulation, we start with constant temperature 1423\,K (sound speed of 3\,km\,s$^{-1}$) and uniform number fractions of $y($H$_2)=10^{-5}$, $y($H$^+)=0.1$ and $y($CO$)=10^{-8}$.
We assume the rest of the carbon is in the form of C$^+$, that helium is neutral, and that the metal, M, is in the form of M$^+$.
The simulation is then run so that it evolves chemically and thermally towards equilibrium for 4\,Myr.
The dense regions have reached equilibrium by this time, but the lowest density gas is still evolving slowly.

\subsection{Physical state of the gas}
\label{sec:fractal:phase}
\begin{figure*}
\centering
\includegraphics[trim = 0mm 17mm 20mm 1mm, clip, height=4.7cm]{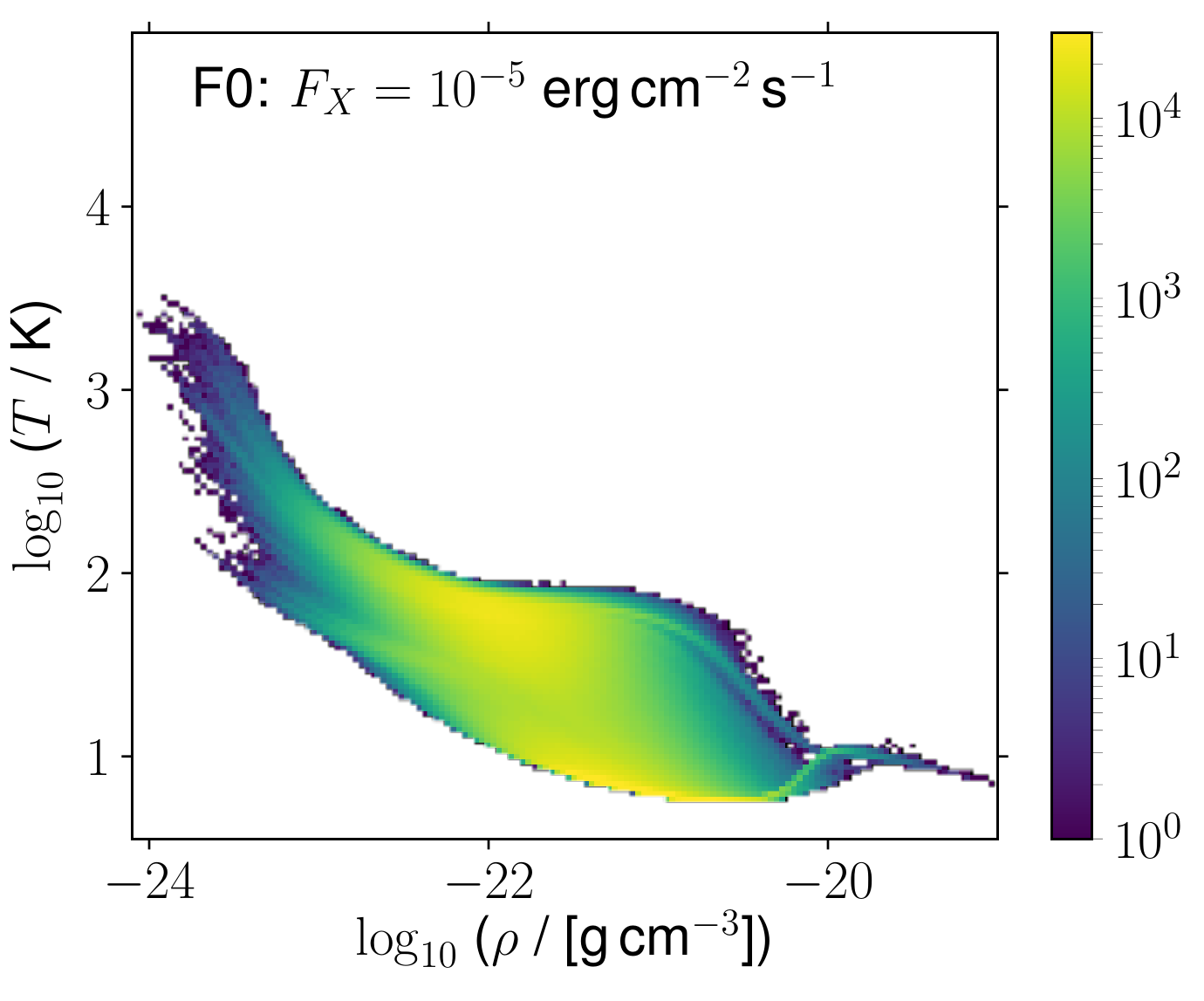}
\includegraphics[trim = 15mm 17mm 20mm 1mm, clip, height=4.7cm]{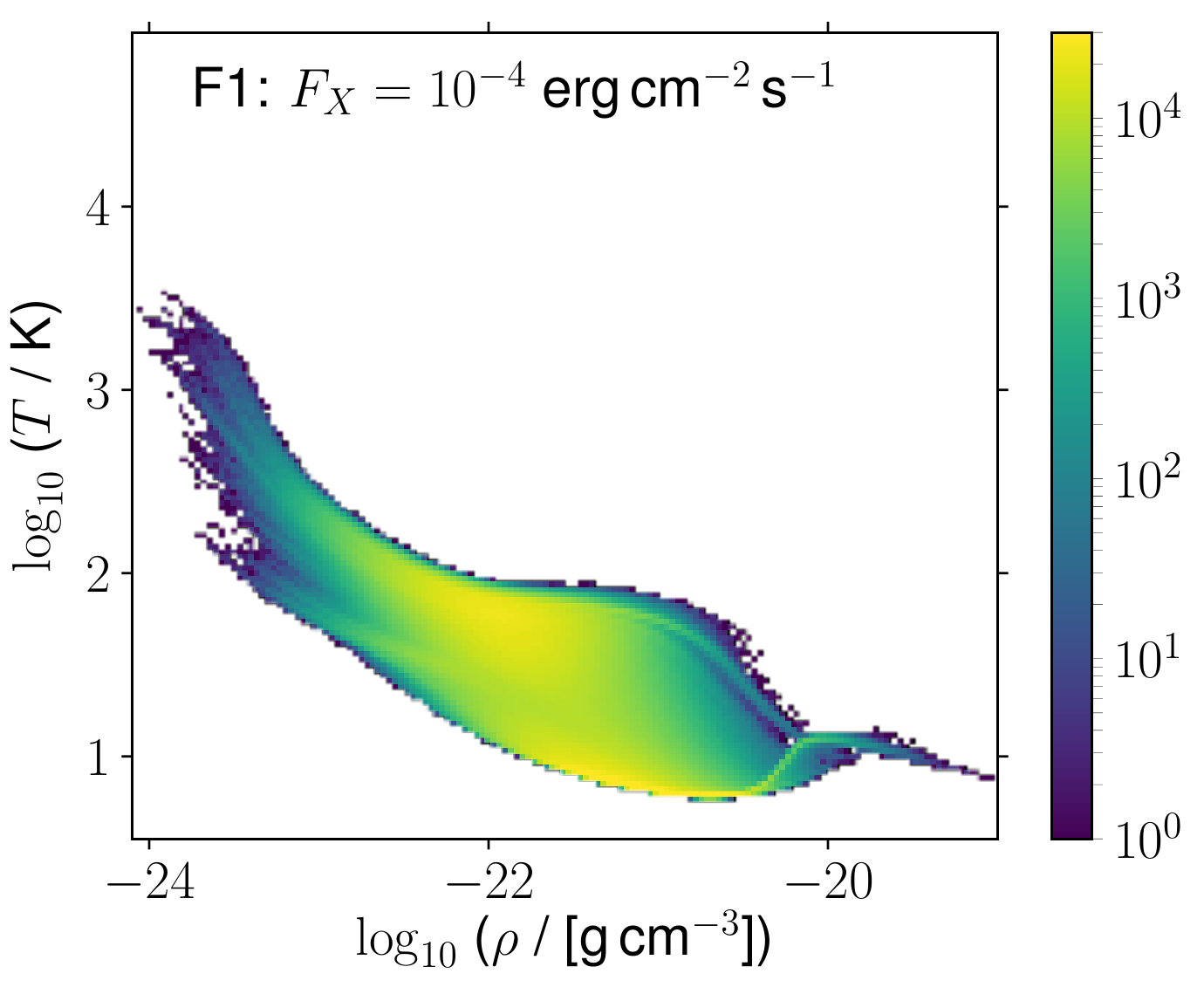}
\includegraphics[trim = 15mm 17mm 2mm 1mm, clip, height=4.7cm]{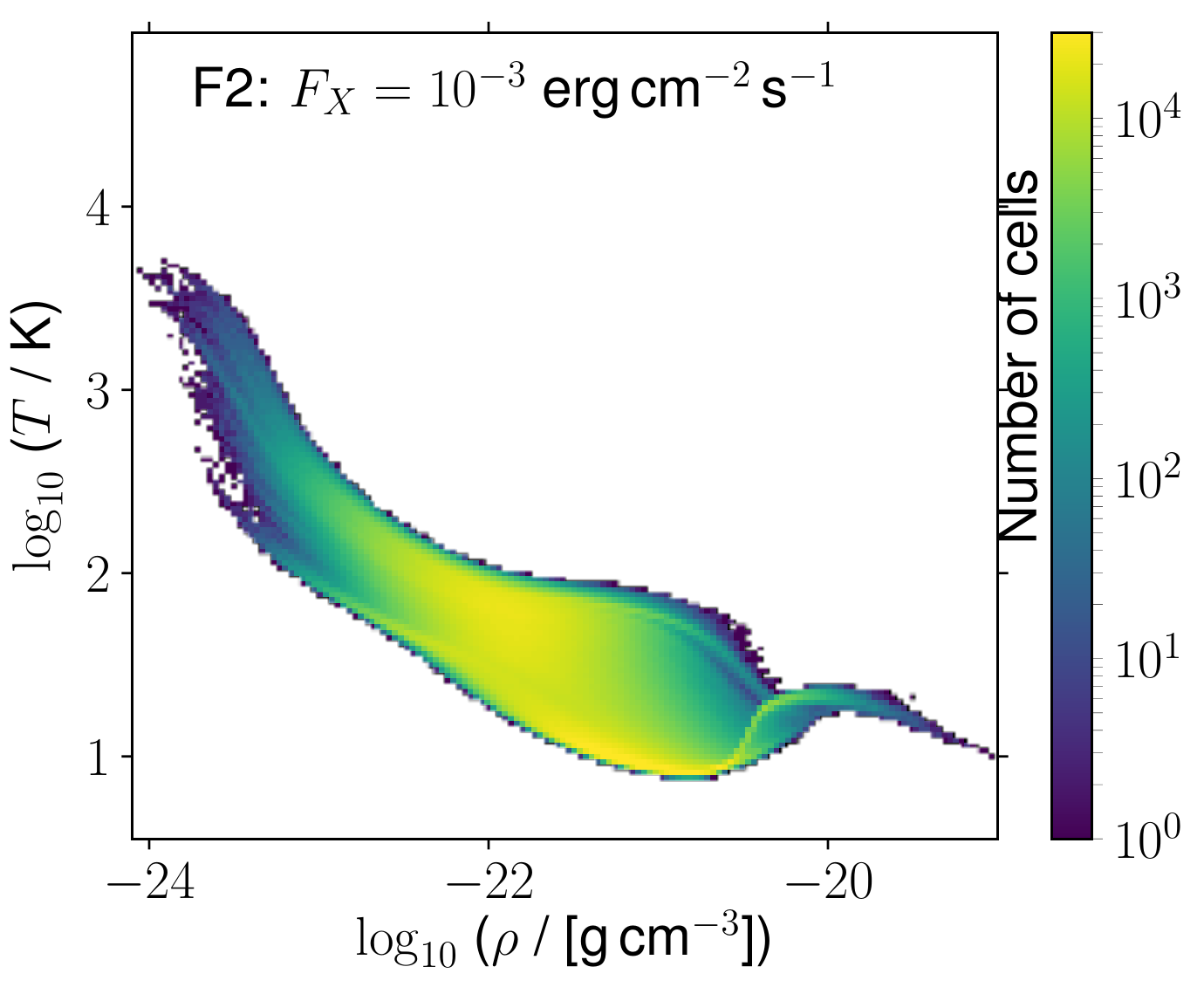}\\
\includegraphics[trim = 0mm 17mm 20mm 1mm, clip, height=4.7cm]{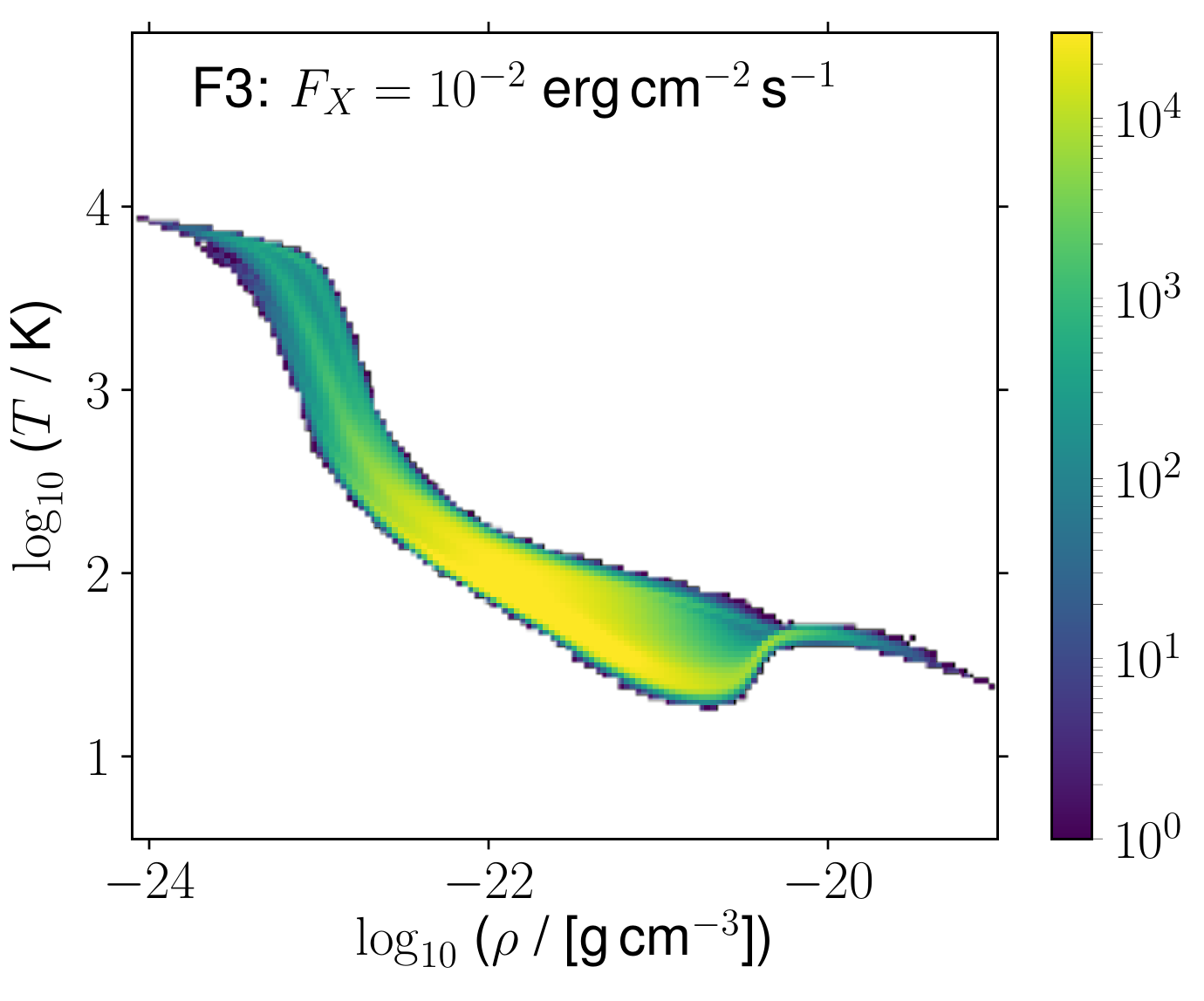}
\includegraphics[trim = 15mm 17mm 20mm 1mm, clip, height=4.7cm]{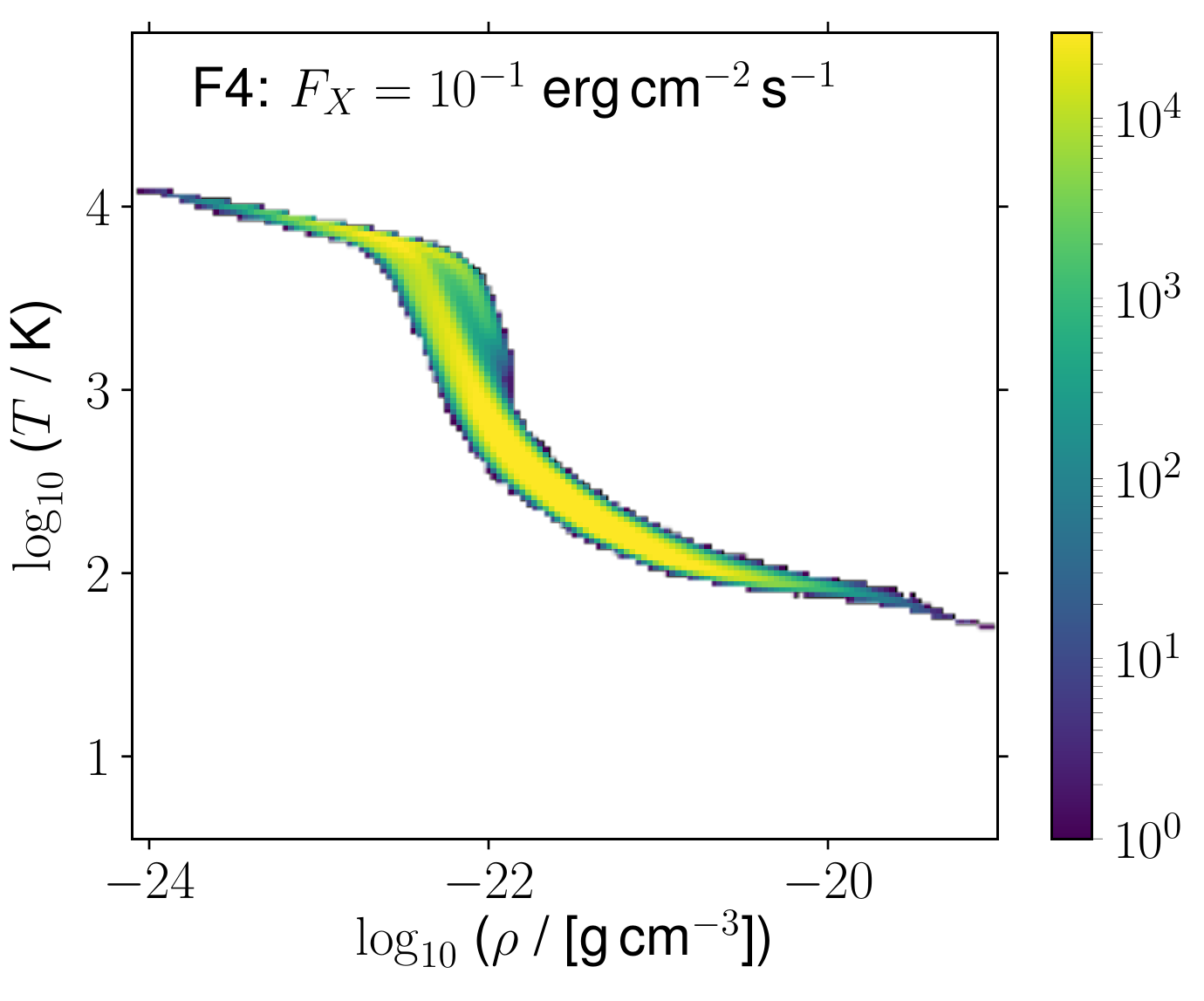}
\includegraphics[trim = 15mm 17mm 2mm 1mm, clip, height=4.7cm]{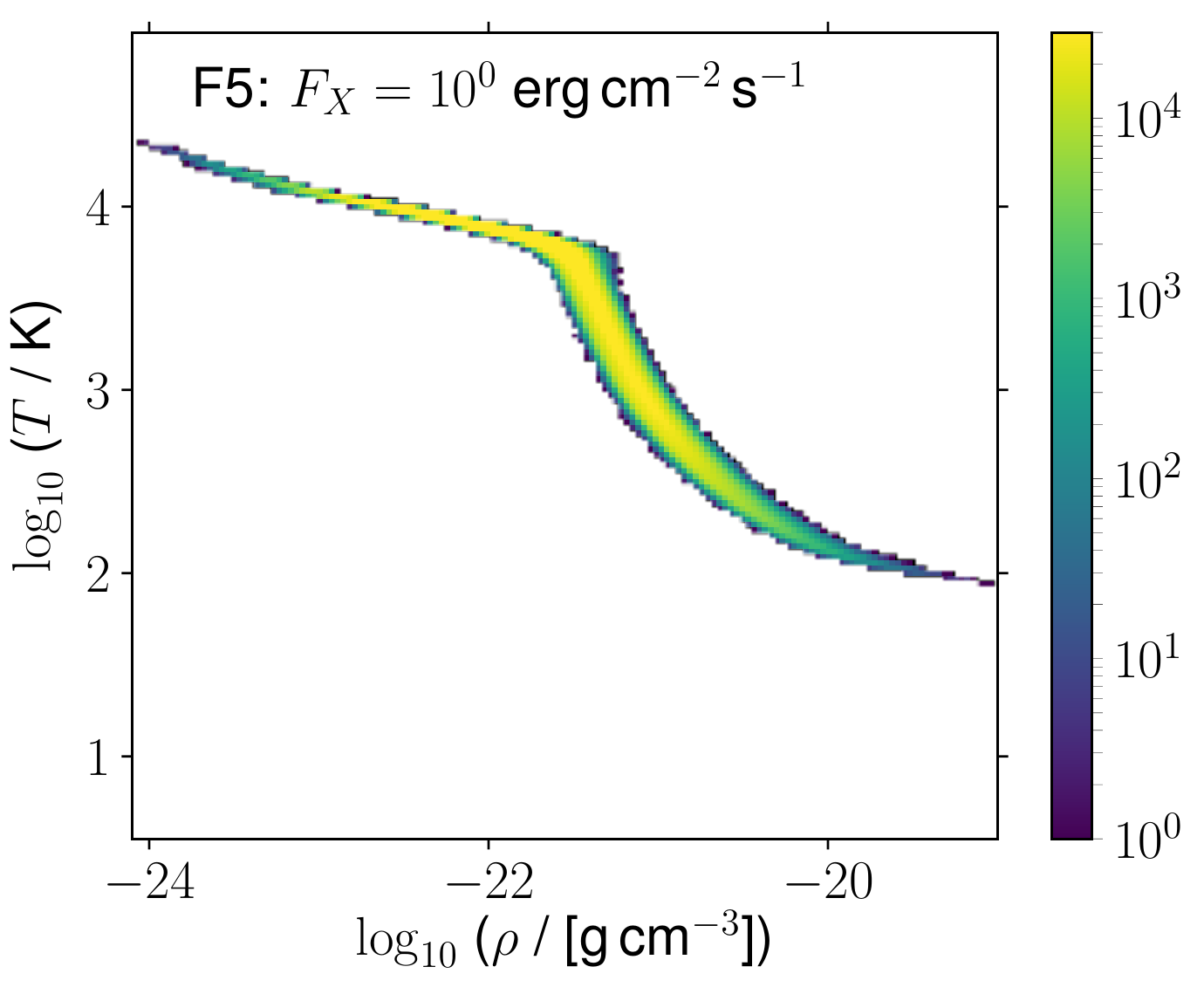}\\
\includegraphics[trim = 0mm 0mm 20mm 1mm, clip, height=5.5cm]{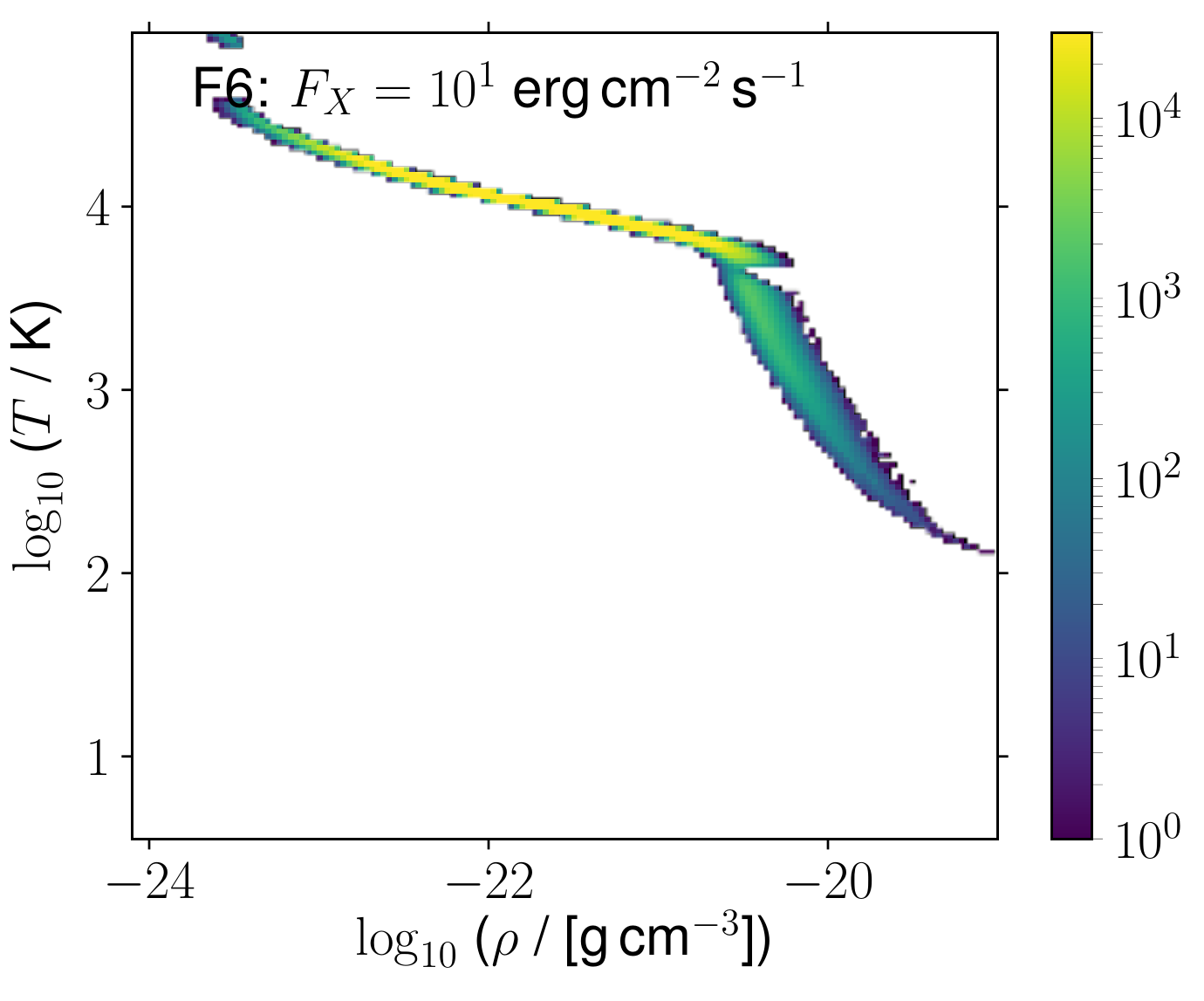}
\includegraphics[trim = 15mm 0mm 20mm 1mm, clip, height=5.5cm]{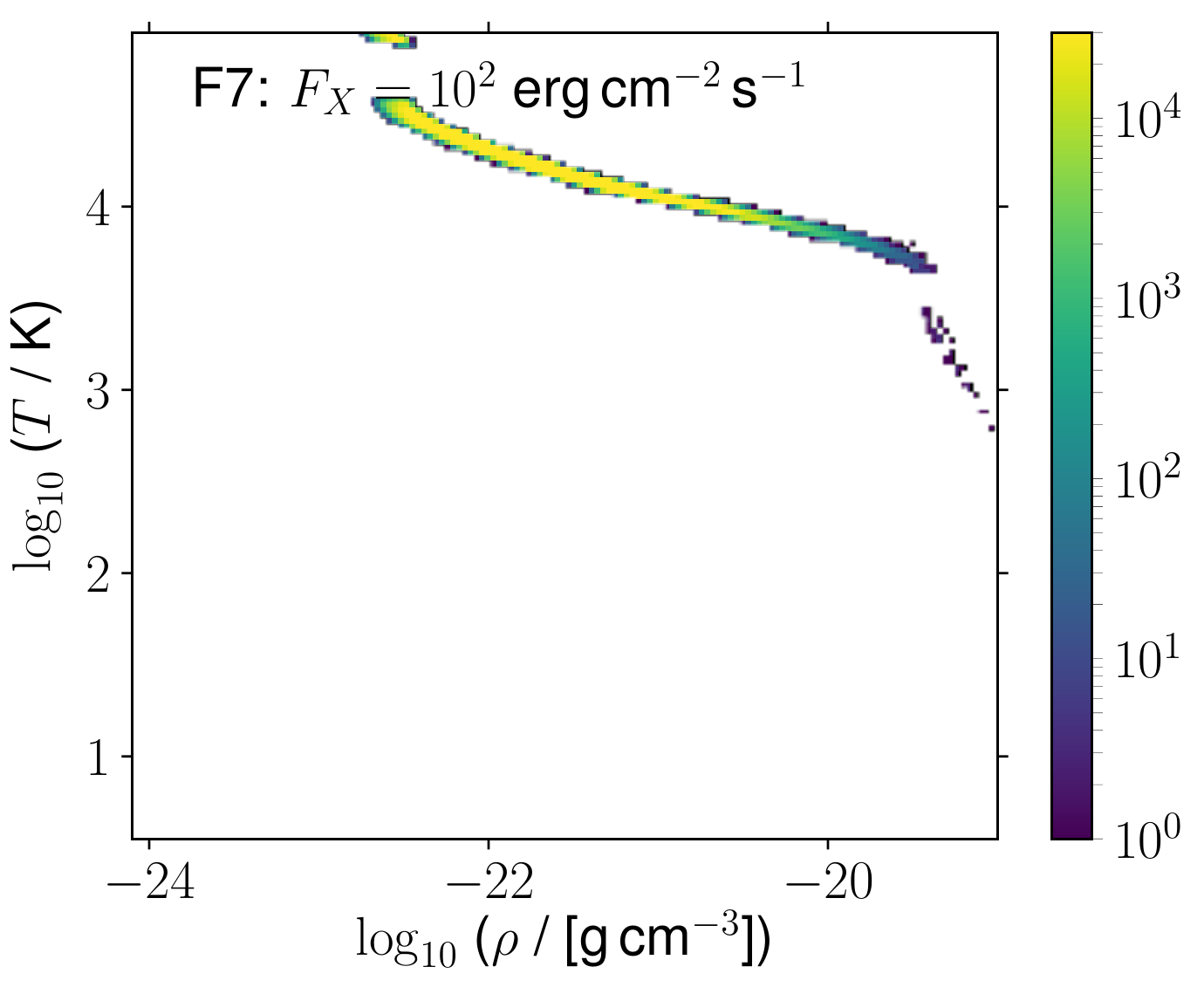}
\includegraphics[trim = 15mm 0mm 2mm 1mm, clip, height=5.5cm]{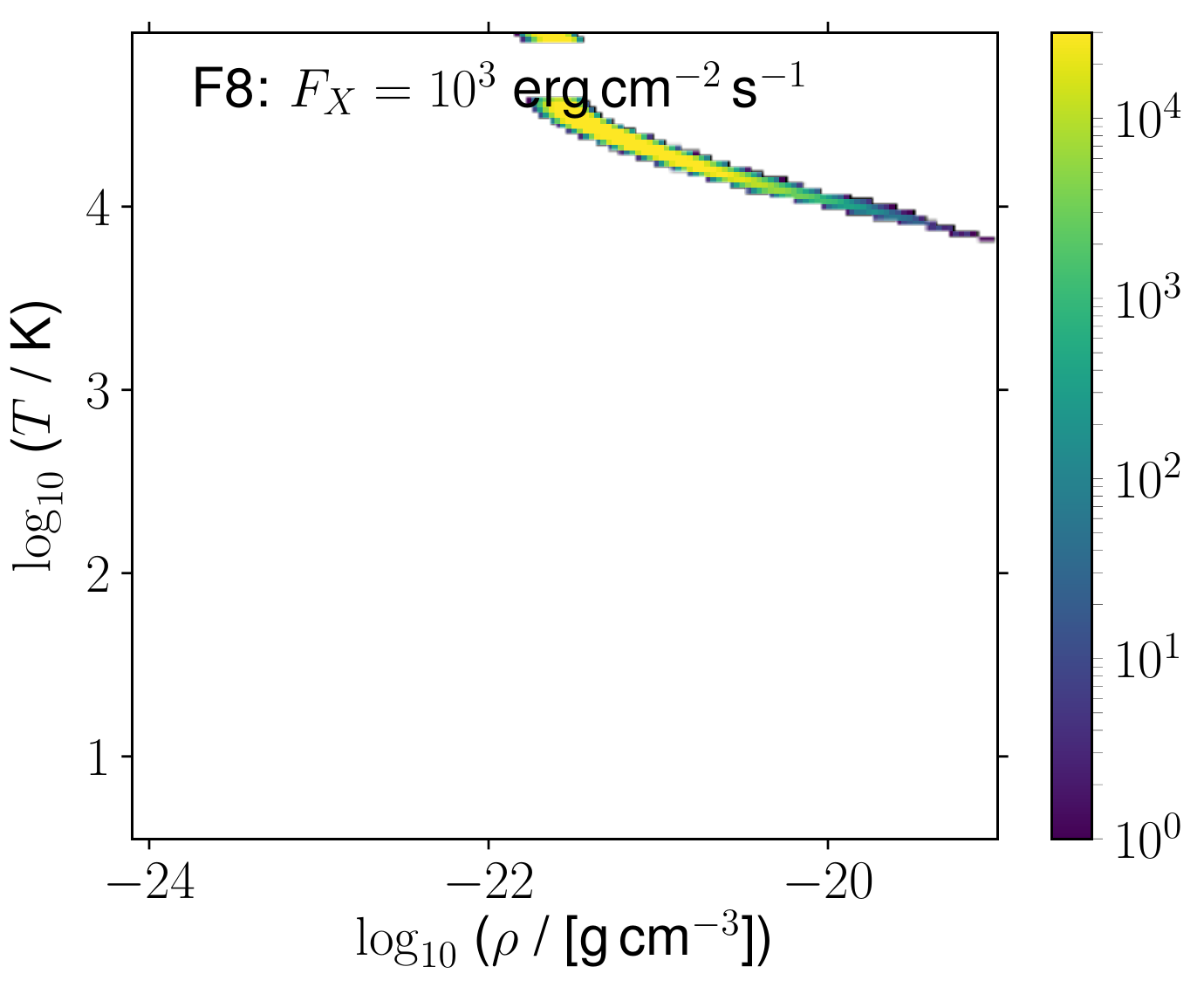}
\caption{
  Volume-weighted (unnormalised) temperature-density probability distribution function of the fractal cloud,
  where the logarithmic colour scale indicates the number of grid cells at a given point in the parameter space.
  Each panel has a different X-ray irradiating flux, with increasing X-ray flux from left to right and top to bottom (F0-F8, see Table~\ref{tab:sims}).
  There are $256^3$ cells in total, each with volume $10^{-3}$\,pc$^3$.
  For weak X-ray irradiation, cells with different extinction can have different equilibrium temperatures for a given density, whereas with strong X-ray irradiation the temperature is almost entirely determined by density alone.
}
\label{fig:rhotemp}
\end{figure*}

\begin{figure*}
\centering
\includegraphics[trim = 0mm 17mm 20mm 1mm, clip, height=4.7cm]{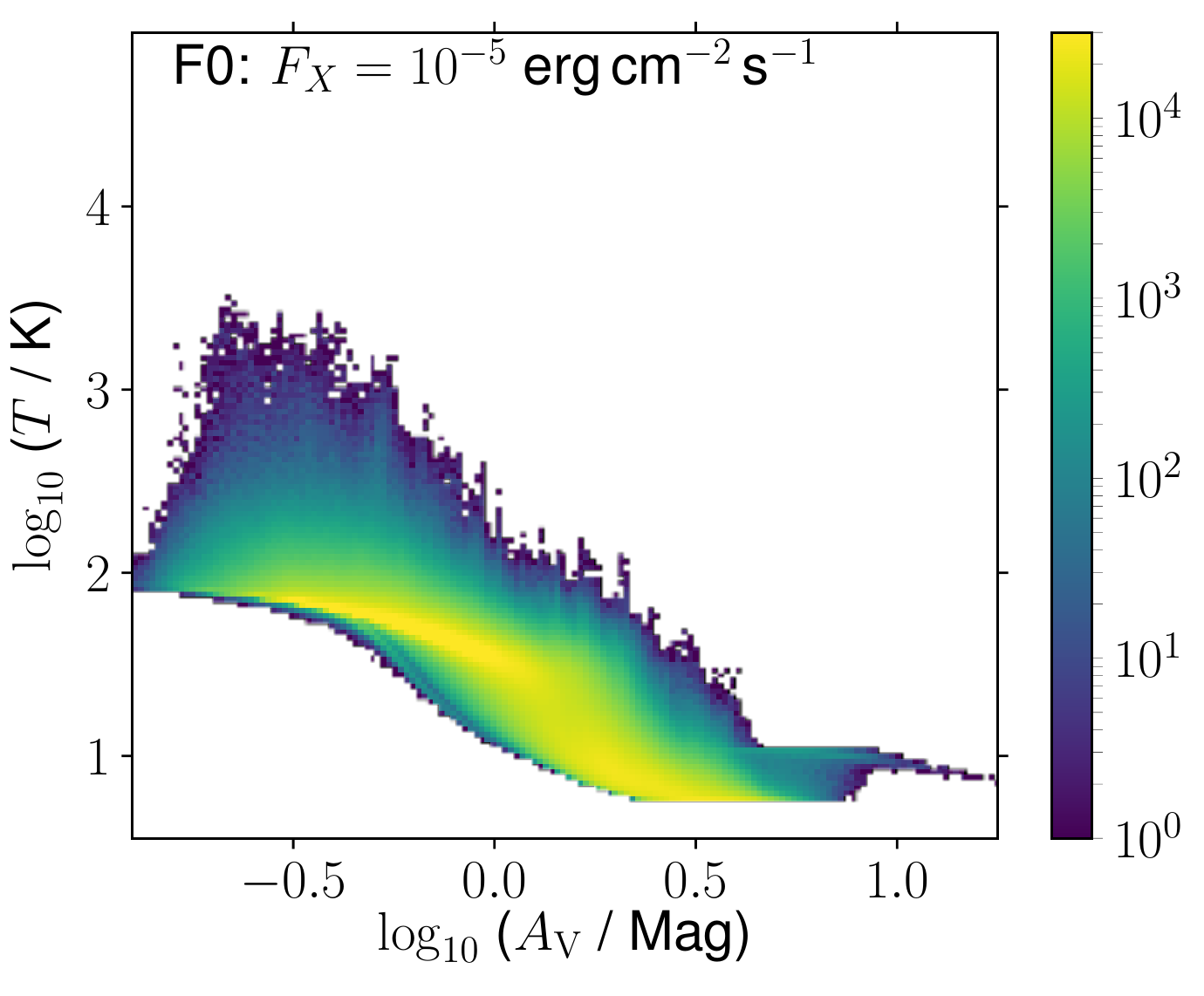}
\includegraphics[trim = 15mm 17mm 20mm 1mm, clip, height=4.7cm]{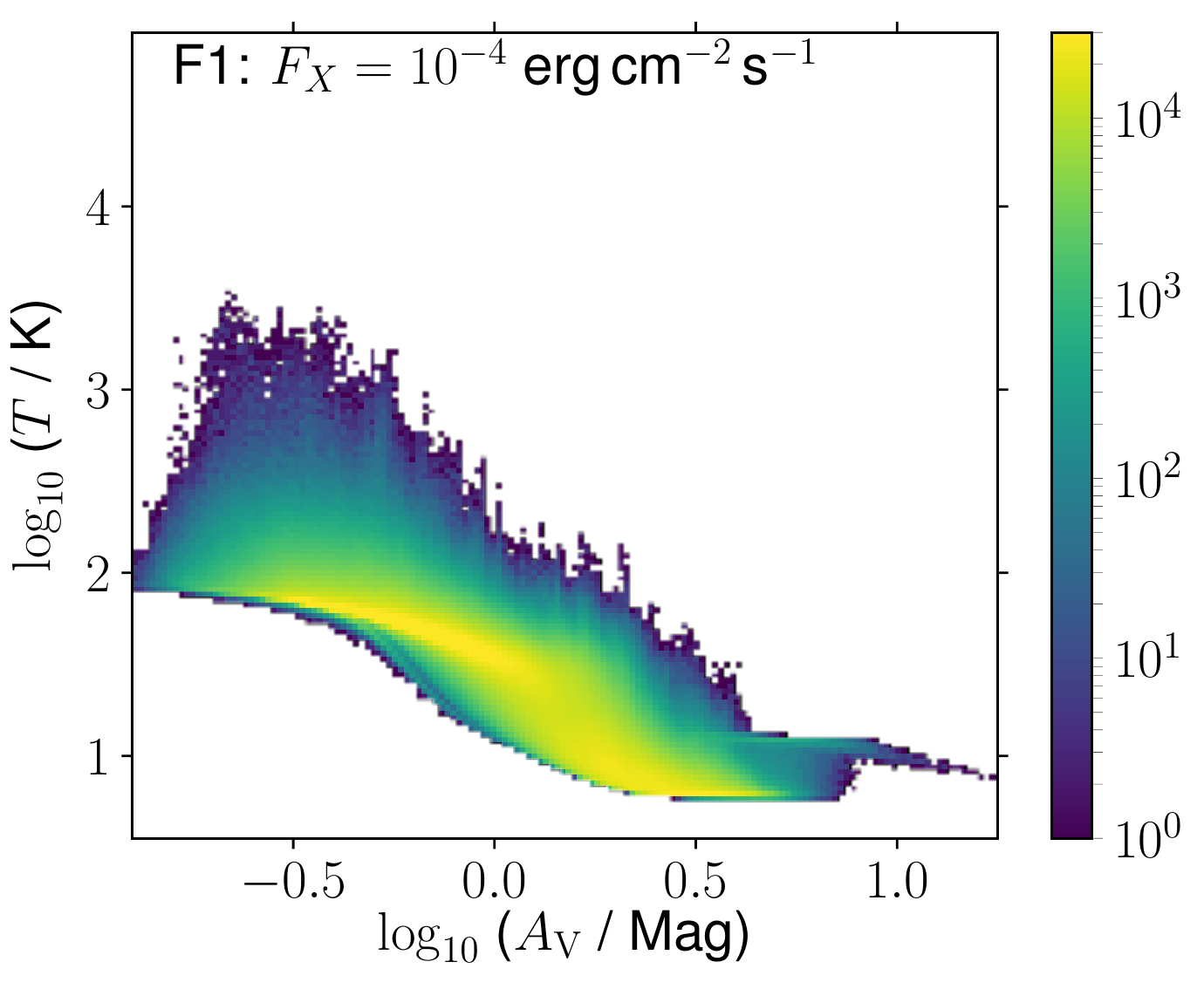}
\includegraphics[trim = 15mm 17mm 2mm 1mm, clip, height=4.7cm]{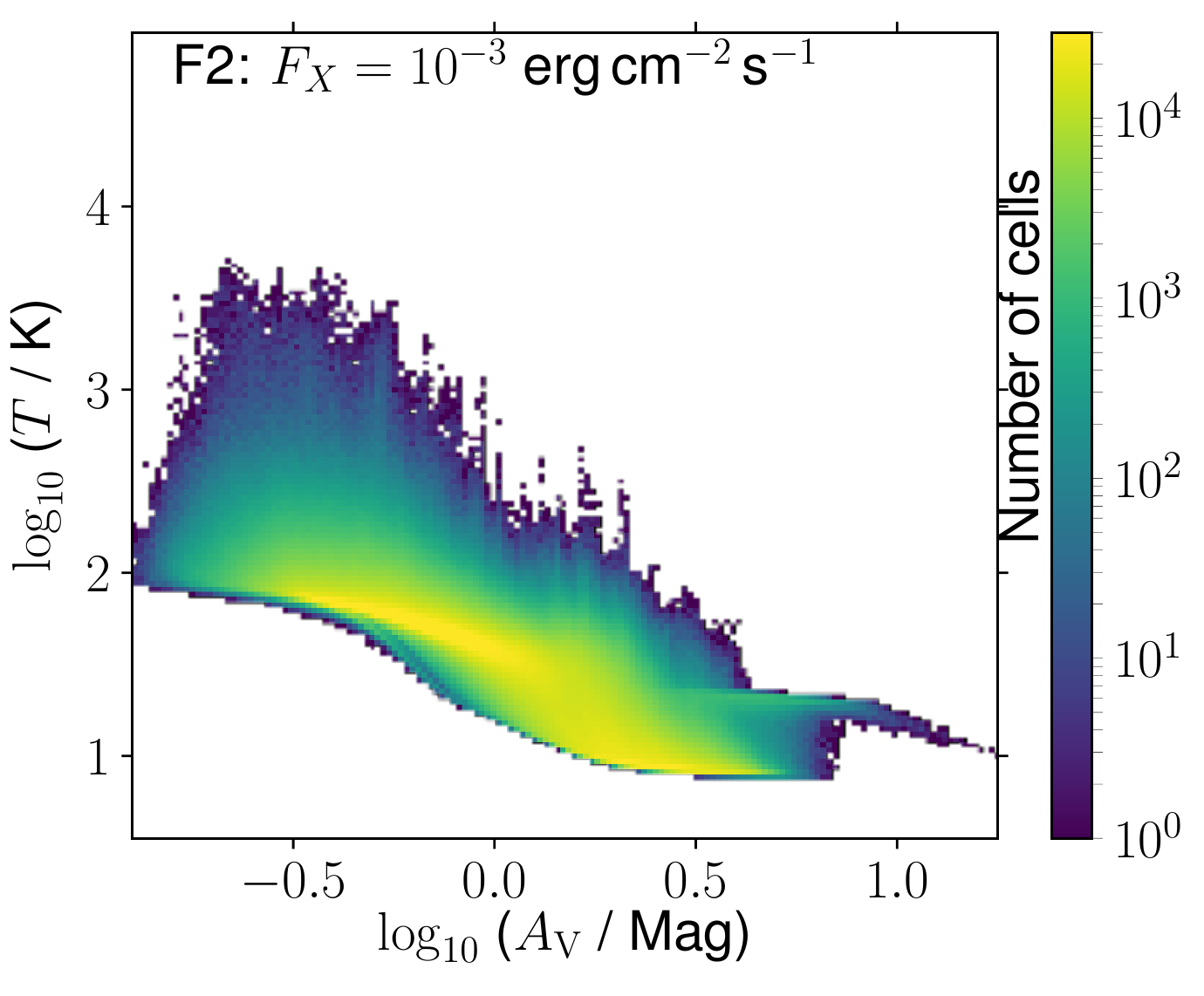}\\
\includegraphics[trim = 0mm 17mm 20mm 1mm, clip, height=4.7cm]{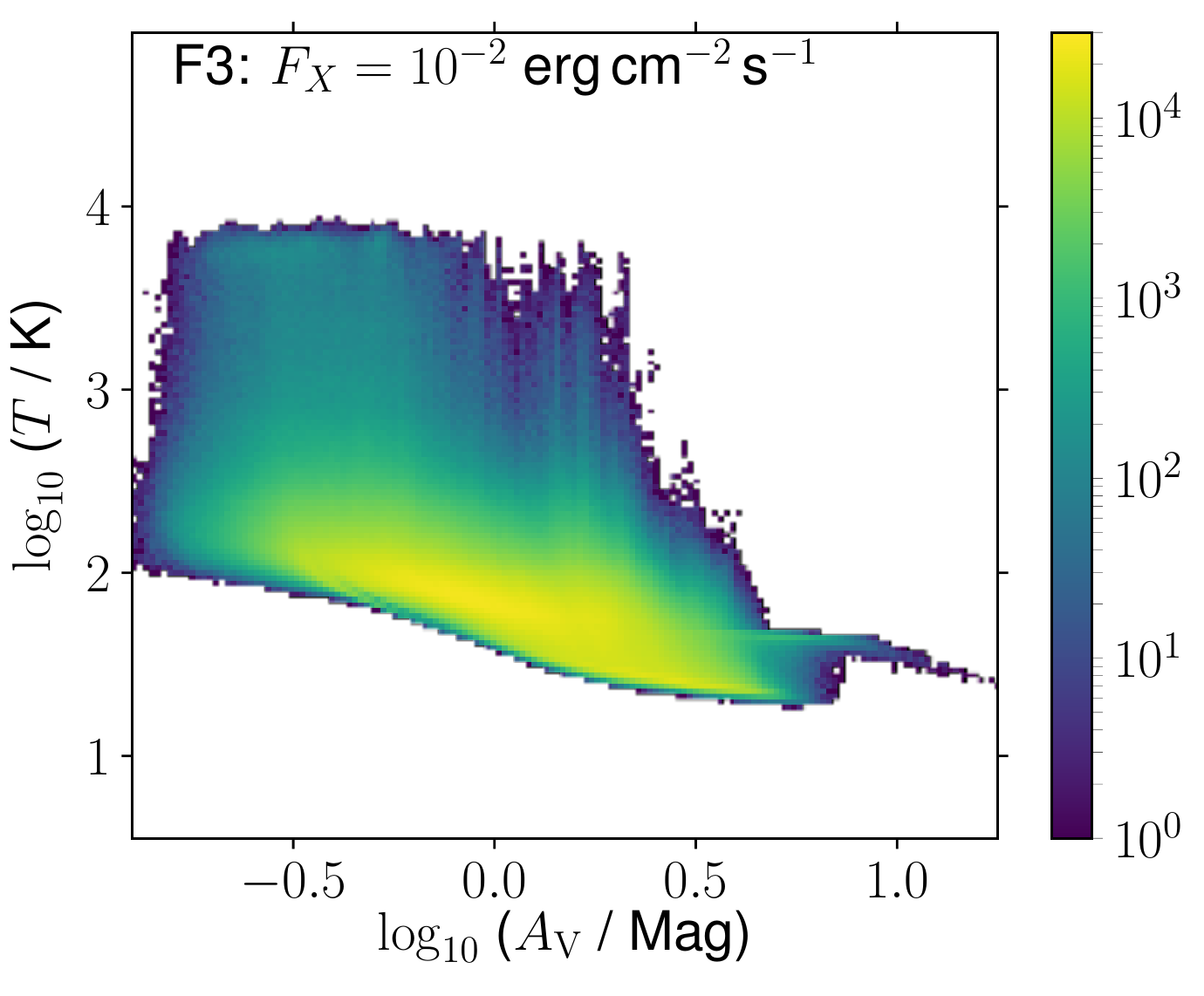}
\includegraphics[trim = 15mm 17mm 20mm 1mm, clip, height=4.7cm]{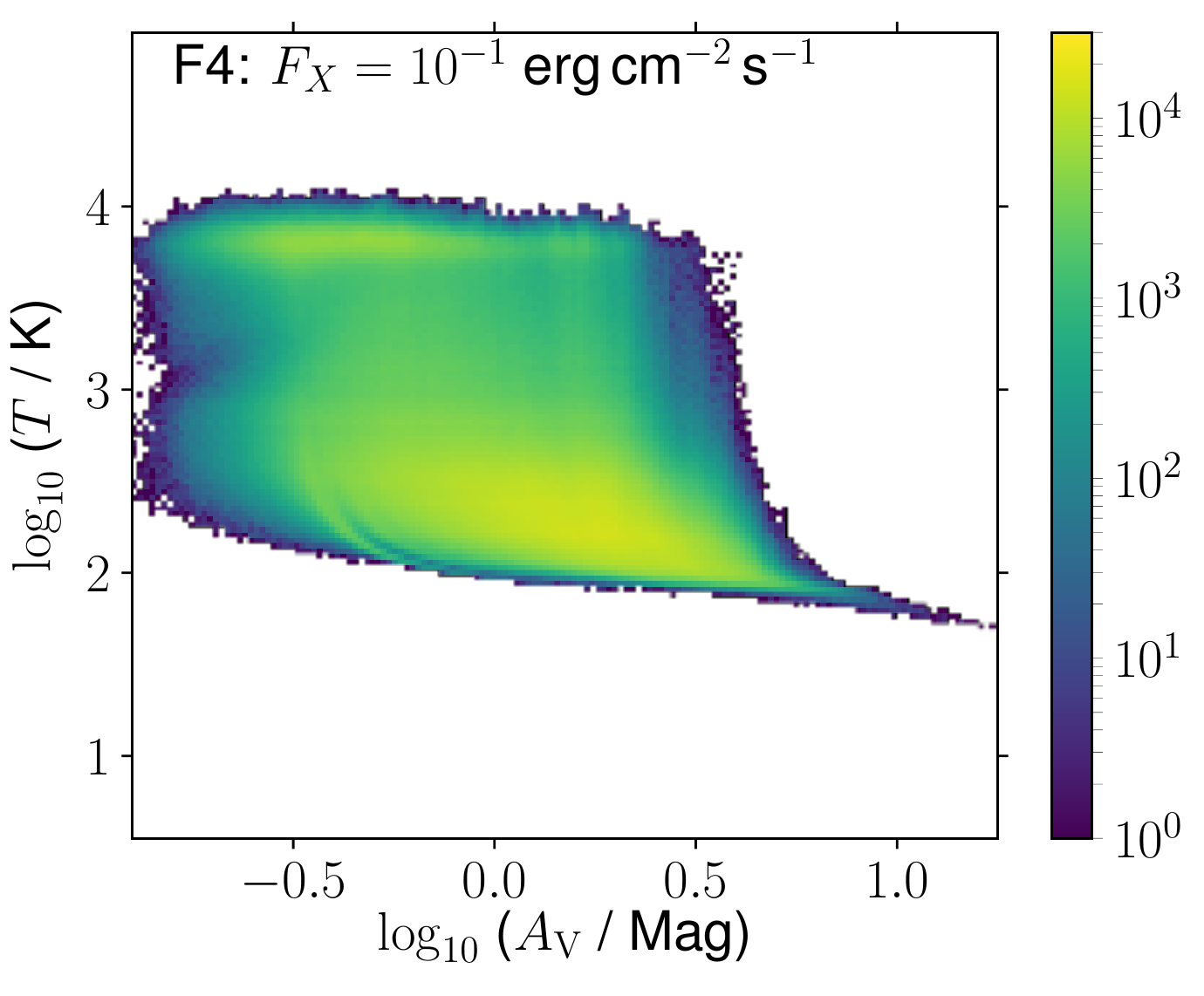}
\includegraphics[trim = 15mm 17mm 2mm 1mm, clip, height=4.7cm]{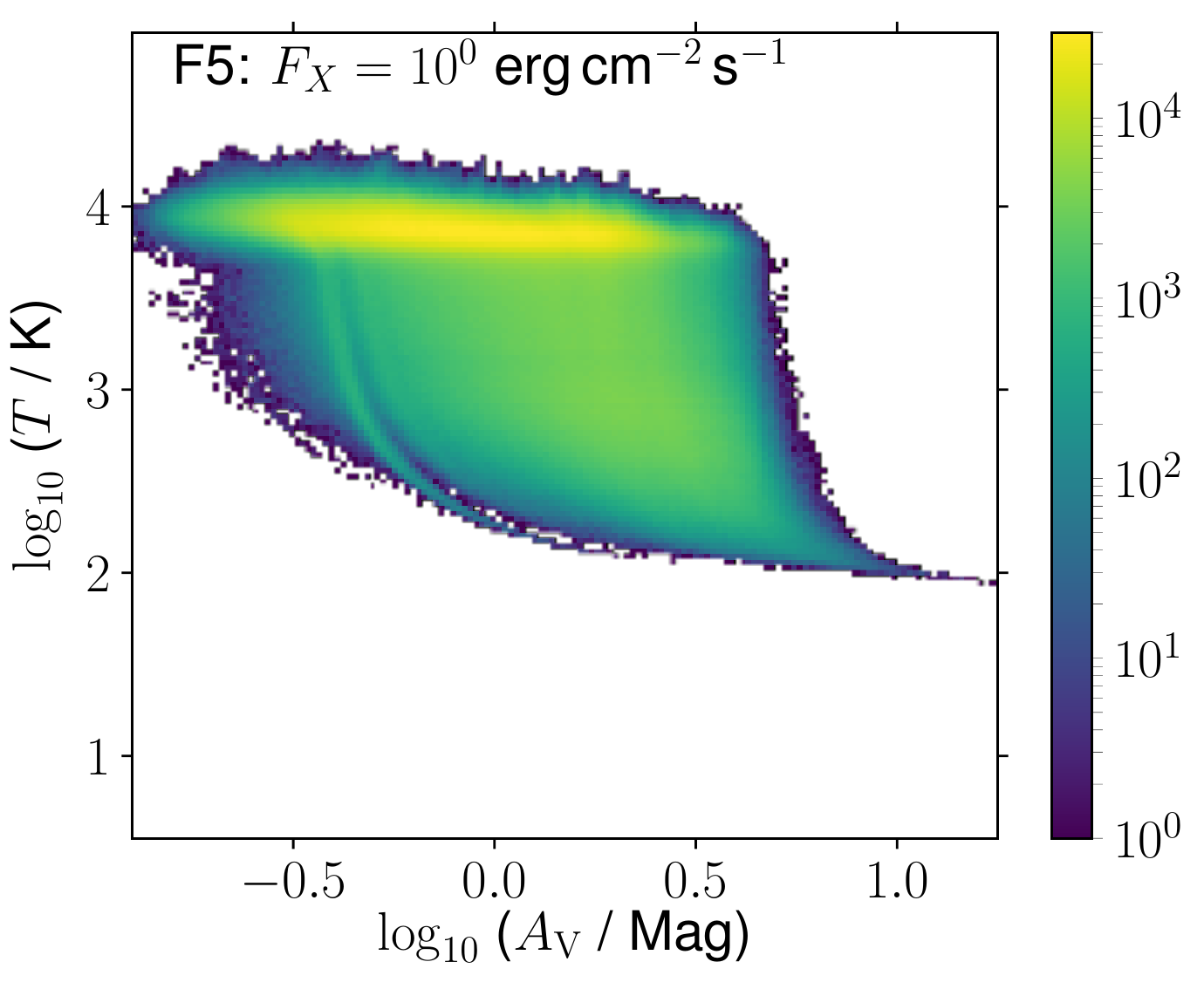}\\
\includegraphics[trim = 0mm 0mm 20mm 1mm, clip, height=5.5cm]{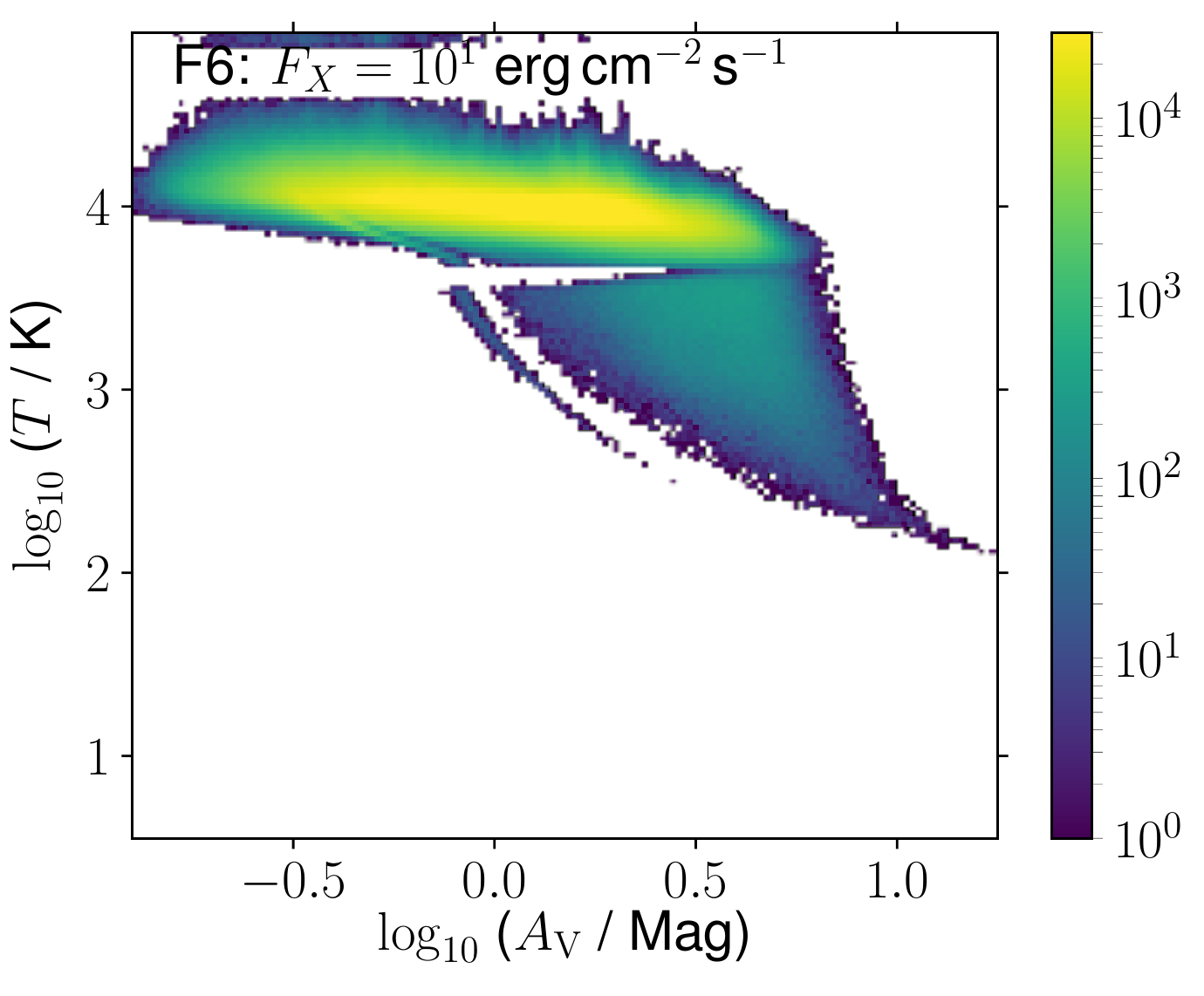}
\includegraphics[trim = 15mm 0mm 20mm 1mm, clip, height=5.5cm]{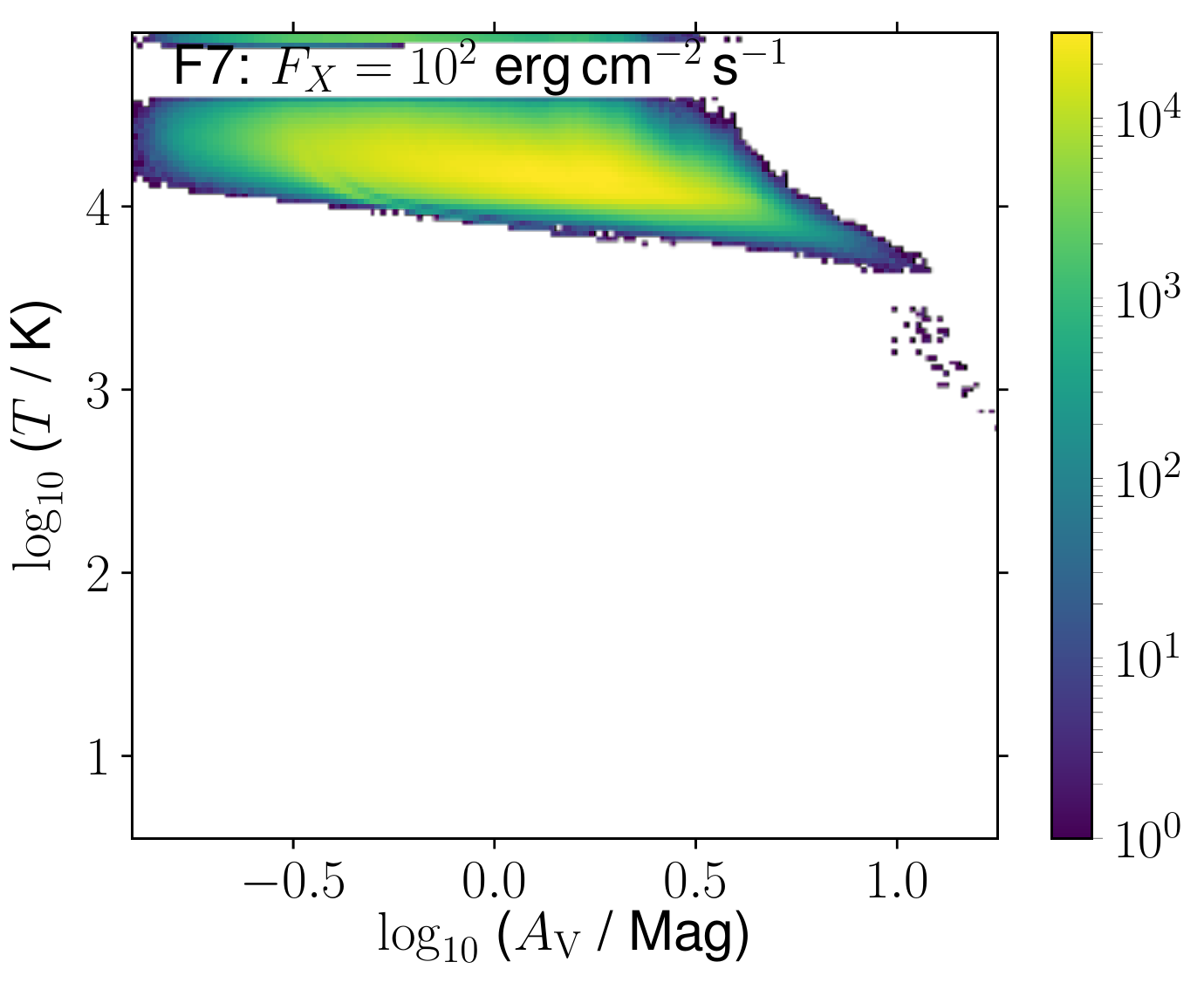}
\includegraphics[trim = 15mm 0mm 2mm 1mm, clip, height=5.5cm]{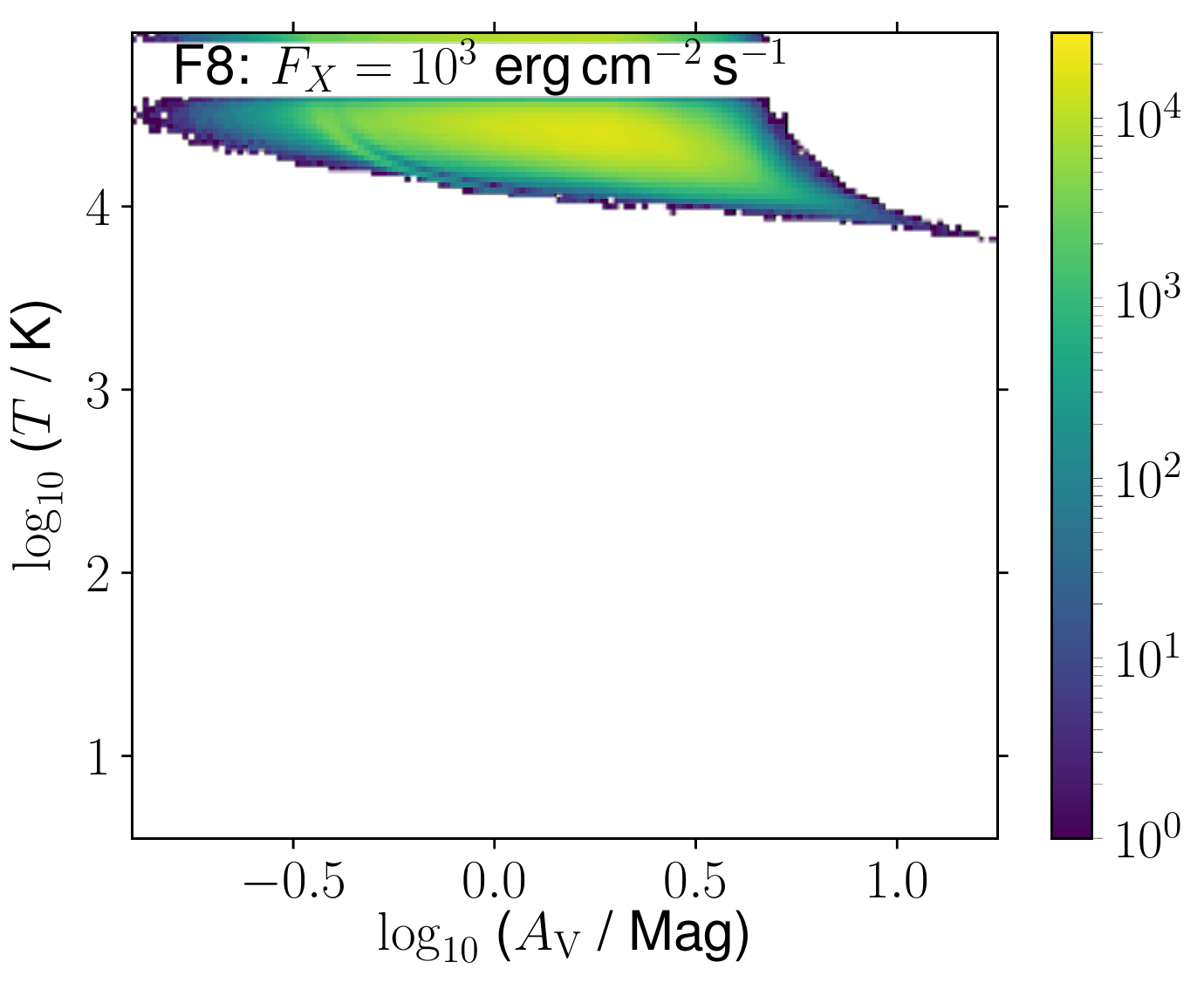}
\caption{
  Volume-weighted (unnormalised) probability distribution function in the plane of local extinction, $A_V$, and temperature, $T$, for the fractal cloud,
  where the logarithmic colour scale indicates the number of grid cells at a given point.
  Each panel has a different X-ray irradiating flux, with increasing X-ray flux from left to right and top to bottom (F0-F8).
  The $A_V$ value is calculated from the angle-averaged attenuation factor of the cell (equation~\ref{eqn:av}).
}
\label{fig:coltemp}
\end{figure*}

Fig.~\ref{fig:rhotemp} plots the location of the grid cells in the density--temperature plane for simulations F0-F8; 
effectively an unnormalised, volume-weighted, probability distribution function (PDF) in density and temperature.
Brighter colours indicate regions with more cells.
Similarly, Fig.~\ref{fig:coltemp} plots the same in the extinction--temperature plane.
It is important to note that different cells in our 3D simulations experience different UV extinction factors and so the equilibrium temperature depends on both density and location.
Once chemical and thermal equilibrium has been reached, the cells all sit on a surface in the space of density, temperature and UV extinction, and Figs.~\ref{fig:rhotemp} and \ref{fig:coltemp} are projections of this surface onto two different planes.
The scatter in these plots arises from this projection and not from the gas being out of equilibrium.
For larger X-ray flux the UV field has decreasing importance and so the effect of extinction on equilibrium temperature starts to drop out.

The extinction, $A_V$, is calculated using equation~\ref{eqn:attenuation}, but for the UV ISRF rather than X-ray radiation field.
This is
\begin{equation}
  \langle A_V \rangle = -\frac{1}{2.5}\log \frac{1}{N_\mathrm{pix}}\sum_{i=1}^{N_\mathrm{pix}} \exp \left(-2.5 A_V^i \right) \;,
  \label{eqn:av}
\end{equation}
where $A_V^i$ is the visual extinction along ray $i$, and $N_\mathrm{pix}$ is the number of rays used to sample all directions in 3D space (here $N_\mathrm{pix}=48$, see section~\ref{sec:xrays}).
Due to the non-linear nature of this equation, the resulting average $\langle A_V \rangle$ is dependent on the radiation energy at which the average is taken, i.e., dependent on the numerical multiplier that here is 2.5, appropriate for the UV ISRF.
Using the attenuation from one of the X-ray energy bins, or indeed the visual attenuation (a numerical multiplier of unity) gives a different mean value.
This shows the importance of 3D simulations: for a 1D calculation the extinction is a single number, but for 3D simulations the weighting of different rays is wavelength dependent, and so the mean UV or X-ray extinction is not necessarily consistent with what one expects given the mean optical extinction.

There is very little difference between F0 and F1 in Fig.~\ref{fig:rhotemp}, because the X-ray field is weak and cannot affect the chemistry or thermal state of the gas to any significant extent
(a run with zero X-ray flux is almost identical to F0 in these plots).
Almost all of the gas is in the temperature range 7-100\,K, and there is a relatively weak correlation between temperature and density (multiple temperatures are found for gas at a given density).
In contrast, there is a strong correlation between temperature and extinction, $A_V$, for these simulations (Fig.~\ref{fig:coltemp}), with temperature decreasing strongly with increasing extinction and most cells following a single curve in the plane.

Simulation F2 is a transitional case, where the X-ray field has a noticeable effect on the gas temperature but where the temperature is still strongly correlated with $A_V$.
The minimum temperature at large column density (where X-ray heating is effective) is increased to $>10$\,K with respect to F0 and F1, but the temperature at low column density (where UV heating is effective) is similar to F0 and F1.
There is similar energy in both the UV and X-ray fields ($F_X\approx10^{-3}$\,erg\,cm$^{-2}$\,s$^{-1}$) and so both have similar levels of influence.
The majority of the UV energy is deposited at $A_V<1$ near the cloud surface, whereas the X-ray energy penetrates beyond $A_V=10$ and so it acts on the whole cloud.

The thermodynamics of the remaining simulations are all dominated by the X-ray radiation field.
The mean temperatures of F3 and F5 are 100\,K and 8\,000\,K, respectively, with very little dependence on extinction (Fig.~\ref{fig:coltemp}).
Simulation F4 has significant quantities of gas at all temperatures from 100\,K to 8\,000\,K, regardless of $A_V$.
This is because the cloud is optically thin to X-rays in the higher energy bins ($>1$\,keV), and so the heating rate of a cell depends on the cell density to a much greater extent than the cell's $A_V$.
Fig.~\ref{fig:rhotemp} reflects this, showing very tight correlations between gas density and temperature for F4-F8.
For F5 ($4\pi J_{X}=1$\,erg\,cm$^{-2}$\,s$^{-1}$) there are two regimes, where gas with $\rho\lesssim10^{-21}$\,g\,cm$^{-3}$ is at $T\sim10^4$\,K, whereas higher density gas has progressively lower temperature.
For F4 the dividing line is $\rho\sim10^{-22}$\,g\,cm$^{-3}$, and for F3 it is about $\rho\sim10^{-23}$\,g\,cm$^{-3}$.
This reflects the fact that the cooling rate increases dramatically at $T\sim10^4$\,K, with Lyman-$\alpha$ and forbidden-line cooling becoming very strong.
The cooling rate scales with $n_\mathrm{H}^2$ whereas X-ray heating scales with $n_\mathrm{H}$, and so the density at which the Lyman-$\alpha$ and forbidden-line cooling equals the heating rate should scale with $4\pi J_{X}$.
At higher densities the temperature decreases with increasing density.

Simulations F6, F7 and F8 have sufficiently strong X-ray fields that the heating rate is stronger than the Lyman-$\alpha$ cooling rate, and so much of the gas becomes highly ionized with $T>10^4$\,K.
With such high temperatures the molecules in these simulations are destroyed, and the chemistry network that we use is no longer well-suited to the physical conditions because we do not include higher ionization stages of important coolants such as C, N, O, Fe, etc.
The empty region in the plots for F6-F8 at $4.6\lesssim \log T \lesssim4.8$ is an artifact of this limitation of the network.
For $T\gg10^4$\,K we assume cooling appropriate for collisional ionization equilibrium \citep[interpolated from a table; see][]{WalGirNaa15}, which is not satisfied for X-ray irradiated gas, and so the cooling rate has an incorrect temperature dependence.
For sufficiently large X-ray heating rates this leads to runaway heating, and so  we set the net heating rate to zero for $T>10^5$\,K in these simulations, because we are not interested in the coronal gas that X-ray heating can produce.
Simulation F6 also has a gap around $T\sim10^{3.5}$\,K, which is a manifestation of the chemo-thermal instability seen in MS05 models 2 and 4.
The gap is also seen in the $T$-$y(\mathrm{H}_2)$ plane.

\subsection{Chemical state of the gas}
\label{sec:fractal:chem}

\begin{figure*}
\centering
\includegraphics[trim = 0mm 17mm 20mm 1mm, clip, height=4.7cm]{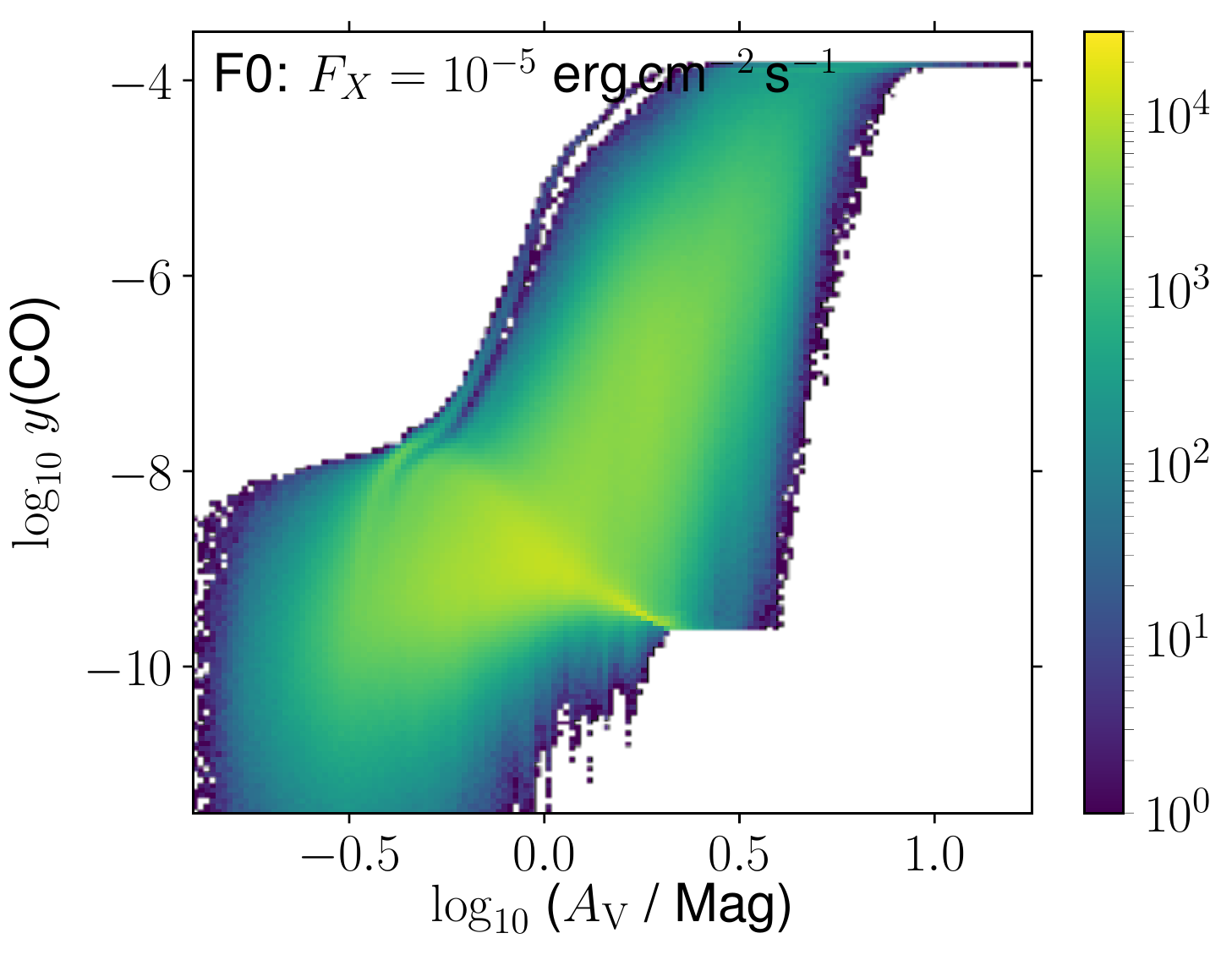}
\includegraphics[trim = 22mm 17mm 20mm 1mm, clip, height=4.7cm]{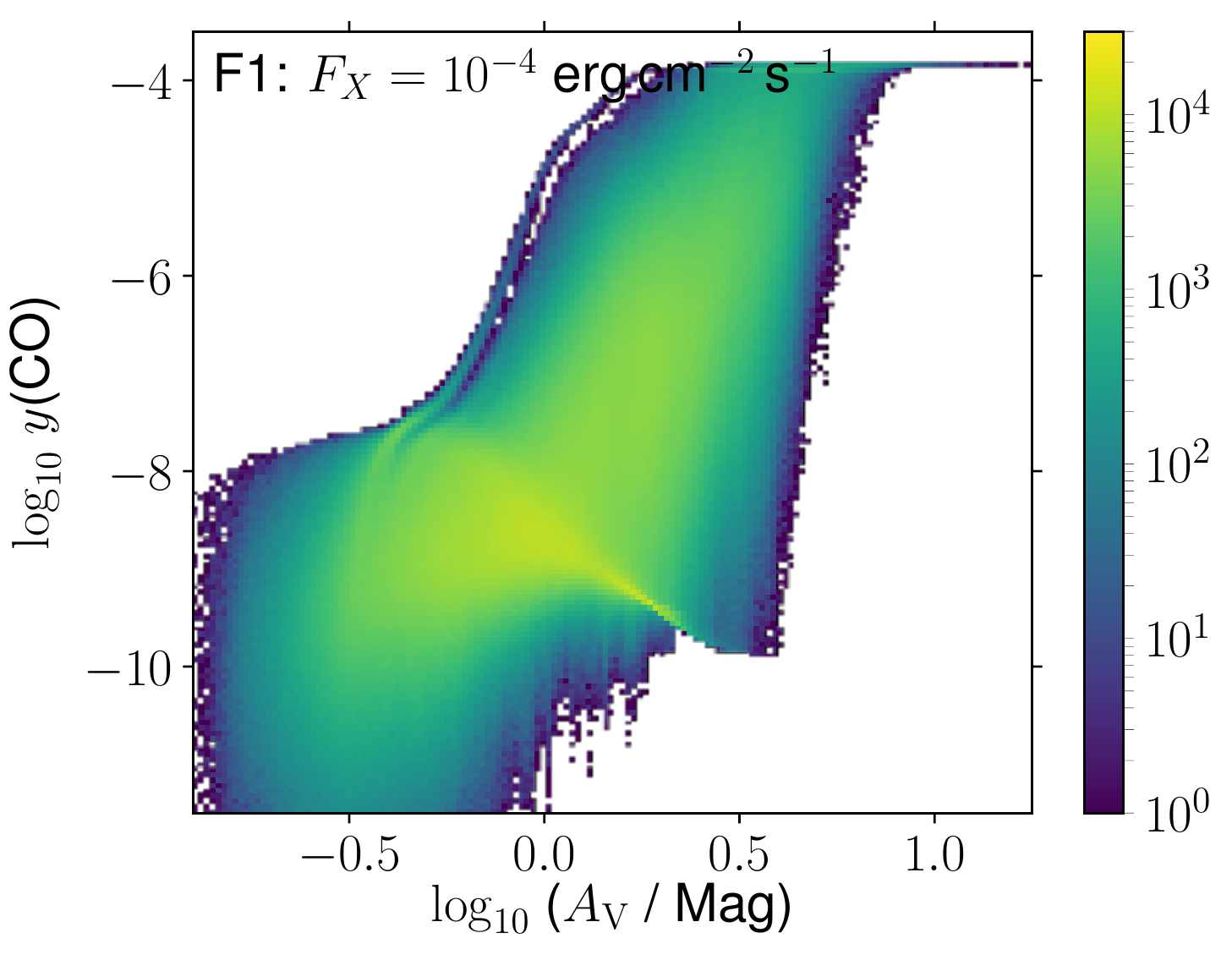}
\includegraphics[trim = 22mm 17mm 2mm 1mm, clip, height=4.7cm]{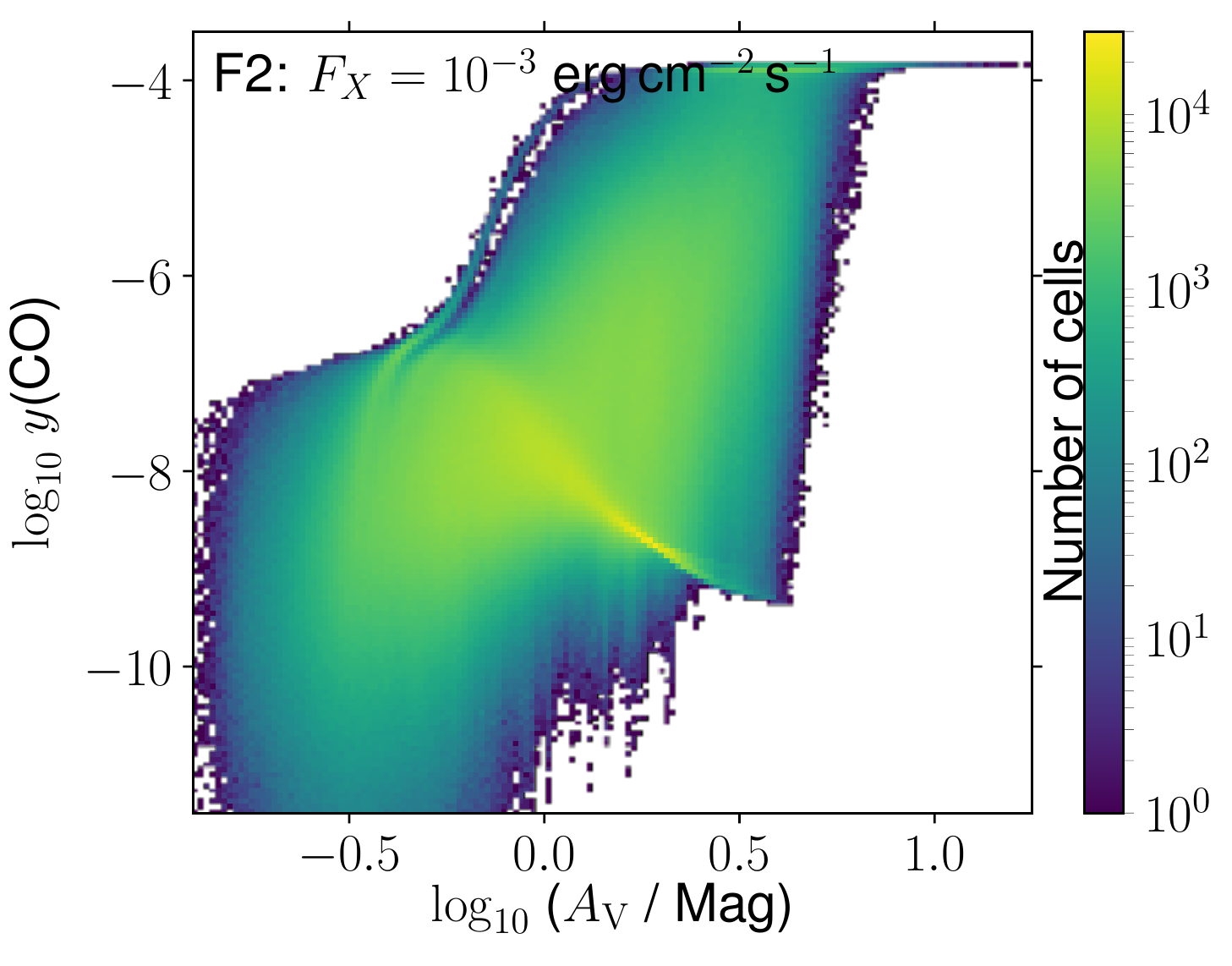}\\
\includegraphics[trim = 0mm 0mm 20mm 1mm, clip, height=5.5cm]{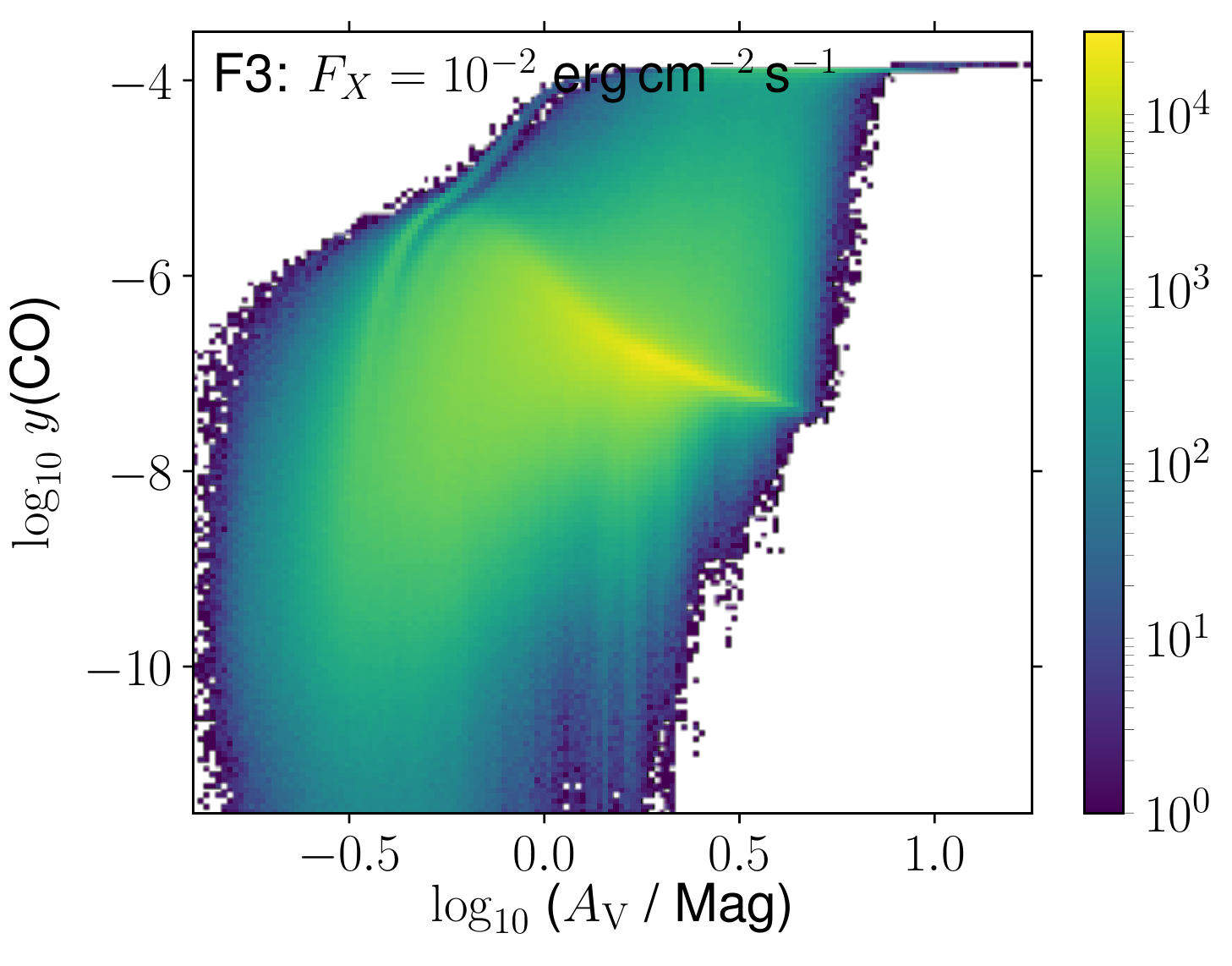}
\includegraphics[trim = 22mm 0mm 20mm 1mm, clip, height=5.5cm]{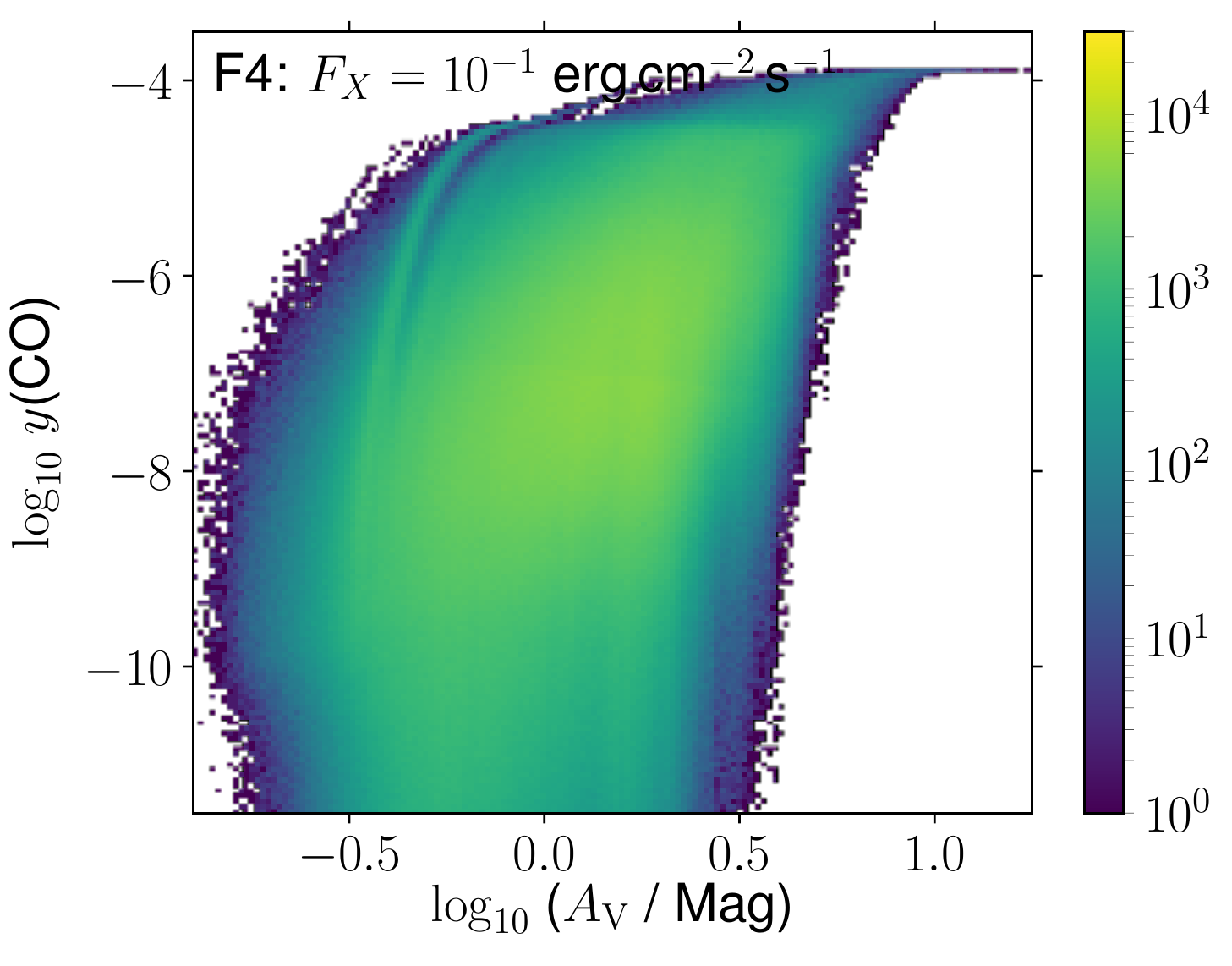}
\includegraphics[trim = 22mm 0mm 2mm 1mm, clip, height=5.5cm]{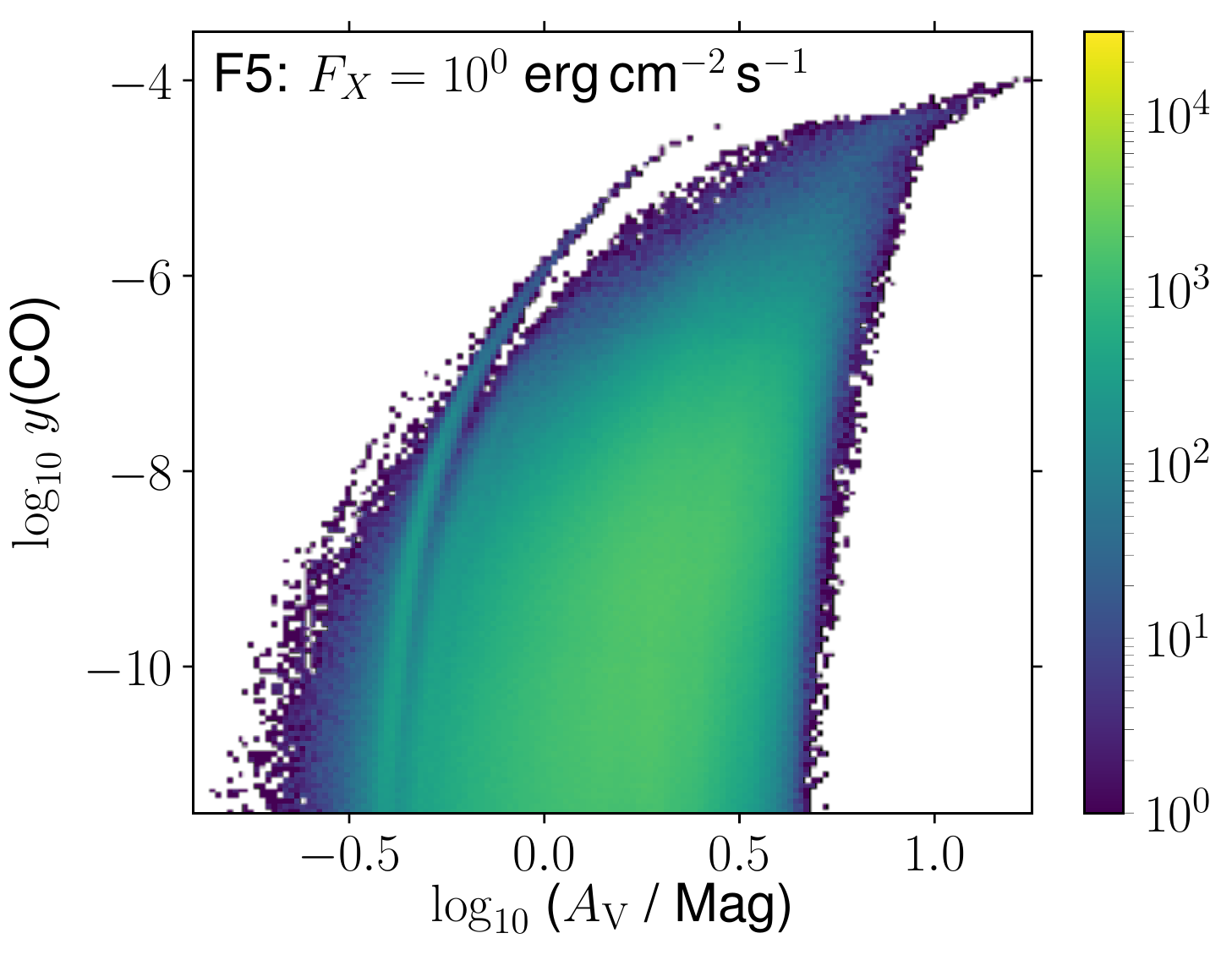}\\
\caption{
  Volume-weighted (unnormalised) probability distribution function in the plane of CO number fraction relative to H, $y(\mathrm{CO})$, and local extinction $A_V$ for the fractal cloud,
  where the logarithmic colour scale indicates the number of grid cells at a given point.
  Each panel has a different X-ray irradiating flux, with increasing X-ray flux from left to right and top to bottom (F0-F5).
  Simulations F6, F7 and F8 have so little CO that they are not shown.
  The $A_V$ value of each cell is calculated from the angle-averaged attenuation factor of the cell (equation~\ref{eqn:av}).
}
\label{fig:co_col}
\end{figure*}

\begin{figure*}
\centering
\includegraphics[trim = 0mm 16mm 20mm 1mm, clip, height=4.7cm]{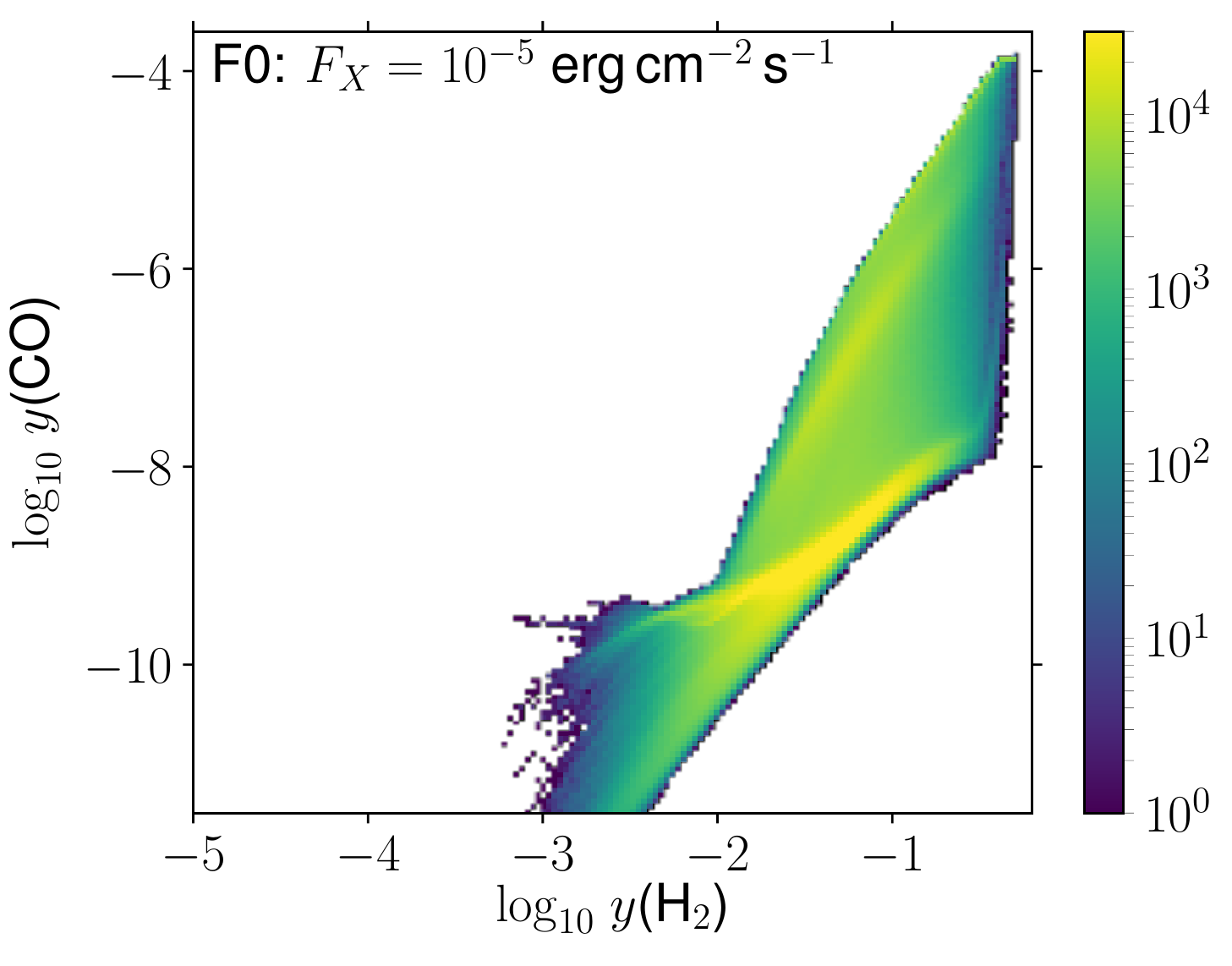}
\includegraphics[trim = 22mm 16mm 20mm 1mm, clip, height=4.7cm]{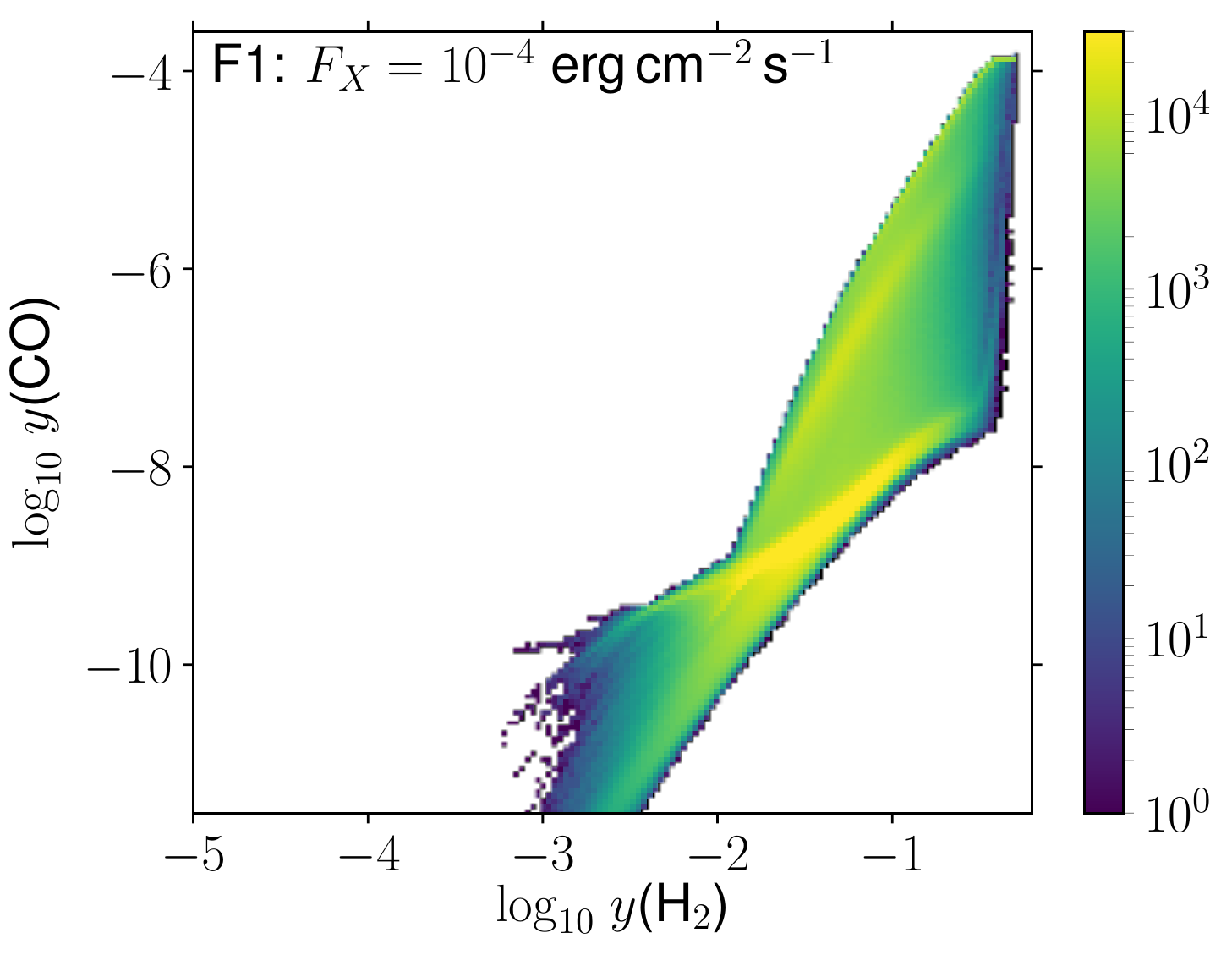}
\includegraphics[trim = 22mm 16mm 2mm 1mm, clip, height=4.7cm]{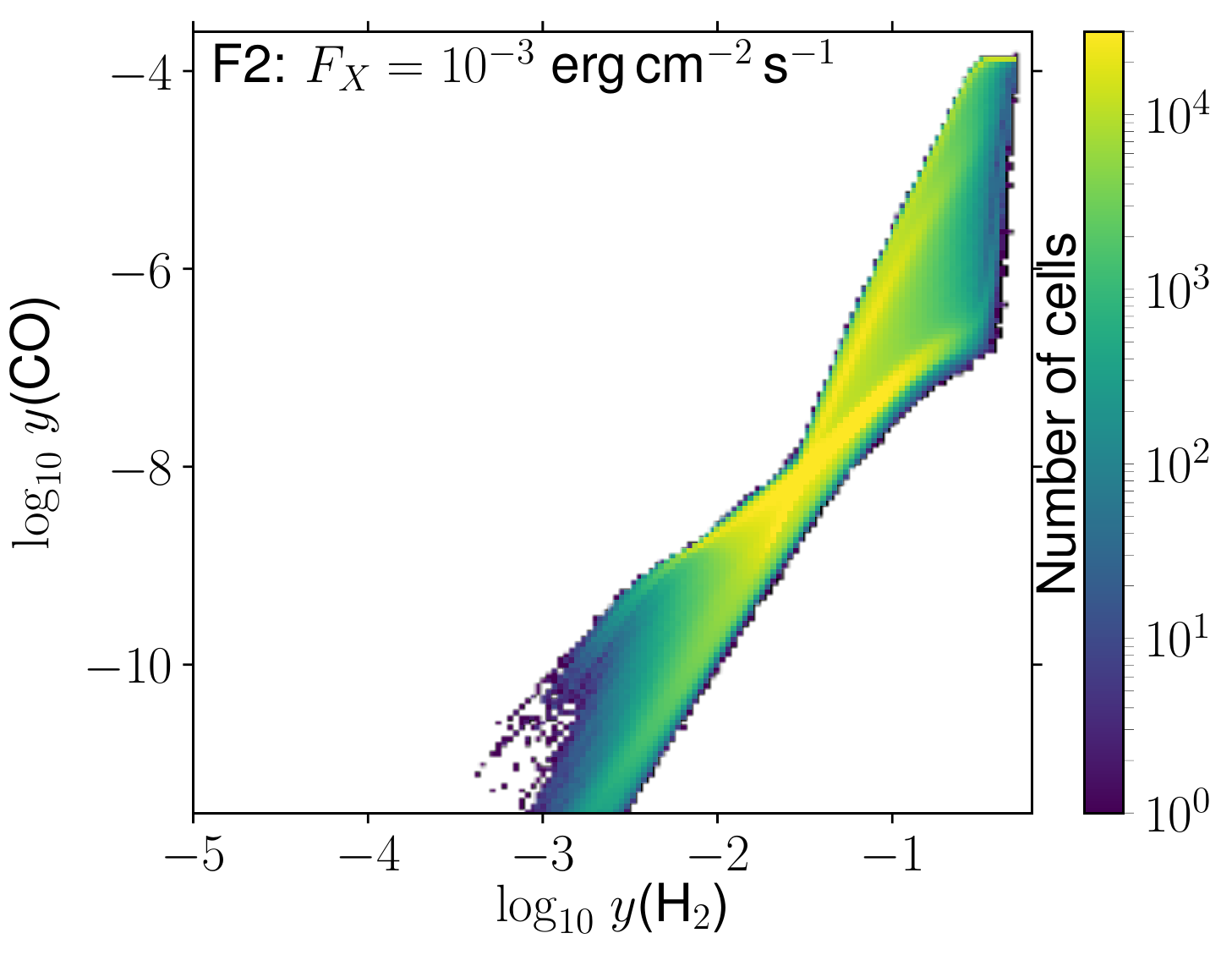}\\
\includegraphics[trim = 0mm 0mm 20mm 1mm, clip, height=5.5cm]{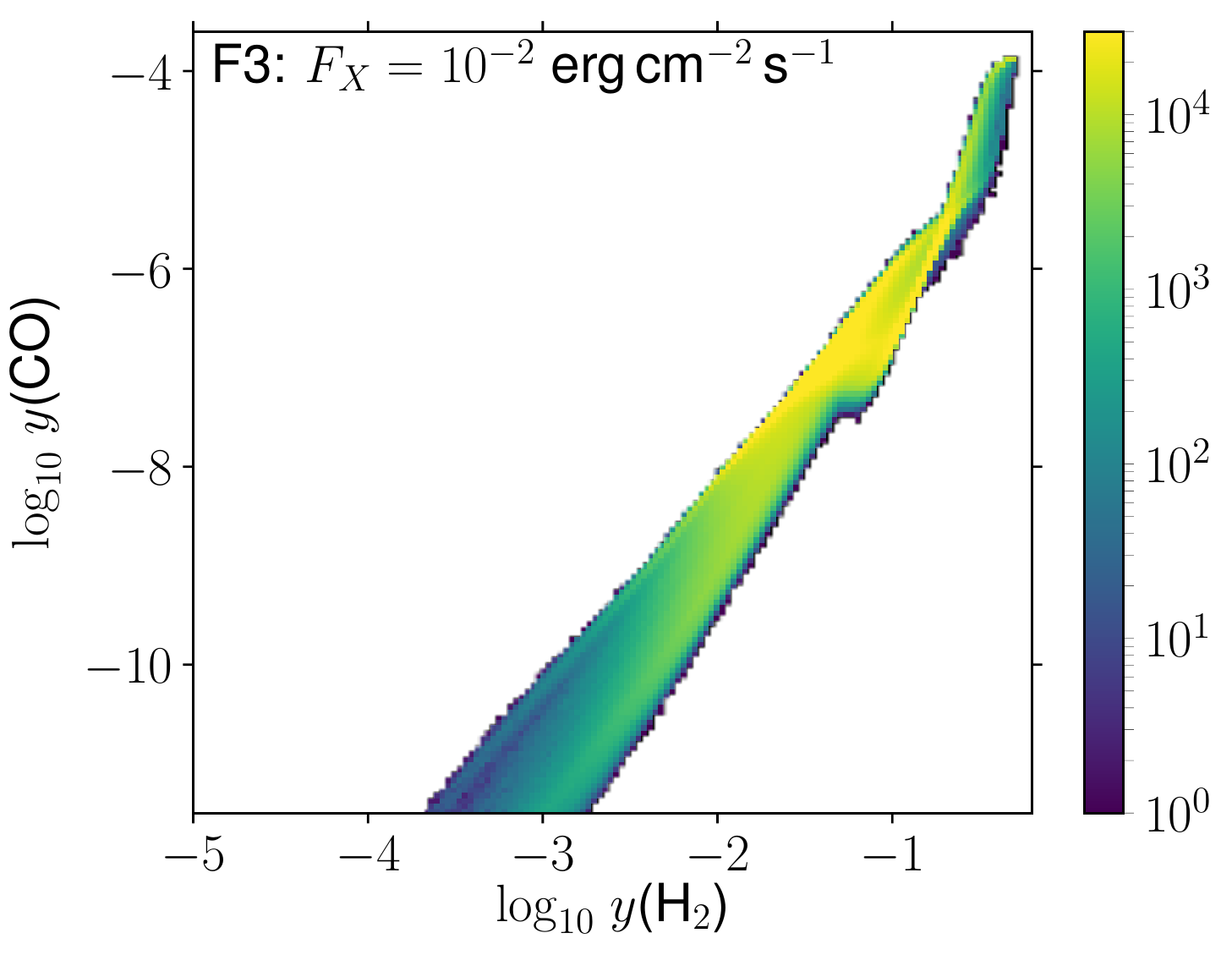}
\includegraphics[trim = 22mm 0mm 20mm 1mm, clip, height=5.5cm]{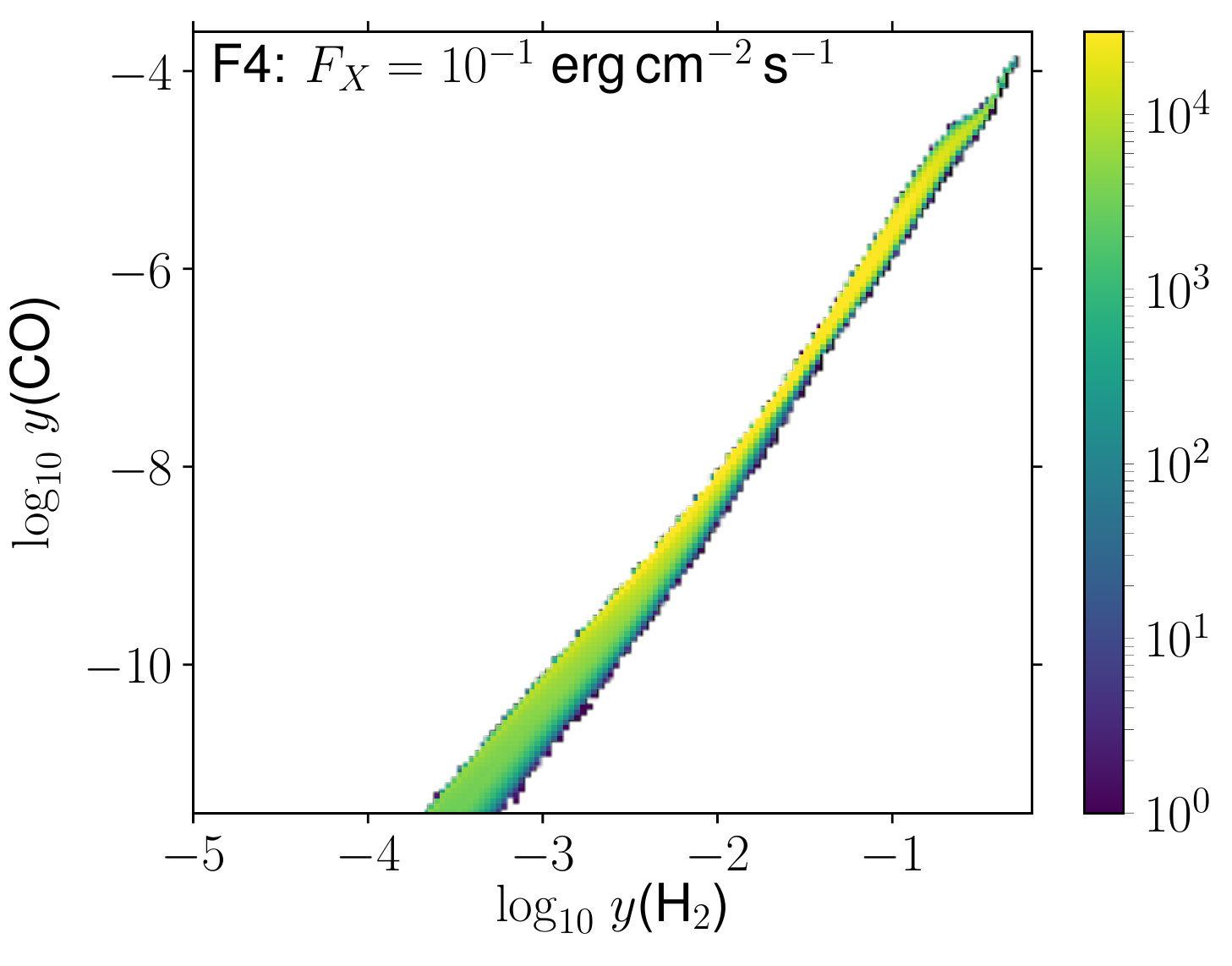}
\includegraphics[trim = 22mm 0mm 2mm 1mm, clip, height=5.5cm]{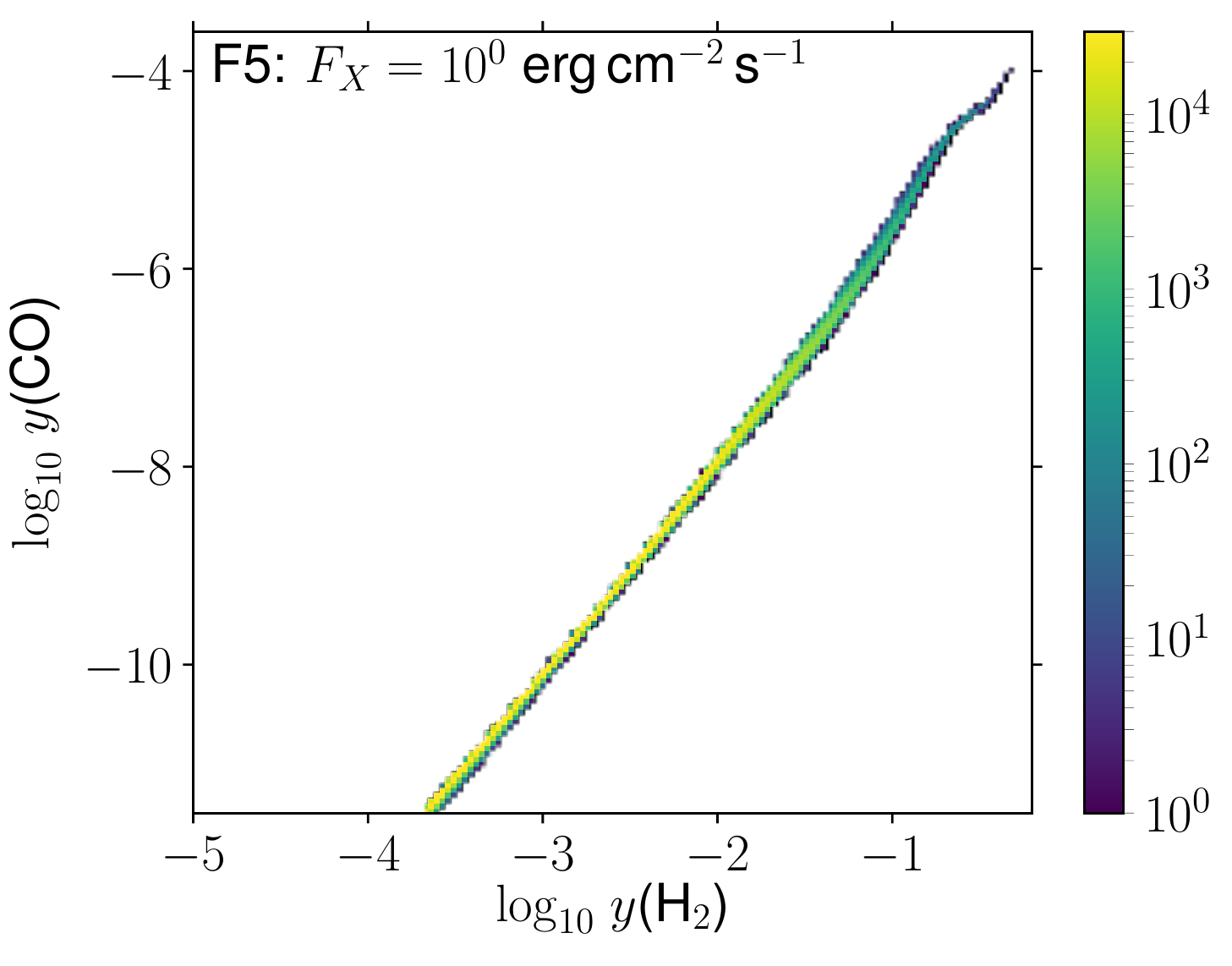}\\
\caption{
  Volume-weighted (unnormalised) probability distribution function in the plane of CO number fraction, $y(\mathrm{CO})$, and H$_2$ number fraction, $y(\mathrm{H}_2)$, for the fractal cloud, where the logarithmic colour scale indicates the number of grid cells at a given point.
  Each panel has a different X-ray irradiating flux, with increasing X-ray flux from left to right and top to bottom (F0-F5).
  Simulations F6-F8 had very little CO and so are not plotted here.
}
\label{fig:co_h2}
\end{figure*}

Fig.~\ref{fig:co_col} plots the CO abundance, $y(\mathrm{CO})$, in each cell as a function of $A_V$ for simulations F0-F5 (F6-F8 have very little CO).
Simulations F0 and F1 are showing what is typically found in PDR simulations, where the molecular fraction increases with column density, and increases dramatically once the column density is sufficient for self-shielding \citep[see e.g.][]{TieHol85, RoeAbeBel07}.

In the remaining simulations (F2-F5) we see the increasingly strong effect of X-ray ionization and heating.
As well as a general decrease in CO abundance at all column densities, the highly molecular gas at large column density progressively decreases with increasing flux, and disappears almost completely for  F5.
The CO abundance as a function of the H$_2$ abundance is plotted in Fig.~\ref{fig:co_h2}, again only for simulations F0-F5.
As the X-ray flux increases, the correlation between $y(\mathrm{CO})$ and $y(\mathrm{H}_2)$ gets stronger, and the overall CO abundance decreases.
The correlation of CO abundance with H$_2$ abundance is stronger than that with $A_V$, and this is again because the hard X-rays can penetrate to large $A_V$.
They are not strongly attenuated by the cloud that we simulate here, and so the thermal and chemical properties of a cell are set much more by the gas density than by the extinction.
The CO abundance increases with the square of the H$_2$ abundance.

\subsection{Column density maps of CO and H$_2$}
\label{sec:fractal:colmaps}

\begin{figure*}
\begin{tabular}{ccc}
\includegraphics[trim = 0mm 16mm 0mm 0mm, clip, width=6.0cm]{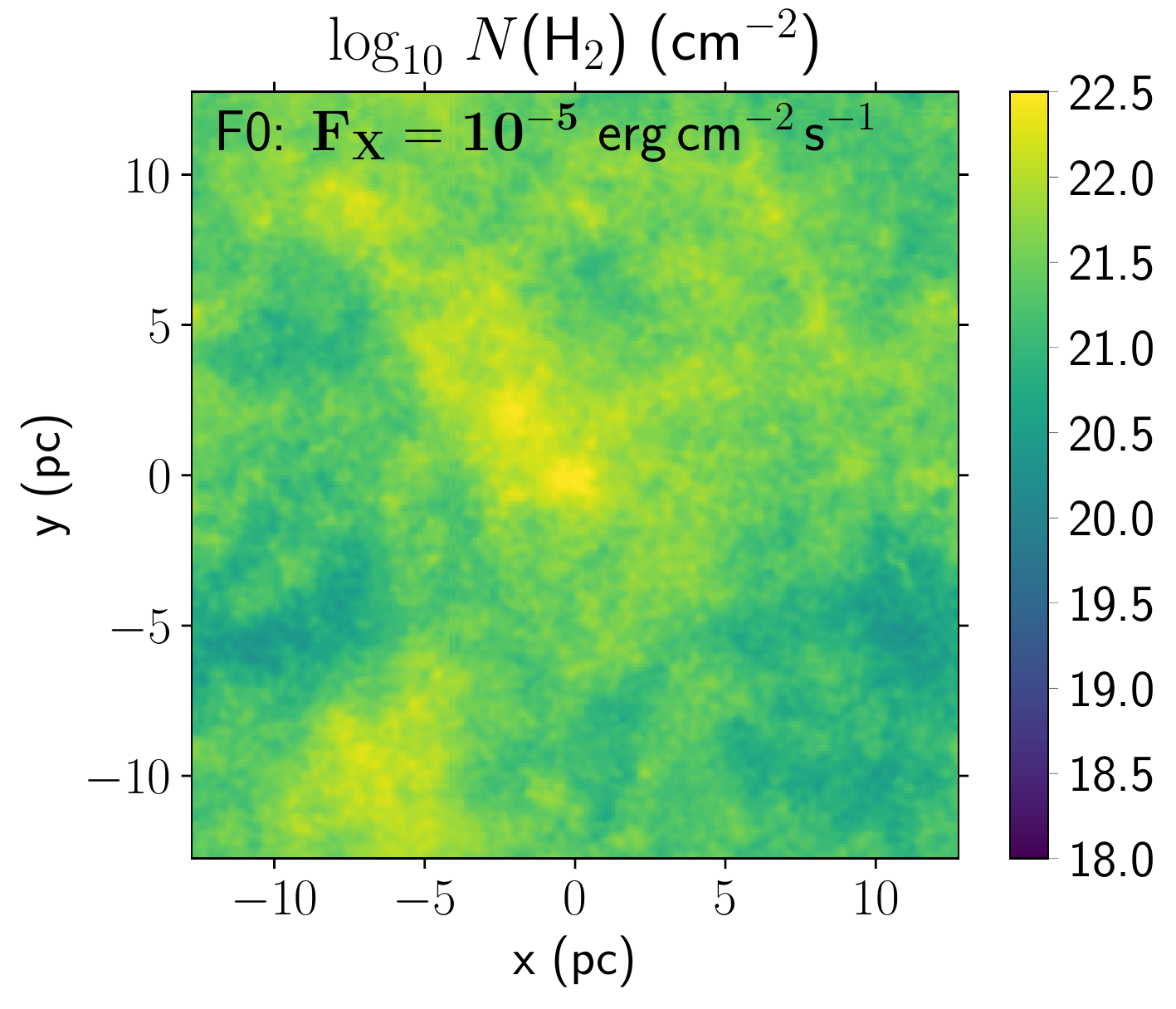} &
\includegraphics[trim = 12mm 16mm 0mm 0mm, clip, width=5.3cm]{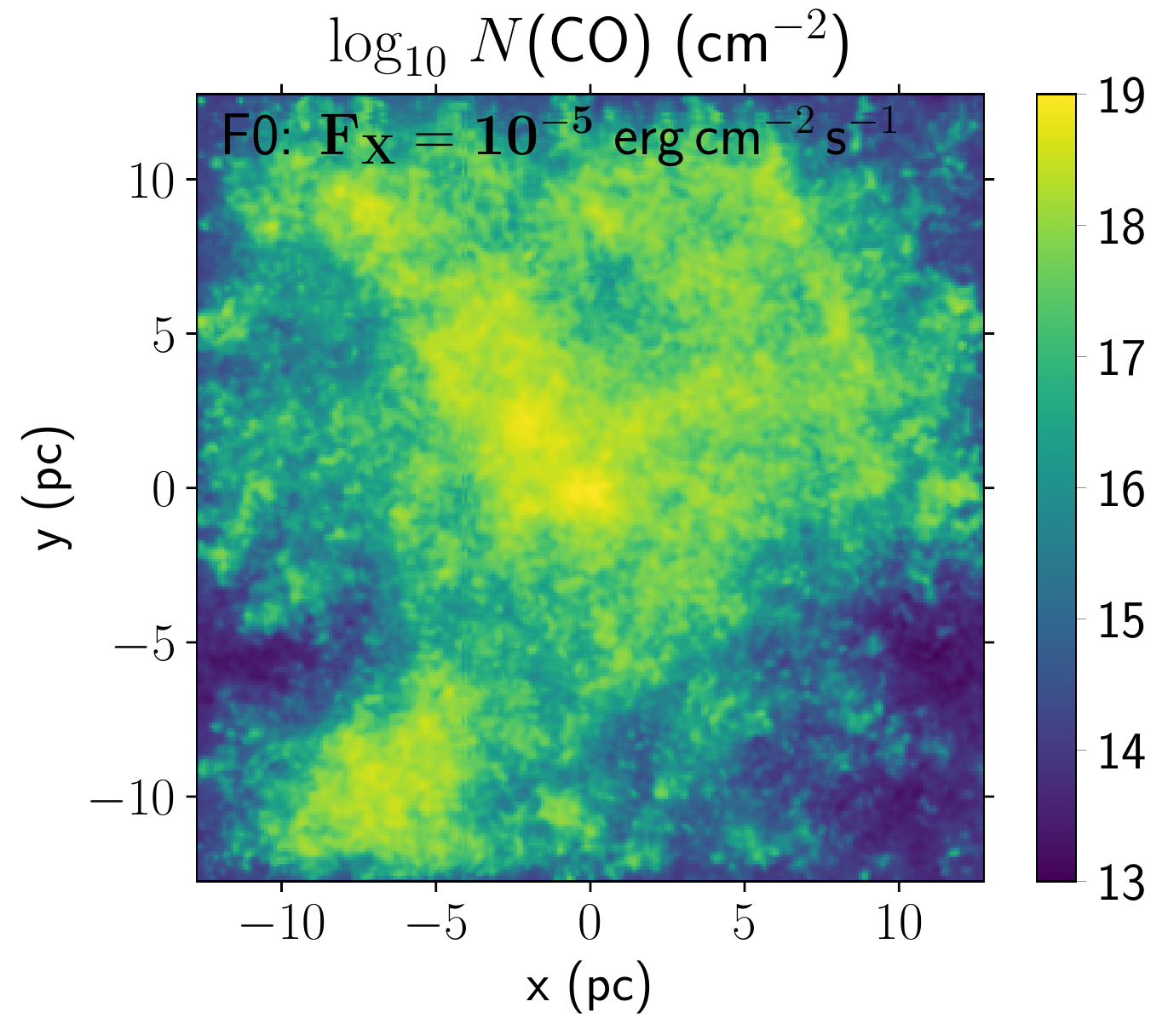} &
\includegraphics[trim = 12mm 16mm 0mm 0mm, clip, width=5.3cm]{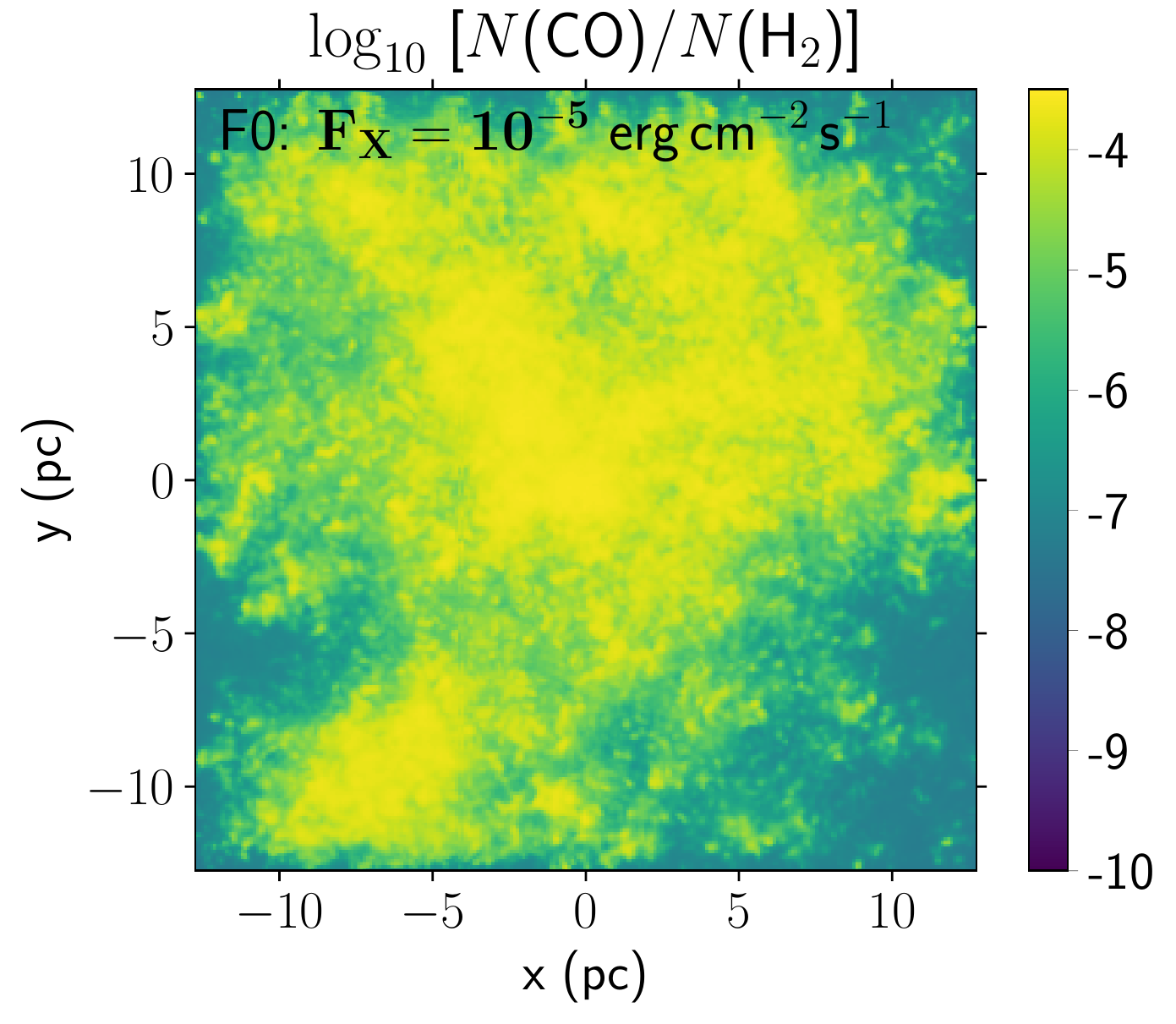} \\

\includegraphics[trim = 0mm 16mm 0mm 9.25mm, clip, width=6.0cm]{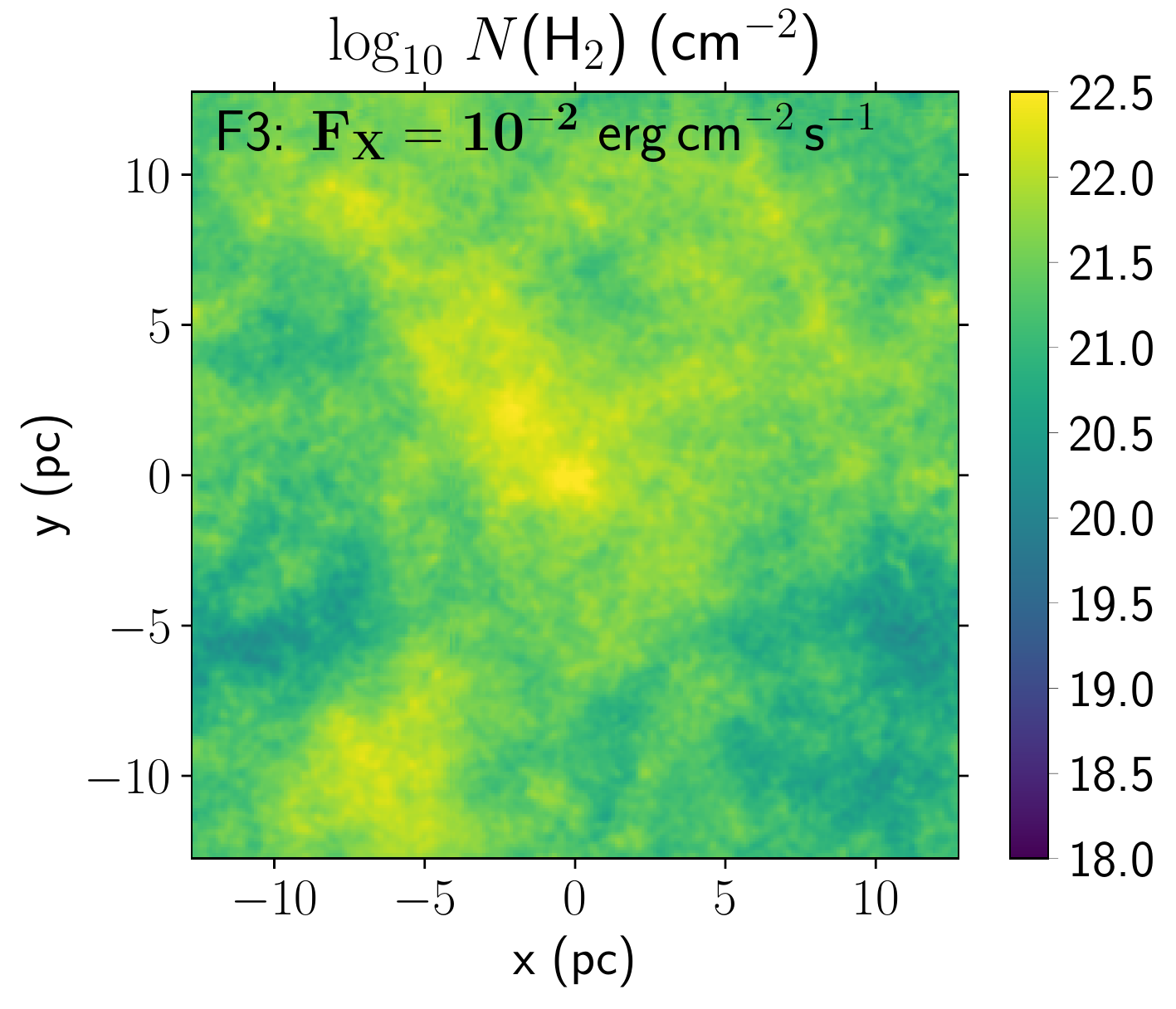} &
\includegraphics[trim = 12mm 16mm 0mm 9.25mm, clip, width=5.3cm]{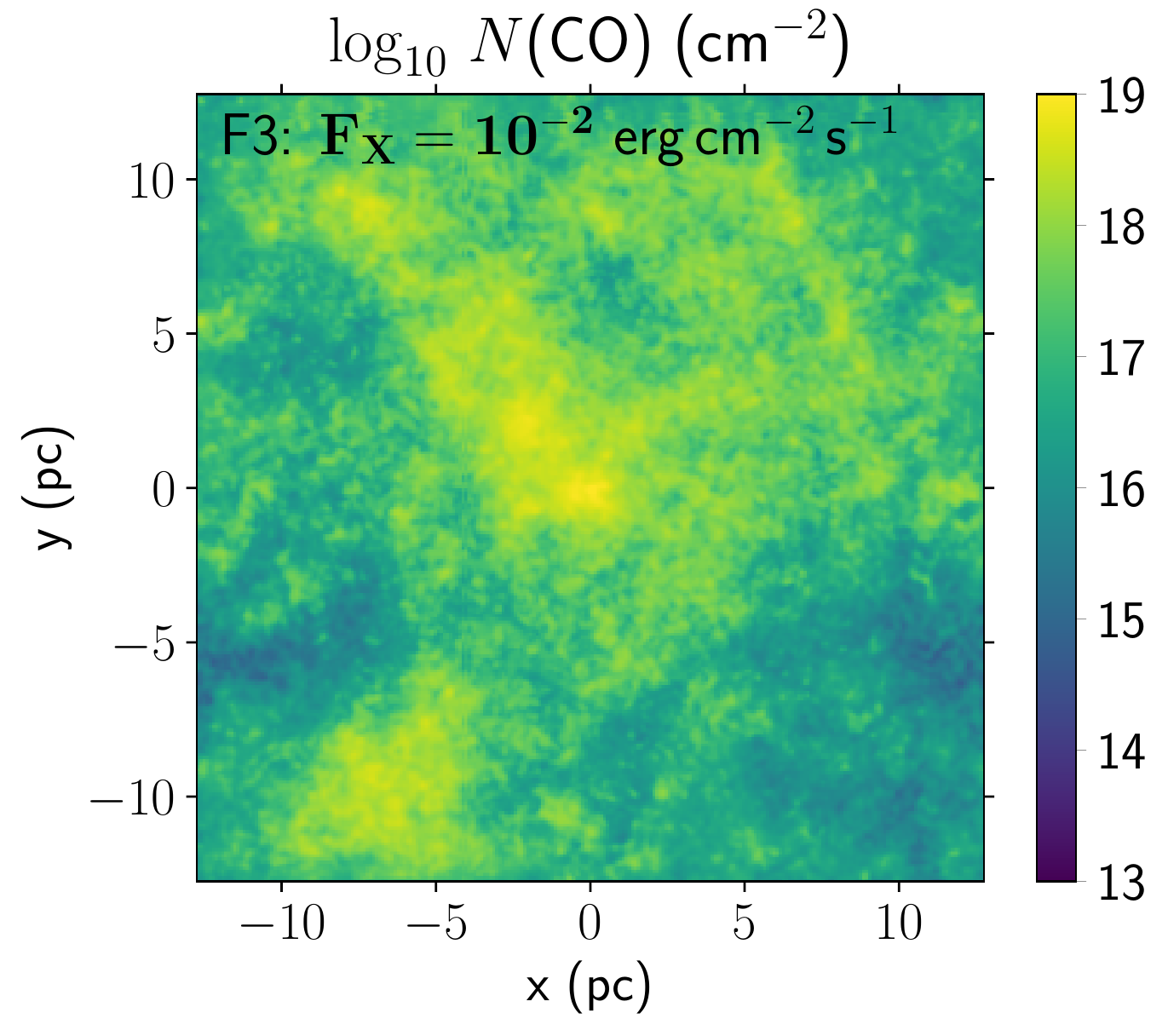} &
\includegraphics[trim = 12mm 16mm 0mm 9.25mm, clip, width=5.3cm]{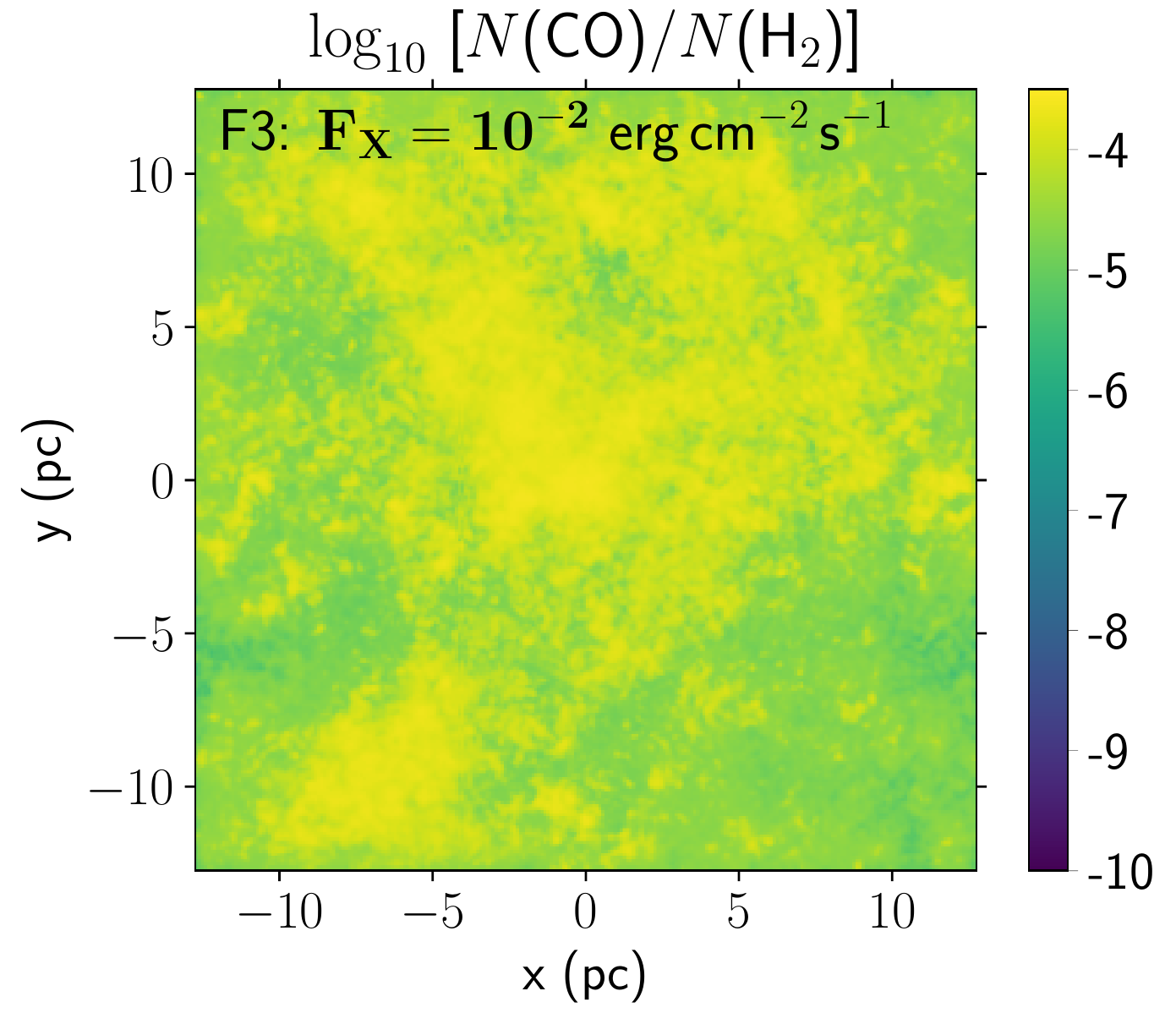} \\

\includegraphics[trim = 0mm 16mm 0mm 9.25mm, clip, width=6.0cm]{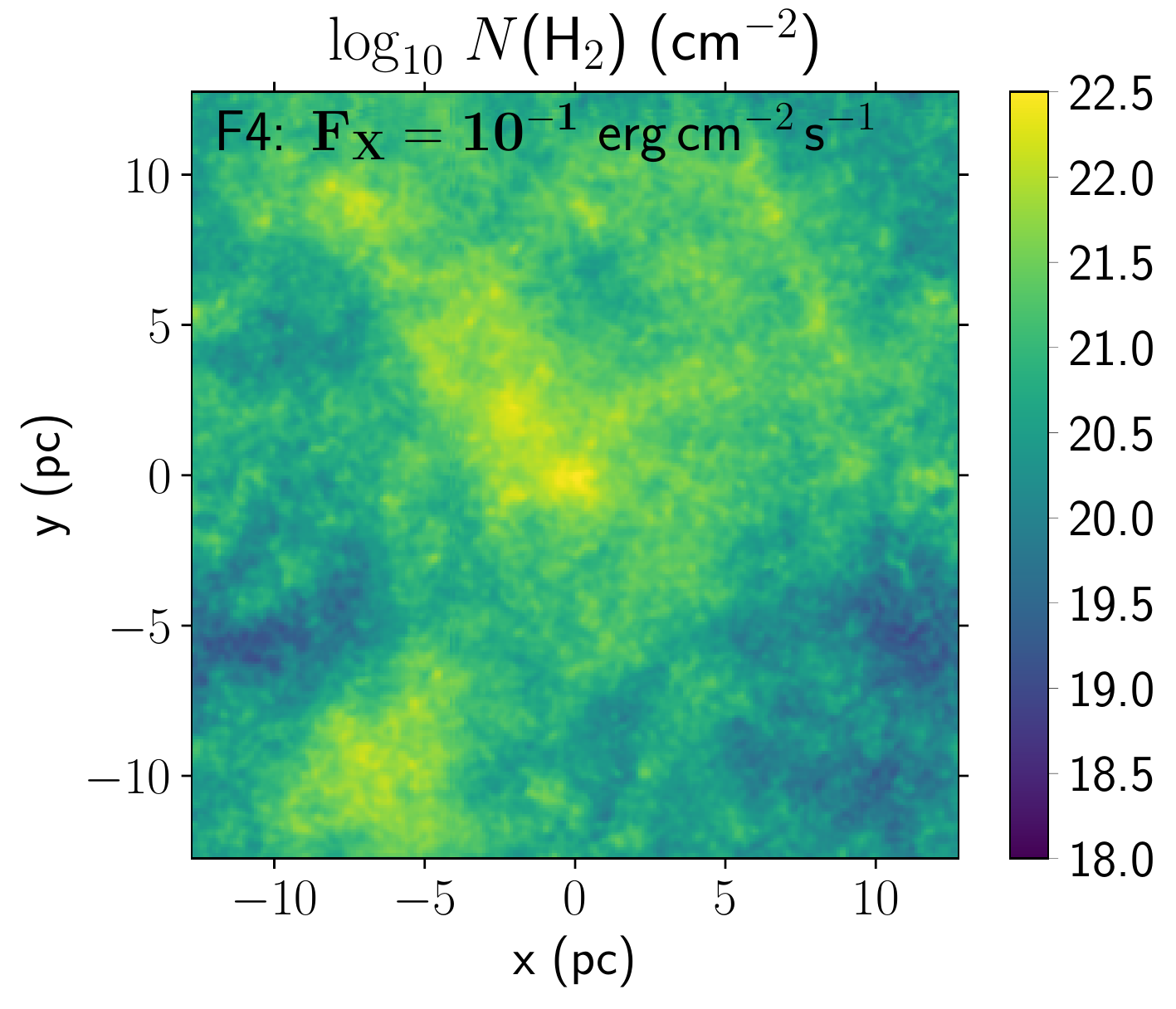} &
\includegraphics[trim = 12mm 16mm 0mm 9.25mm, clip, width=5.3cm]{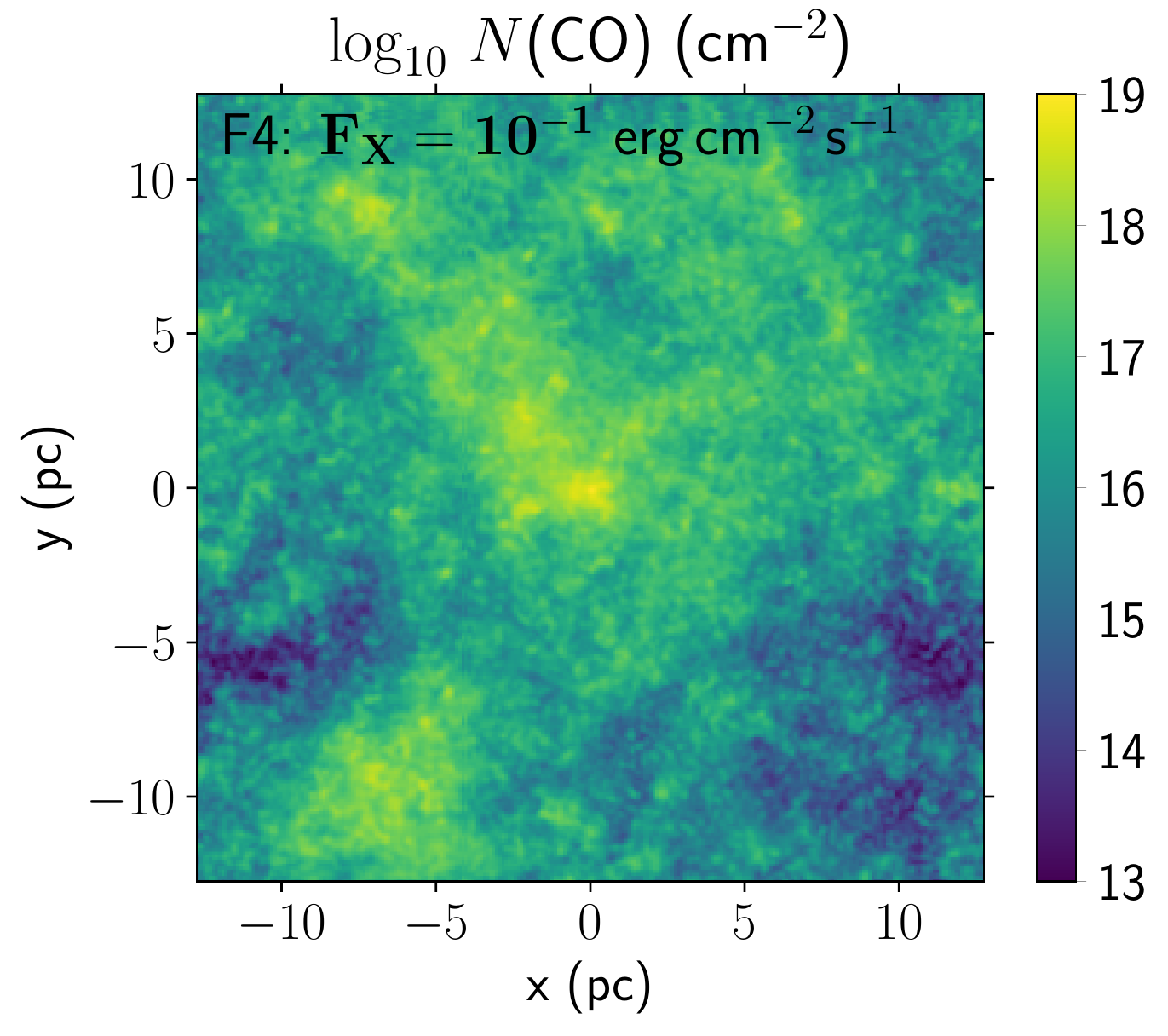} &
\includegraphics[trim = 12mm 16mm 0mm 9.25mm, clip, width=5.3cm]{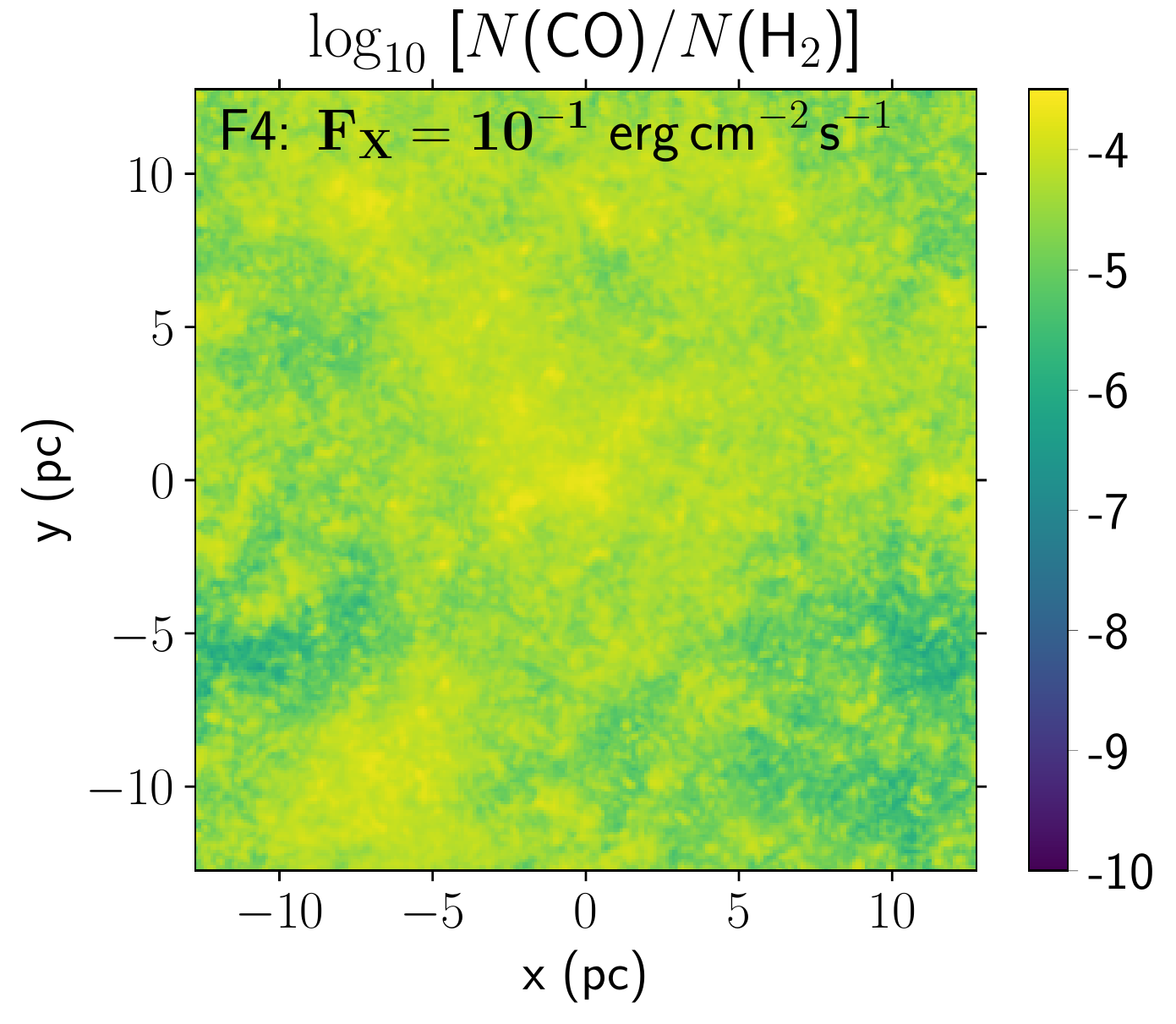} \\

\includegraphics[trim = 0mm 0mm 0mm 9.25mm, clip, width=6.0cm]{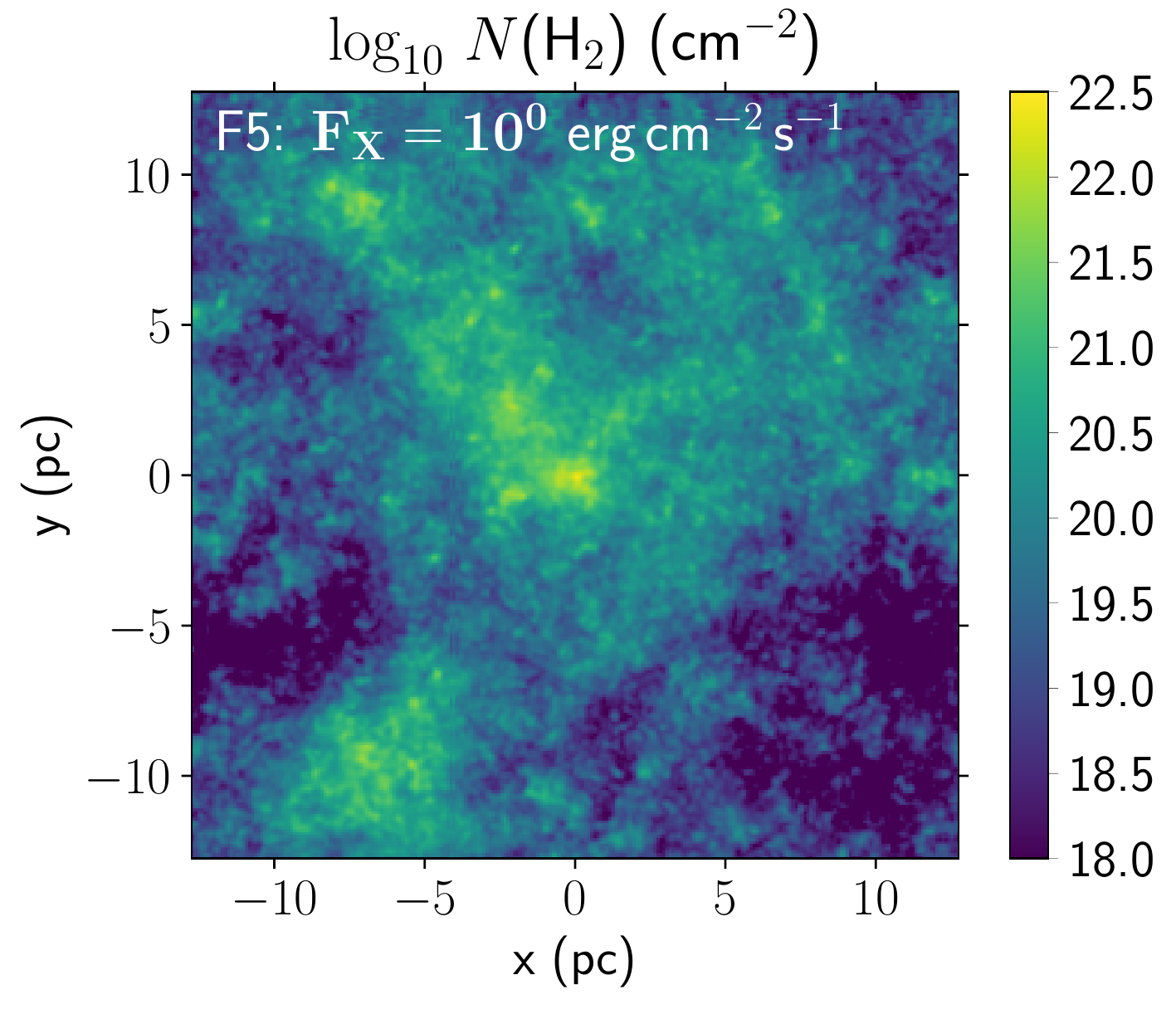} &
\includegraphics[trim = 12mm 0mm 0mm 9.25mm, clip, width=5.3cm]{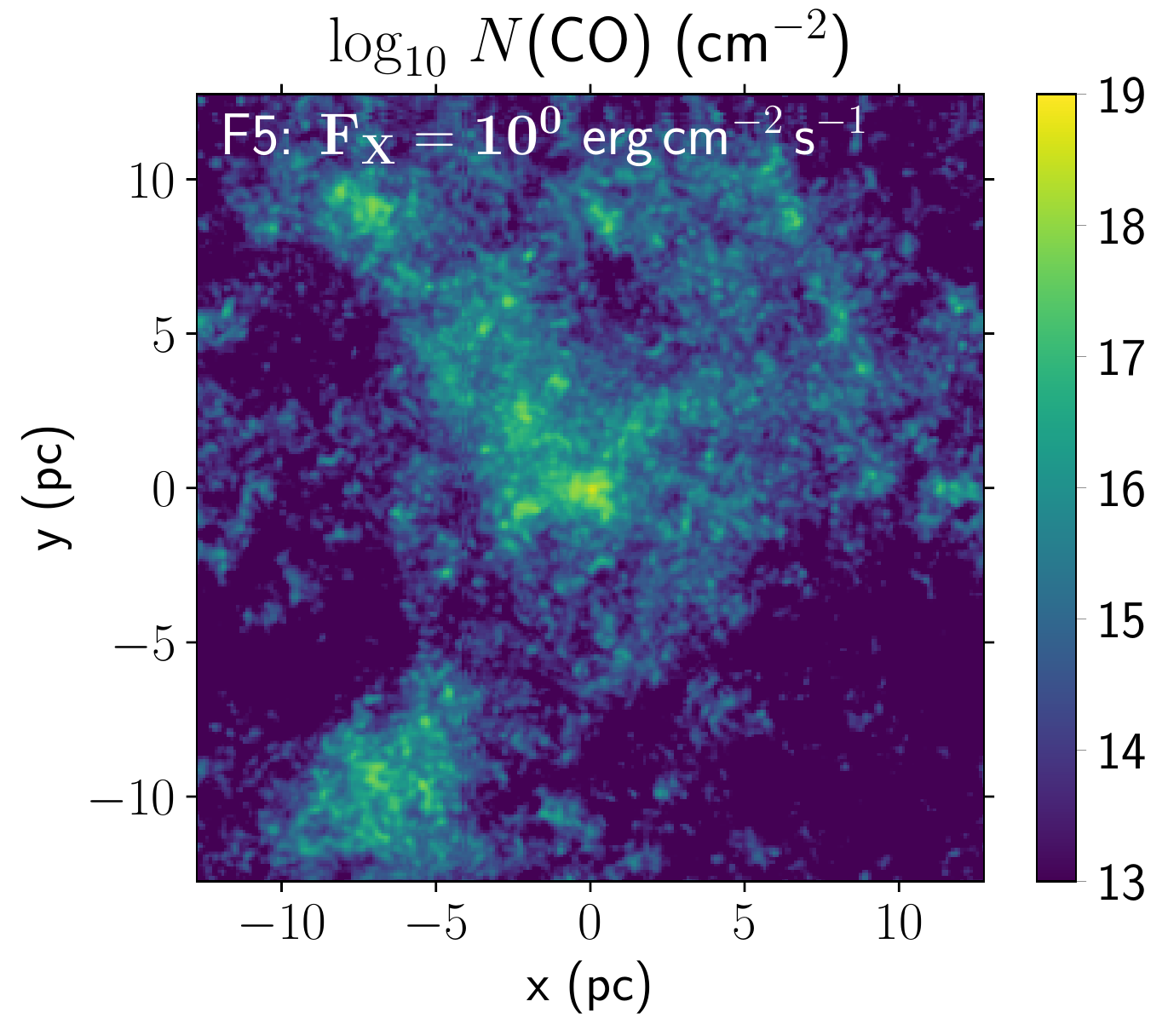} &
\includegraphics[trim = 12mm 0mm 0mm 9.25mm, clip, width=5.3cm]{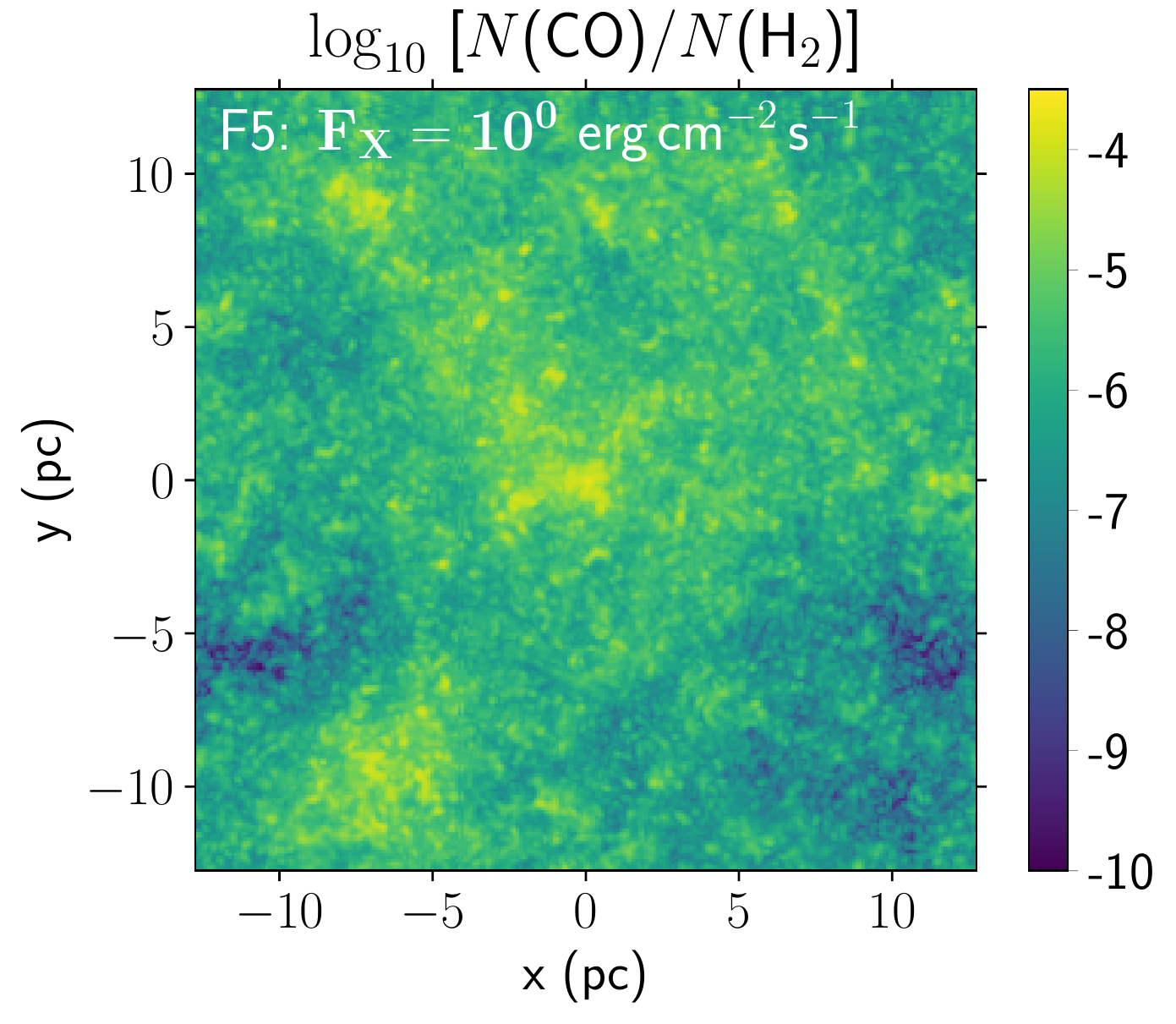} \\

\end{tabular}
\caption{
  Column density of H$_2$ (left), CO (centre), and the column-density ratio $N$(CO)/$N$(H$_2$) (right), for the fractal cloud irradiated with an external X-ray radiation field of $10^{-5}$ erg\,cm$^{-2}$\,s$^{-1}$ (thermal spectrum, $kT=1$\,keV) (run F0; top row), $10^{-2}$ erg\,cm$^{-2}$\,s$^{-1}$ (run F3; 2$^{nd}$ row), $10^{-1}$ erg\,cm$^{-2}$\,s$^{-1}$ (run F4; 3$^{rd}$ row), and $1$ erg\,cm$^{-2}$\,s$^{-1}$ (run F5; bottom row) for 4 Myr.
  CO is more effectively destroyed by the incident X-ray field than H$_2$, leading to a decreasing CO-to-H$_2$ ratio with increasing $4\pi J_{X}$.
  }
\label{fig:frac_F0_F5}
\end{figure*}

In Fig.~\ref{fig:frac_F0_F5} we show the column density of H$_2$ and CO, and the column-density ratio of the two, for simulations F0, F3, F4 and F5.
Runs F1 and F2 are not shown because they are similar to F0, and F6-F8 are also not shown because they have very little CO (F7 has no cells with $y($CO$)>3\times10^{-8}$, F6 has only a handful with $y($CO$)>10^{-6}$).

Visual inspection of these figures shows that CO and H$_2$ start to be depleted for $4\pi J_{X} \gtrsim 10^{-1}$\,erg\,cm$^{-2}$\,s$^{-1}$ (F4) and are mostly destroyed for $4\pi J_{X}\gtrsim1$\,erg\,cm$^{-2}$\,s$^{-1}$ (F5).
CO also is destroyed more completely than H$_2$ for large X-ray fluxes: 
the mass ratio of CO to H$_2$ in the simulation box decreases from about $10^{-3}$ for F0-F4 to $3.7\times10^{-4}$ for F5, $1.1\times10^{-4}$ for F6, $1.4\times10^{-5}$ for F7, and F8 has no CO.
In simulation F3 the densest regions still have large CO column densities and, counter-intuitively, the lowest column density regions at the edges of the simulation box have more CO in F3 than in F0.
The effect of X-rays is to raise the gas and dust temperatures (speeding up most reactions) and to increase the abundance of electrons and ions that are required for the formation of CO.

\begin{figure}
\centering
\includegraphics[height=6.5cm]{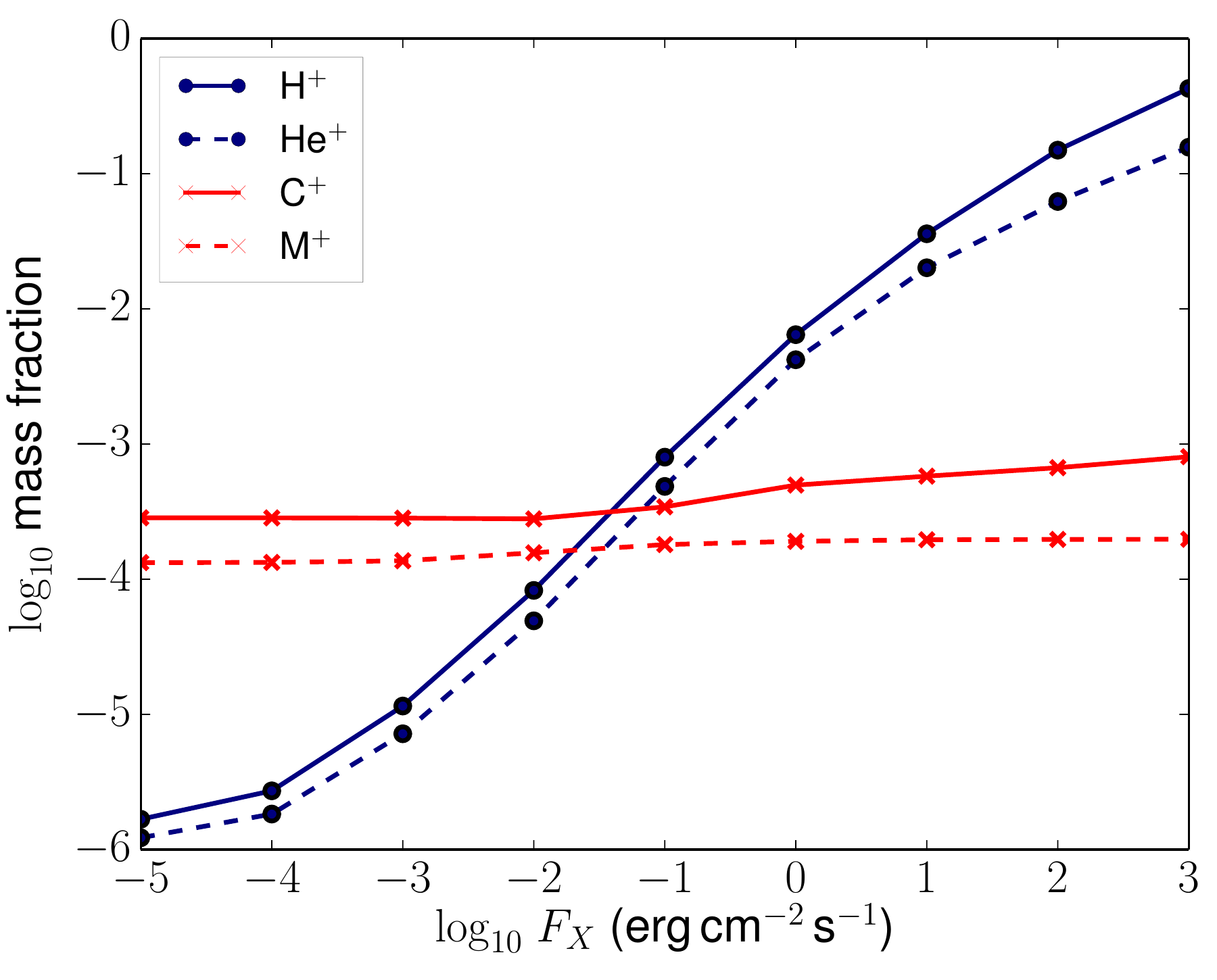}
\includegraphics[height=6.5cm]{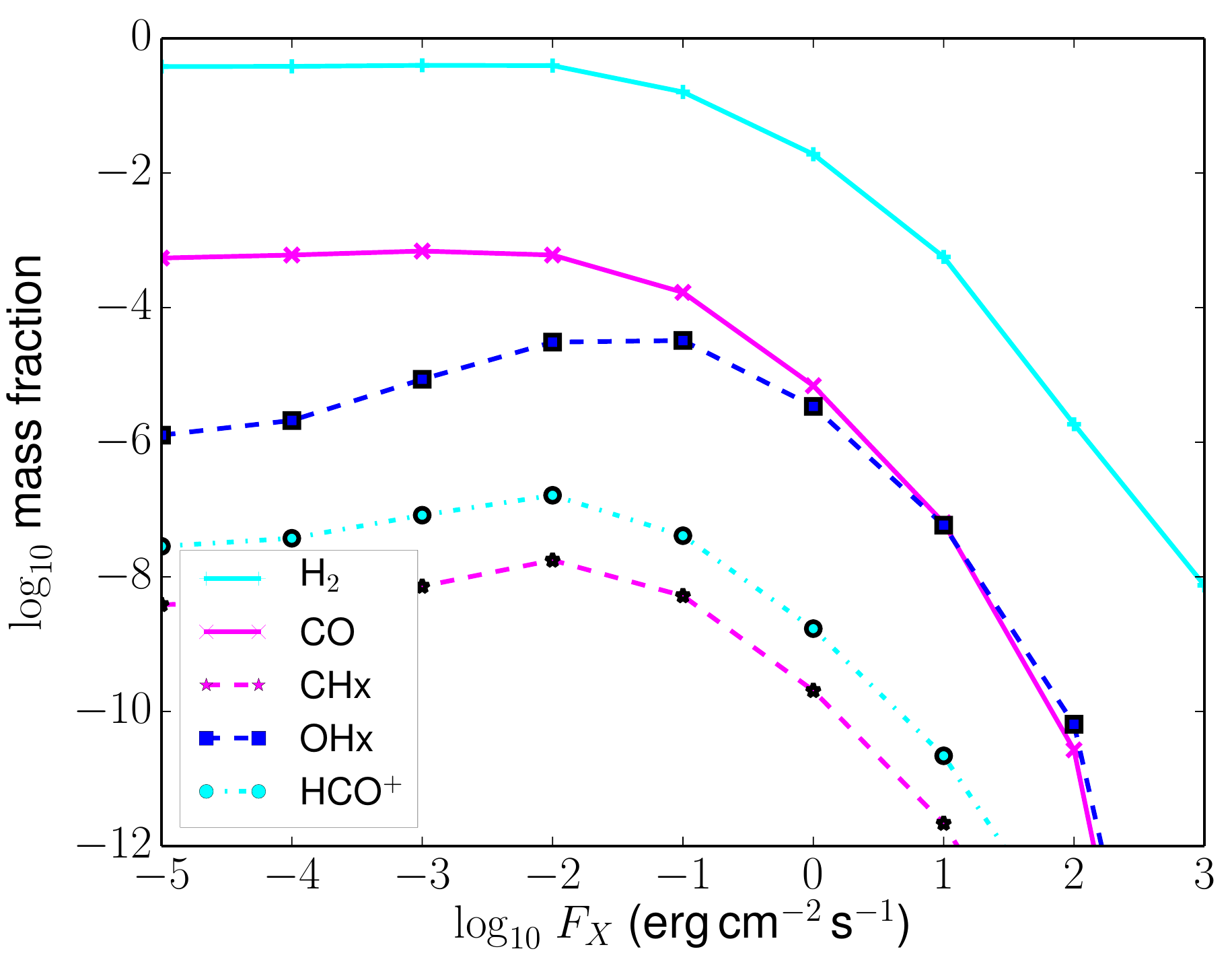}
\includegraphics[height=6.5cm]{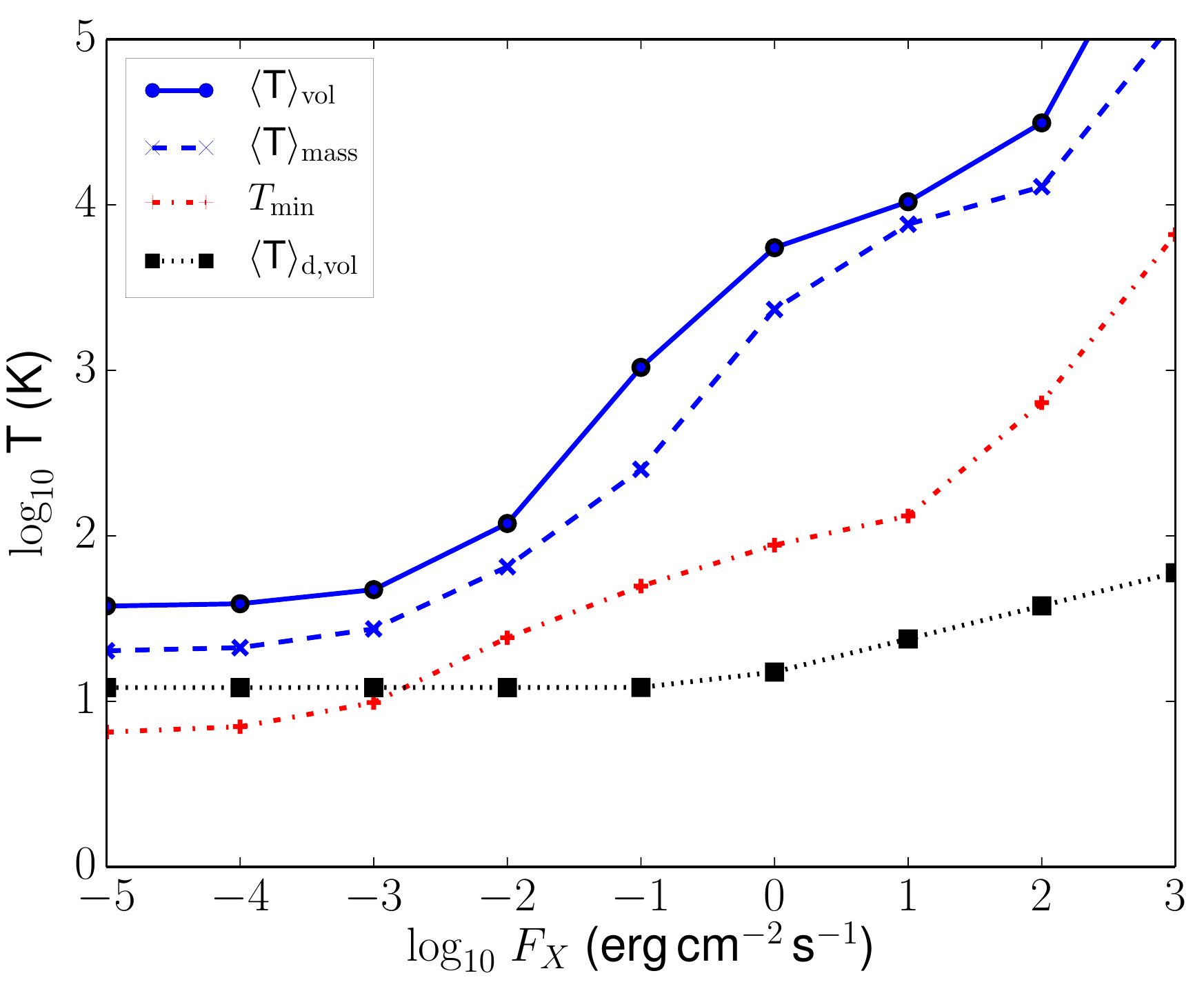}
\caption{
  Change in the mass fraction of ionic species (top panel), molecular species (middle panel), and temperature evolution (bottom panel) as a function of the incident X-ray flux on a fractal molecular cloud.
  These are the mass-fractions of all gas in the simulation domain, after $4\times10^6$ years of evolution to chemical equilibrium.
  The volume-weighted ($\langle T\rangle_\mathrm{vol}$) and mass-weighted ($\langle T\rangle_\mathrm{mass}$) mean temperatures are plotted, together with the minimum gas temperature, $T_\mathrm{min}$, and volume-weighted mean dust temperature, $\langle T\rangle_\mathrm{d,vol}$.}
\label{fig:3d_ion_mol}
\end{figure}

Fig.~\ref{fig:3d_ion_mol} shows the total mass fractions of various chemical species in the simulation domain for simulations F0-F8, again at $t=4$\,Myr, with the X-ray flux on the $x$-axis.
For low fluxes, the CO mass fraction actually increases slightly with increasing X-ray flux (already seen in Fig.~\ref{fig:frac_F0_F5} and discussed above), along with CH$_\mathrm{x}$, OH$_\mathrm{x}$ and HCO$^+$.
All molecular species are destroyed with increasing flux following similar trends and beginning at the same flux value: $4\pi J_{X}>10^{-2}$\,erg\,cm$^{-2}$\,s$^{-1}$ (F3).
H$_2$ is more resistant for large X-ray fluxes than any other molecule, surviving at trace levels up to the highest X-ray fluxes, whereas CO and the other molecular species are completely destroyed for $4\pi J_{X}>10^{2}$\,erg\,cm$^{-2}$\,s$^{-1}$ (F7).
The reason for this can be seen in the temperature panel, where the mean temperature approaches $10^4$\,K for $4\pi J_{X}>1$\,erg\,cm$^{-2}$\,s$^{-1}$, and the minimum temperature jumps from $\sim10^2$\,K to nearly $10^4$\,K between $4\pi J_{X}=10$ and $10^3$\,erg\,cm$^{-2}$\,s$^{-1}$.
Most of the destroyed CO goes into increasing the C$^+$ abundance, but this has a small effect on the total electron abundance because most electrons are produced from H$^+$ and He$^+$ for $4\pi J_{X}>10^{-2}$\,erg\,cm$^{-2}$\,s$^{-1}$
Neutral carbon also decreases in abundance with increasing $4\pi J_{X}$, albeit with a much weaker dependence on $4\pi J_{X}$ than CO.

\section{Flaring X-ray sources}
\label{sec:flare}

\subsection{Effect of increasing the X-ray irradiation}
\label{sec:flare_on}

Here we study the effects of a strong X-ray radiation field that is switched on for a given length of time and then switched off (i.e.~a flare) to see how the chemistry of a molecular cloud responds.
We take as initial conditions the cloud in simulation F2, where the chemistry and thermodynamics have been allowed to relax towards equilibrium for 4\,Myr.
We then increase the X-ray flux instantaneously by a factor of $10^5$, from $4\pi J_{X}=10^{-3}$ erg\,cm$^{-2}$\,s$^{-1}$ to $10^{2}$ erg\,cm$^{-2}$\,s$^{-1}$.
This large flux is similar to models 2 and 4 in MS05, who chose this value because it is typical of the cloud irradiation near AGN (it is also what is used in our simulation F7).
Because the speed of light is considered to be infinite, this affects all parts of the simulation instantaneously, heating, ionizing atoms and dissociating molecules.
Note that we find strong chemical and thermal effects on timescales shorter than the light-crossing-time of the simulation domain (i.e.~100 years).
The actual thermal and chemical effects we see on the cloud are robust, but the time-lag would be slightly different if we had a greater level of realism in modelling the radiative transfer.

\begin{figure}
\centering
\includegraphics[width=0.49\textwidth]{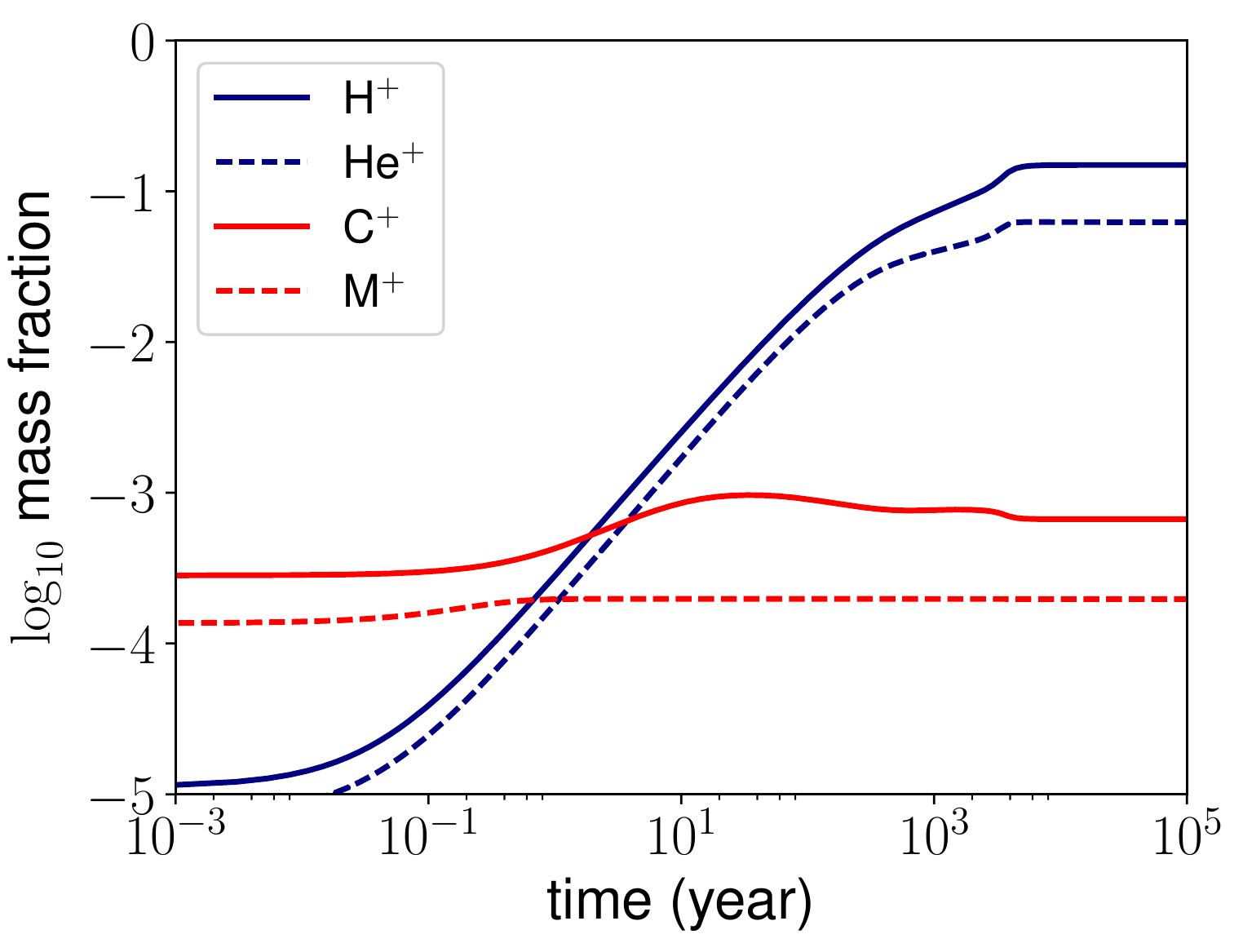}\\
\includegraphics[width=0.49\textwidth]{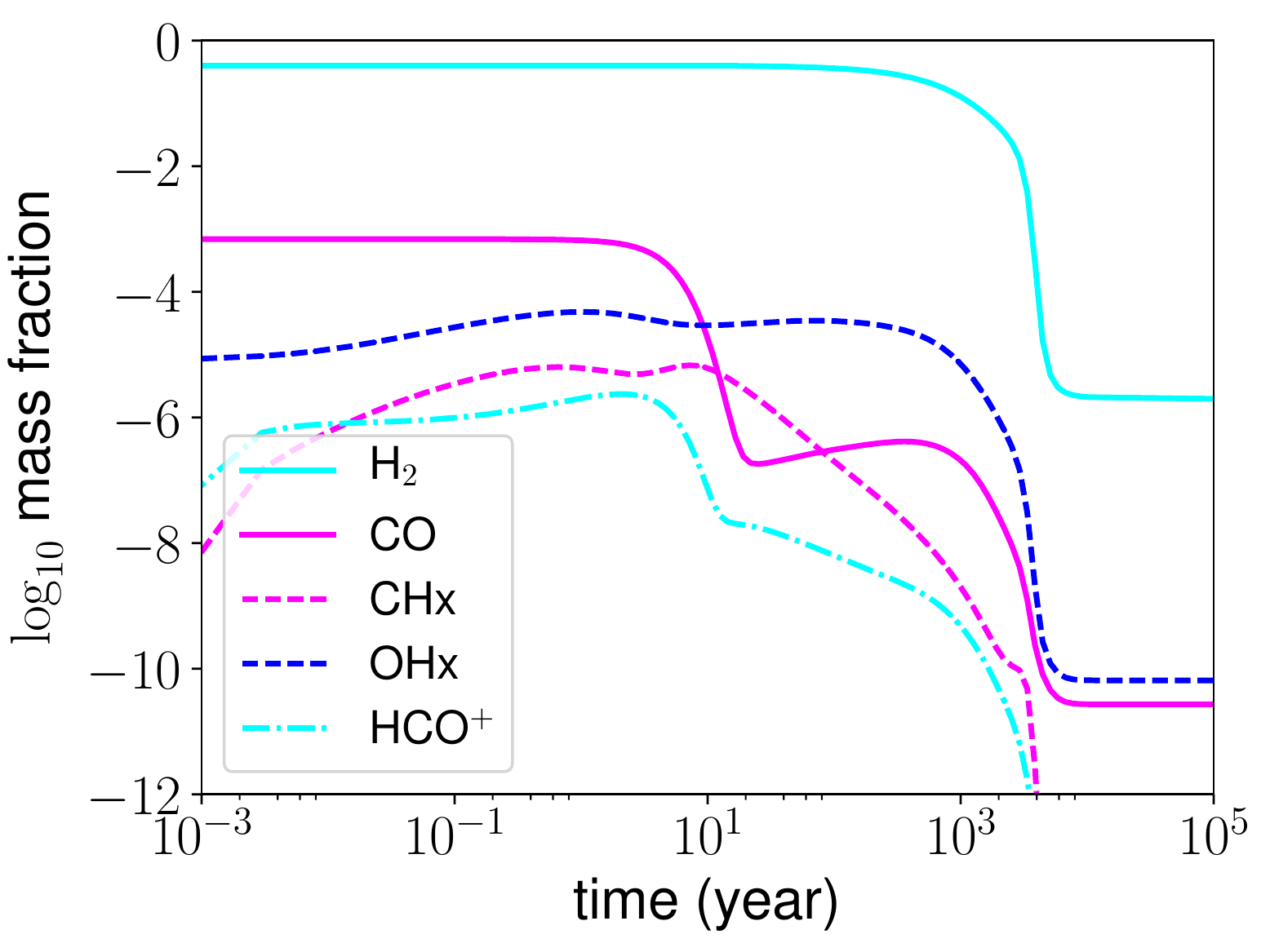}\\
\includegraphics[width=0.49\textwidth]{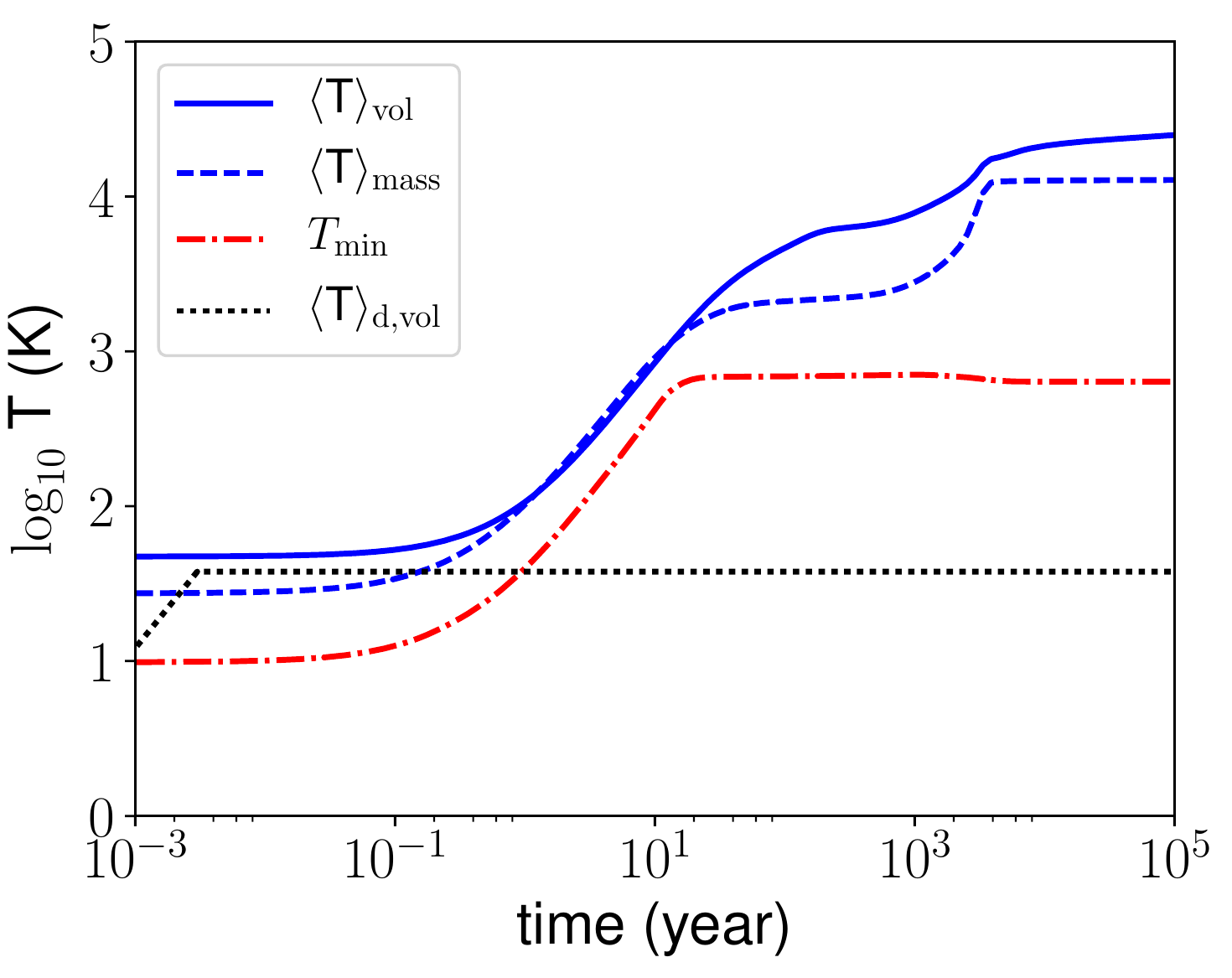}
\caption{
  Evolution of the mass fraction of various ionic species (top panel), molecular species (middle panel), and temperature (bottom panel) over time, measured from when the X-ray flux is increased by a factor of $10^5$.
  The volume-weighted ($\langle T\rangle_\mathrm{vol}$) and mass-weighted ($\langle T\rangle_\mathrm{mass}$) mean temperatures are plotted in the bottom panel, together with the minimum gas temperature, $T_\mathrm{min}$, and volume-weighted mean dust temperature, $\langle T\rangle_\mathrm{d,vol}$.
  }
\label{fig:flare_ion_mol}
\end{figure}

The evolution of the mass fractions of ions and molecules as a function of time, as well as the mean temperature, are plotted in Fig.~\ref{fig:flare_ion_mol}.
The mass fractions of H$^+$ and He$^+$ increase rapidly because of the dramatically increased ionization rate until they saturate at their equilibrium values after about $3\times10^3$ years.
Carbon goes from being partially ionized to almost fully ionized throughout the whole simulation after about 10 years, and the equilibrium mass fraction of C$^+$ at $t>10^3$ years is slightly larger than, but comparable to, that of neutral carbon.
The metal (M) is almost fully ionized in the initial conditions, and so its ionization state doesn't change much.

The results for the molecules are more interesting and subtle.
The middle panel of Fig.~\ref{fig:flare_ion_mol} shows that CO is very rapidly destroyed between 1 and 20 years after the X-ray flare switches on, and after about 20 years its rate of destruction decreases noticeably.
HCO$^+$ follows the same trend, whereas CH$_\mathrm{x}$ and OH$_\mathrm{x}$ are destroyed more gradually.
H$_2$ is almost unaffected for 100 years, and is significantly destroyed only after $10^3$ years.
This means that an X-ray flare can destroy almost all of the CO in a molecular cloud, while leaving the H$_2$ unaffected if it is shorter than $\sim10^3$ years.

This surprising result can be explained by looking at the temperature dependence of the various creation and destruction reactions for CO.
The mass-weighted mean temperature shows a rapid rise from $\approx30$\,K initally to $\approx100$\,K after 1 year to $\approx1000$\,K after 10 years.
This increase in temperature affects the dominant creation and destruction reaction rates for CO in a different way to H$_2$,
with the result that the CO abundance is much more sensitive to cloud heating than the H$_2$ abundance for $T\lesssim1000$\,K.

At early times the main creation reaction is through HCO$^+$ + e$^-$ (Table~\ref{tab:reactions}, \#38), and destruction is through H$_3^+$ (Table~\ref{tab:reactions}, \#24).
This pair of reactions is circular, however, and largely just convert CO to HCO$^+$ and back again, rather than reducing the overall quantity of CO.
The H$_3^+$ destruction rate is constant, whereas the destruction through locally generated FUV by fast electrons (Table~\ref{tab:photoreactions}, \#74) increases with temperature, so as the gas heats up, the FUV destruction becomes dominant after 1 year.
The creation rate (\#38) decreases as $T$ increases, so there is a phase of runaway CO destruction as long as these two (\#38 and \#74) are the dominant rates and $T$ is increasing with time.
During this phase the HCO$^+$ abundance decreases because it is being converted to CO through reaction \#38 whereas the reverse reaction (\#24) is no longer effective.
The abundances of CH$_\mathrm{x}$ and OH$_\mathrm{x}$ are not so dramatically affected because the FUV destruction reactions (\#75 and \#76 in Table~\ref{tab:photoreactions}) are independent of temperature, unlike the CO destruction rate.

After about 10 years, the HCO$^+$ creation channel for CO  (\#38) becomes too small, and the main creation rates are the constant rate from CH$_\mathrm{x}$ + O  (\#36) and OH$_\mathrm{x}$ + C  (\#37).
This slows down the CO destruction because after 10 years $T$ remains relatively constant, and so the FUV destruction rate (\#74) scales with the decreasing CO abundance.
Only after $>100$ years does the CO + He$^+$ destruction reaction (\#34) become the main one, by which stage most CO is already destroyed.

For CR ionization of molecular clouds, \citet{BisPapVit15} found that He$^+$ is the main destruction agent of CO, which superficially appears in conflict with our result.
The resolution to this seems to be that at late times in our flare simulation He$^+$ \textit{is} the main destruction channel, but most of the CO has already been destroyed through other reaction channels by the time He$^+$ becomes important.
This highlights an important difference between equilibrium and non-equilibrium chemistry.

We do assume that the rotational temperature of CO molecules (which is what determines the UV dissociation rate) is the same as the kinetic temperature.
In fact the rotational temperature lags behind rapid changes in the kinetic temperature, but the timescale is $\ll1$\,yr for the gas densities in the cloud that we simulate.

The reason H$_2$ is so much more robust than other molecular species is that it is not destroyed by the FUV radiation that the non-thermal electrons excite. 
Indeed the excitation of H$_2$ molecules is the main source of this locally generated FUV field.
Once the H$^+$ mass fraction increases to the point that the electron fraction reaches $\sim0.1$, most of the absorbed X-ray energy goes into Coulomb heating \citep{DalYanLiu99} and the gas temperature rises above $10^3$\,K in most of the cloud mass.
The rate of collisional dissociation of H$_2$ from collisions with H atoms increases hugely from $T=1000$\,K to $T=5000$\,K, and this is what ultimately destroys the H$_2$.
When the H$_2$ mass fraction decreases, this reduces the cooling rate and the temperature increases, further decreasing the H$_2$ fraction in a runaway process until a new equilibrium temperature is reached.

\subsection{Relaxation once the flare switches off}
\label{sec:flare_off}
\begin{figure*}
\begin{tabular}{ccc}
\raisebox{7.5\normalbaselineskip}[0pt][0pt]{\rotatebox{90}{1 year flare}} &
\includegraphics[width=0.45\textwidth]{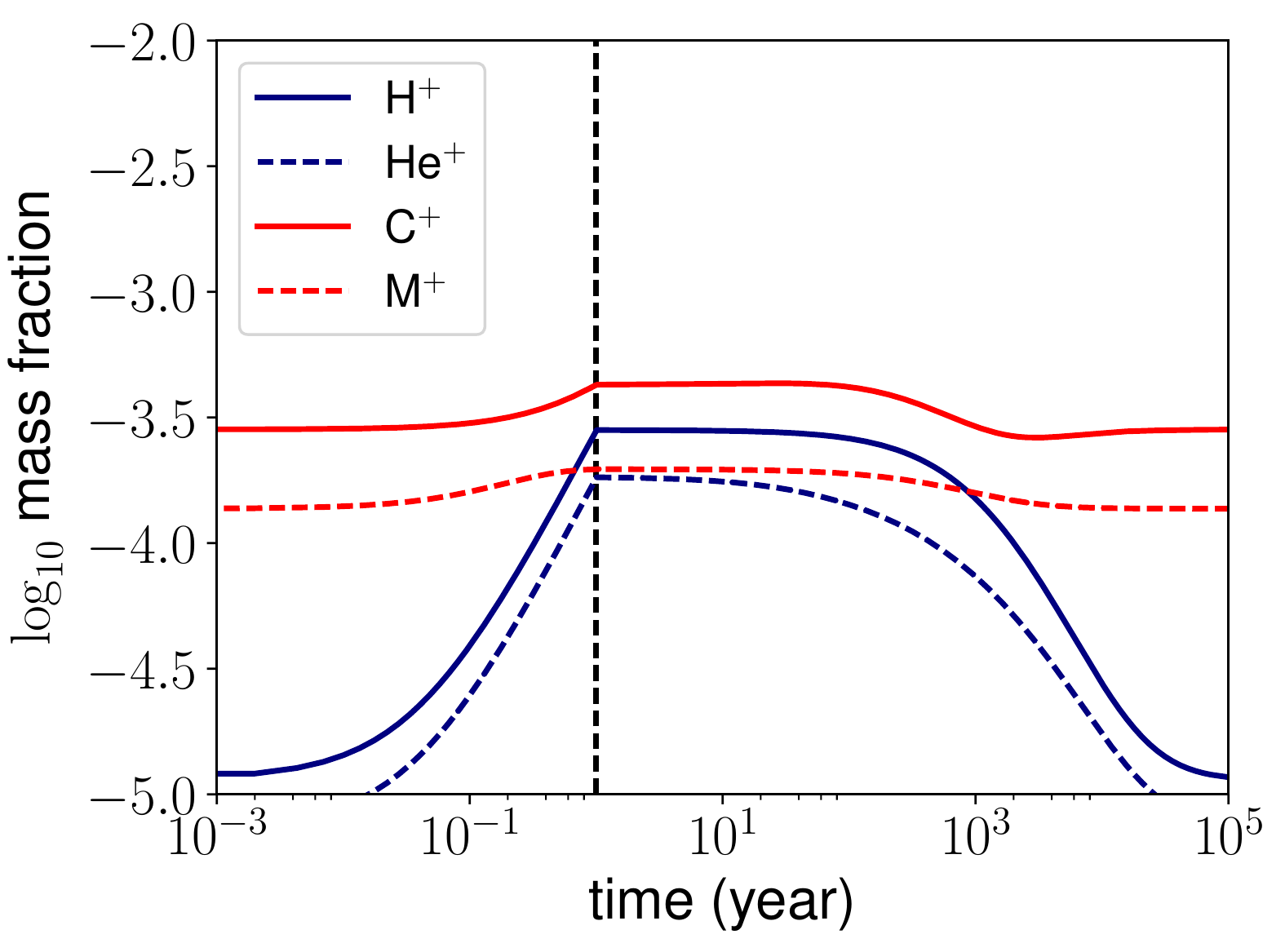}&
\includegraphics[width=0.45\textwidth]{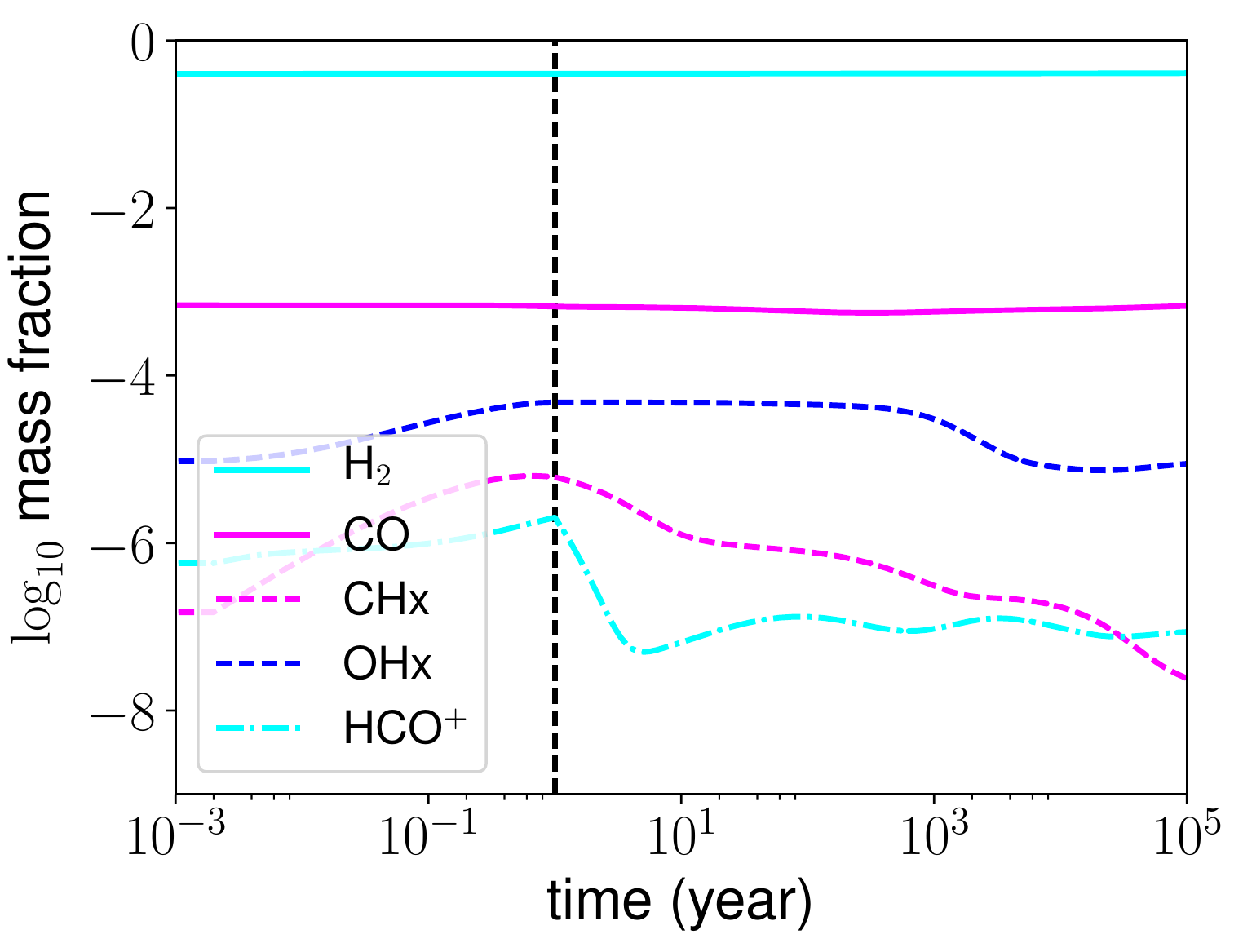}\\

\raisebox{7.\normalbaselineskip}[0pt][0pt]{\rotatebox{90}{10 year flare}} &
\includegraphics[width=0.45\textwidth]{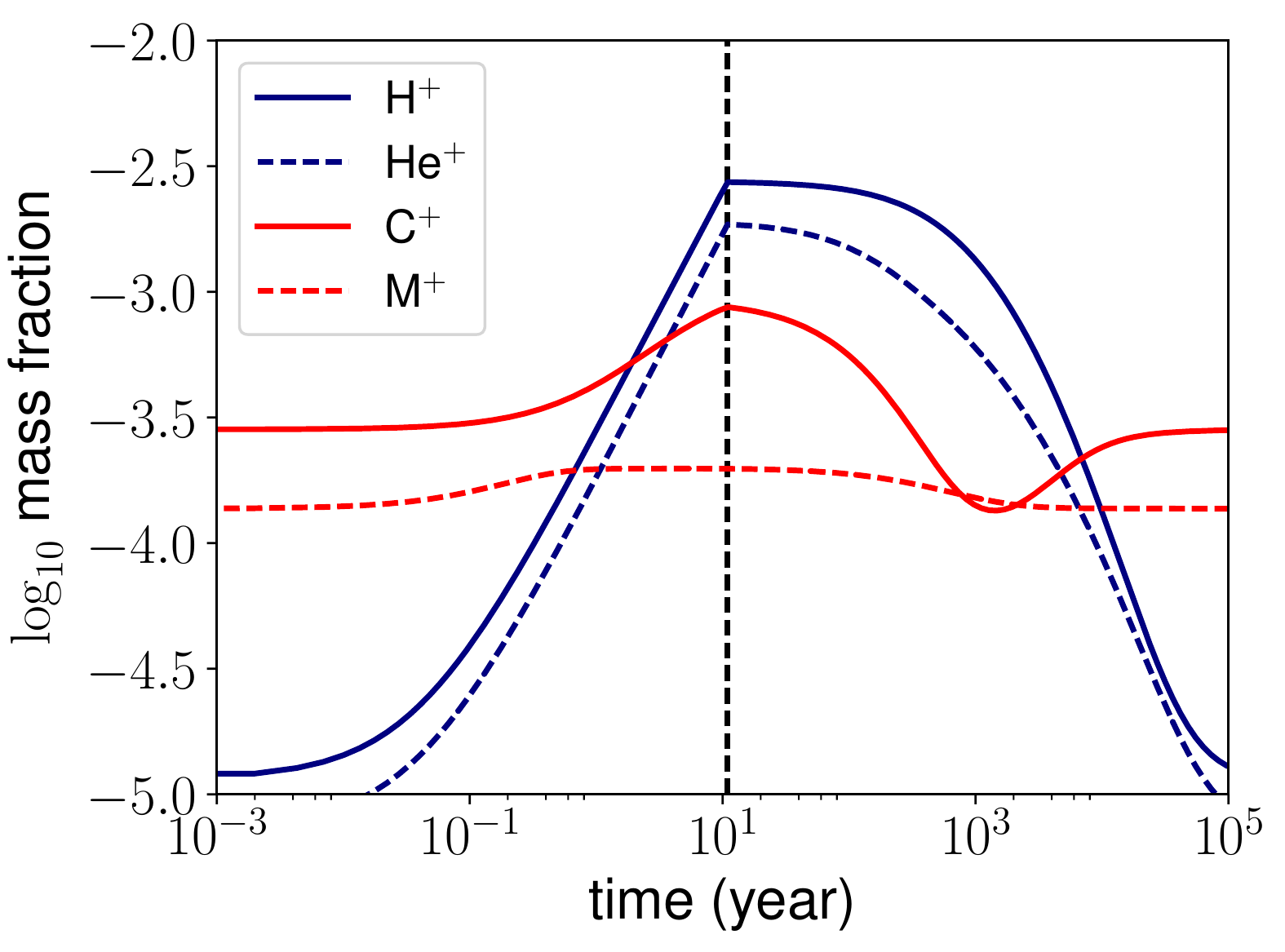} &
\includegraphics[width=0.45\textwidth]{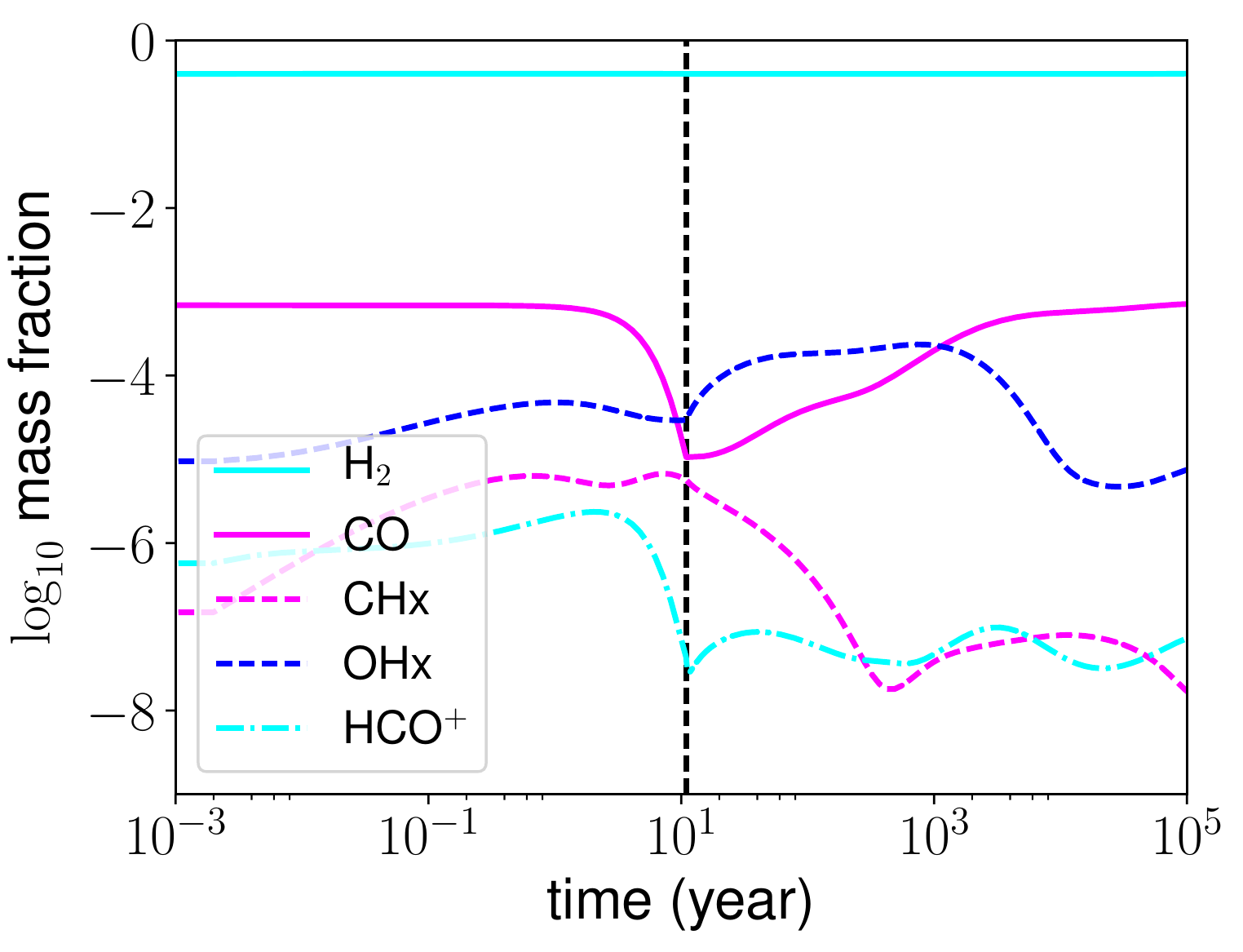} \\

\raisebox{7.\normalbaselineskip}[0pt][0pt]{\rotatebox{90}{25 year flare}} &
\includegraphics[width=0.45\textwidth]{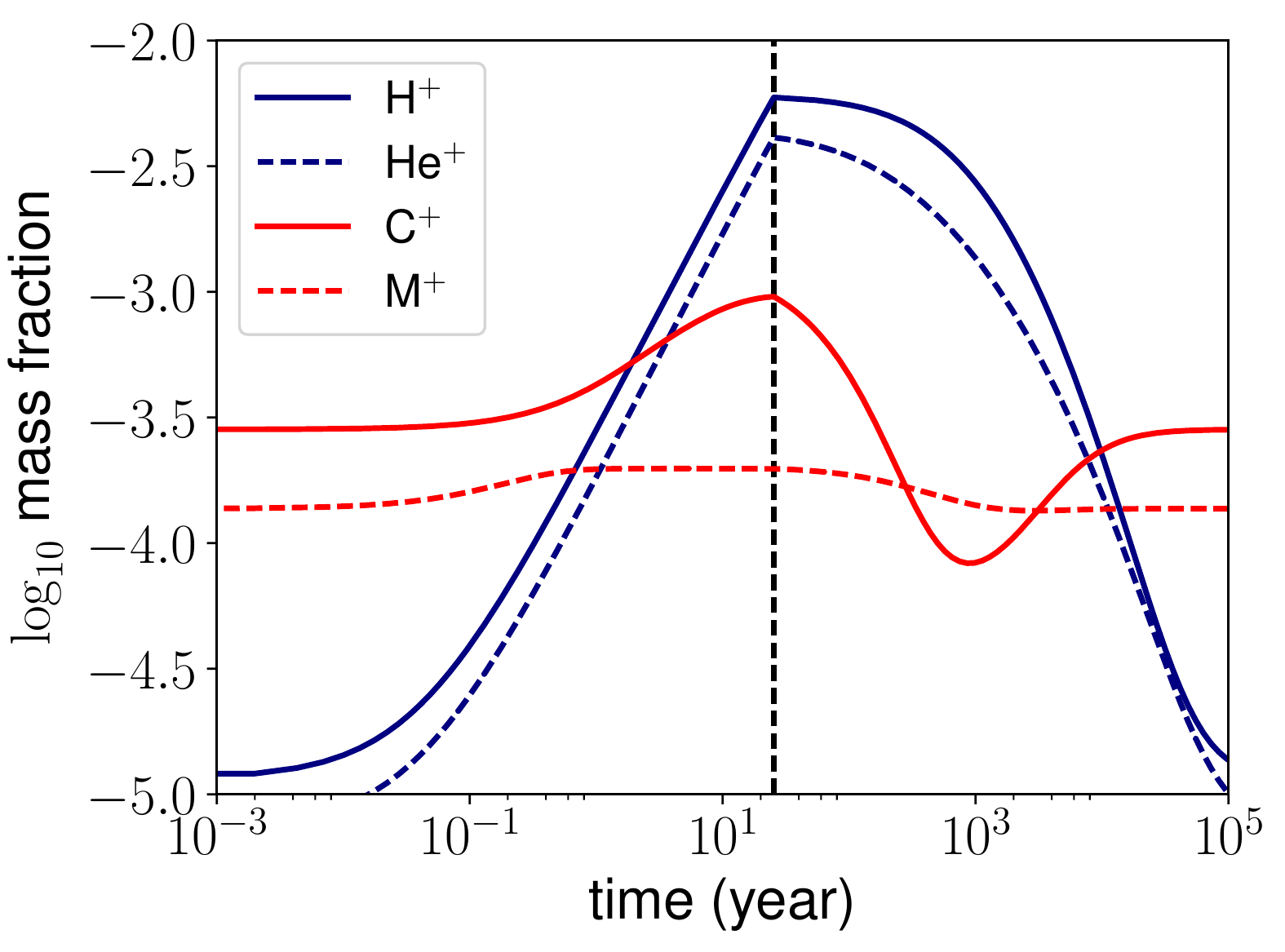} &
\includegraphics[width=0.45\textwidth]{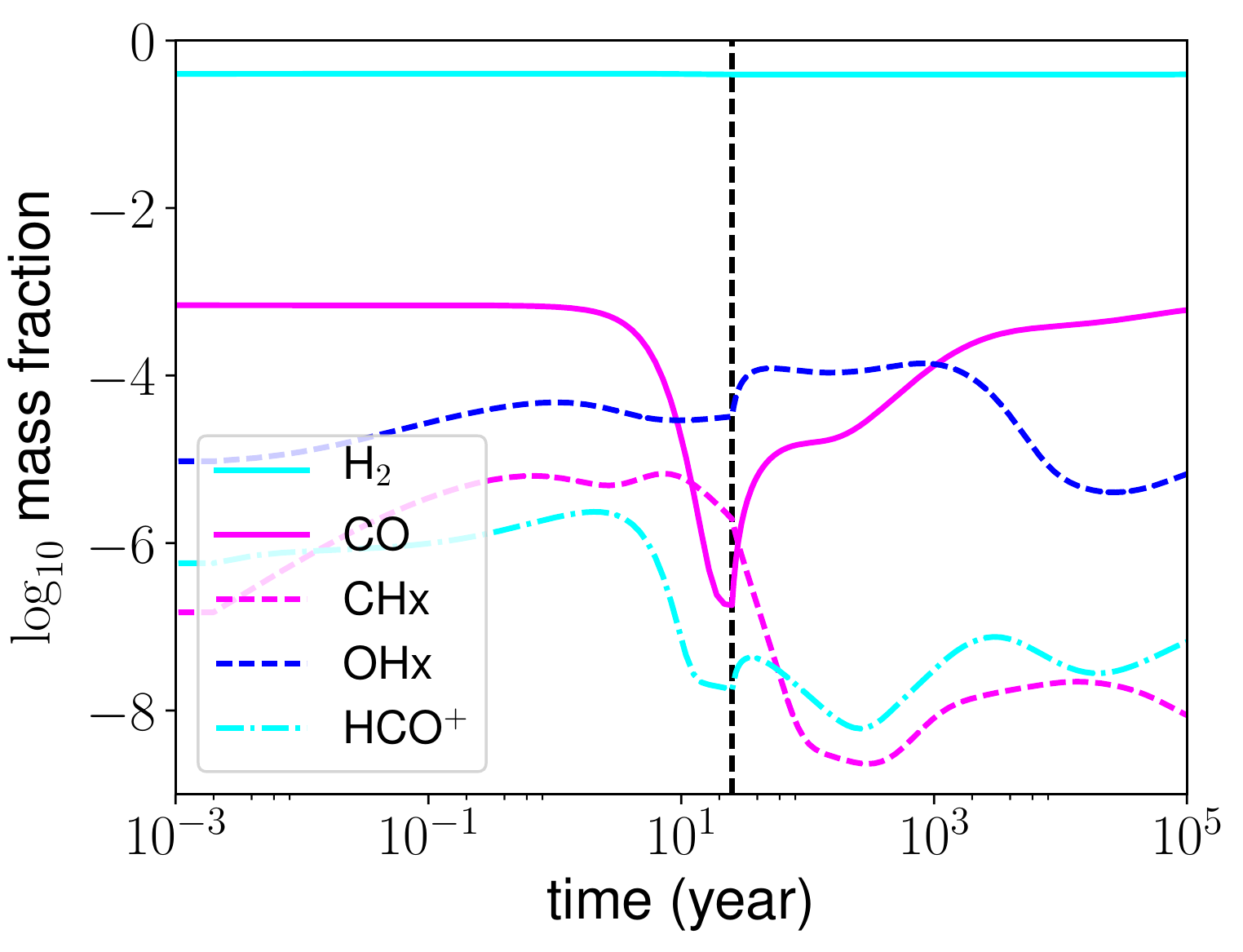} \\
\end{tabular}
\caption{
  Evolution of the total mass fractions of various ionic species (left column) and molecular species (right column) over time, measured from when the X-ray flare is switched on, for the case where the flare duration was 1 year (top row), 10 years (middle row), and 25 years (bottom row).
  The vertical line shows when the flare was switched off in each simulation.
  While H$_2$ is largely unaffected, CO is effectively destroyed by the X-ray flare if it lasts for 10 years or longer. 
  }
\label{fig:flare_all}
\end{figure*}

We now consider what happens if the increased X-ray irradiation switches off after a certain time; here we take 1, 10, 25 and 100 years as examples.
We restart from the flare simulation of the previous subsection but decrease the X-ray irradiation to $4\pi J_{X}=10^{-3}$ erg\,cm$^{-2}$\,s$^{-1}$ (model F2).
This decrease is again instantaneous, and takes effect everywhere in the domain because of the infinite-speed-of-light approximation.
The gas then cools and molecules reform.
The global evolution of the ions and molecules is plotted for these three flare durations in Fig.~\ref{fig:flare_all}, where the top panel shows the results of the 1 year flare, the middle panel the 10 year flare, and the bottom panel depicts the results of the 25 year flare.

If the duration of the flare is only 1 year, then the gas temperature has not increased dramatically and the molecular species have not been significantly affected by the X-rays (see also bottom panel of Fig.~\ref{fig:flare_ion_mol}), and so not too much changes after the flare is switched off.
Fig.~\ref{fig:flare_ion_mol} shows that most of the CO and HCO$^+$ are already destroyed after 10 years, so for a flare duration of 10 years or longer we  see significant evolution during and after the flare in Fig.~\ref{fig:flare_all}.

After the flare the ionic mass fractions decrease over $10^2$-$10^4$ years, and reach equilibrium in about $10^5$ years in all cases.
The molecular evolution is somewhat more complicated, but the trend is that CO starts to reform immediately, and is approaching its equilibrium mass fraction after $10^5$ years.
For shorters flares the recovery is faster:  for a 10 year flare the CO mass fractions reaches half of its pre-flare equilibrium value after 1750 years; for a 25 year flare it takes 4000 years; and for the 100 year flare (not shown) 31\,000 years.
H$_2$ remains constant because it was not destroyed by the flare.
This result raises the possibility that molecular clouds with negligible CO abundance may exist near X-ray sources simply because X-ray flares efficiently destroy CO but not H$_2$.
Since it takes $10^3 - 10^5$ years to reform the CO, we expect that molecular clouds near centres of galaxies that are occasionally active, and clouds hosting young massive star clusters with X-ray binaries, can have out-of-equilibrium CO-to-H$_2$ ratios for much of their lifetime (see sec.~\ref{sec:discussion}).

\section{Discussion}
\label{sec:discussion}

We have shown that a gas cloud exposed to an X-ray flare with radiation energy density of $E_\mathrm{rad}\sim3\times10^{-9}$\,erg\,cm$^{-3}$ will suffer catastropic CO destruction for flares of duration 10 years or longer, and that the flare duration must be $\gtrsim1000$ years to significantly destroy the H$_2$.
Also, gas clouds irradiated by a constant X-ray energy density $E_\mathrm{rad}\gtrsim3\times10^{-13}$\,erg\,cm$^{-3}$ (F3) show significant heating and chemical effects, and X-rays dominate over CRs as the main heating agent (assuming the CR flux does not scale with X-ray flux).
If $E_\mathrm{rad}\gtrsim3\times10^{-12}$\,erg\,cm$^{-3}$ (F4) then X-rays begin to significantly destroy CO and H$_2$.
It is useful to discuss where such conditions arise, ignoring for now the issue of attenuation and focusing purely on the dilution due to the inverse-square law.
The energy density at a distance $d$ from a point source with luminosity $L_\mathrm{x}$ is given by
\begin{equation}
E_\mathrm{rad} = \frac{L_\mathrm{x}}{4\pi c d^2} = 
2.8\times10^{-9} \frac{L_\mathrm{x}}{10^{40}\,\mathrm{erg\,s}^{-1}} \left(\frac{1\,\mathrm{pc}}{d}\right)^2 \;\mathrm{erg\,cm}^{-3}\;.
\end{equation}

The Galactic Centre today has an X-ray luminosity of $L_\mathrm{x}\lesssim10^{35}$\,erg\,s$^{-1}$, implying that only clouds within a small fraction of a parsec have significant CO depletion from the current X-ray emission of Sgr A$^\star$.
During the flare from 100 years ago, the luminosity was 4 orders of magnitude larger, but still only clouds within $\lesssim0.5$\,pc of Sgr A$^\star$ would have been affected as strongly as the cloud we simulate.
Our results for the simulations with X-ray fields of differing strength show that clouds close to Sgr A$^\star$ (0.5-10\,pc) would have some CO destruction, with the effect decreasing with distance.
For $d\gtrsim10$\,pc ($E_\mathrm{rad}\lesssim3\times10^{-12}$\,erg\,cm$^{-3}$, comparable to simulation F4 or weaker) the CO abundance should actually be enhanced because of the X-ray heating and production of free electrons.
Our results imply that the clouds in the circumnuclear disk around Sgr A$^\star$ could have been significantly affected by X-rays, but the clouds in the 100-pc molecular ring would have remained largely unaffected, given the luminosity estimates of the flare obtained from X-ray reflection \citep{PonTerGol10}.

Active Galactic Nuclei (AGN) can have $L_\mathrm{x}>10^{43}$\,erg\,s$^{-1}$, for which gas clouds up to 30\,pc (larger for higher $L_\mathrm{x}$) from the black hole should have their CO completely destroyed by X-ray radiation, unless they are optically thick to hard X-rays.
CO should be depleted out to $d\gtrsim1000$\,pc, and for sources that emit with this luminosity for thousands of years the H$_2$ should also be depleted, again with stronger depletion closer to the source.

The class of ultraluminous X-ray sources (ULX) have $L_\mathrm{x}\sim10^{39}-10^{42}$\,erg\,s$^{-1}$ \citep{SwaGhoTen04}, and it is thought that some of these are powered by pulsars, stellar-mass black holes, and possibly intermediate-mass black holes for the most luminous of them \citep{MezRobSut13, EarRobHei16, MezCivFab16, Bac16}.
The pulsars and stellar-mass black holes are associated with high-mass star formation, and hence with molecular clouds.
This, together with the variable nature of ULX sources \citep{Bac16} suggests that we should see strong effects of X-rays on the chemistry and temperature of molecular clouds in the vicinity of ULX, out to tens of parsecs from the source.
For the most luminous ULXs, this radius is 300-1000\,pc, a significant fraction of the volume of a dwarf galaxy.


Fig.~\ref{fig:3d_ion_mol} shows that simulation F3, with $E_\mathrm{rad}\sim3\times10^{-13}$\,erg\,cm$^{-3}$ (Table~\ref{tab:sims}), divides the lower-flux simulations where X-rays have little effect, from the high-flux simulations where X-rays have a big impact on the chemistry and thermal state of the molecular cloud.
This is a few times less than the energy density of the ISM in the Galactic plane in CRs, magnetic fields and turbulent kinetic energy \citep[$\sim1$\,eV\,cm$^{-3}$;][]{Cox05}.
Our results imply, therefore, that X-rays will dominate the chemistry/thermodynamics of molecular clouds if the X-ray energy density is comparable to or exceeds that of CRs.
This claim is of course dependent on energy and environment, because the interaction cross-sections of both X-rays and CRs are strongly energy-dependent.
Furthermore, sources of CRs are invariably also sources of X-rays, but the scaling of energy density with respect to distance from the source is not the same for CRs and X-rays, because CRs diffuse whereas X-rays stream freely until they are absorbed.
Absorption cross-sections of CRs are very uncertain, but it should still be possible to use the code we have developed to constrain the conditions under which X-rays deposit more energy in molecular clouds than CRs, and vice versa.

The local far-UV ISRF has $E_\mathrm{FUV}\sim5\times10^{-14}$\,erg\,cm$^{-3}$ \citep{Dra78} (or $4\pi J_\mathrm{FUV}\approx1.5\times10^{-3}$\,erg\,cm$^{-2}$\,s$^{-1}$) which is significantly smaller than the ISM energy density in CRs and the X-ray energy density in simulation F4.
The ISRF can significantly affect ISM chemistry with a smaller energy density than X-rays (or CRs) because it has a larger absorption cross-section, and so a larger heating rate per unit energy density, but it consequently can only affect the outer (low-extinction) layers of a molecular cloud \citep[cf.][]{MeiSpa05}.
Fig.~\ref{fig:coltemp} shows that the low-extinction part of the cloud is only significantly affected by the X-rays for Simulations F3 and above, with $E_\mathrm{rad}\gtrsim3.3\times10^{-13}$\,erg\,cm$^{-3}$.
This reflects that only a small fraction of the X-ray radiation is absorbed in the low-extinction part of the cloud, so the X-rays must have a significantly larger energy density than FUV in order to have a comparable effect at low column densities.
In contrast, the high-extinction part of the cloud is already heated by X-rays for a flux 10 times lower (F2) because (i) here it is not competing with the FUV but only with CRs, and (ii) the majority of the X-ray radiation is deposited here.

Our 1D test calculations in Section~\ref{sec:ms05} showed that H$_2$ is a significant coolant when dense clouds are strongly irradiated by X-rays, supported by the \textsc{Cloudy} calculations in Section~\ref{sec:cloudy}.
The 3D simulations of an X-ray flare show (see Fig.~\ref{fig:flare_ion_mol}) that molecular gas is heated to $T\sim10^3$\,K in about 10 years, into the temperature regime where H$_2$ cooling becomes effective.
We therefore expect that this hot H$_2$ gas would emit in the infrared and be observable with upcoming observatories such as the \textit{James Webb Space Telescope} \citep{GarMatCla06, Kal18}.
Our simulations predict that CO is destroyed on a similar (10-20 year) timescale to gas heating, and so it should be possible to observe CO emission decreasing on the same timescale as H$_2$ emission switches on after a bright flare near a molecular cloud.

\citet{GloMac07b} showed that turbulent motions in molecular clouds can significantly speed up the formation of H$_2$ and other molecules.
We cannot address this with the static simulations presented here, but future calculations with a turbulent cloud will study whether CO can re-form more quickly than indicated by our results.

Our results should also have application to protoplanetary disks, where  \citet{CleBerObe17} showed that time-dependent X-ray irradiation can modify the observable HCO$^+$ signature in the disk.
Low-mass protostars typically have strong X-ray emission and variability on account of the strong surface magnetic fields, and this radiation field strongly affects the properties of protostellar disks \citep{GlaNajIge97}.
The time-dependent effects of the X-ray irradiation have not yet been investigated in great detail.

A limitation of our work is that we use the infinite speed-of-light approximation, whereas the chemical and thermal properties of the molecular cloud that we model are changing on a timescale less than the light travel time across the cloud for the model of an X-ray flare.
If we tracked the photon front propagating through a cloud then the heating, dissociation and ionization would sweep through the cloud rather than happen simultaneously at all places.
The same chemical and thermal evolution would still occur, but there would be time offsets between different parts of the cloud depending on when they were first exposed to the X-ray flare.
How this would appear to an observer is very dependent on the angle between the photon propagation direction and the observer's line of sight.
If the photon front were propagating directly towards the observer then nothing would look different, whereas if it were propagating at right angles then we could potentially see different molecular and atomic transitions switch on and off in a wave moving across a cloud as more of the cloud gets heated by X-rays.
The long-term evolution of the cloud, which is perhaps the most interesting result we have obtained, would not look any different because the timescales for recombination and for CO to re-form are much longer than the light-crossing time of a cloud.

\section{Conclusions}
\label{sec:conclusions}

This paper presents a new implementation of hydrogen and carbon non-equilibrium chemistry when exposed to a (potentially time-varying) X-ray radiation field.
The chemical network is relatively small, so that it can be integrated efficiently enough for use in 3D magnetohydrodynamic simulations of molecular clouds and the ISM.
Comparison of 1D test calculations using the new network and more complex XDR/PDR codes such as \textsc{cloudy} shows that the gas temperature and abundances of the most abundant species agree satisfactorily.
Species with typically low abundance, namely CH$_\mathrm{x}$, OH$_\mathrm{x}$, and HCO$^+$, show poor agreement with \textsc{cloudy}, probably reflecting their status in our network as \emph{helper molecules} whose main purpose is to obtain the correct abundances of C$^+$, C, and CO.
The chemical network is coupled to the \textsc{TreeRay/Optical depth} solver \citep{WunWalDin18} for radiative transfer of the far-UV ISRF, modified to include X-ray radiative transfer, and implemented in the simulation code \textsc{Flash}.

The first application of the code was to study the equilibrium chemical and thermal state of a fractal molecular cloud when exposed to X-ray radiation of different intensities.
UV radiation acts only on the surface layers of a molecular cloud, but hard X-rays can penetrate deep into the whole volume of the simulated cloud, and so have a much stronger effect.
X-ray energy densities of $3\times10^{-16}-3\times10^{-14}$\,erg\,cm$^{-3}$ had limited effects on the cloud other than a small increase in the minimum temperature and also an increase in the CO to H$_2$ ratio (on account of the increased ion and electron abundances induced by the X-rays).
A radiation field with $E_\mathrm{rad}=3\times10^{-13}$\,erg\,cm$^{-3}$ increased the mean cloud temperature to nearly 100\,K, and provided sufficient ionizations that H$^+$ and He$^+$ became the main source of electrons (instead of C$^+$ and M$^+$, which have much lower overall abundance).
The CO abundance for this X-ray radiation field is elevated compared with a zero flux case because of the increased electron abundance.
Still stronger radiation fields increased the mean temperature to $10^3-10^4$\,K or above, and the ionized fractions of H and He to 10\% or more.

For weak X-ray irradiation the gas temperature and molecular abundances are strongly correlated with the local extinction at a given point in the cloud because the UV radiation field is stronger than the X-ray field.
For stronger irradiation this correlation disappears and the chemical and thermal properties of the gas depend almost entirely on gas density.

We studied the time-dependent response of the fractal cloud to a sudden increase in X-ray radiation intensity for a duration of 1 to 100 years, followed by a sudden decrease back to the original intensity.
This is a crude model of an X-ray flare from a variable source, such as Sgr A$^\star$ in the Galactic Centre, or an AGN or ultra-luminous X-ray source.
In one year the mass-weighted-mean gas temperature increased from $\sim30$\,K to $\gtrsim10^2$\,K, and the ionization fraction of H and He increased by more than an order of magnitude.
The abundances of molecular species do not change on this short timescale, however.

After a flare of 10 year duration, the gas temperature increased to $10^3$\,K, and H$^+$ fraction to $\sim0.01$, and the molecular species start to be affected.
The CO abundance decreases by more than an order of magnitude, whereas the H$_2$ abundance is unchanged.
For a flare of 25 years duration or more, the effects on the cloud are similar, with the temperature and H$^+$ fraction even larger and the CO almost completely destroyed, but H$_2$ again unaffected.
The temperature increase means that H$_2$ may become a major coolant in the molecular cloud and should emit brightly in the infrared.
It takes hundreds to thousands of years after the flare for the CO to re-form and reach a value close to its pre-flare abundance.
The main agent of CO destruction is the locally generated FUV radiation field, produced by H atoms and H$_2$ molecules that are excited by collisions with high-energy, non-thermal secondary electrons.
Only once the CO abundance is already very low does the He$^+$ destruction channel become important.

As a function of time, the CO-to-H$_2$ abundance decreases dramatically for flares of duration a few years or more.
Our main result is that CO is destroyed almost 100 times more rapidly than H$_2$, because of the different destruction channels of these molecules.
Our results show that some molecular clouds that have been exposed to recent intense X-ray radiation should be still out of chemical equilibrium, and we predict that some of these clouds will still have fully molecular hydrogen, but will contain very little CO.
These CO-dark clouds should remain deficient in CO for about $10^3$ years after a flare (depending on gas density, shorter for higher-density gas).
Depending on the frequency and intensity of X-ray flares, a molecular cloud near a flaring source could be permanently deficient in CO  but still be fully molecular as far as hydrogen is concerned.

For Galactic Centre clouds at $\gtrsim10$\,pc from Sgr A$^\star$ the irradiation from the strong X-ray flare about 100 years ago was not sufficiently strong to destroy CO, and in fact we predict that the CO abundance may actually have been enhanced by the X-ray irradiation.
Only for clouds within a parsec of Sgr A$^\star$ would significant CO destruction have occured.

\section*{Acknowledgements}
We are grateful to the referee for very useful comments and suggestions that have improved this paper.
We thank B.~Godard for a very useful discussion about modeling the X-ray absorption,  T.~Millar for discussions on the CO dissociation rate in UMIST12, R.~Meijerink for discussions on comparison with his results, and E.~Pellegrini for suggesting to use \textsc{Cloudy} for the comparison and providing preliminary results.
JM acknowledges funding from a Royal Society-Science Foundation Ireland University Research Fellowship (14/RS-URF/3219).
This research was funded by the ERC starting grant No. 679852 ``RADFEEDBACK''. We further acknowledge support by the Deutsche Forschungsgemeinschaft via priority program 1573, Physics of the Interstellar Medium.
SW thanks the Bonn-Cologne Graduate School, which is funded through the German Excellence Initiative.
DS acknowledges funding by the Deutsche Forschungsgemeinschaft (DFG)
via the Collaborative Research Center SFB 956 ``Conditions and Impact of
Star Formation'' (subproject C5).
SCOG acknowledges support from the Deutsche Forschungsgemeinschaft via SFB 881, ``The Milky Way System'' (sub-projects B1, B2 and B8), and from the European Research Council under the European Community's Seventh Framework Programme (FP7/2007 - 2013) via the ERC Advanced Grant ``STARLIGHT: Formation of the First Stars'' (project number 339177).
RW acknowledges support by the Albert Einstein Centre for Gravitation and Astrophysics via the Czech Science Foundation grant 14-37086G and by the institutional project RVO:67985815 of the Academy of Sciences of the Czech Republic.
The authors acknowledge the DJEI/DES/SFI/HEA Irish Centre for High-End Computing (ICHEC) for the provision of computational facilities and support.
The software used in this work was in part developed by the DOE-supported ASC/Alliance Center for Astrophysical Thermonuclear Flashes at the University of Chicago.
This research has made use of NASA's Astrophysics Data System.


\bibliographystyle{mnras}
\bibliography{./refs} 

\begin{thebibliography}{}
\makeatletter
\relax
\def\mn@urlcharsother{\let\do\@makeother \do\$\do\&\do\#\do\^\do\_\do\%\do\~}
\def\mn@doi{\begingroup\mn@urlcharsother \@ifnextchar [ {\mn@doi@}
  {\mn@doi@[]}}
\def\mn@doi@[#1]#2{\def\@tempa{#1}\ifx\@tempa\@empty \href
  {http://dx.doi.org/#2} {doi:#2}\else \href {http://dx.doi.org/#2} {#1}\fi
  \endgroup}
\def\mn@eprint#1#2{\mn@eprint@#1:#2::\@nil}
\def\mn@eprint@arXiv#1{\href {http://arxiv.org/abs/#1} {{\tt arXiv:#1}}}
\def\mn@eprint@dblp#1{\href {http://dblp.uni-trier.de/rec/bibtex/#1.xml}
  {dblp:#1}}
\def\mn@eprint@#1:#2:#3:#4\@nil{\def\@tempa {#1}\def\@tempb {#2}\def\@tempc
  {#3}\ifx \@tempc \@empty \let \@tempc \@tempb \let \@tempb \@tempa \fi \ifx
  \@tempb \@empty \def\@tempb {arXiv}\fi \@ifundefined
  {mn@eprint@\@tempb}{\@tempb:\@tempc}{\expandafter \expandafter \csname
  mn@eprint@\@tempb\endcsname \expandafter{\@tempc}}}

\bibitem[\protect\citeauthoryear{{Abel}, {Anninos}, {Zhang}  \&
  {Norman}}{{Abel} et~al.}{1997}]{AbeAnnZha97}
{Abel} T.,  {Anninos} P.,  {Zhang} Y.,   {Norman} M.~L.,  1997, \mn@doi [New
  Astronomy] {10.1016/S1384-1076(97)00010-9}, \href
  {http://adsabs.harvard.edu/abs/1997NewA....2..181A} {2, 181}

\bibitem[\protect\citeauthoryear{{Abrahamsson}, {Krems}  \&
  {Dalgarno}}{{Abrahamsson} et~al.}{2007}]{akd07}
{Abrahamsson} E.,  {Krems} R.~V.,   {Dalgarno} A.,  2007, \mn@doi [\apj]
  {10.1086/509631}, \href {http://adsabs.harvard.edu/abs/2007ApJ...654.1171A}
  {654, 1171}

\bibitem[\protect\citeauthoryear{{Ackermann} et~al.,}{{Ackermann}
  et~al.}{2011}]{AckAjeAll11}
{Ackermann} M.,  et~al., 2011, \mn@doi [\apj] {10.1088/0004-637X/743/2/171},
  \href {http://adsabs.harvard.edu/abs/2011ApJ...743..171A} {743, 171}

\bibitem[\protect\citeauthoryear{{Aharonian}}{{Aharonian}}{2013}]{Aha13}
{Aharonian} F.~A.,  2013, \mn@doi [Astroparticle Physics]
  {10.1016/j.astropartphys.2012.08.007}, \href
  {http://adsabs.harvard.edu/abs/2013APh....43...71A} {43, 71}

\bibitem[\protect\citeauthoryear{{Asplund}, {Grevesse}, {Sauval}  \&
  {Scott}}{{Asplund} et~al.}{2009}]{AspGreSau09}
{Asplund} M.,  {Grevesse} N.,  {Sauval} A.~J.,   {Scott} P.,  2009, \mn@doi
  [\araa] {10.1146/annurev.astro.46.060407.145222}, \href
  {http://adsabs.harvard.edu/abs/2009ARA%26A..47..481A} {47, 481}

\bibitem[\protect\citeauthoryear{{Bachetti}}{{Bachetti}}{2016}]{Bac16}
{Bachetti} M.,  2016, \mn@doi [Astronomische Nachrichten]
  {10.1002/asna.201612312}, \href
  {http://ukads.nottingham.ac.uk/abs/2016AN....337..349B} {337, 349}

\bibitem[\protect\citeauthoryear{{Baczynski}, {Glover}  \&
  {Klessen}}{{Baczynski} et~al.}{2015}]{bac15}
{Baczynski} C.,  {Glover} S.~C.~O.,   {Klessen} R.~S.,  2015, \mn@doi [\mnras]
  {10.1093/mnras/stv1906}, \href
  {http://adsabs.harvard.edu/abs/2015MNRAS.454..380B} {454, 380}

\bibitem[\protect\citeauthoryear{{Badnell}}{{Badnell}}{2006}]{Bad06}
{Badnell} N.~R.,  2006, \mn@doi [\apjs] {10.1086/508465}, \href
  {http://adsabs.harvard.edu/abs/2006ApJS..167..334B} {167, 334}

\bibitem[\protect\citeauthoryear{{Badnell} et~al.,}{{Badnell}
  et~al.}{2003}]{BadOMuSum03}
{Badnell} N.~R.,  et~al., 2003, \mn@doi [\aap] {10.1051/0004-6361:20030816},
  \href {http://adsabs.harvard.edu/abs/2003A%26A...406.1151B} {406, 1151}

\bibitem[\protect\citeauthoryear{{Bakes} \& {Tielens}}{{Bakes} \&
  {Tielens}}{1994}]{bt94}
{Bakes} E.~L.~O.,  {Tielens} A.~G.~G.~M.,  1994, \mn@doi [\apj]
  {10.1086/174188}, \href {http://adsabs.harvard.edu/abs/1994ApJ...427..822B}
  {427, 822}

\bibitem[\protect\citeauthoryear{{Barlow}}{{Barlow}}{1984}]{Bar84}
{Barlow} S.~E.,  1984, PhD thesis, UNIVERSITY OF COLORADO AT BOULDER.

\bibitem[\protect\citeauthoryear{{Barnes} \& {Hut}}{{Barnes} \&
  {Hut}}{1986}]{Barnes1986}
{Barnes} J.,  {Hut} P.,  1986, \mn@doi [\nat] {10.1038/324446a0}, \href
  {http://adsabs.harvard.edu/abs/1986Natur.324..446B} {324, 446}

\bibitem[\protect\citeauthoryear{{Bell}, {Berrington}  \& {Thomas}}{{Bell}
  et~al.}{1998}]{bbt98}
{Bell} K.~L.,  {Berrington} K.~A.,   {Thomas} M.~R.~J.,  1998, \mn@doi [\mnras]
  {10.1046/j.1365-8711.1998.01364.x}, \href
  {http://adsabs.harvard.edu/abs/1998MNRAS.293L..83B} {293, L83}

\bibitem[\protect\citeauthoryear{{Bisbas}, {Papadopoulos}  \& {Viti}}{{Bisbas}
  et~al.}{2015}]{BisPapVit15}
{Bisbas} T.~G.,  {Papadopoulos} P.~P.,   {Viti} S.,  2015, \mn@doi [\apj]
  {10.1088/0004-637X/803/1/37}, \href
  {http://adsabs.harvard.edu/abs/2015ApJ...803...37B} {803, 37}

\bibitem[\protect\citeauthoryear{{Bisbas}, {van Dishoeck}, {Papadopoulos},
  {Sz{\H u}cs}, {Bialy}  \& {Zhang}}{{Bisbas} et~al.}{2017}]{BisVanPap17}
{Bisbas} T.~G.,  {van Dishoeck} E.~F.,  {Papadopoulos} P.~P.,  {Sz{\H u}cs} L.,
   {Bialy} S.,   {Zhang} Z.-Y.,  2017, \mn@doi [\apj]
  {10.3847/1538-4357/aa696d}, \href
  {http://adsabs.harvard.edu/abs/2017ApJ...839...90B} {839, 90}

\bibitem[\protect\citeauthoryear{{Black}}{{Black}}{1981}]{black81}
{Black} J.~H.,  1981, \mn@doi [\mnras] {10.1093/mnras/197.3.553}, \href
  {http://adsabs.harvard.edu/abs/1981MNRAS.197..553B} {197, 553}

\bibitem[\protect\citeauthoryear{{Black} \& {Dalgarno}}{{Black} \&
  {Dalgarno}}{1977}]{bd77}
{Black} J.~H.,  {Dalgarno} A.,  1977, \mn@doi [\apjs] {10.1086/190455}, \href
  {http://adsabs.harvard.edu/abs/1977ApJS...34..405B} {34, 405}

\bibitem[\protect\citeauthoryear{{Brian} \& {Mitchell}}{{Brian} \&
  {Mitchell}}{1990}]{BriMit90}
{Brian} J.,  {Mitchell} A.,  1990, \mn@doi [\physrep]
  {10.1016/0370-1573(90)90159-Y}, \href
  {https://ui.adsabs.harvard.edu/#abs/1990PhR...186..215B} {186, 215}

\bibitem[\protect\citeauthoryear{{Brown}, {Byrne}  \& {Hindmarsh}}{{Brown}
  et~al.}{1989}]{Brown1989}
{Brown} P.~N.,  {Byrne} G.~D.,   {Hindmarsh} A.~C.,  1989, SIAM J. Sci. Stat.
  Comput., 10, 1038ff

\bibitem[\protect\citeauthoryear{{Burton}, {Hollenbach}  \& {Tielens}}{{Burton}
  et~al.}{1990}]{bht90}
{Burton} M.~G.,  {Hollenbach} D.~J.,   {Tielens} A.~G.~G.~M.,  1990, \mn@doi
  [\apj] {10.1086/169516}, \href
  {http://adsabs.harvard.edu/abs/1990ApJ...365..620B} {365, 620}

\bibitem[\protect\citeauthoryear{{Carty}, {Goddard}, {K{\"o}hler}, {Sims}  \&
  {Smith}}{{Carty} et~al.}{2006}]{CarGodKoh06}
{Carty} D.,  {Goddard} A.,  {K{\"o}hler} S.~P.~K.,  {Sims} I.~R.,   {Smith}
  I.~W.~M.,  2006, \mn@doi [Journal of Physical Chemistry A]
  {10.1021/jp054429u}, \href
  {http://adsabs.harvard.edu/abs/2006JPCA..110.3101C} {110, 3101}

\bibitem[\protect\citeauthoryear{{Caselli}, {Walmsley}, {Terzieva}  \&
  {Herbst}}{{Caselli} et~al.}{1998}]{CasWalTer98}
{Caselli} P.,  {Walmsley} C.~M.,  {Terzieva} R.,   {Herbst} E.,  1998, \mn@doi
  [\apj] {10.1086/305624}, \href
  {http://adsabs.harvard.edu/abs/1998ApJ...499..234C} {499, 234}

\bibitem[\protect\citeauthoryear{{Cen}}{{Cen}}{1992}]{cen92}
{Cen} R.,  1992, \mn@doi [\apjs] {10.1086/191630}, \href
  {http://adsabs.harvard.edu/abs/1992ApJS...78..341C} {78, 341}

\bibitem[\protect\citeauthoryear{{Churazov}, {Khabibullin}, {Sunyaev}  \&
  {Ponti}}{{Churazov} et~al.}{2017a}]{Churazov2017a}
{Churazov} E.,  {Khabibullin} I.,  {Sunyaev} R.,   {Ponti} G.,  2017a, \mn@doi
  [\mnras] {10.1093/mnras/stw2750}, \href
  {http://adsabs.harvard.edu/abs/2017MNRAS.465...45C} {465, 45}

\bibitem[\protect\citeauthoryear{{Churazov}, {Khabibullin}, {Ponti}  \&
  {Sunyaev}}{{Churazov} et~al.}{2017b}]{Churazov2017b}
{Churazov} E.,  {Khabibullin} I.,  {Ponti} G.,   {Sunyaev} R.,  2017b, \mn@doi
  [\mnras] {10.1093/mnras/stx443}, \href
  {http://adsabs.harvard.edu/abs/2017MNRAS.468..165C} {468, 165}

\bibitem[\protect\citeauthoryear{{Churazov}, {Khabibullin}, {Sunyaev}  \&
  {Ponti}}{{Churazov} et~al.}{2017c}]{Churazov2017c}
{Churazov} E.,  {Khabibullin} I.,  {Sunyaev} R.,   {Ponti} G.,  2017c, \mn@doi
  [\mnras] {10.1093/mnras/stx1855}, \href
  {http://adsabs.harvard.edu/abs/2017MNRAS.471.3293C} {471, 3293}

\bibitem[\protect\citeauthoryear{{Clark}, {Glover}  \& {Klessen}}{{Clark}
  et~al.}{2012}]{Clark2012}
{Clark} P.~C.,  {Glover} S.~C.~O.,   {Klessen} R.~S.,  2012, \mn@doi [\mnras]
  {10.1111/j.1365-2966.2011.20087.x}, \href
  {http://adsabs.harvard.edu/abs/2012MNRAS.420..745C} {420, 745}

\bibitem[\protect\citeauthoryear{{Clark}, {Glover}, {Ragan}, {Shetty}  \&
  {Klessen}}{{Clark} et~al.}{2013}]{ClaGloRag13}
{Clark} P.~C.,  {Glover} S.~C.~O.,  {Ragan} S.~E.,  {Shetty} R.,   {Klessen}
  R.~S.,  2013, \mn@doi [\apjl] {10.1088/2041-8205/768/2/L34}, \href
  {http://adsabs.harvard.edu/abs/2013ApJ...768L..34C} {768, L34}

\bibitem[\protect\citeauthoryear{{Cleeves}, {Bergin}, {{\"O}berg}, {Andrews},
  {Wilner}  \& {Loomis}}{{Cleeves} et~al.}{2017}]{CleBerObe17}
{Cleeves} L.~I.,  {Bergin} E.~A.,  {{\"O}berg} K.~I.,  {Andrews} S.~M.,
  {Wilner} D.~J.,   {Loomis} R.~A.,  2017, preprint, \href
  {http://adsabs.harvard.edu/abs/2017arXiv170600833C} {} (\mn@eprint {arXiv}
  {1706.00833})

\bibitem[\protect\citeauthoryear{{Cox}}{{Cox}}{2005}]{Cox05}
{Cox} D.~P.,  2005, \mn@doi [\araa] {10.1146/annurev.astro.43.072103.150615},
  \href {http://adsabs.harvard.edu/abs/2005ARAandA..43..337C} {43, 337}

\bibitem[\protect\citeauthoryear{{Dalgarno} \& {McCray}}{{Dalgarno} \&
  {McCray}}{1972}]{DalMcC72}
{Dalgarno} A.,  {McCray} R.~A.,  1972, \mn@doi [\araa]
  {10.1146/annurev.aa.10.090172.002111}, \href
  {http://adsabs.harvard.edu/abs/1972ARA%26A..10..375D} {10, 375}

\bibitem[\protect\citeauthoryear{{Dalgarno}, {Yan}  \& {Liu}}{{Dalgarno}
  et~al.}{1999}]{DalYanLiu99}
{Dalgarno} A.,  {Yan} M.,   {Liu} W.,  1999, \mn@doi [\apjs] {10.1086/313267},
  \href {http://adsabs.harvard.edu/abs/1999ApJS..125..237D} {125, 237}

\bibitem[\protect\citeauthoryear{{Draine}}{{Draine}}{1978}]{Dra78}
{Draine} B.~T.,  1978, \mn@doi [\apjs] {10.1086/190513}, \href
  {http://adsabs.harvard.edu/abs/1978ApJS...36..595D} {36, 595}

\bibitem[\protect\citeauthoryear{{Dufton} \& {Kingston}}{{Dufton} \&
  {Kingston}}{1991}]{dk91}
{Dufton} P.~L.,  {Kingston} A.~E.,  1991, \mn@doi [\mnras]
  {10.1093/mnras/248.4.827}, \href
  {http://adsabs.harvard.edu/abs/1991MNRAS.248..827D} {248, 827}

\bibitem[\protect\citeauthoryear{{Earnshaw} et~al.,}{{Earnshaw}
  et~al.}{2016}]{EarRobHei16}
{Earnshaw} H.~M.,  et~al., 2016, \mn@doi [\mnras] {10.1093/mnras/stv2945},
  \href {http://ukads.nottingham.ac.uk/abs/2016MNRAS.456.3840E} {456, 3840}

\bibitem[\protect\citeauthoryear{{Ferland}, {Peterson}, {Horne}, {Welsh}  \&
  {Nahar}}{{Ferland} et~al.}{1992}]{FerPetHor92}
{Ferland} G.~J.,  {Peterson} B.~M.,  {Horne} K.,  {Welsh} W.~F.,   {Nahar}
  S.~N.,  1992, \mn@doi [\apj] {10.1086/171063}, \href
  {http://adsabs.harvard.edu/abs/1992ApJ...387...95F} {387, 95}

\bibitem[\protect\citeauthoryear{{Ferland} et~al.,}{{Ferland}
  et~al.}{2013}]{FerPorVan13}
{Ferland} G.~J.,  et~al., 2013, \rmxaa, \href
  {http://adsabs.harvard.edu/abs/2013RMxAA..49..137F} {49, 137}

\bibitem[\protect\citeauthoryear{{Ferland} et~al.,}{{Ferland}
  et~al.}{2017}]{FerChaGuz17}
{Ferland} G.~J.,  et~al., 2017, \rmxaa, \href
  {http://adsabs.harvard.edu/abs/2017RMxAA..53..385F} {53, 385}

\bibitem[\protect\citeauthoryear{{Flower}}{{Flower}}{2001}]{fl01}
{Flower} D.~R.,  2001, \mn@doi [Journal of Physics B Atomic Molecular Physics]
  {10.1088/0953-4075/34/13/315}, \href
  {http://adsabs.harvard.edu/abs/2001JPhB...34.2731F} {34, 2731}

\bibitem[\protect\citeauthoryear{{Flower} \& {Launay}}{{Flower} \&
  {Launay}}{1977}]{fl77}
{Flower} D.~R.,  {Launay} J.~M.,  1977, \mn@doi [Journal of Physics B Atomic
  Molecular Physics] {10.1088/0022-3700/10/18/024}, \href
  {http://adsabs.harvard.edu/abs/1977JPhB...10.3673F} {10, 3673}

\bibitem[\protect\citeauthoryear{{Fryxell} et~al.,}{{Fryxell}
  et~al.}{2000}]{Fryxell2000}
{Fryxell} B.,  et~al., 2000, \mn@doi [\apjs] {10.1086/317361}, \href
  {http://adsabs.harvard.edu/abs/2000ApJS..131..273F} {131, 273}

\bibitem[\protect\citeauthoryear{{Gardner} et~al.,}{{Gardner}
  et~al.}{2006}]{GarMatCla06}
{Gardner} J.~P.,  et~al., 2006, \mn@doi [\ssr] {10.1007/s11214-006-8315-7},
  \href {http://adsabs.harvard.edu/abs/2006SSRv..123..485G} {123, 485}

\bibitem[\protect\citeauthoryear{{Gatto} et~al.,}{{Gatto}
  et~al.}{2017}]{Gatto2017}
{Gatto} A.,  et~al., 2017, \mn@doi [\mnras] {10.1093/mnras/stw3209}, \href
  {http://adsabs.harvard.edu/abs/2017MNRAS.466.1903G} {466, 1903}

\bibitem[\protect\citeauthoryear{{Girichidis} et~al.,}{{Girichidis}
  et~al.}{2016}]{Girichidis2016}
{Girichidis} P.,  et~al., 2016, \mn@doi [\mnras] {10.1093/mnras/stv2742}, \href
  {http://adsabs.harvard.edu/abs/2016MNRAS.456.3432G} {456, 3432}

\bibitem[\protect\citeauthoryear{{Glassgold}, {Najita}  \& {Igea}}{{Glassgold}
  et~al.}{1997}]{GlaNajIge97}
{Glassgold} A.~E.,  {Najita} J.,   {Igea} J.,  1997, \mn@doi [\apj]
  {10.1086/303952}, \href {http://adsabs.harvard.edu/abs/1997ApJ...480..344G}
  {480, 344}

\bibitem[\protect\citeauthoryear{{Glover}}{{Glover}}{2003}]{Glo03}
{Glover} S.~C.~O.,  2003, \mn@doi [\apj] {10.1086/345684}, \href
  {http://adsabs.harvard.edu/abs/2003ApJ...584..331G} {584, 331}

\bibitem[\protect\citeauthoryear{{Glover} \& {Abel}}{{Glover} \&
  {Abel}}{2008}]{ga08}
{Glover} S.~C.~O.,  {Abel} T.,  2008, \mn@doi [\mnras]
  {10.1111/j.1365-2966.2008.13224.x}, \href
  {http://adsabs.harvard.edu/abs/2008MNRAS.388.1627G} {388, 1627}

\bibitem[\protect\citeauthoryear{{Glover} \& {Clark}}{{Glover} \&
  {Clark}}{2012}]{GloCla12}
{Glover} S.~C.~O.,  {Clark} P.~C.,  2012, \mn@doi [\mnras]
  {10.1111/j.1365-2966.2011.20260.x}, \href
  {http://adsabs.harvard.edu/abs/2012MNRAS.421..116G} {421, 116}

\bibitem[\protect\citeauthoryear{{Glover} \& {Jappsen}}{{Glover} \&
  {Jappsen}}{2007}]{gj07}
{Glover} S.~C.~O.,  {Jappsen} A.-K.,  2007, \mn@doi [\apj] {10.1086/519445},
  \href {http://adsabs.harvard.edu/abs/2007ApJ...666....1G} {666, 1}

\bibitem[\protect\citeauthoryear{{Glover} \& {Mac Low}}{{Glover} \& {Mac
  Low}}{2007a}]{GloMac07a}
{Glover} S.~C.~O.,  {Mac Low} M.-M.,  2007a, \mn@doi [\apjs] {10.1086/512238},
  \href {http://adsabs.harvard.edu/abs/2007ApJS..169..239G} {169, 239}

\bibitem[\protect\citeauthoryear{{Glover} \& {Mac Low}}{{Glover} \& {Mac
  Low}}{2007b}]{GloMac07b}
{Glover} S.~C.~O.,  {Mac Low} M.-M.,  2007b, \mn@doi [\apj] {10.1086/512227},
  \href {http://adsabs.harvard.edu/abs/2007ApJ...659.1317G} {659, 1317}

\bibitem[\protect\citeauthoryear{{Glover} \& {Savin}}{{Glover} \&
  {Savin}}{2009}]{GloSav09}
{Glover} S.~C.~O.,  {Savin} D.~W.,  2009, \mn@doi [\mnras]
  {10.1111/j.1365-2966.2008.14156.x}, \href
  {http://adsabs.harvard.edu/abs/2009MNRAS.393..911G} {393, 911}

\bibitem[\protect\citeauthoryear{{Glover}, {Federrath}, {Mac Low}  \&
  {Klessen}}{{Glover} et~al.}{2010}]{g10}
{Glover} S.~C.~O.,  {Federrath} C.,  {Mac Low} M.-M.,   {Klessen} R.~S.,  2010,
  \mn@doi [\mnras] {10.1111/j.1365-2966.2009.15718.x}, \href
  {http://adsabs.harvard.edu/abs/2010MNRAS.404....2G} {404, 2}

\bibitem[\protect\citeauthoryear{{Gnat} \& {Ferland}}{{Gnat} \&
  {Ferland}}{2012}]{gf12}
{Gnat} O.,  {Ferland} G.~J.,  2012, \mn@doi [\apjs]
  {10.1088/0067-0049/199/1/20}, \href
  {http://adsabs.harvard.edu/abs/2012ApJS..199...20G} {199, 20}

\bibitem[\protect\citeauthoryear{{Goldsmith} \& {Langer}}{{Goldsmith} \&
  {Langer}}{1978}]{gl78}
{Goldsmith} P.~F.,  {Langer} W.~D.,  1978, \mn@doi [\apj] {10.1086/156206},
  \href {http://adsabs.harvard.edu/abs/1978ApJ...222..881G} {222, 881}

\bibitem[\protect\citeauthoryear{{Gong}, {Ostriker}  \& {Wolfire}}{{Gong}
  et~al.}{2017}]{GonOstWol17}
{Gong} M.,  {Ostriker} E.~C.,   {Wolfire} M.~G.,  2017, \mn@doi [\apj]
  {10.3847/1538-4357/aa7561}, \href
  {http://adsabs.harvard.edu/abs/2017ApJ...843...38G} {843, 38}

\bibitem[\protect\citeauthoryear{{G{\'o}rski}, {Hivon}, {Banday}, {Wandelt},
  {Hansen}, {Reinecke}  \& {Bartelmann}}{{G{\'o}rski}
  et~al.}{2005}]{Gorski2005}
{G{\'o}rski} K.~M.,  {Hivon} E.,  {Banday} A.~J.,  {Wandelt} B.~D.,  {Hansen}
  F.~K.,  {Reinecke} M.,   {Bartelmann} M.,  2005, \mn@doi [\apj]
  {10.1086/427976}, \href {http://adsabs.harvard.edu/abs/2005ApJ...622..759G}
  {622, 759}

\bibitem[\protect\citeauthoryear{{Gredel}, {Lepp}  \& {Dalgarno}}{{Gredel}
  et~al.}{1987}]{GreLepDal87}
{Gredel} R.,  {Lepp} S.,   {Dalgarno} A.,  1987, \mn@doi [\apjl]
  {10.1086/185073}, \href {http://adsabs.harvard.edu/abs/1987ApJ...323L.137G}
  {323, L137}

\bibitem[\protect\citeauthoryear{{Gredel}, {Lepp}, {Dalgarno}  \&
  {Herbst}}{{Gredel} et~al.}{1989}]{GreLepDal89}
{Gredel} R.,  {Lepp} S.,  {Dalgarno} A.,   {Herbst} E.,  1989, \mn@doi [\apj]
  {10.1086/168117}, \href
  {https://ui.adsabs.harvard.edu/#abs/1989ApJ...347..289G} {347, 289}

\bibitem[\protect\citeauthoryear{{H.~E.~S.~S.~Collaboration}, {:}, {Abdalla},
  {Abramowski}, {Aharonian}  \& et al.}{{H.~E.~S.~S.~Collaboration}
  et~al.}{2017}]{HESS17}
{H.~E.~S.~S.~Collaboration} {:} {Abdalla} H.,  {Abramowski} A.,  {Aharonian}
  F.,   et al. 2017, \aap, in press, arXiv:1706.04535, \href
  {http://adsabs.harvard.edu/abs/2017arXiv170604535H} {}

\bibitem[\protect\citeauthoryear{{Heays}, {Bosman}  \& {van Dishoeck}}{{Heays}
  et~al.}{2017}]{HeaBosVan17}
{Heays} A.~N.,  {Bosman} A.~D.,   {van Dishoeck} E.~F.,  2017, \mn@doi [\aap]
  {10.1051/0004-6361/201628742}, \href
  {https://ui.adsabs.harvard.edu/#abs/2017A&A...602A.105H} {602, A105}

\bibitem[\protect\citeauthoryear{{Hocuk} \& {Spaans}}{{Hocuk} \&
  {Spaans}}{2010}]{HocSpa10}
{Hocuk} S.,  {Spaans} M.,  2010, \mn@doi [\aap] {10.1051/0004-6361/201015055},
  \href {http://adsabs.harvard.edu/abs/2010A%26A...522A..24H} {522, A24}

\bibitem[\protect\citeauthoryear{{Hollenbach} \& {McKee}}{{Hollenbach} \&
  {McKee}}{1979}]{HolMcK79}
{Hollenbach} D.,  {McKee} C.~F.,  1979, \mn@doi [\apjs] {10.1086/190631}, \href
  {http://adsabs.harvard.edu/abs/1979ApJS...41..555H} {41, 555}

\bibitem[\protect\citeauthoryear{{Hollenbach} \& {McKee}}{{Hollenbach} \&
  {McKee}}{1989}]{hm89}
{Hollenbach} D.,  {McKee} C.~F.,  1989, \mn@doi [\apj] {10.1086/167595}, \href
  {http://adsabs.harvard.edu/abs/1989ApJ...342..306H} {342, 306}

\bibitem[\protect\citeauthoryear{{Hummer} \& {Storey}}{{Hummer} \&
  {Storey}}{1998}]{HumSto98}
{Hummer} D.~G.,  {Storey} P.~J.,  1998, \mn@doi [\mnras]
  {10.1046/j.1365-8711.1998.2970041073.x}, \href
  {http://adsabs.harvard.edu/abs/1998MNRAS.297.1073H} {297, 1073}

\bibitem[\protect\citeauthoryear{{Johnson}, {Burke}  \& {Kingston}}{{Johnson}
  et~al.}{1987}]{joh87}
{Johnson} C.~T.,  {Burke} P.~G.,   {Kingston} A.~E.,  1987, \mn@doi [Journal of
  Physics B Atomic Molecular Physics] {10.1088/0022-3700/20/11/022}, \href
  {http://adsabs.harvard.edu/abs/1987JPhB...20.2553J} {20, 2553}

\bibitem[\protect\citeauthoryear{{Kalirai}}{{Kalirai}}{2018}]{Kal18}
{Kalirai} J.,  2018, \mn@doi [Contemporary Physics]
  {10.1080/00107514.2018.1467648}, \href
  {https://ui.adsabs.harvard.edu/#abs/2018ConPh..59..251K} {59, 251}

\bibitem[\protect\citeauthoryear{{Karpas}, {Anicich}  \& {Huntress}}{{Karpas}
  et~al.}{1979}]{KarAniHun79}
{Karpas} Z.,  {Anicich} V.,   {Huntress} W.~T.,  1979, \mn@doi [\jcp]
  {10.1063/1.437823}, \href {http://adsabs.harvard.edu/abs/1979JChPh..70.2877K}
  {70, 2877}

\bibitem[\protect\citeauthoryear{{Keenan}, {Lennon}, {Johnson}  \&
  {Kingston}}{{Keenan} et~al.}{1986}]{k86}
{Keenan} F.~P.,  {Lennon} D.~J.,  {Johnson} C.~T.,   {Kingston} A.~E.,  1986,
  \mn@doi [\mnras] {10.1093/mnras/220.3.571}, \href
  {http://adsabs.harvard.edu/abs/1986MNRAS.220..571K} {220, 571}

\bibitem[\protect\citeauthoryear{{Kim}, {Theard}  \& {Huntress}}{{Kim}
  et~al.}{1975}]{KimTheHun75}
{Kim} J.~K.,  {Theard} L.~P.,   {Huntress} Jr. W.~T.,  1975, \mn@doi [Chemical
  Physics Letters] {10.1016/0009-2614(75)85252-3}, \href
  {http://adsabs.harvard.edu/abs/1975CPL....32..610K} {32, 610}

\bibitem[\protect\citeauthoryear{{Koyama}, {Maeda}, {Sonobe}, {Takeshima},
  {Tanaka}  \& {Yamauchi}}{{Koyama} et~al.}{1996}]{KoyMaeSon96}
{Koyama} K.,  {Maeda} Y.,  {Sonobe} T.,  {Takeshima} T.,  {Tanaka} Y.,
  {Yamauchi} S.,  1996, \mn@doi [\pasj] {10.1093/pasj/48.2.249}, \href
  {http://adsabs.harvard.edu/abs/1996PASJ...48..249K} {48, 249}

\bibitem[\protect\citeauthoryear{{Krivonos} et~al.,}{{Krivonos}
  et~al.}{2017}]{Krivonos2017}
{Krivonos} R.,  et~al., 2017, \mn@doi [\mnras] {10.1093/mnras/stx585}, \href
  {http://adsabs.harvard.edu/abs/2017MNRAS.468.2822K} {468, 2822}

\bibitem[\protect\citeauthoryear{{Lepp} \& {Dalgarno}}{{Lepp} \&
  {Dalgarno}}{1996}]{LepDal96}
{Lepp} S.,  {Dalgarno} A.,  1996, \aap, \href
  {http://adsabs.harvard.edu/abs/1996A%26A...306L..21L} {306, L21}

\bibitem[\protect\citeauthoryear{{Lepp} \& {McCray}}{{Lepp} \&
  {McCray}}{1983}]{LepMcC83}
{Lepp} S.,  {McCray} R.,  1983, \mn@doi [\apj] {10.1086/161062}, \href
  {http://adsabs.harvard.edu/abs/1983ApJ...269..560L} {269, 560}

\bibitem[\protect\citeauthoryear{{Lepp} \& {Shull}}{{Lepp} \&
  {Shull}}{1983}]{LepShu83}
{Lepp} S.,  {Shull} J.~M.,  1983, \mn@doi [\apj] {10.1086/161149}, \href
  {http://adsabs.harvard.edu/abs/1983ApJ...270..578L} {270, 578}

\bibitem[\protect\citeauthoryear{{Liszt}}{{Liszt}}{2003}]{Lis03}
{Liszt} H.,  2003, \mn@doi [\aap] {10.1051/0004-6361:20021660}, \href
  {https://ui.adsabs.harvard.edu/#abs/2003A&A...398..621L} {398, 621}

\bibitem[\protect\citeauthoryear{{Mac Low} \& {Shull}}{{Mac Low} \&
  {Shull}}{1986}]{MacShu86}
{Mac Low} M.-M.,  {Shull} J.~M.,  1986, \mn@doi [\apj] {10.1086/164017}, \href
  {http://adsabs.harvard.edu/abs/1986ApJ...302..585M} {302, 585}

\bibitem[\protect\citeauthoryear{{Mackey} \& {Lim}}{{Mackey} \&
  {Lim}}{2010}]{MacLim10}
{Mackey} J.,  {Lim} A.~J.,  2010, \mn@doi [\mnras]
  {10.1111/j.1365-2966.2009.16181.x}, \href
  {http://adsabs.harvard.edu/abs/2010MNRAS.403..714M} {403, 714}

\bibitem[\protect\citeauthoryear{{Maloney}, {Hollenbach}  \&
  {Tielens}}{{Maloney} et~al.}{1996}]{MalHolTie96}
{Maloney} P.~R.,  {Hollenbach} D.~J.,   {Tielens} A.~G.~G.~M.,  1996, \mn@doi
  [\apj] {10.1086/177532}, \href
  {http://adsabs.harvard.edu/abs/1996ApJ...466..561M} {466, 561}

\bibitem[\protect\citeauthoryear{{Martin}, {Schwarz}  \& {Mandy}}{{Martin}
  et~al.}{1996}]{MarSchMan96}
{Martin} P.~G.,  {Schwarz} D.~H.,   {Mandy} M.~E.,  1996, \mn@doi [\apj]
  {10.1086/177053}, \href {http://adsabs.harvard.edu/abs/1996ApJ...461..265M}
  {461, 265}

\bibitem[\protect\citeauthoryear{{Martin}, {Keogh}  \& {Mandy}}{{Martin}
  et~al.}{1998}]{MarKeoMan98}
{Martin} P.~G.,  {Keogh} W.~J.,   {Mandy} M.~E.,  1998, \mn@doi [\apj]
  {10.1086/305665}, \href {http://adsabs.harvard.edu/abs/1998ApJ...499..793M}
  {499, 793}

\bibitem[\protect\citeauthoryear{{Mathis}, {Rumpl}  \& {Nordsieck}}{{Mathis}
  et~al.}{1977}]{MatRumNor77}
{Mathis} J.~S.,  {Rumpl} W.,   {Nordsieck} K.~H.,  1977, \mn@doi [\apj]
  {10.1086/155591}, \href {http://adsabs.harvard.edu/abs/1977ApJ...217..425M}
  {217, 425}

\bibitem[\protect\citeauthoryear{{McCall} et~al.,}{{McCall}
  et~al.}{2004}]{McCHunSay04}
{McCall} B.~J.,  et~al., 2004, \mn@doi [\pra] {10.1103/PhysRevA.70.052716},
  \href {http://adsabs.harvard.edu/abs/2004PhRvA..70e2716M} {70, 052716}

\bibitem[\protect\citeauthoryear{{McElroy}, {Walsh}, {Markwick}, {Cordiner},
  {Smith}  \& {Millar}}{{McElroy} et~al.}{2013}]{McEWalMar12}
{McElroy} D.,  {Walsh} C.,  {Markwick} A.~J.,  {Cordiner} M.~A.,  {Smith} K.,
  {Millar} T.~J.,  2013, \mn@doi [\aap] {10.1051/0004-6361/201220465}, \href
  {http://adsabs.harvard.edu/abs/2013A%26A...550A..36M} {550, A36}

\bibitem[\protect\citeauthoryear{{Meijerink} \& {Spaans}}{{Meijerink} \&
  {Spaans}}{2005}]{MeiSpa05}
{Meijerink} R.,  {Spaans} M.,  2005, \mn@doi [\aap]
  {10.1051/0004-6361:20042398}, \href
  {http://adsabs.harvard.edu/abs/2005A%26A...436..397M} {436, 397}

\bibitem[\protect\citeauthoryear{{Meijerink}, {Spaans}  \&
  {Israel}}{{Meijerink} et~al.}{2006}]{MeiSpaIsr06}
{Meijerink} R.,  {Spaans} M.,   {Israel} F.~P.,  2006, \mn@doi [\apjl]
  {10.1086/508938}, \href {http://adsabs.harvard.edu/abs/2006ApJ...650L.103M}
  {650, L103}

\bibitem[\protect\citeauthoryear{{Meijerink}, {Spaans}, {Loenen}  \& {van der
  Werf}}{{Meijerink} et~al.}{2011}]{MeiSpaLoe11}
{Meijerink} R.,  {Spaans} M.,  {Loenen} A.~F.,   {van der Werf} P.~P.,  2011,
  \mn@doi [\aap] {10.1051/0004-6361/201015136}, \href
  {http://adsabs.harvard.edu/abs/2011A%26A...525A.119M} {525, A119}

\bibitem[\protect\citeauthoryear{{Mezcua}, {Roberts}, {Sutton}  \&
  {Lobanov}}{{Mezcua} et~al.}{2013}]{MezRobSut13}
{Mezcua} M.,  {Roberts} T.~P.,  {Sutton} A.~D.,   {Lobanov} A.~P.,  2013,
  \mn@doi [\mnras] {10.1093/mnras/stt1794}, \href
  {http://ukads.nottingham.ac.uk/abs/2013MNRAS.436.3128M} {436, 3128}

\bibitem[\protect\citeauthoryear{{Mezcua}, {Civano}, {Fabbiano}, {Miyaji}  \&
  {Marchesi}}{{Mezcua} et~al.}{2016}]{MezCivFab16}
{Mezcua} M.,  {Civano} F.,  {Fabbiano} G.,  {Miyaji} T.,   {Marchesi} S.,
  2016, \mn@doi [\apj] {10.3847/0004-637X/817/1/20}, \href
  {http://ukads.nottingham.ac.uk/abs/2016ApJ...817...20M} {817, 20}

\bibitem[\protect\citeauthoryear{{Micic}, {Glover}, {Federrath}  \&
  {Klessen}}{{Micic} et~al.}{2012}]{MicGloFed12}
{Micic} M.,  {Glover} S.~C.~O.,  {Federrath} C.,   {Klessen} R.~S.,  2012,
  \mn@doi [\mnras] {10.1111/j.1365-2966.2012.20477.x}, \href
  {http://adsabs.harvard.edu/abs/2012MNRAS.421.2531M} {421, 2531}

\bibitem[\protect\citeauthoryear{{Mills}, {G{\"u}sten}, {Requena-Torres}  \&
  {Morris}}{{Mills} et~al.}{2013}]{Mills2013}
{Mills} E.~A.~C.,  {G{\"u}sten} R.,  {Requena-Torres} M.~A.,   {Morris} M.~R.,
  2013, \mn@doi [\apj] {10.1088/0004-637X/779/1/47}, \href
  {http://adsabs.harvard.edu/abs/2013ApJ...779...47M} {779, 47}

\bibitem[\protect\citeauthoryear{{Molaro}, {Khatri}  \& {Sunyaev}}{{Molaro}
  et~al.}{2016}]{MolKhaSun16}
{Molaro} M.,  {Khatri} R.,   {Sunyaev} R.~A.,  2016, \mn@doi [\aap]
  {10.1051/0004-6361/201527760}, \href
  {http://adsabs.harvard.edu/abs/2016A%26A...589A..88M} {589, A88}

\bibitem[\protect\citeauthoryear{{Morrison} \& {McCammon}}{{Morrison} \&
  {McCammon}}{1983}]{MorMcC83}
{Morrison} R.,  {McCammon} D.,  1983, \mn@doi [\apj] {10.1086/161102}, \href
  {http://adsabs.harvard.edu/abs/1983ApJ...270..119M} {270, 119}

\bibitem[\protect\citeauthoryear{{Moser} et~al.,}{{Moser}
  et~al.}{2017}]{Moser2016}
{Moser} L.,  et~al., 2017, \mn@doi [\aap] {10.1051/0004-6361/201628385}, 603,
  A68

\bibitem[\protect\citeauthoryear{{Nelson} \& {Langer}}{{Nelson} \&
  {Langer}}{1997}]{NelLan97}
{Nelson} R.~P.,  {Langer} W.~D.,  1997, \apj, \href
  {http://adsabs.harvard.edu/abs/1997ApJ...482..796N} {482, 796}

\bibitem[\protect\citeauthoryear{{Nelson} \& {Langer}}{{Nelson} \&
  {Langer}}{1999}]{NelLan99}
{Nelson} R.~P.,  {Langer} W.~D.,  1999, \mn@doi [\apj] {10.1086/307823}, \href
  {http://adsabs.harvard.edu/abs/1999ApJ...524..923N} {524, 923}

\bibitem[\protect\citeauthoryear{{Neufeld} \& {Kaufman}}{{Neufeld} \&
  {Kaufman}}{1993}]{nk93}
{Neufeld} D.~A.,  {Kaufman} M.~J.,  1993, \mn@doi [\apj] {10.1086/173388},
  \href {http://adsabs.harvard.edu/abs/1993ApJ...418..263N} {418, 263}

\bibitem[\protect\citeauthoryear{{Neufeld}, {Lepp}  \& {Melnick}}{{Neufeld}
  et~al.}{1995}]{nlm95}
{Neufeld} D.~A.,  {Lepp} S.,   {Melnick} G.~J.,  1995, \mn@doi [\apjs]
  {10.1086/192211}, \href {http://adsabs.harvard.edu/abs/1995ApJS..100..132N}
  {100, 132}

\bibitem[\protect\citeauthoryear{{Odaka}, {Aharonian}, {Watanabe}, {Tanaka},
  {Khangulyan}  \& {Takahashi}}{{Odaka} et~al.}{2011}]{OdaAhaWat11}
{Odaka} H.,  {Aharonian} F.,  {Watanabe} S.,  {Tanaka} Y.,  {Khangulyan} D.,
  {Takahashi} T.,  2011, \mn@doi [\apj] {10.1088/0004-637X/740/2/103}, \href
  {http://adsabs.harvard.edu/abs/2011ApJ...740..103O} {740, 103}

\bibitem[\protect\citeauthoryear{{Osterbrock}}{{Osterbrock}}{1989}]{Ost89}
{Osterbrock} D.~E.,  1989, {Astrophysics of gaseous nebulae and active galactic
  nuclei}.
University Science Books, Mill Valley, CA

\bibitem[\protect\citeauthoryear{{Palla}, {Salpeter}  \& {Stahler}}{{Palla}
  et~al.}{1983}]{PalSalSta83}
{Palla} F.,  {Salpeter} E.~E.,   {Stahler} S.~W.,  1983, \mn@doi [\apj]
  {10.1086/161231}, \href {http://adsabs.harvard.edu/abs/1983ApJ...271..632P}
  {271, 632}

\bibitem[\protect\citeauthoryear{{Panoglou}, {Cabrit}, {Pineau Des For{\^e}ts},
  {Garcia}, {Ferreira}  \& {Casse}}{{Panoglou} et~al.}{2012}]{PanCabPin12}
{Panoglou} D.,  {Cabrit} S.,  {Pineau Des For{\^e}ts} G.,  {Garcia} P.~J.~V.,
  {Ferreira} J.,   {Casse} F.,  2012, \mn@doi [\aap]
  {10.1051/0004-6361/200912861}, \href
  {http://adsabs.harvard.edu/abs/2012A%26A...538A...2P} {538, A2}

\bibitem[\protect\citeauthoryear{{Pequignot}}{{Pequignot}}{1990}]{p90}
{Pequignot} D.,  1990, \aap, \href
  {http://adsabs.harvard.edu/abs/1990A%26A...231..499P} {231, 499}

\bibitem[\protect\citeauthoryear{{Pequignot}}{{Pequignot}}{1996}]{p96}
{Pequignot} D.,  1996, \aap, \href
  {http://adsabs.harvard.edu/abs/1996A%26A...313.1026P} {313, 1026}

\bibitem[\protect\citeauthoryear{{Peters} et~al.,}{{Peters}
  et~al.}{2017}]{Peters2017}
{Peters} T.,  et~al., 2017, \mn@doi [\mnras] {10.1093/mnras/stw3216}, \href
  {http://adsabs.harvard.edu/abs/2017MNRAS.466.3293P} {466, 3293}

\bibitem[\protect\citeauthoryear{{Petuchowski}, {Dwek}, {Allen}  \&
  {Nuth}}{{Petuchowski} et~al.}{1989}]{PetDweAll89}
{Petuchowski} S.~J.,  {Dwek} E.,  {Allen} Jr. J.~E.,   {Nuth} III J.~A.,  1989,
  \mn@doi [\apj] {10.1086/167601}, \href
  {http://adsabs.harvard.edu/abs/1989ApJ...342..406P} {342, 406}

\bibitem[\protect\citeauthoryear{{Pfrommer}, {Pakmor}, {Schaal}, {Simpson}  \&
  {Springel}}{{Pfrommer} et~al.}{2017}]{PfrPakSch17}
{Pfrommer} C.,  {Pakmor} R.,  {Schaal} K.,  {Simpson} C.~M.,   {Springel} V.,
  2017, \mn@doi [\mnras] {10.1093/mnras/stw2941}, \href
  {http://adsabs.harvard.edu/abs/2017MNRAS.465.4500P} {465, 4500}

\bibitem[\protect\citeauthoryear{{Ponti}, {Terrier}, {Goldwurm}, {Belanger}  \&
  {Trap}}{{Ponti} et~al.}{2010}]{PonTerGol10}
{Ponti} G.,  {Terrier} R.,  {Goldwurm} A.,  {Belanger} G.,   {Trap} G.,  2010,
  \mn@doi [\apj] {10.1088/0004-637X/714/1/732}, \href
  {http://adsabs.harvard.edu/abs/2010ApJ...714..732P} {714, 732}

\bibitem[\protect\citeauthoryear{{Ponti} et~al.,}{{Ponti}
  et~al.}{2015}]{Ponti2015}
{Ponti} G.,  et~al., 2015, \mn@doi [\mnras] {10.1093/mnras/stv1331}, \href
  {http://adsabs.harvard.edu/abs/2015MNRAS.453..172P} {453, 172}

\bibitem[\protect\citeauthoryear{{Prasad} \& {Huntress}}{{Prasad} \&
  {Huntress}}{1980}]{PraHun80}
{Prasad} S.~S.,  {Huntress} Jr. W.~T.,  1980, \mn@doi [\apjs] {10.1086/190665},
  \href {http://adsabs.harvard.edu/abs/1980ApJS...43....1P} {43, 1}

\bibitem[\protect\citeauthoryear{{Prasad} \& {Tarafdar}}{{Prasad} \&
  {Tarafdar}}{1983}]{PraTar83}
{Prasad} S.~S.,  {Tarafdar} S.~P.,  1983, \mn@doi [\apj] {10.1086/160896},
  \href {http://adsabs.harvard.edu/abs/1983ApJ...267..603P} {267, 603}

\bibitem[\protect\citeauthoryear{{Punsly} \& {Rodriguez}}{{Punsly} \&
  {Rodriguez}}{2013}]{Punsly2013}
{Punsly} B.,  {Rodriguez} J.,  2013, \mn@doi [\apj]
  {10.1088/0004-637X/764/2/173}, \href
  {http://adsabs.harvard.edu/abs/2013ApJ...764..173P} {764, 173}

\bibitem[\protect\citeauthoryear{{R{\"o}llig} et~al.,}{{R{\"o}llig}
  et~al.}{2007}]{RoeAbeBel07}
{R{\"o}llig} M.,  et~al., 2007, \mn@doi [\aap] {10.1051/0004-6361:20065918},
  \href {http://adsabs.harvard.edu/abs/2007A%26A...467..187R} {467, 187}

\bibitem[\protect\citeauthoryear{{Roueff}}{{Roueff}}{1990}]{r90}
{Roueff} E.,  1990, \aap, \href
  {http://adsabs.harvard.edu/abs/1990A%26A...234..567R} {234, 567}

\bibitem[\protect\citeauthoryear{{Roueff} \& {Le Bourlot}}{{Roueff} \& {Le
  Bourlot}}{1990}]{rlb90}
{Roueff} E.,  {Le Bourlot} J.,  1990, \aap, \href
  {http://adsabs.harvard.edu/abs/1990A%26A...236..515R} {236, 515}

\bibitem[\protect\citeauthoryear{{Savin}, {Krsti{\'c}}, {Haiman}  \&
  {Stancil}}{{Savin} et~al.}{2004}]{SavKrsPre04}
{Savin} D.~W.,  {Krsti{\'c}} P.~S.,  {Haiman} Z.,   {Stancil} P.~C.,  2004,
  \mn@doi [\apj] {10.1086/421108}, \href
  {https://ui.adsabs.harvard.edu/#abs/2004ApJ...606L.167S} {606, L167}

\bibitem[\protect\citeauthoryear{{Schauer}, {Jefferts}, {Barlow}  \&
  {Dunn}}{{Schauer} et~al.}{1989}]{SchJefBar89}
{Schauer} M.~M.,  {Jefferts} S.~R.,  {Barlow} S.~E.,   {Dunn} G.~H.,  1989,
  \mn@doi [\jcp] {10.1063/1.456748}, \href
  {http://adsabs.harvard.edu/abs/1989JChPh..91.4593S} {91, 4593}

\bibitem[\protect\citeauthoryear{{Schroder}, {Staemmler}, {Smith}, {Flower}  \&
  {Jaquet}}{{Schroder} et~al.}{1991}]{sch91}
{Schroder} K.,  {Staemmler} V.,  {Smith} M.~D.,  {Flower} D.~R.,   {Jaquet} R.,
   1991, \mn@doi [Journal of Physics B Atomic Molecular Physics]
  {10.1088/0953-4075/24/10/007}, \href
  {http://adsabs.harvard.edu/abs/1991JPhB...24.2487S} {24, 2487}

\bibitem[\protect\citeauthoryear{{Seifried} et~al.,}{{Seifried}
  et~al.}{2017}]{SeiWalGir17}
{Seifried} D.,  et~al., 2017, \mn@doi [\mnras] {10.1093/mnras/stx2343}, \href
  {http://adsabs.harvard.edu/abs/2017MNRAS.472.4797S} {472, 4797}

\bibitem[\protect\citeauthoryear{{Shadmehri} \& {Elmegreen}}{{Shadmehri} \&
  {Elmegreen}}{2011}]{Shadmehri2011}
{Shadmehri} M.,  {Elmegreen} B.~G.,  2011, \mn@doi [\mnras]
  {10.1111/j.1365-2966.2010.17481.x}, \href
  {http://adsabs.harvard.edu/abs/2011MNRAS.410..788S} {410, 788}

\bibitem[\protect\citeauthoryear{{Shang}, {Glassgold}, {Shu}  \&
  {Lizano}}{{Shang} et~al.}{2002}]{ShaGlaShu02}
{Shang} H.,  {Glassgold} A.~E.,  {Shu} F.~H.,   {Lizano} S.,  2002, \mn@doi
  [\apj] {10.1086/324197}, \href
  {http://adsabs.harvard.edu/abs/2002ApJ...564..853S} {564, 853}

\bibitem[\protect\citeauthoryear{{Shapiro} \& {Kang}}{{Shapiro} \&
  {Kang}}{1987}]{ShaKan87}
{Shapiro} P.~R.,  {Kang} H.,  1987, \mn@doi [\apj] {10.1086/165350}, \href
  {http://adsabs.harvard.edu/abs/1987ApJ...318...32S} {318, 32}

\bibitem[\protect\citeauthoryear{{Shull} \& {van Steenberg}}{{Shull} \& {van
  Steenberg}}{1985}]{ShuSte85}
{Shull} J.~M.,  {van Steenberg} M.~E.,  1985, \mn@doi [\apj] {10.1086/163605},
  \href {http://adsabs.harvard.edu/abs/1985ApJ...298..268S} {298, 268}

\bibitem[\protect\citeauthoryear{{Silva} \& {Viegas}}{{Silva} \&
  {Viegas}}{2002}]{sv02}
{Silva} A.~I.,  {Viegas} S.~M.,  2002, \mn@doi [\mnras]
  {10.1046/j.1365-8711.2002.04956.x}, \href
  {http://adsabs.harvard.edu/abs/2002MNRAS.329..135S} {329, 135}

\bibitem[\protect\citeauthoryear{{Spitzer} \& {Tomasko}}{{Spitzer} \&
  {Tomasko}}{1968}]{SpiTom68}
{Spitzer} Jr. L.,  {Tomasko} M.~G.,  1968, \mn@doi [\apj] {10.1086/149610},
  \href {http://adsabs.harvard.edu/abs/1968ApJ...152..971S} {152, 971}

\bibitem[\protect\citeauthoryear{{Stancil}, {Lepp}  \& {Dalgarno}}{{Stancil}
  et~al.}{1998}]{StaLepDal98}
{Stancil} P.~C.,  {Lepp} S.,   {Dalgarno} A.,  1998, \mn@doi [\apj]
  {10.1086/306473}, \href {http://adsabs.harvard.edu/abs/1998ApJ...509....1S}
  {509, 1}

\bibitem[\protect\citeauthoryear{{Stancil}, {Schultz}, {Kimura}, {Gu}, {Hirsch}
   \& {Buenker}}{{Stancil} et~al.}{1999}]{StaSchKim99}
{Stancil} P.~C.,  {Schultz} D.~R.,  {Kimura} M.,  {Gu} J.-P.,  {Hirsch} G.,
  {Buenker} R.~J.,  1999, \mn@doi [\aaps] {10.1051/aas:1999419}, \href
  {http://adsabs.harvard.edu/abs/1999A%26AS..140..225S} {140, 225}

\bibitem[\protect\citeauthoryear{{Stutzki}, {Bensch}, {Heithausen}, {Ossenkopf}
   \& {Zielinsky}}{{Stutzki} et~al.}{1998}]{Stutzki1998}
{Stutzki} J.,  {Bensch} F.,  {Heithausen} A.,  {Ossenkopf} V.,   {Zielinsky}
  M.,  1998, \aap, \href {http://adsabs.harvard.edu/abs/1998A%26A...336..697S}
  {336, 697}

\bibitem[\protect\citeauthoryear{{Sunyaev} \& {Churazov}}{{Sunyaev} \&
  {Churazov}}{1998}]{Sunyaev1998}
{Sunyaev} R.,  {Churazov} E.,  1998, \mn@doi [\mnras]
  {10.1046/j.1365-8711.1998.01684.x}, \href
  {http://adsabs.harvard.edu/abs/1998MNRAS.297.1279S} {297, 1279}

\bibitem[\protect\citeauthoryear{{Sunyaev}, {Markevitch}  \&
  {Pavlinsky}}{{Sunyaev} et~al.}{1993}]{Sunyaev1993}
{Sunyaev} R.~A.,  {Markevitch} M.,   {Pavlinsky} M.,  1993, \mn@doi [\apj]
  {10.1086/172542}, \href {http://adsabs.harvard.edu/abs/1993ApJ...407..606S}
  {407, 606}

\bibitem[\protect\citeauthoryear{{Swartz}, {Ghosh}, {Tennant}  \&
  {Wu}}{{Swartz} et~al.}{2004}]{SwaGhoTen04}
{Swartz} D.~A.,  {Ghosh} K.~K.,  {Tennant} A.~F.,   {Wu} K.,  2004, \mn@doi
  [\apjs] {10.1086/422842}, \href
  {http://ukads.nottingham.ac.uk/abs/2004ApJS..154..519S} {154, 519}

\bibitem[\protect\citeauthoryear{{Tielens} \& {Hollenbach}}{{Tielens} \&
  {Hollenbach}}{1985}]{TieHol85}
{Tielens} A.~G.~G.~M.,  {Hollenbach} D.,  1985, \mn@doi [\apj]
  {10.1086/163111}, \href {http://adsabs.harvard.edu/abs/1985ApJ...291..722T}
  {291, 722}

\bibitem[\protect\citeauthoryear{{Trevisan} \& {Tennyson}}{{Trevisan} \&
  {Tennyson}}{2002}]{TreTen02}
{Trevisan} C.~S.,  {Tennyson} J.,  2002, \mn@doi [Plasma Physics and Controlled
  Fusion] {10.1088/0741-3335/44/7/315}, \href
  {http://adsabs.harvard.edu/abs/2002PPCF...44.1263T} {44, 1263}

\bibitem[\protect\citeauthoryear{{Vaupr{\'e}}, {Hily-Blant}, {Ceccarelli},
  {Dubus}, {Gabici}  \& {Montmerle}}{{Vaupr{\'e}} et~al.}{2014}]{VauHilCec14}
{Vaupr{\'e}} S.,  {Hily-Blant} P.,  {Ceccarelli} C.,  {Dubus} G.,  {Gabici} S.,
    {Montmerle} T.,  2014, \mn@doi [\aap] {10.1051/0004-6361/201424036}, \href
  {http://adsabs.harvard.edu/abs/2014A%26A...568A..50V} {568, A50}

\bibitem[\protect\citeauthoryear{{Vissapragada}, {Buzard}, {Miller},
  {O'Connor}, {de Ruette}, {Urbain}  \& {Savin}}{{Vissapragada}
  et~al.}{2016}]{VisBuzMil16}
{Vissapragada} S.,  {Buzard} C.~F.,  {Miller} K.~A.,  {O'Connor} A.~P.,  {de
  Ruette} N.,  {Urbain} X.,   {Savin} D.~W.,  2016, \mn@doi [\apj]
  {10.3847/0004-637X/832/1/31}, \href
  {http://adsabs.harvard.edu/abs/2016ApJ...832...31V} {832, 31}

\bibitem[\protect\citeauthoryear{{Voronov}}{{Voronov}}{1997}]{Vor97}
{Voronov} G.~S.,  1997, \mn@doi [Atomic Data and Nuclear Data Tables]
  {10.1006/adnd.1997.0732}, \href
  {http://adsabs.harvard.edu/abs/1997ADNDT..65....1V} {65, 1}

\bibitem[\protect\citeauthoryear{{Wakelam} et~al.,}{{Wakelam}
  et~al.}{2010}]{WakSmiHer10}
{Wakelam} V.,  et~al., 2010, \mn@doi [\ssr] {10.1007/s11214-010-9712-5}, \href
  {http://adsabs.harvard.edu/abs/2010SSRv..156...13W} {156, 13}

\bibitem[\protect\citeauthoryear{{Wakelam} et~al.,}{{Wakelam}
  et~al.}{2015}]{WakLoiHer15}
{Wakelam} V.,  et~al., 2015, \mn@doi [\apjs] {10.1088/0067-0049/217/2/20},
  \href {http://adsabs.harvard.edu/abs/2015ApJS..217...20W} {217, 20}

\bibitem[\protect\citeauthoryear{{Walch}, {Whitworth}, {Bisbas}, {W{\"u}nsch}
  \& {Hubber}}{{Walch} et~al.}{2012}]{Walch2012}
{Walch} S.~K.,  {Whitworth} A.~P.,  {Bisbas} T.,  {W{\"u}nsch} R.,   {Hubber}
  D.,  2012, \mn@doi [\mnras] {10.1111/j.1365-2966.2012.21767.x}, \href
  {http://adsabs.harvard.edu/abs/2012MNRAS.427..625W} {427, 625}

\bibitem[\protect\citeauthoryear{{Walch} et~al.,}{{Walch}
  et~al.}{2015}]{WalGirNaa15}
{Walch} S.,  et~al., 2015, \mn@doi [\mnras] {10.1093/mnras/stv1975}, \href
  {http://adsabs.harvard.edu/abs/2015MNRAS.454..238W} {454, 238}

\bibitem[\protect\citeauthoryear{{Walls}, {Chernyakova}, {Terrier}  \&
  {Goldwurm}}{{Walls} et~al.}{2016}]{WalCheTer16}
{Walls} M.,  {Chernyakova} M.,  {Terrier} R.,   {Goldwurm} A.,  2016, \mn@doi
  [\mnras] {10.1093/mnras/stw2181}, \href
  {http://adsabs.harvard.edu/abs/2016MNRAS.463.2893W} {463, 2893}

\bibitem[\protect\citeauthoryear{{Weingartner} \& {Draine}}{{Weingartner} \&
  {Draine}}{2001}]{WeiDra01}
{Weingartner} J.~C.,  {Draine} B.~T.,  2001, \mn@doi [\apj] {10.1086/324035},
  \href {http://adsabs.harvard.edu/abs/2001ApJ...563..842W} {563, 842}

\bibitem[\protect\citeauthoryear{{Wernli}, {Valiron}, {Faure}, {Wiesenfeld},
  {Jankowski}  \& {Szalewicz}}{{Wernli} et~al.}{2006}]{w06}
{Wernli} M.,  {Valiron} P.,  {Faure} A.,  {Wiesenfeld} L.,  {Jankowski} P.,
  {Szalewicz} K.,  2006, \mn@doi [\aap] {10.1051/0004-6361:20053919}, \href
  {http://adsabs.harvard.edu/abs/2006A%26A...446..367W} {446, 367}

\bibitem[\protect\citeauthoryear{{Wilms}, {Allen}  \& {McCray}}{{Wilms}
  et~al.}{2000}]{WilAllMcC00}
{Wilms} J.,  {Allen} A.,   {McCray} R.,  2000, \mn@doi [\apj] {10.1086/317016},
  \href {http://adsabs.harvard.edu/abs/2000ApJ...542..914W} {542, 914}

\bibitem[\protect\citeauthoryear{{Wilson} \& {Bell}}{{Wilson} \&
  {Bell}}{2002}]{wb02}
{Wilson} N.~J.,  {Bell} K.~L.,  2002, \mn@doi [\mnras]
  {10.1046/j.1365-8711.2002.05982.x}, \href
  {http://adsabs.harvard.edu/abs/2002MNRAS.337.1027W} {337, 1027}

\bibitem[\protect\citeauthoryear{{Wolfire}, {McKee}, {Hollenbach}  \&
  {Tielens}}{{Wolfire} et~al.}{2003}]{w03}
{Wolfire} M.~G.,  {McKee} C.~F.,  {Hollenbach} D.,   {Tielens} A.~G.~G.~M.,
  2003, \mn@doi [\apj] {10.1086/368016}, \href
  {http://adsabs.harvard.edu/abs/2003ApJ...587..278W} {587, 278}

\bibitem[\protect\citeauthoryear{{Woodall}, {Ag{\'u}ndez}, {Markwick-Kemper}
  \& {Millar}}{{Woodall} et~al.}{2007}]{WooAguMar07}
{Woodall} J.,  {Ag{\'u}ndez} M.,  {Markwick-Kemper} A.~J.,   {Millar} T.~J.,
  2007, \mn@doi [\aap] {10.1051/0004-6361:20064981}, \href
  {http://adsabs.harvard.edu/abs/2007A%26A...466.1197W} {466, 1197}

\bibitem[\protect\citeauthoryear{{W{\"u}nsch}, {Walch}, {Dinnbier}  \&
  {Whitworth}}{{W{\"u}nsch} et~al.}{2018}]{WunWalDin18}
{W{\"u}nsch} R.,  {Walch} S.,  {Dinnbier} F.,   {Whitworth} A.,  2018, \mn@doi
  [\mnras] {10.1093/mnras/sty015}, \href
  {http://adsabs.harvard.edu/abs/2018MNRAS.475.3393W} {475, 3393}

\bibitem[\protect\citeauthoryear{{Yan}}{{Yan}}{1997}]{Yan97}
{Yan} M.,  1997, PhD thesis, Harvard University

\bibitem[\protect\citeauthoryear{{Zanchet}, {Bussery-Honvault}, {Jorfi}  \&
  {Honvault}}{{Zanchet} et~al.}{2009}]{ZanBusJor09}
{Zanchet} A.,  {Bussery-Honvault} B.,  {Jorfi} M.,   {Honvault} P.,  2009,
  \mn@doi [Physical Chemistry Chemical Physics (Incorporating Faraday
  Transactions)] {10.1039/B903829A}, \href
  {http://adsabs.harvard.edu/abs/2009PCCP...11.6182Z} {11, 6182}

\bibitem[\protect\citeauthoryear{{de Avillez} \& {Breitschwerdt}}{{de Avillez}
  \& {Breitschwerdt}}{2012a}]{DeABre12}
{de Avillez} M.~A.,  {Breitschwerdt} D.,  2012a, \mn@doi [\apjl]
  {10.1088/2041-8205/756/1/L3}, \href
  {http://adsabs.harvard.edu/abs/2012ApJ...756L...3D} {756, L3}

\bibitem[\protect\citeauthoryear{{de Avillez} \& {Breitschwerdt}}{{de Avillez}
  \& {Breitschwerdt}}{2012b}]{DeAMigBre12}
{de Avillez} M.~A.,  {Breitschwerdt} D.,  2012b, \mn@doi [\apj]
  {10.1088/2041-8205/756/1/L3}, \href
  {https://ui.adsabs.harvard.edu/#abs/2012ApJ...756L...3D} {756, L3}

\bibitem[\protect\citeauthoryear{{de Ruette}, {Miller}, {O'Connor}, {Urbain},
  {Buzard}, {Vissapragada}  \& {Savin}}{{de Ruette} et~al.}{2016}]{DeRMilOCo16}
{de Ruette} N.,  {Miller} K.~A.,  {O'Connor} A.~P.,  {Urbain} X.,  {Buzard}
  C.~F.,  {Vissapragada} S.,   {Savin} D.~W.,  2016, \mn@doi [\apj]
  {10.3847/0004-637X/816/1/31}, \href
  {http://adsabs.harvard.edu/abs/2016ApJ...816...31D} {816, 31}

\makeatother
\end{thebibliography}

\appendix

\section{Chemical network} \label{app:network}

\begin{table*}
  \centering
  \caption{
    Collisional reactions used in the new chemical network for modelling X-ray-irradiated gas.
    GOW17 refers to \citet{GonOstWol17}.
    }
  \label{tab:reactions}
  \begin{tabular}{llllp{4cm}} 
    \hline
    ID & Reaction & Type & Note & Reference \\
    \hline
    %
    1 & H  +  e $\rightarrow$ H$^+$ + 2e & Collisional ionization & polynomial fit & \citet{AbeAnnZha97} \\ 
    2 & He  +  e $\rightarrow$ He$^+$ + 2e & Collisional ionization & Not in GOW17 & \citet{AbeAnnZha97} \\ 
    3 & C  +  e $\rightarrow$ C$^+$ + 2e & Collisional ionization & Not in GOW17 & \citet{Vor97} \\ 
    4 & M  +  e $\rightarrow$ M$^+$ + 2e & Collisional ionization & Not in GOW17 & \citet{Vor97} \\ 
    5 & H$_2$ + e $\rightarrow$  2H + e & Collisional dissociation & Not in GOW17 & \citet{TreTen02} \\ 
    6 & H$_2$ + H $\rightarrow$  3H     & Collisional dissociation & similar to GOW17 & \citet{LepShu83, MacShu86, MarSchMan96} \\  
    7 & H$_2$ + H$_2$ $\rightarrow$  H$_2$ + 2H     & Collisional dissociation & similar to GOW17 & \citet{MarKeoMan98, ShaKan87, PalSalSta83} \\ 
    \hline
    8 & H$^+$ + e $\rightarrow$ H + $\gamma$ & Radiative recomb. & Same as GOW17 & \citet{FerPetHor92} \\ 
    9 & He$^+$ + e $\rightarrow$ He + $\gamma$  & Radiative+dielec.\ recomb. & &  \citet{Ost89, HumSto98, Bad06} \\ 
    10 & C$^+$ + e $\rightarrow$ C + $\gamma$  & Radiative recomb. & Same as GOW17 & \citet{BadOMuSum03, Bad06}  \\ 
    11 & M$^+$ + e $\rightarrow$ M + $\gamma$  & Radiative recomb. & Similar to GOW17 & \citet{NelLan99} \\ 
    \hline
    12 & H$^+$ + e $\rightarrow$ H  & Grain-assisted recomb. & Same as GOW17 &  \citet{WeiDra01} \\ 
    13 & He$^+$ + e $\rightarrow$ He  & Grain-assisted recomb. & Same as GOW17  &  \citet{WeiDra01} \\
    14 & C$^+$ + e $\rightarrow$ C  & Grain-assisted recomb. & Same as GOW17  &  \citet{WeiDra01} \\
    15 & M$^+$ + e $\rightarrow$ M  & Grain-assisted recomb. & Same as GOW17  &  \citet{WeiDra01} \\
    \hline
    %
    16 & H + H $\rightarrow$  H$_2$                  & Grain-assisted H$_2$ form.  & Similar to GOW17 & \citet{HolMcK79} \\
    17 & H$_2^+$ + H $\rightarrow$ H$_2$ + H$^+$     & Charge ex.  & Same as GOW17 &  \citet{KarAniHun79}  \\
    18 & H$_2$ + H$^+$ $\rightarrow$ H$_2^+$ + H     & Charge ex.  & Not in GOW17 & \citet{SavKrsPre04}  \\

    19 & H$_3^+$ + M $\rightarrow$ M$^+$ + H$_2$ + H & Dissociative charge ex. & not in GOW17 &  \citet{NelLan99} \\
    20 & H$_3^+$ + e $\rightarrow$ H$_2$ + H         & Dissociative recomb.   & Same as GOW17 & \citet{McCHunSay04, WooAguMar07} \\
    21 & H$_3^+$ + C $\rightarrow$ CH$_\mathrm{x}$ + H$_2$       & Formation of CH$_\mathrm{x}$    & Same as GOW17 & \citet{VisBuzMil16, GonOstWol17} \\
    22 & H$_3^+$ + O $\rightarrow$ OH$_\mathrm{x}$ + H$_2$       & Formation of OH$_\mathrm{x}$    & Same as GOW17 & \citet{DeRMilOCo16, GonOstWol17} \\
    23 & H$_3^+$ + O + e $\rightarrow$ O + 3H        & Pseudo-reaction     & Same as GOW17 & \citet{DeRMilOCo16, GonOstWol17}  \\
    24 & H$_3^+$ + CO $\rightarrow$ HCO$^+$ + H$_2$  & Proton transfer  & Same as GOW17 &  \citet{KimTheHun75}  \\
    25 & CH$_\mathrm{x}$     + H  $\rightarrow$ H$_2$ + C        & Exchange reaction  &  Same as GOW17  & \citet{WakSmiHer10, GonOstWol17}  \\
    \hline
    %
    %
    26 & He$^+$ + H$_2$    $\rightarrow$  He + H + H$^+$  & Dissociative charge ex.  & Same as GOW17 &   \citet{SchJefBar89} \\
    27 & He$^+$ + H$_2$    $\rightarrow$ He + H$_2^+$  & Charge ex.  & Same as GOW17 & \citet{Bar84}  \\
    28 & O$^+$ + H$_2$     $\rightarrow$ OH$_\mathrm{x}$ + H       & Formation of OH$_\mathrm{x}$    & Same as GOW17 & \citet{GonOstWol17}  \\
    29 & O$^+$ + H$_2$ + e $\rightarrow$ O + 2H        & H$_2$ destruction & Same as GOW17 & \citet{GonOstWol17}  \\
    30 & C$^+$ + H$_2$     $\rightarrow$ CH$_\mathrm{x}$ + H       & Formation of CH$_\mathrm{x}$    &  Same as GOW17 & \citet{WakSmiHer10}  \\
    31 & C$^+$ + H$_2$ + e $\rightarrow$ C + 2H        & H$_2$ Destruction   & Same as GOW17 & \citet{WakSmiHer10}  \\
    32 & H$_2$ + H$_2^+$   $\rightarrow$ H$_3^+$ + H   & Formation of H$_3^+$ & Same as GOW17 & \citet{StaLepDal98} \\
    33 & C + H$_2$         $\rightarrow$ CH$_\mathrm{x}$           & Radiative association  & Not in GOW17 & \citet{PraHun80} \\
    \hline
    %
    %
    34 & He$^+$ + CO $\rightarrow$ He + C$^+$ + O  & Dissociative charge ex. & GOW17 differs & \citet{PetDweAll89} \\
    35 & C$^+$ + OH$_\mathrm{x}$ $\rightarrow$ HCO$^+$         & HCO$^+$ formation &  Same as GOW17 & \citet{WakSmiHer10} \\
    36 & O     + CH$_\mathrm{x}$ $\rightarrow$ CO + H          & CO formation  & Same as GOW17  & \citet{WakSmiHer10} \\
    37 & C     + OH$_\mathrm{x}$ $\rightarrow$ CO + H          & CO formation  & Same as GOW17  & \citet{ZanBusJor09, WakSmiHer10} \\
    38 & HCO$^+$ + e $\rightarrow$ CO + H          & CO formation  &  GOW17 rate similar & \citet{BriMit90, McEWalMar12} \\
    39 & OH$_\mathrm{x}$     + O $\rightarrow$ 2O + H          & OH$_\mathrm{x}$ destruction &  Same as GOW17  & \citet{CarGodKoh06}  \\
    40 & OH$_\mathrm{x}$ + He$^+$ $\rightarrow$ O$^+$ + He + H & Dissociative charge ex. &  Same as GOW17  & \citet{WakSmiHer10}  \\
    41 & O$^+$ + H $\rightarrow$ O + H$^+$     & Charge ex.  & Equilibrium & \citet{StaSchKim99} \\
    42 & H$^+$ + O $\rightarrow$ H + O$^+$     & Charge ex.  & Equilibrium & \citet{StaSchKim99} \\
    43 & H$_2^+$ + e $\rightarrow$ 2H  & Dissociative recomb. & Not in GOW17 & \citet{AbeAnnZha97}. \\
    \hline
  \end{tabular}
\end{table*}

\begin{table*}
  \centering
  \caption{
    Cosmic-ray, X-ray, and photo-reactions used in the new chemical network.
    GOW17 refers to \citet{GonOstWol17}.
    }
  \label{tab:photoreactions}
  \begin{tabular}{llllp{4cm}} 
    \hline
    ID & Reaction & Type & Note & Reference \\
    \hline
     %
    %
    \hline
    44 & H$_2$   + FUV $\rightarrow$ 2H          &  Photodiss.      & same as GOW17 &   \citet{HeaBosVan17} \\
    45 & HCO$^+$ + FUV $\rightarrow$ CO + H$^+$  &  Photodiss.      & not in GOW17 & \citet{HeaBosVan17} \\
    46 & CO      + FUV $\rightarrow$ C + O       &  Photodiss.      & same as GOW17 & \citet{HeaBosVan17} \\ 
    47 & C       + FUV $\rightarrow$ C$^+$ + e   &  Photoioniz.     & same as GOW17 & \citet{HeaBosVan17}  \\ 
    48 & M       + FUV $\rightarrow$ M$^+$ + e   &  Photoioniz.     & same as GOW17 & \citet{HeaBosVan17} \\ 
    49 & OH$_\mathrm{x}$   + FUV $\rightarrow$ O + H         &  Photodiss. & same as GOW17 & \citet{HeaBosVan17} \\
    50 & CH$_\mathrm{x}$   + FUV $\rightarrow$ C + H         &  Photodiss. & same as GOW17 & \citet{HeaBosVan17} \\
    \hline
    51 & H     + CR $\rightarrow$ H$^+$ + e      & Cosmic-ray ioniz.   & $\zeta_\mathrm{H}=3\times10^{-17}$\,s$^{-1}$ per H & \citet{WalGirNaa15}  \\
    52 & He    + CR $\rightarrow$ He$^+$ + e     & Cosmic-ray ioniz.   & same as GOW17 &  \citet{g10} \\
    53 & C     + CR $\rightarrow$ C$^+$ + e      & Cosmic-ray ioniz.   & within 1\% of GOW17 &  \citet{Lis03} \\
    54 & H$_2$ + CR $\rightarrow$ H$^+$ + H + e  & Cosmic-ray ioniz.   & $0.037\zeta_\mathrm{H}$ per H$_2$ &  \citet{MicGloFed12} \\
    55 & H$_2$ + CR $\rightarrow$ 2H             & Cosmic-ray diss.    & $0.21\zeta_\mathrm{H}$ per H$_2$ &   \citet{MicGloFed12} \\
    56 & H$_2$ + CR $\rightarrow$ H$_2^+$ + e    & Cosmic-ray ioniz.   & as GOW17; $2\zeta_\mathrm{H}$ per H$_2$ &   \citet{GonOstWol17} \\
    57 & CO    + CR (+H) $\rightarrow$ HCO$^+$   & Pseudoreaction, via CO$^+$ &  same as GOW17 & \citet{g10}  \\
    58 & CO    + CR $\rightarrow$ C + O          & Cosmic-ray diss.    & 10$\zeta_\mathrm{H}y$(CO)  & \citet{WakLoiHer15} \\
    \hline
    59 & C  + CRPHOT $\rightarrow$ C$^+$ + e      & Ioniz. by CR-induced FUV &  Similar to GOW17 & \citet{GreLepDal87, McEWalMar12} \\
    60 & CO + CRPHOT $\rightarrow$ C + O          & Diss. by CR-induced FUV  & GOW17 rate differs & \citet{GreLepDal87, McEWalMar12} \\
    61 & M  + CRPHOT $\rightarrow$ M$^+$ + e      & Ioniz. by CR-induced FUV & Similar to GOW17 & \citet{McEWalMar12} reference Rawlings (1992, priv.\ comm.) \\
    \hline
    %
    %
    62 & H     + XR $\rightarrow$ H$^+$ + e      & Secondary ioniz.  & Fitted from table & \citet{DalYanLiu99} \\
    63 & He    + XR $\rightarrow$ He$^+$ + e     & Secondary ioniz.  & Fitted from table & \citet{DalYanLiu99} \\
    64 & C     + XR $\rightarrow$ C$^+$ + e      & Secondary ioniz.  & 3.92$\times$ rate for H   & MS05 \\
    65 & M     + XR $\rightarrow$ M$^+$ + e      & Secondary ioniz.  & 6.67$\times$ rate for H   & MS05 \\
    66 & H$_2$ + XR $\rightarrow$ H$_2^+$ + e    & Secondary ioniz.  & Fitted from table & \citet{DalYanLiu99} \\
    67 & H$_2$ + XR $\rightarrow$ 2H             & Secondary diss.   & Fitted from table & \citet{DalYanLiu99} \\
    68 & CO    + XR $\rightarrow$ C$^+$ + O + e  & Secondary ioniz.  & 3.92$\times$ rate for H   & MS05 \\
    69 & CH$_\mathrm{x}$   + XR $\rightarrow$ C$^+$ + H + e  & Secondary ioniz.  & 3.92$\times$ rate for H   & MS05  \\
    70 & OH$_\mathrm{x}$   + XR $\rightarrow$ O + H$^+$ + e  & Secondary ioniz.  & 2.97$\times$ rate for H   & MS05  \\
    71 & HCO$^+$ + XR $\rightarrow$ C$^+$ + H$^+$ + O + e & Secondary ioniz.  & 3.92$\times$ rate for H   & MS05 \\
    \hline
    72 & C     + XRPHOT $\rightarrow$ C$^+$ + e        & Ioniz. by XR-induced FUV    & Eqn.~\ref{eqn:xrfuv}; Tab.~\ref{tab:crphot} & See Tab.~\ref{tab:crphot} \\
    73 & M     + XRPHOT $\rightarrow$ M$^+$ + e        & Ioniz. by XR-induced FUV    & Eqn.~\ref{eqn:xrfuv}; Tab.~\ref{tab:crphot} & See Tab.~\ref{tab:crphot} \\
    74 & CO    + XRPHOT $\rightarrow$ C + O            & Ioniz. by XR-induced FUV    & Eqn.~\ref{eqn:crphot} & \citet{GreLepDal87,McEWalMar12}  \\
    75 & CH$_\mathrm{x}$   + XRPHOT $\rightarrow$ C + H            & Diss. by XR-induced FUV  & Eqn.~\ref{eqn:xrfuv}; Tab.~\ref{tab:crphot} & See Tab.~\ref{tab:crphot} \\
    76 & OH$_\mathrm{x}$   + XRPHOT $\rightarrow$ O + H            & Diss. by XR-induced FUV  & Eqn.~\ref{eqn:xrfuv}; Tab.~\ref{tab:crphot} & See Tab.~\ref{tab:crphot} \\
    \hline
  \end{tabular}
\end{table*}

The collisional reactions considered are listed in Table~\ref{tab:reactions} and photo/CR/X-ray reactions in Table~\ref{tab:photoreactions}.
The reaction network is a superset of the NL99 \citet{GloCla12} network, with most additions taken from \citet{GonOstWol17}.
The extra reactions included are numbers \#13, \#14, \#15, \#18, \#25, \#28, \#29, \#31, \#39, \#40, \#41, \#56, \#60, plus the X-ray photoreactions \#62-76.

\begin{figure*}
\begin{tabular}{ccc}
&$4\pi J_{X}=1.6$ erg\,cm$^{-2}$\,s$^{-1}$ & $4\pi J_{X}=160$ erg\,cm$^{-2}$\,s$^{-1}$ \\
\raisebox{7.5\normalbaselineskip}[0pt][0pt]{\rotatebox{90}{$n_\mathrm{H}=10^3$ cm$^{-3}$}} &
\includegraphics[trim = 0mm 11mm 11mm 1mm, clip, height=6.0cm]{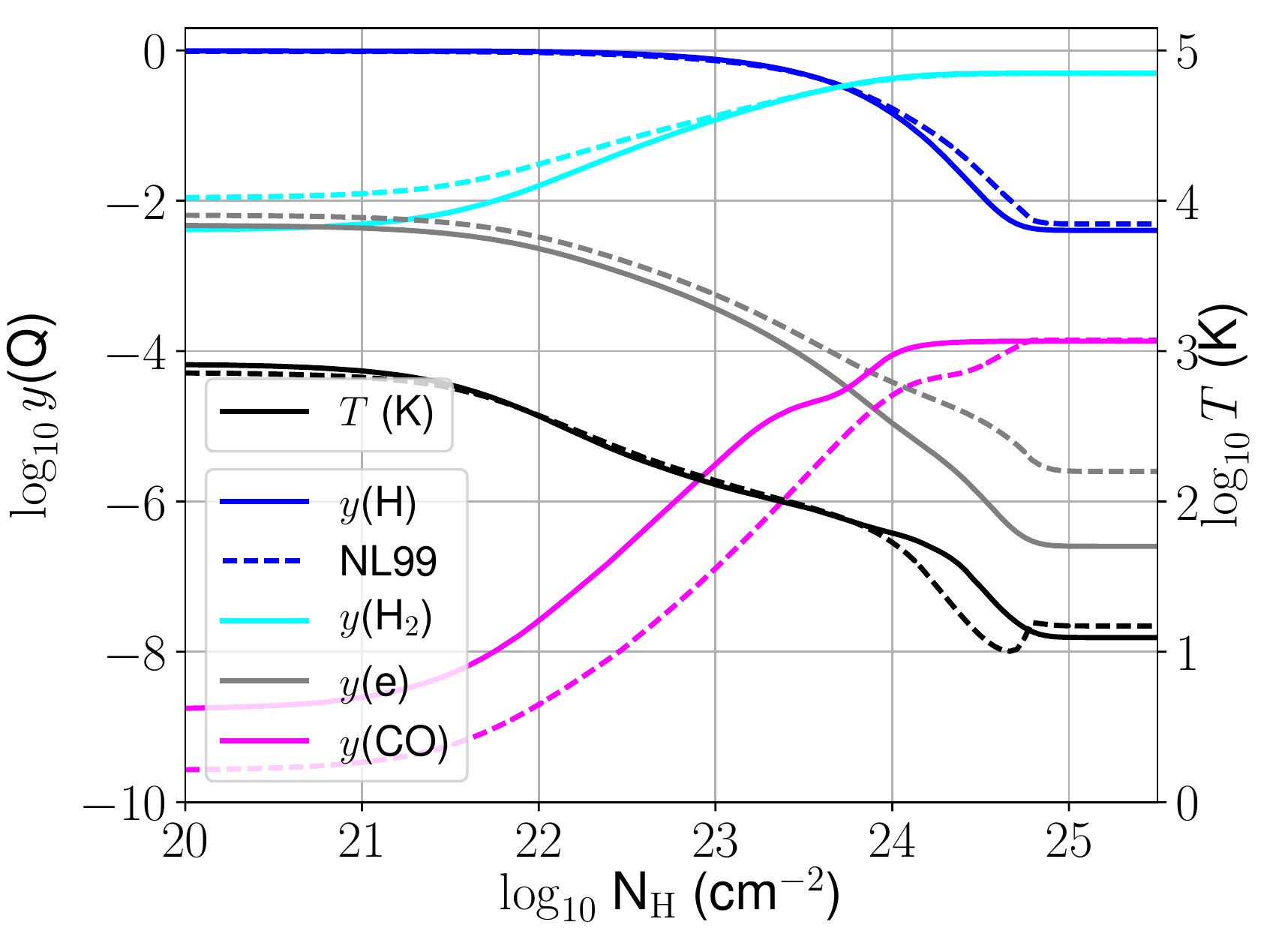} &
\includegraphics[trim = 16mm 11mm 0mm 1mm, clip,  height=6.0cm]{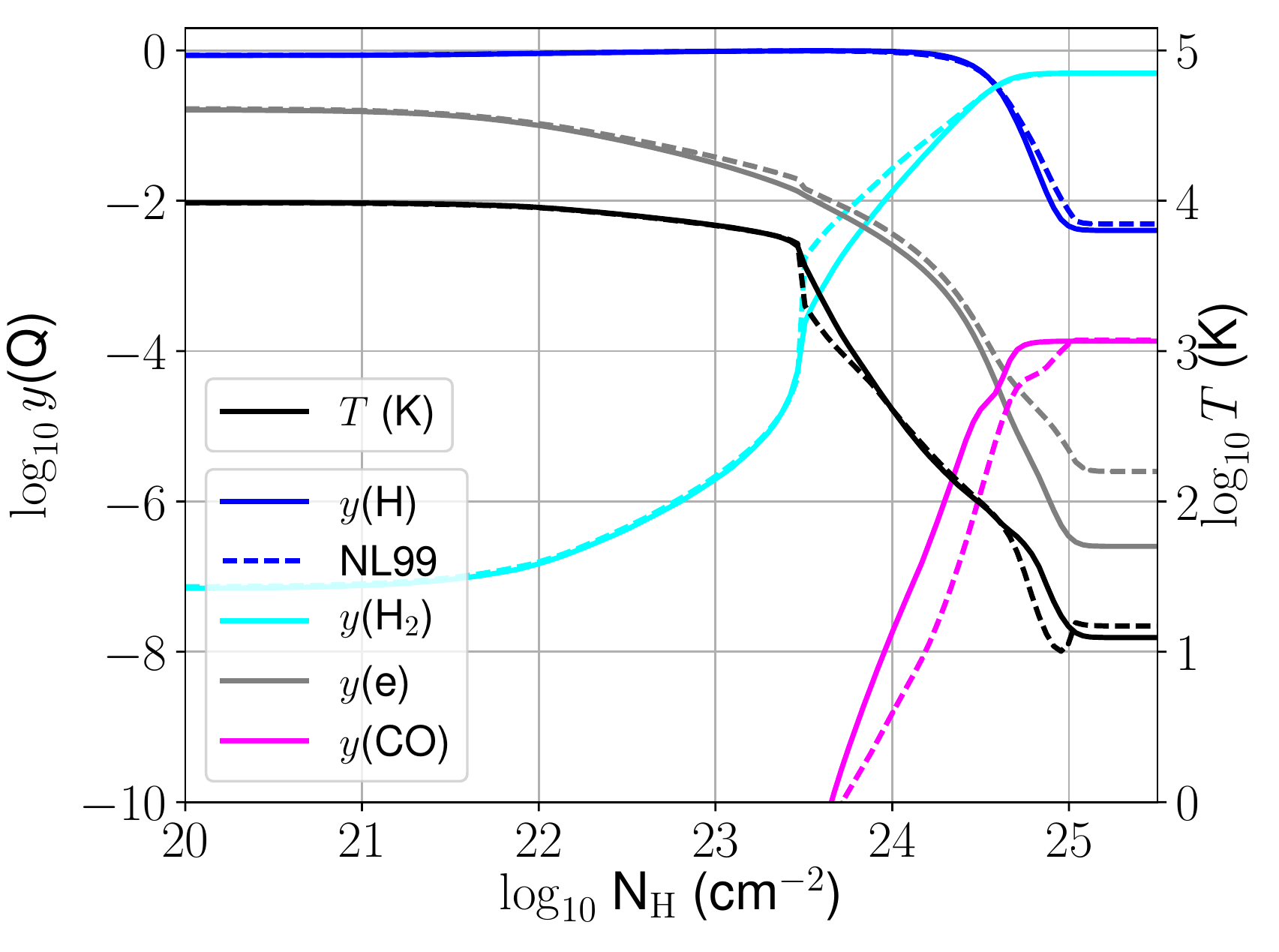} \\
\raisebox{7.5\normalbaselineskip}[0pt][0pt]{\rotatebox{90}{$n_\mathrm{H}=10^{5.5}$ cm$^{-3}$}} &
\includegraphics[trim = 0mm 0mm 11mm 0mm, clip, height=6.5cm]{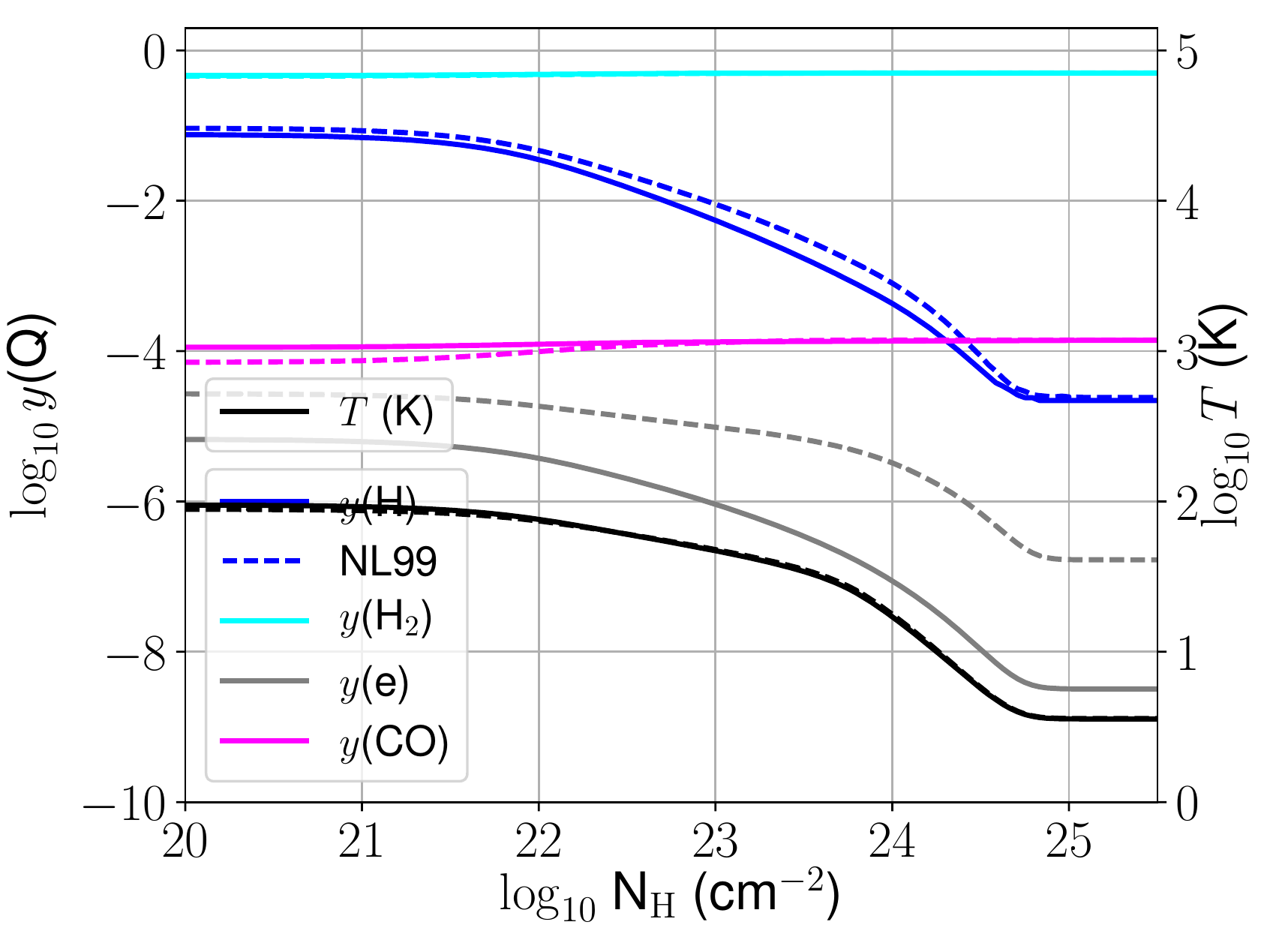} & 
\includegraphics[trim = 16mm 0mm 0mm 0mm, clip, height=6.5cm]{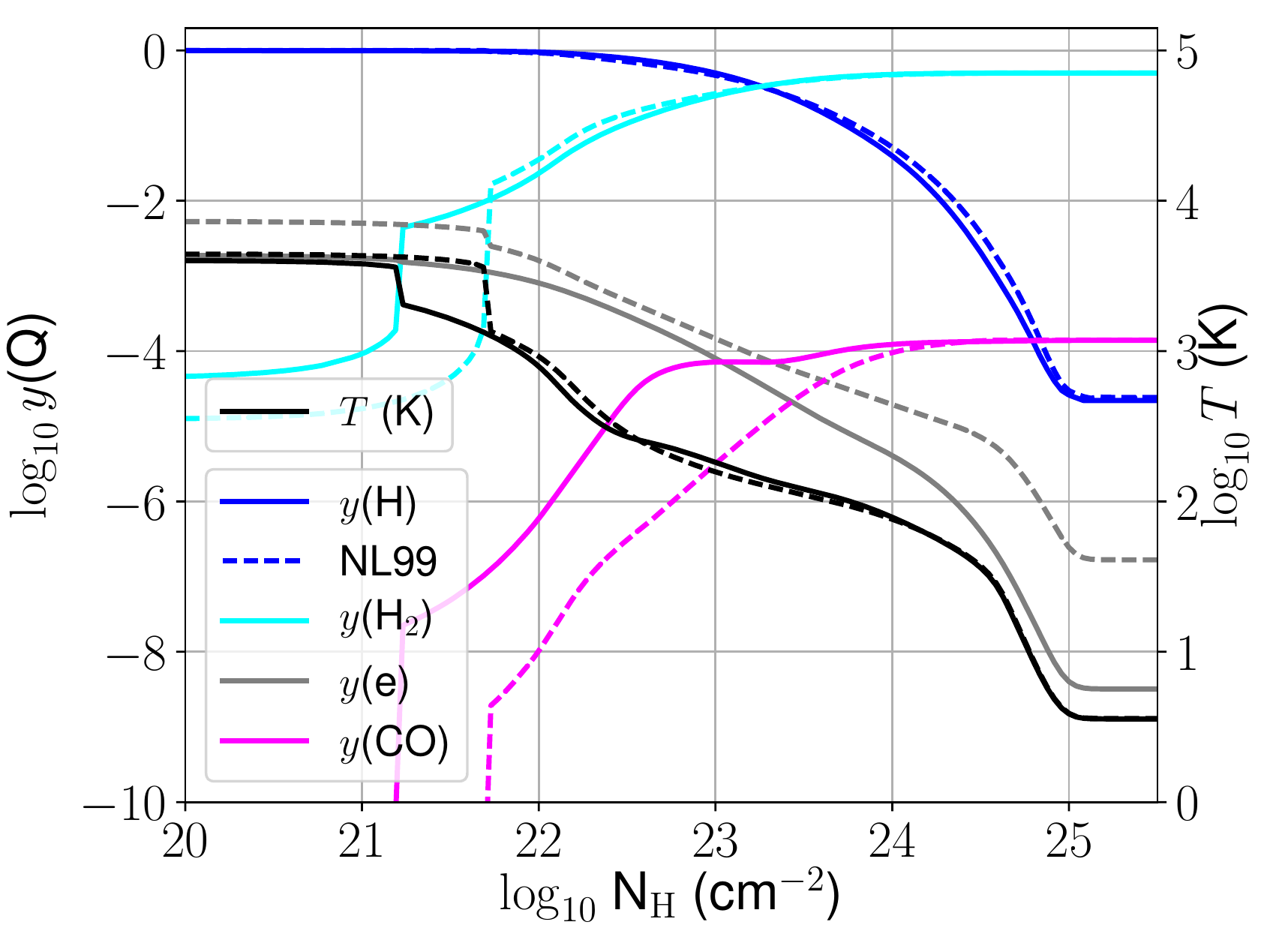} \\
\end{tabular}
  \caption{
  Abundances of H$_2$, CO, H, electrons, and gas temperature for models 1 (upper left), 2 (upper right), 3 (lower left) and 4 (lower right) calculated using the original NL99 network (dashed lines) and including the \citet{GonOstWol17} additions (solid lines).
  The results are plotted as a function of column density of hydrogen.
  The left-hand vertical axis shows the fractional abundance whereas the right-hand vertical axis shows the temperature scale.
}
  \label{fig:gongAB}
\end{figure*}

\begin{figure*}
\begin{tabular}{ccc}
&$4\pi J_{X}=1.6$ erg\,cm$^{-2}$\,s$^{-1}$ & $4\pi J_{X}=160$ erg\,cm$^{-2}$\,s$^{-1}$ \\
\raisebox{7.5\normalbaselineskip}[0pt][0pt]{\rotatebox{90}{$n_\mathrm{H}=10^3$ cm$^{-3}$}} &
\includegraphics[trim = 0mm 12mm 3mm 2mm, clip, height=5.5cm]{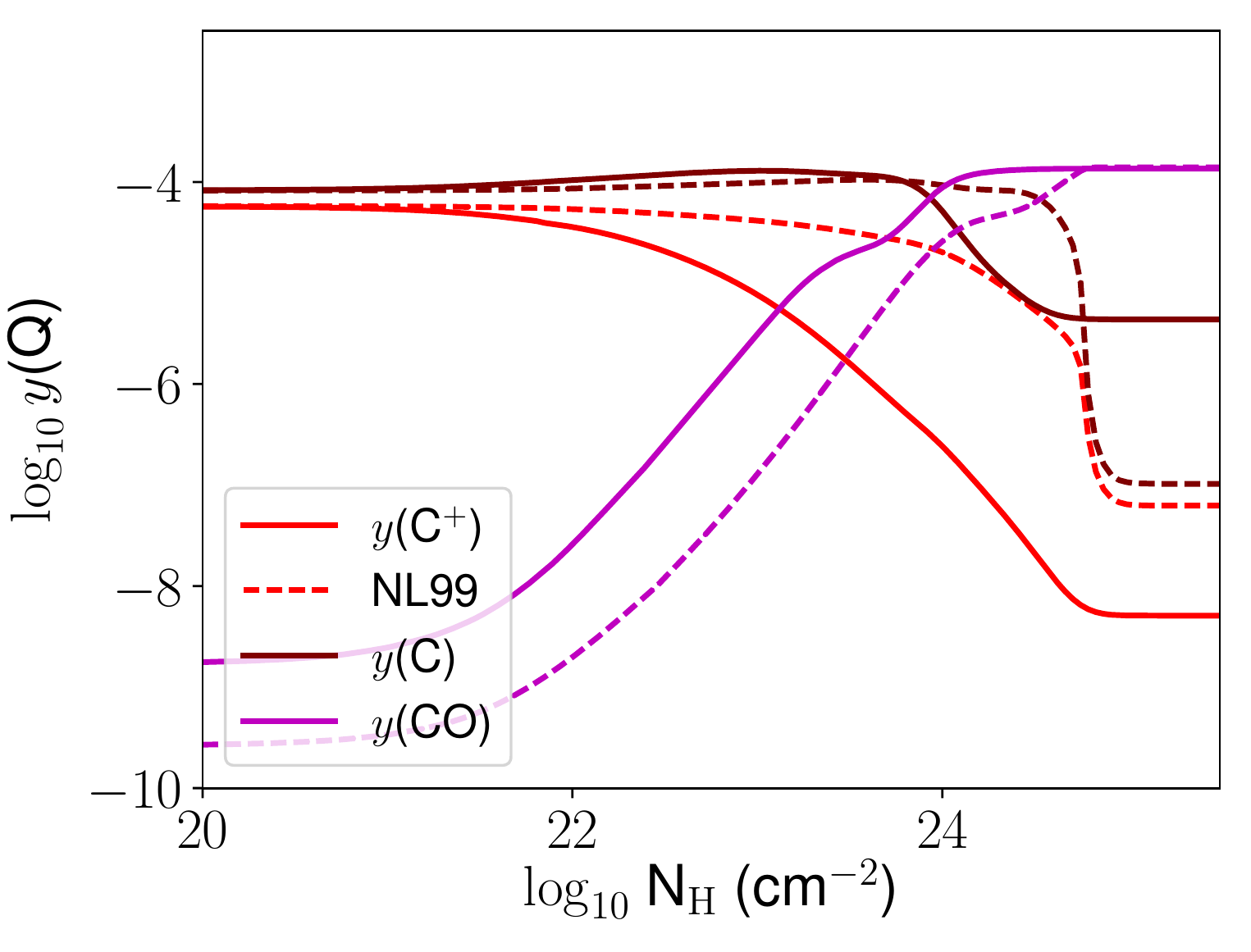} &
\includegraphics[trim = 16mm 12mm 0mm 2mm, clip,  height=5.5cm]{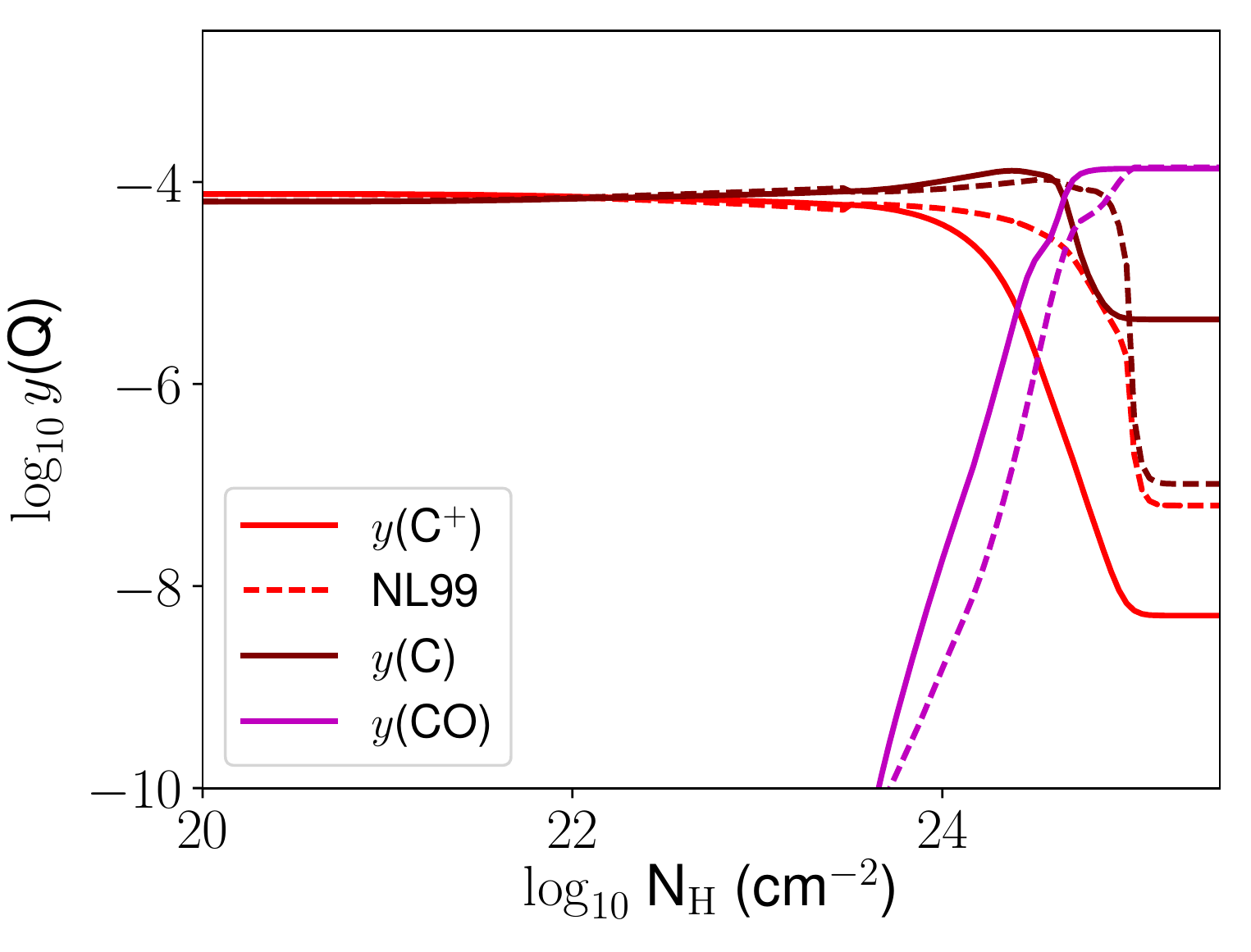} \\
\raisebox{7.5\normalbaselineskip}[0pt][0pt]{\rotatebox{90}{$n_\mathrm{H}=10^{5.5}$ cm$^{-3}$}} &
\includegraphics[trim = 0mm 0mm 3mm 2mm, clip, height=6.3cm]{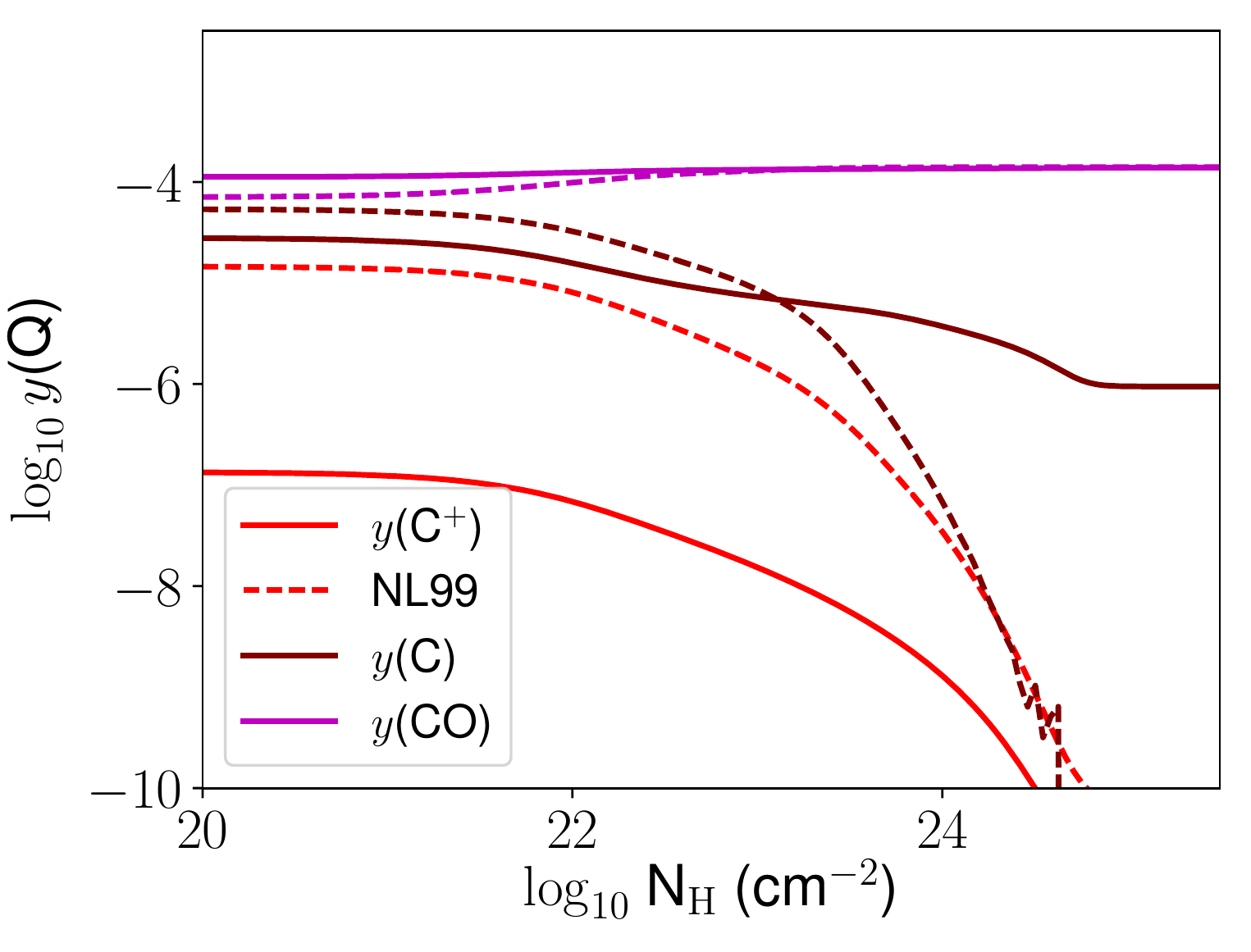} & 
\includegraphics[trim = 16mm 0mm 0mm 2mm, clip, height=6.3cm]{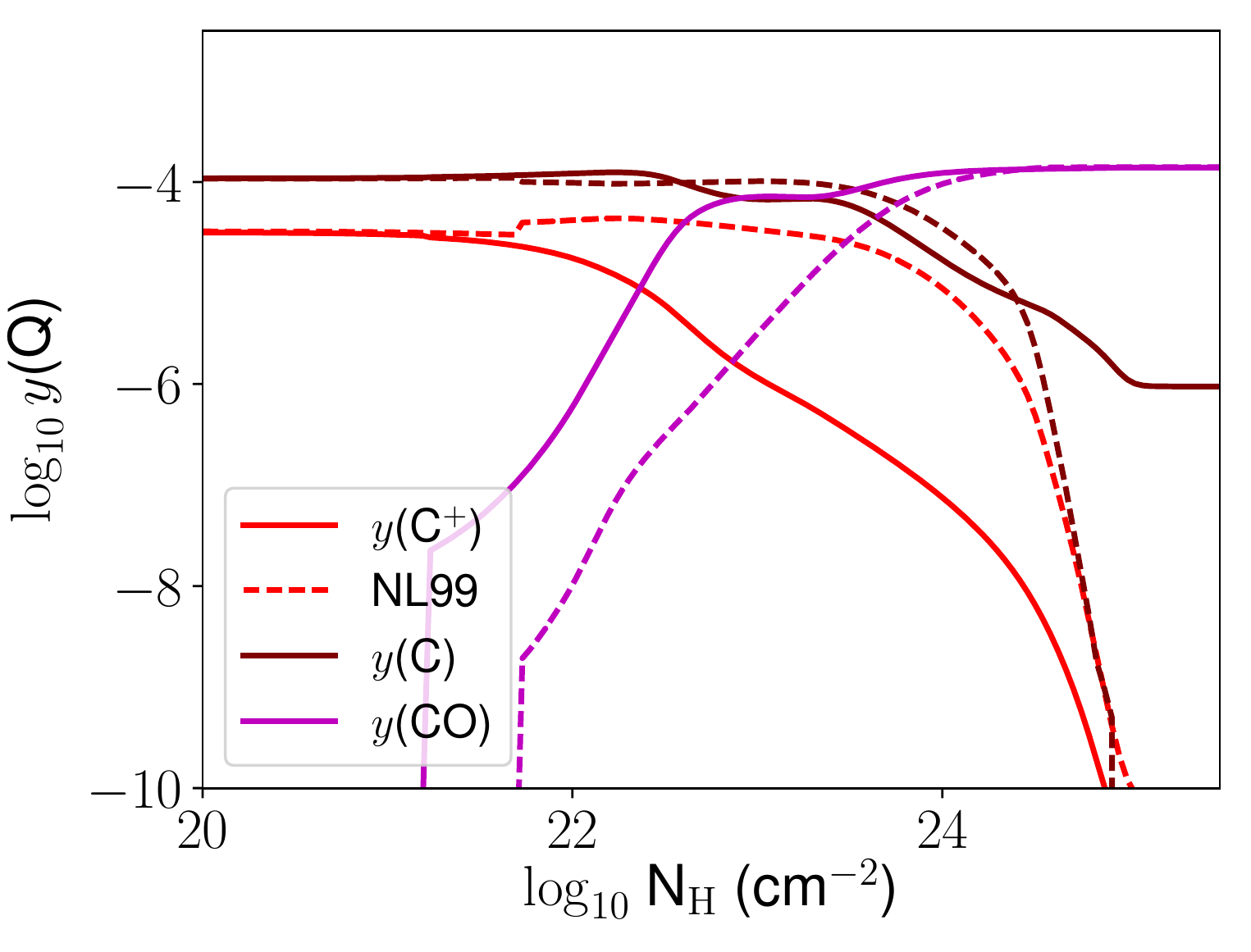} \\
\end{tabular}
  \caption{
  Abundances of carbon-bearing species for models 1 (upper left), 2 (upper right), 3 (lower left) and 4 (lower right) calculated using the original NL99 network (dashed lines) and including the \citet{GonOstWol17} additions (solid lines).
  The results are plotted as a function of column density of hydrogen.
}
  \label{fig:gongC}
\end{figure*}

The results of 1D simulations of the MS05 models 1-4, calculated with and without these additional reactions, are plotted in Figs.~\ref{fig:gongAB} and~\ref{fig:gongC}.
The abundances of H$_2$, CO, H, electrons, and gas temperature are shown in Fig.~\ref{fig:gongAB}, and abundances of carbon-bearing species in Fis.~\ref{fig:gongC}.

The main difference apparent from Fig.~\ref{fig:gongAB} is that $y$(CO) has a very different relationship with column density for the two sets of reactions.
The gas temperature is not strongly affected, except for models 2 and 4, which have a strong chemo-thermal instability for the original NL99 network.
This is weaker when using the updated network.
Looking at the carbon chemistry in Fig.~\ref{fig:gongC}, the updated network has consistently lower C$^+$ abundance for all calculations.
The original NL99 network produces results much closer to those of MS05; in fact the C$^+$ abundance showed the largest discrepancy between our results and MS05 in section~\ref{sec:ms05}.
The neutral C abundance is higher using the updated network except in the region of column density where C$^+$ and CO co-exist, for which the updated network typically has lower neutral C abundance.
At very high column density, the neutral C abundance is much higher with the updated network.
CO forms more rapidly with increasing column density using the updated network; this is in much better agreement with the \textsc{Cloudy} results in Section~\ref{sec:cloudy}.

\section{Heating and Cooling Rates}
\label{sec:cooling}
  We model the thermal evolution of the gas in our simulations using a cooling function based largely on the one developed by \citet{g10}  and \citet{GloCla12}, but updated to account for the effects of X-ray heating, as detailed in Section~2.3 of the current paper.
  A full list of the processes included in the cooling function is given in Table~\ref{cool_model}, along with the sources for the rates used.
  For a few processes, we also give additional details below.

\begin{table*}
\centering
\begin{tabular}{ll}
\hline
Process & Reference(s) \\
\hline
{\bf Radiative cooling:} & \\
C fine structure lines & Atomic data -- \citet{sv02} \\
& Collisional rates (H) -- \citet{akd07} \\
& Collisional rates (H$_{2}$) -- \citet{sch91} \\
& Collisional rates (e$^{-}$) -- \citet{joh87} \\
& Collisional rates (H$^{+}$) -- \citet{rlb90} \\
C$^{+}$ fine structure lines &  Atomic data -- \citet{sv02} \\
&  Collisional rates (H$_{2}$) -- \citet{fl77}  \\
&  Collisional rates (H, $T < 2000 \: {\rm K}$) -- \citet{hm89} \\
&  Collisional rates (H, $T > 2000 \: {\rm K}$) -- \citet{k86} \\
&  Collisional rates (e$^{-}$) -- \citet{wb02} \\
O fine structure lines &  Atomic data -- \citet{sv02} \\
& Collisional rates (H) -- \citet{akd07} \\
& Collisional rates (H$_{2}$) -- see \citet{gj07} \\ 
& Collisional rates (e$^{-}$) -- \citet{bbt98} \\
& Collisional rates (H$^{+}$) -- \citet{p90,p96} \\
Si fine structure lines & All data -- \citet{hm89} \\
Si$^{+}$ fine structure lines & Atomic data -- \citet{sv02} \\
& Collisional rates (H) -- \citet{r90} \\
& Collisional rates (e$^{-}$) -- \citet{dk91} \\
H$_{2}$ rovibrational lines & \citet{ga08} \\
CO rovibrational lines & \citet{nk93,nlm95} \\
Gas-grain energy transfer & \citet{hm89} \\
Atomic resonance lines & Hydrogen -- \citet{black81,cen92} \\
& Helium and metals -- \citet{gf12} \\
Atomic metastable transitions & \citet{hm89,bac15} \\
Compton cooling & \citet{cen92} \\
\hline
{\bf Chemical cooling:} & \\
H collisional ionisation& See Table A1 \\
H$_{2}$ collisional dissociation & See Table A1 \\
H$^{+}$ recombination & \citet{FerPetHor92,w03} \\
\hline
{\bf Heating:} & \\
Photoelectric effect & \citet{bt94,w03} \\
H$_{2}$ photoionisation & \citet{MeiSpa05} \\
H$_{2}$ photodissociation & \citet{bd77} \\ 
UV pumping of H$_{2}$ & \citet{bht90}  \\
H$_{2}$ formation on dust grains & \citet{hm89} \\
X-ray Coulomb heating & See Section 2.3.2 \\
Cosmic ray ionisation & \citet{gl78}  \\
\hline
\end{tabular}
\caption{Processes included in our thermal model. \label{cool_model}}
\end{table*}

\subsection*{Fine structure cooling} 
  We model atomic fine structure cooling from neutral C, O and Si atoms and C$^{+}$ and Si$^{+}$ ions by directly solving for the
fine structure level populations, with the assumption that the populations of any electronically-excited states are zero. This 
assumption allows us to model C$^{+}$ and Si$^{+}$ as two-level systems and C, O and Si as three-level systems, allowing us
to write down analytical expressions for the cooling rate from each species in a relatively simple fashion. We do not account for 
any external sources of radiation other than the cosmic microwave background. The sources for the data used in the level 
population calculations are listed in Table~\ref{cool_model}, and a more detailed discussion of our approach can be found in \citet{gj07}. 
Note that we use the Si and Si$^{+}$ cooling rates as a proxy for the cooling coming from the species represented by M and M$^{+}$,
which include not only Si but also other low ionization potential metals such as Mg or Fe. This simplification is somewhat inaccurate,
but in practice this is unlikely to be important as the fine structure cooling is typically dominated by C$^{+}$ and O in regions with low
A$_{\rm V}$ and by C in regions with high A$_{\rm V}$.

\subsection*{CO rovibrational line cooling}
We model CO cooling using the cooling tables given in \citet{nk93} and \citet{nlm95}, which are based on a large velocity gradient (LVG)
calculation of the CO level populations as a function of the H$_{2}$ number density, CO number density temperature, and local velocity gradient.
The lowest temperature included in these tables is 10~K, but to allow us to handle very cold molecular gas we have extended them down to 5~K
using collisional data from \citet{fl01} and \citet{w06}, as described in Appendix A of \citet{GloCla12}. The LVG calculation in \citet{nk93} and \citet{nlm95} assumes
that CO is excited primarily by collisions with H$_{2}$. However, in our cooling function, we also account for collisions with atomic hydrogen and
with electrons, using the procedure described in Section C.4 of \citet{MeiSpa05}.

\bsp	
\label{lastpage}
\end{document}